# Analytical Aberration Theory for Plane-symmetric Optical Systems and its Application in the Analysis of Distortion in Freeform Spectrometers

by

Yuxuan Liu

Submitted in Partial Fulfillment of the

Requirements for the Degree

Doctor of Philosophy

Supervised by Professor Jannick Rolland

The Institute of Optics

Arts, Science and Engineering

Edmund A. Hajim School of Engineering and Applied Sciences

University of Rochester

Rochester, New York

2024

# Dedication

This work is dedicated to my family who raised me and supported me throughout my educational life.

# Table of Contents

# Biographical Sketch

Yuxuan Liu was born in Luoyang, China. He attended Fudan University in Shanghai from 2011 until 2015, completing a Bachelor of Science degree in Physics. During his undergraduate studies, he was involved in a college funded project on designing and constructing a modular spectrometer for physics teaching purposes. During this experience, he grew a strong interest in optical engineering. In 2015, he moved to the United States to continue his graduate study in optical engineering at Rose-Hulman Institute of Technology, where he designed and constructed the optics and control systems for a microscope based on magneto-optic effects.

In 2017, Yuxuan entered the doctoral program at The Institute of Optics to further his interest in optical engineering. He joined the research group of Professor Jannick P. Rolland, where he researched freeform optical design and aberration theory. Supervised by Professor Jannick P. Rolland and Doctor Aaron Bauer, he completed a design project on a freeform hyperspectral imager under The Center for Freeform Optics. In 2021, Yuxuan was awarded the Robert S. Hilbert Memorial Optical Design Competition Winner for his hyperspectral imager design.

During the design project, a key question was raised about the sources of distortion in the freeform spectrometer, which led Yuxuan to deeper research on analytical aberration theories for freeform plane-symmetric systems. He also completed an internship at Apple, CA, as an Exploratory Design Intern in 2022.

The following peer-reviewed publications were a result of work conducted during the doctoral study:

Yuxuan Liu, Aaron Bauer, Thierry Viard and Jannick P. Rolland, "Freeform hyperspectral imager design in a CubeSat format," Opt. Express **29**(22), 35915-35928 (2021).

Yuxuan Liu, Jessica Steidle, and Jannick P. Rolland, "Intrinsic aberration coefficients for plane-symmetric optical systems consisting of spherical surfaces," J. Opt. Soc. Am. A **40**, 378-387 (2023).

Yuxuan Liu, and Jannick P. Rolland, "Analytical aberration theory for plane-symmetric optical systems and its application in analysis of distortion in freeform spectrometers," (In Preparation)

## Acknowledgements

Working towards my PhD degree has been an important and influential life chapter for me. I want to express my gratefulness to many people who supported and positively influenced me during this journey.

First, I want to thank my advisor, Professor Jannick Rolland, for all her mentorship on my research. I am grateful for the wonderful research opportunities and resources she provided that helped me tremendously throughout my time in her group. Each meeting we had helped me improve not only in academic but also as a person. The experience I had during my PhD program will surely have a positive and long-lasting impact on my future career.

I would also like to thank Doctor Aaron Bauer. As the co-PI on my design project, Doctor Aaron Bauer has provided me with guidance and advice on almost all things academically related, from preparing for a presentation to finding possible design strategies. His help has been most valuable to me during my work on freeform design.

I thank the group members, Jonathan Papa, Nick Takaki, Changsik Yoon, Eric Schiesser, Wooyoun Kim, Di Xu, and Romita Chaudhuri, for all the rewarding discussions we had. There were great learning experiences and also fun chats. I want to especially thank Jessica Steidle for the collaborations we had on the derivation work of fifth-order aberration coefficients for plane-symmetric systems. She also helped proofread my dissertation, for which I am very grateful.

I want to thank my close friends in Rochester for the good time we had together. Kejia Zhang has been an amazing inspiration to me in both life and work. I thank her for her immense and selfless support through the years. I am grateful to have her in my life. I want to also thank Jiacheng Zhao, a great friend with a great personality, who shares my passion for optics and video games. I wish you a great future in your study and career.

I want to thank the teachers who mentored me in different stages of my education. I would not have reached this far without them. I would like to thank Professor Maarij Syed at Rose-Hulman Institute of Technology, my advisor during my Master's program in optical engineering. He gave me a warm welcome as I was a newcomer to the United States and helped me immensely during my thesis project. Working with him was a pleasant and rewarding experience. I want to also thank Professor Yongkang Le at Fudan University. Leading the university physics teaching labs, he has done so much for his students, organizing learning groups, holding invited talks, and mentoring student projects. He has showed me the intricacy and beautifulness of instrumentation physics when I was an undergraduate student. Influenced by him, I have found and chosen this road of optical engineering.

More importantly, I want to thank my family, my mom and dad. They held this family together despite hardships in life and raised me to be the person who I am. They showed me how to be a person who shows kindness and takes responsibility. They have supported me with all they have. I can always feel their love from across the ocean. I hope I can find more opportunities to visit them in the future.

# Abstract

The recent history of optical design saw a progressive trend of also designing without rotational symmetry, especially spectrometers due to the use of reflective and diffractive elements in their designs. A freeform hyperspectral imager design in CubeSat format is presented in this work, which has large deviation from rotational symmetry.

Breaking the rotational symmetry in an optical design proposes a challenge for designers to understand the resulting aberration behavior. Paraxial optics and analytical aberration theories derived on the basis of an optical axis do not apply to systems where this optical axis does not exist, preventing people from using them in analysis of the systems.

Based on a generalization of paraxial optics and wavefront aberration expansion applicable to plane-symmetric systems, in this dissertation we derived the aberration coefficients of aberration types in the third group for plane-symmetric systems that include the second order effects of the light beam footprint on optical surfaces. We also expanded the theory to include the contributions from freeform surfaces and induced aberrations. For the application to plane-symmetric spectrometers, the aberration coefficients related to distortion are of special interest.

Comparisons between the distortion results predicted by the aberration coefficients and simulated from real raytracing show good consistency in example systems of a Dyson spectrometer and freeform spectrometers that share the three-mirror double-pass structure, which is similar to the spectrometer component of the hyperspectral imager we designed. In comparison to real raytracing, results of the theory applied to the spectrometer of the hyperspectral imager we designed show the limits of the theory that is limited to the third group of aberrations, while in this design, high-order groups also significantly contribute. In all cases, the analytical formulae of the aberration coefficients contain the information on surface contributions, induced aberrations, and the relation between the aberration behavior and system parameters, which provide valuable insights in the analysis of plane-symmetric systems.

# Contributor and Funding Sources

This work was supervised by a dissertation committee of Professor Jannick P. Rolland (advisor), Professor Jake Bromage, Professor James Fienup, Professor Julie Bentley of The Institute of Optics, and Professor John Lambropoulos of the Department of Mechanical Engineering. The freeform three-mirror double-pass spectrometers shown in Chapters 7-8 were designed by the student in collaboration with Assistant Professor Aaron Bauer. All other work was completed by the student independently and funded by the Center for Freeform Optics.

# List of Tables

# List of Figures

# List of Symbols and Abbreviations

| Symbols and Abbreviations | Definition |
| --- | --- |
| 2D | Two Dimensional |
| AVG | Average |
| MAX | Maximum |
| MIN | Minimum |
| NAT | Nodal Aberration Theory |
| OAR | Optical Axis Ray |
| TMA | Three-Mirror Anastigmat |
| REP | Reference Sphere at the Entrance Pupil in Object Space |
| RSI | Reference Sphere at the Surface in Image Space |
| RSO | Reference Sphere at the Surface in Object Space |
| RXP | Reference Sphere at the Exit Pupil in Image Space |
| SD | Spectral Distortion |
| SysK | A Freeform Optical System Optimized for Keystone Distortion Correction |
| SysS | A Freeform Optical System Optimized for Smile Distortion Correction |
| SysF | A Freeform Optical System that Follows the Final Spectrometer Design in Chapter 7. |
| W | Wavefront Aberration |

# Chapter 1. Introduction

Optical spectrometers, which analyze the intensity of light as a function of wavelength or frequency, are used in a wide range of purposes such as material analysis, environmental monitoring, optical metrology, and research in physics and astronomy. While spectrometers can be designed based on interferometric methods or optical filtering, a widely used type of spectrometer is an imaging spectrometer, consisting of an optical imaging system and a dispersive element such as a prism or grating. In this type of spectrometer, light enters the system through a slit which is imaged onto a detector. Before reaching the detector, the light goes through the dispersive element, forming a spectrum on the detector as shown in Figure 1.1.

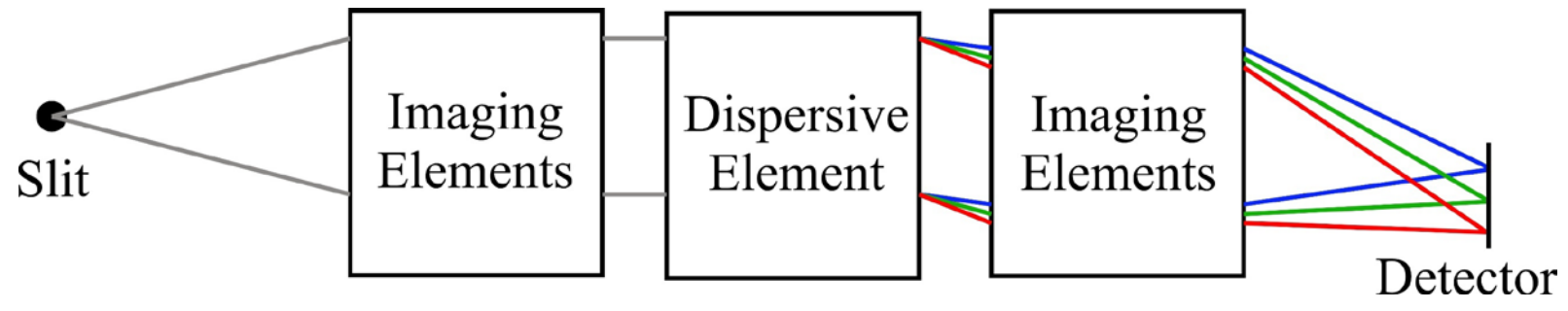

**Figure 1.1. Structure of an imaging spectrometer.**

Most imaging spectrometer designs are based on diffractive gratings. Early designs usually utilize planar reflective gratings and spherical mirrors, such as Czerny-Turner configuration [1, 2] and Ebert-Fastie configuration [3], the general structures of which are shown in Figures 1.2 and 1.3. Both Czerny-Turner and Ebert-Fastie configurations show similar design structures that include one optical surface that collimates the incoming light from the slit and another surface that focuses the light diffracted from the grating. In cases with high dispersion gratings, the gratings are sometimes designed to rotate with a output slit at the spectrum to cover a wider spectral range or acquire with higher spectral resolution [4]. Other modifications made to Czerny-Turner and Ebert-Fastie configurations include the use of curved slits [5, 6] and toroidal or cylindrical surface shapes [7, 8] to reduce astigmatism.

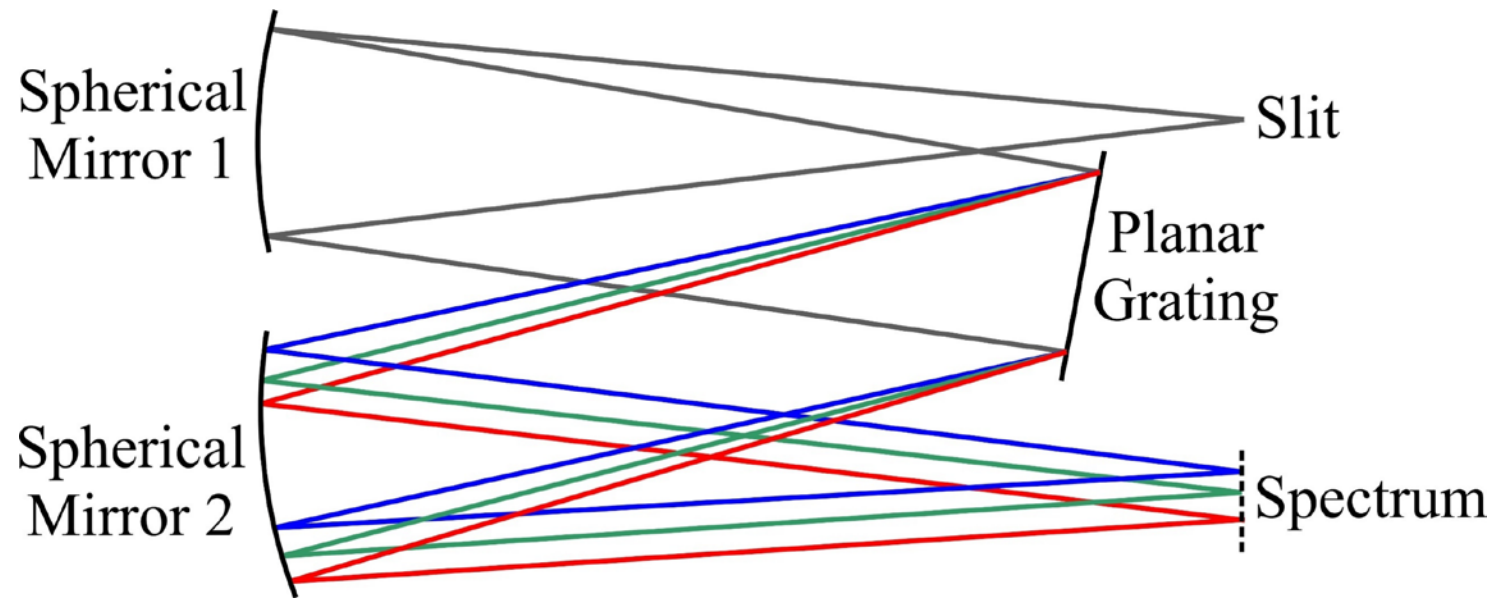

**Figure 1.2. Illustration of Czerny–Turner spectrometer configuration.**

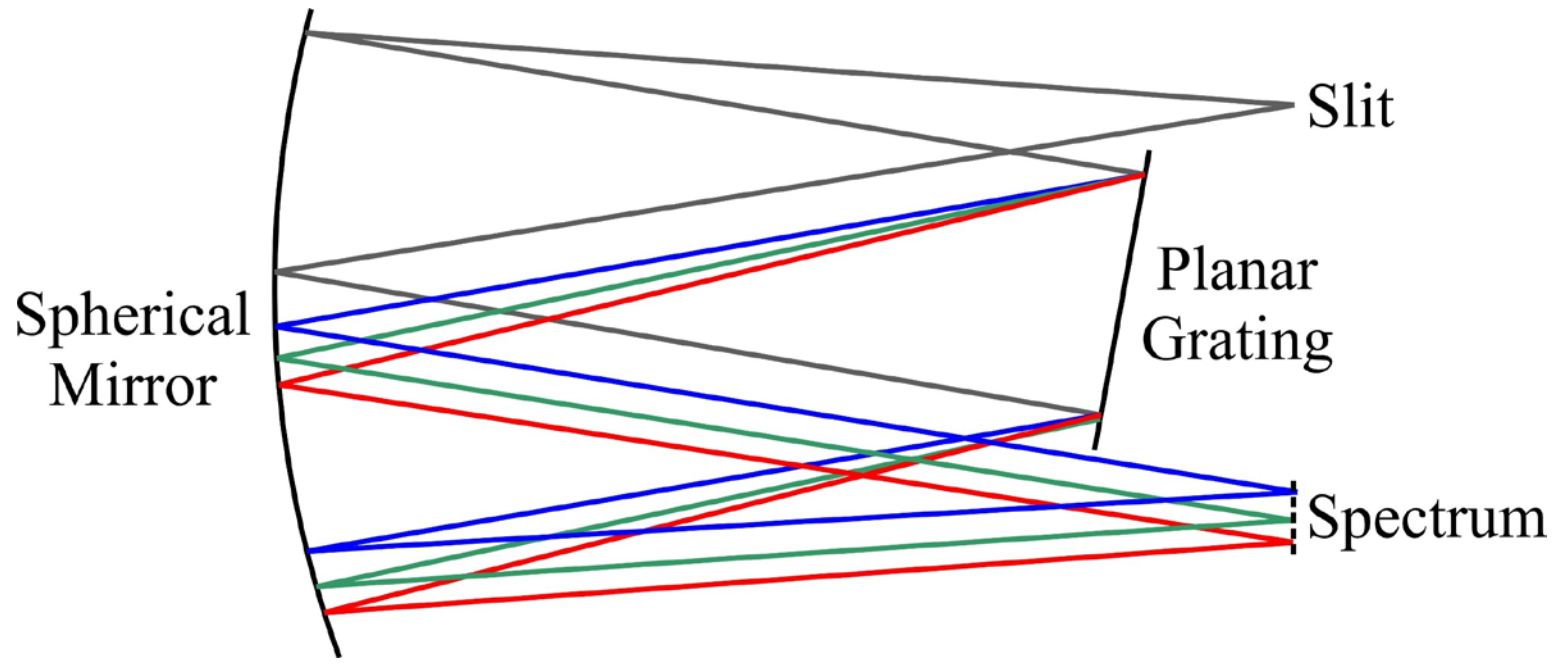

**Figure 1.3. Illustration of Ebert-Fastie spectrometer configuration.**

Other types of widely used imaging spectrometer designs are concentric designs with curved gratings, such as Dyson [9] and Offner [10] spectrometers as shown in Figures 1.4 and 1.5. Both Dyson and Offner designs were first proposed as relays with spherical mirrors [11, 12], and their concentric structures enable the systems to be free of third-order aberrations when used at -1 magnification. Dyson and Offner spectrometers are designed based on the corresponding relays and can achieve compact and fast (low F-number) designs that are widely used in remote sensing applications [13-16].

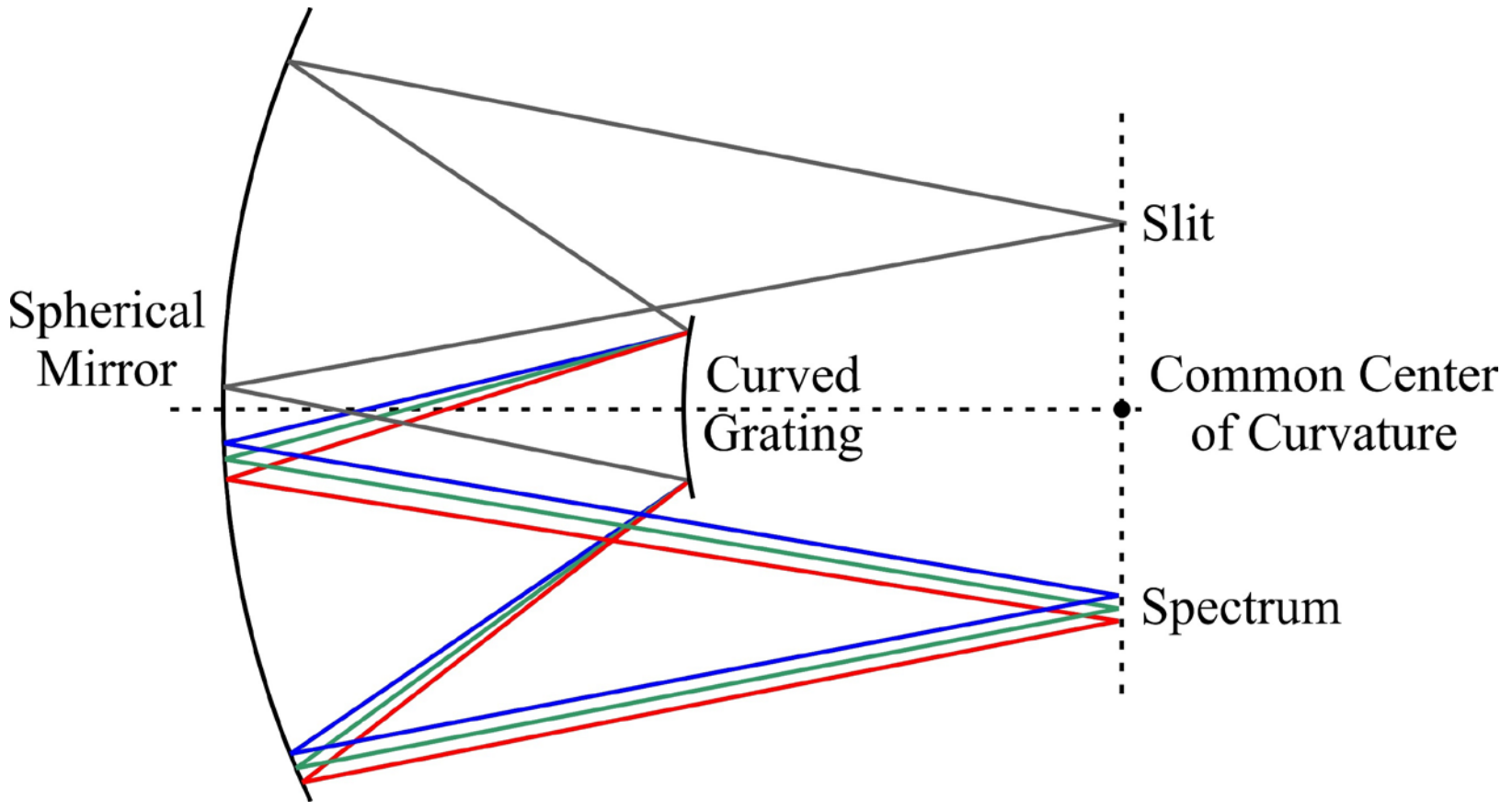

**Figure 1.4. Illustration of Offner spectrometer configuration. The common center of curvature is shared by the spherical mirror and the curved grating.**

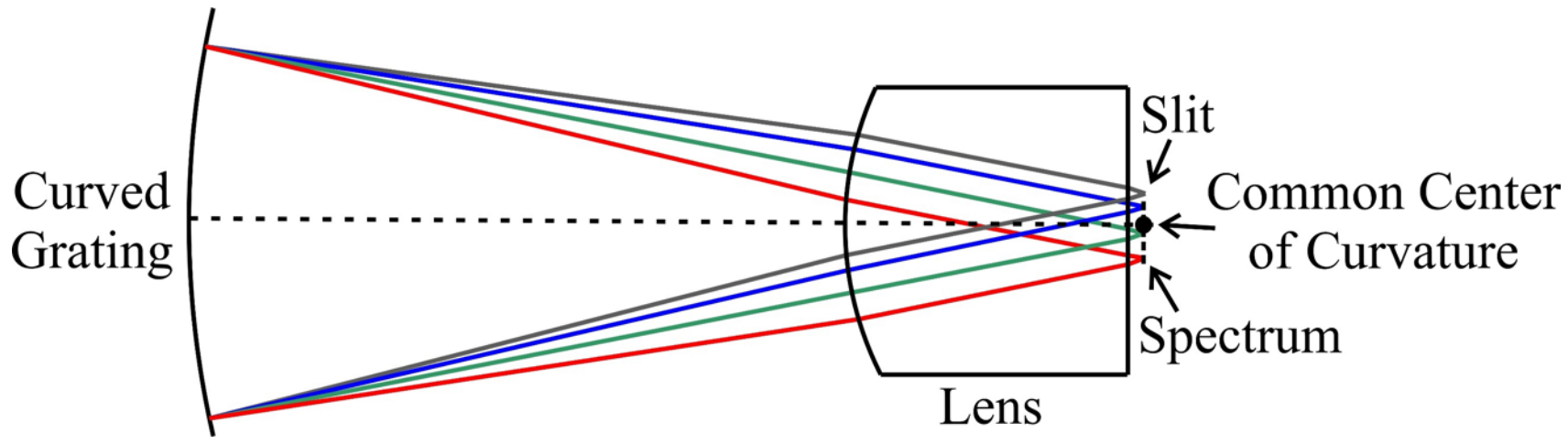

**Figure 1.5. Illustration of Dyson spectrometer configuration. The common center of curvature is shared by the convex side of the lens and the curved grating.**

The spectrometer designs described above show some properties that are common in imaging spectrometer designs. Firstly, reflective components are used in most spectrometer designs, because they do not introduce additional chromatic aberrations. With reflections, the light beams need to be directed to specific space to avoid obscuration. Therefore, in contrast to other optical systems that are designed with rotational symmetry around an optical axis, imaging spectrometer designs usually have geometries where the light beams do not follow rotational symmetry. The involvement of dispersive elements also creates asymmetry between the dispersive and non-dispersive directions and introduces an additional source of aberration contribution, affecting how the imaging aberrations are analyzed in spectrometers. For example, Seidel aberrations, derived for rotationally symmetric systems, can be applied to a class of reflective imaging systems when they are a part of a parent rotationally symmetric systems such as the Offner relay mentioned before with off-axis fields and the unobscured three-mirror anastigmat (TMA) designed by Cook with an off-set aperture [17]. However, the existence of dispersive elements makes the system essentially biased in one direction that cannot be derived from a parent rotationally symmetric system. Although with low dispersion, the imaging quality of the spectrum may share similar characteristics of the non-dispersive image, which still provides valuable insights, the aberration theory for rotationally symmetric systems cannot be fully applied to spectrometers.

Recent developments also saw the utilization of freeform surfaces in optical designs, which enables higher imaging performance in compact form factors and further drives the system structure away from rotational symmetry. Freeform surfaces are not rotationally symmetric by definition [18] and can be used to correct target aberrations that also have no rotational symmetry [19-21]. In recent years, freeform spectrometer designs are also emerging quickly [22, 23]. Reimers *et al*. designed a freeform spectrometer based on Offner-Chrisp geometry and reached a volume 5 times more compact than similar designs with spherical and aspherical surfaces [24]. Zhang *et al*. proposed a freeform spectrometer design consisting of a single concave grating with freeform shape and variable grating line spacing [25]. Due to the asymmetric nature of spectrometer structures, freeform surfaces have been shown to greatly improve the performance and compactness of spectrometer designs. In this work, we also present a freeform hyperspectral imager design in CubeSat format [26] that originally motivated our research into aberration theory. The spectrometer component of the hyperspectral imager design is the main system under investigation.

Another class of optical systems heavily investigated are based on off-axis conics, which are defined as part of conic surfaces not centered around the axes of conics, as shown in Figure 1.6. An important imaging property often leveraged in optical designs based on off-axis conics is the perfect imaging relation between the two foci of a conical surface. In Figure 1.6, the two foci of the ellipsoid, $F_1$ and $F_2$, are by definition perfect image of each other. A design structure based on multiple off-axis conics placed with each pair of consecutive conics sharing a common focus is proposed to ensure stigmatic imaging at the corresponding foci, which is usually referred to as confocal conics [27-29]. Freeform designs are also explored with confocal conics as starting points [20, 30]. Although confocal conics are best-known in rotationally symmetric telescopes such as Cassegrain and Gregorian telescopes, this configuration is effective at correcting field-constant aberrations in designs without rotational symmetry.

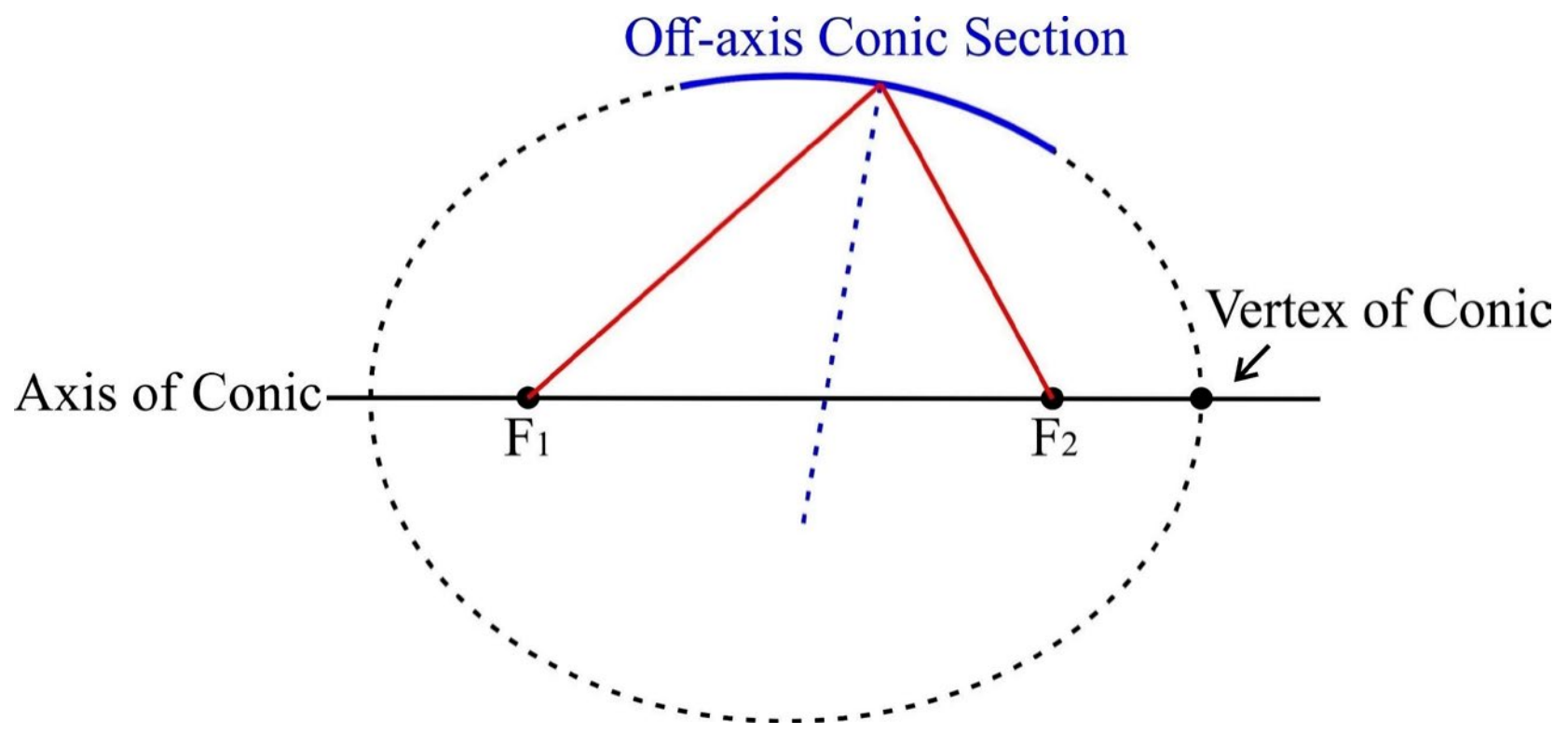

**Figure 1.6. Illustration of off-axis conics. The shown case is based on an ellipsoid.**

With the increasing complexity of optical design, there is also a great need for new knowledge and tools to help understand and analyze imaging qualities and aberrations in optical systems without rotational symmetry. One type of desired tools are aberration theories that can provide insights on the aberration behavior including the magnitude of different aberration types, aberration contributions from different sources, and the relation between aberrations and system parameters. For rotational symmetric systems, the concepts of paraxial optics and Seidel aberrations have been used to guide optical designs for over a century. However, without extension or modification, the same concepts do not apply to many systems described above. Different approaches have also been made to analyze aberrations in systems without rotational symmetry. Tang and Gross utilized a mixed ray-tracing method to calculate surface aberration contribution and differentiate intrinsic and induced (extrinsic) aberration components in symmetry-free optical systems [31, 32]. Caron *et al.* developed a matrix formalism based on a set of generalized ray-tracing equations [33, 34] to derive ray aberration coefficients for reflective optical systems and gave an illustration of ray aberration coefficients for N-mirror confocal systems where recursive relations between N-mirror and N+1-mirror systems were used [35].

One notable aberration theory dedicated to optical systems without rotational symmetry is Nodal Aberration Theory (NAT) developed by Shack and Thompson [36-40]. NAT was initially aimed to express the effect of tilt and decenter of spherical surfaces on the aberration field derived using the symmetry of the aberration field contribution around a line connecting the center of the pupil and the center of curvature [37]. In later development, Fuerschbach *et al.* extended NAT to derive the formulae for the wave aberrations caused by Zernike freeform surfaces [41]. This development enables NAT to be used to analytically calculate aberration contributions from freeform surfaces. Bauer *et al*. then utilized these formulae in a design method based on analyzing the potential of folding geometries to correct aberrations [19].

As discussed before, paraxial optics and Seidel aberrations derived based on an optical axis of rotational symmetry do not generally apply to systems without rotational symmetry. However, the concepts of paraxial optics or first-order optics and wavefront aberration terms based on their dependence on field and pupil vectors are still valuable and can be extended to systems without rotational symmetry. Buchroeder introduced a method to describe optical systems without rotational symmetry following a real ray passing through the vertex of each surface [42]. Paraxial optics can be developed near this real ray similar to that in rotationally symmetric systems. While rotational symmetry is lost, most imaging systems without rotational symmetry maintain a plane symmetry where half of the system is the mirror image of the other half. Sasian built on the vectorial expression of wavefront aberrations and grouped various aberration terms based on the number of vector products in the aberration expression. He derived the approximated intrinsic aberration coefficients in the third group for plane-symmetric (bilaterally-symmetric) systems, which are expressed in system parameters such as radii of curvature, distances between surface vertices, and tilt angles [43].

Some of the work presented here extends the work by Sasian and the wavefront aberration theory by Hopkins [44] to develop an analytical aberration theory. This work removes the approximations used by Sasian to derive accurate intrinsic aberration coefficients. In addition to intrinsic aberrations from spherical surfaces, contributions from freeform surfaces and induced aberrations are also derived. This work also applies the derived aberration coefficients in the prediction of distortion behavior in example freeform spectrometers. Compared with numerical simulations, the analytical theory can provide more physical insights on how system parameters affect aberration behavior, and a computationally more efficient way to evaluate system performance.

In Chapter 2, the description method for plane-symmetric systems is reviewed and illustrated. Chapter 3 illustrates the paraxial optics and the first-order properties in plane-symmetric systems. Chapter 4 reviews the wavefront aberration terms in plane-symmetric systems and their relation to the transverse aberration. The aberration coefficients related to distortion in plane-symmetric imaging spectrometers are also discussed and derived analytically in Chapter 5. While Chapter 5 focuses on one-surface systems with perfectly spherical incoming wavefront, Chapter 6 covers induced aberrations caused by aberrated incoming wavefront. In Chapter 7, we present a hyperspectral imager design in CubeSat format, which motivated this research on aberration theory. The spectrometer component of the hyperspectral imager is the main system to which we apply the aberration theory in the analysis of its distortion behavior. In Chapter 8, example systems, a Dyson spectrometer and multiple freeform spectrometers in the design form of the spectrometer design in Chapter 7, are presented. With the example systems, the distortion results calculated from the aberration coefficients derived in Chapters 5-6 are compared with the results from real raytracing in CODE V.

# Chapter 2. Description of plane-symmetric optical systems

To develop analytical formulae on an optical system's aberration behavior, the first step is to describe the system by parameterizing the locations, shapes and materials of optical surfaces. In this chapter, we describe the parametrization method of plane-symmetric systems on which all following analytical derivations are based.

The optical systems under investigation are limited to plane-symmetric systems, which means a plane of symmetry can always be found so that half of the optical system on one side of the plane of symmetry is the mirror image of the other half. Rotationally symmetric systems are also plane-symmetric, since any plane that includes the optical axis can be treated as a plane of symmetry. Common plane-symmetric systems involve reflective mirrors which are tilted to avoid obscuration as shown in Figure 2.1.

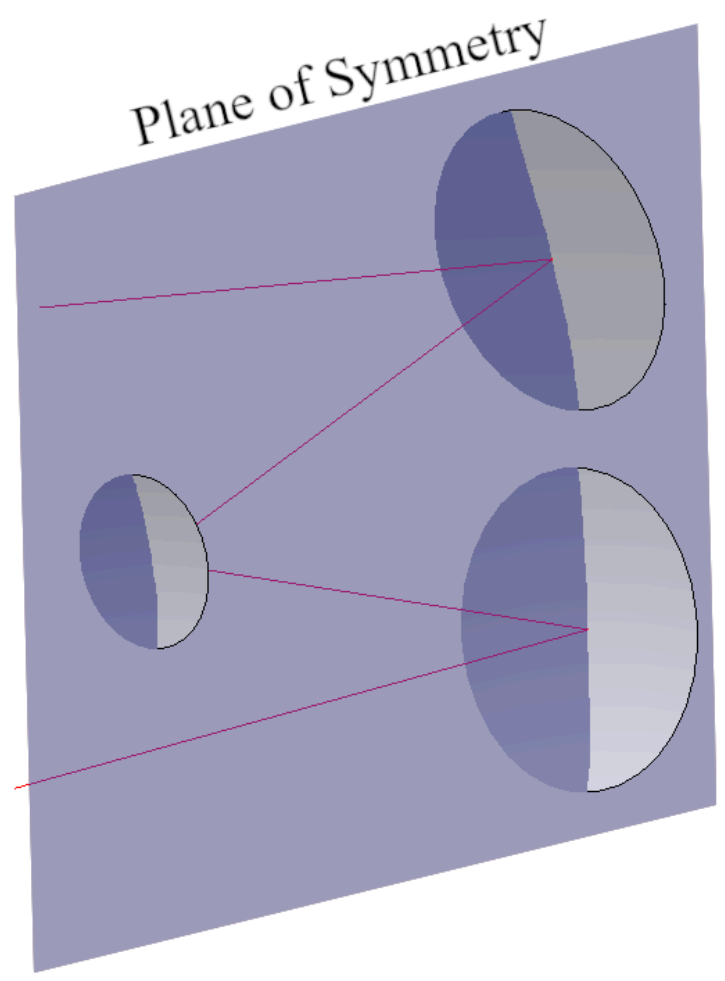

**Figure 2.1. A three-mirror plane-symmetric system. Each mirror is plane-symmetric.**

## 2.1 Parametrization of optical surface location and orientation

In this work, the description method of plane-symmetric systems follows the system parametrization by Sasian [43] that describes a system following an optical axis ray (OAR) connecting the center of the object and the center of the aperture stop. This OAR lies in the plane of symmetry, as shown in Figure 2.2, intersecting each optical surface at its vertex.

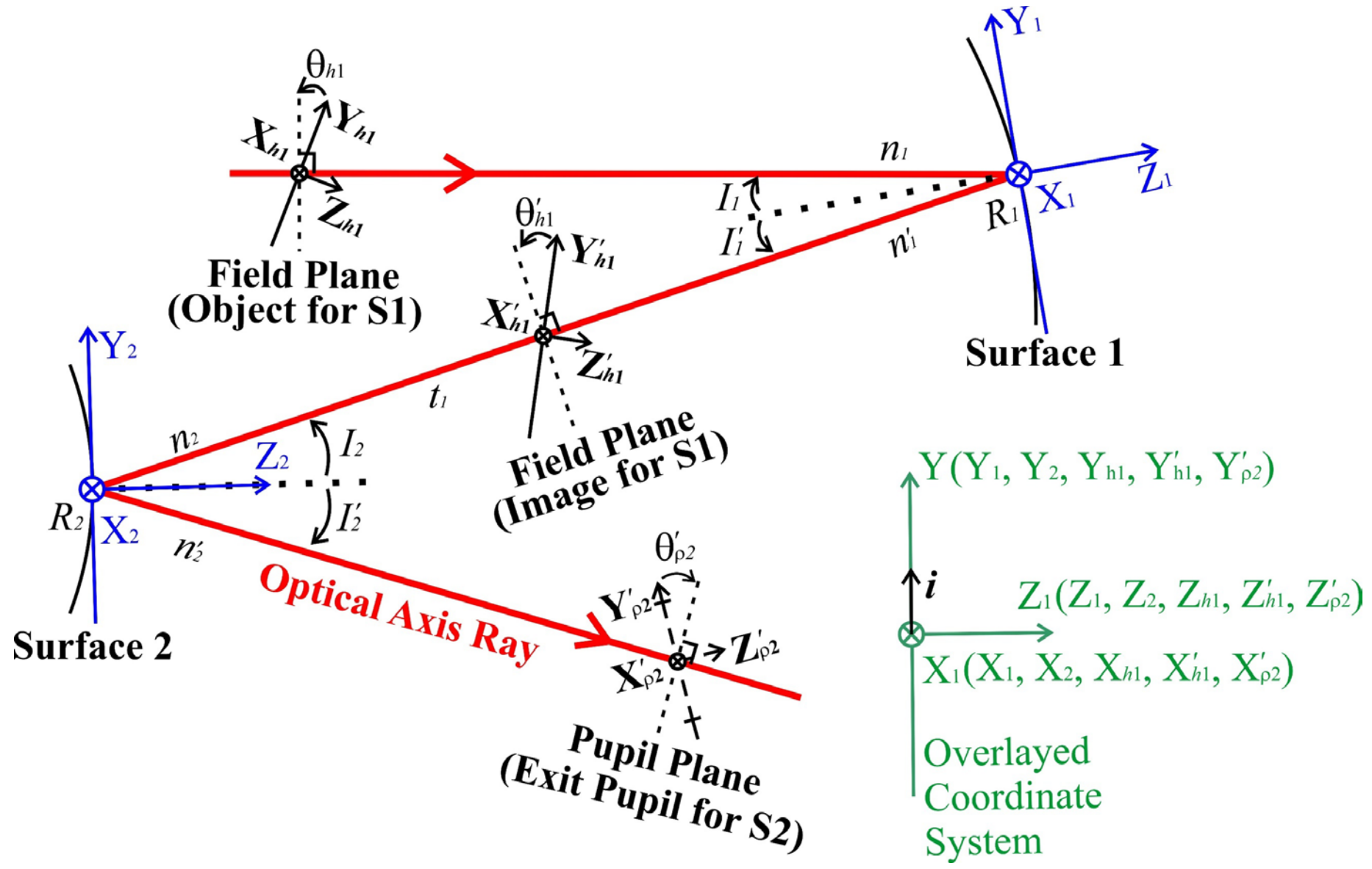

**Figure 2.2. Illustration of the system description in the plane of symmetry (not a real system); in green is the overlayed coordinate system where all local coordinate systems for surfaces, field and pupil planes coincide, with *i* being the common unit vector in the Y direction.**

The system is described sequentially with the location of each surface depending on that of the previous surface by two parameters: the distance between the vertices, $t$, and the incident angle of the OAR at the vertex, $I$. The field planes shown in Figure 2.2 are object or image planes, and pupil planes are entrance and exit pupil planes. The field and pupil planes are defined by paraxial optics, detailed in Section 3.2, and have their origins on the OAR. In addition, they have tilt angles along the OAR, $\theta_h$ for field planes and $\theta_\rho$ for pupil planes, as shown in Figure 2.2. Parameter $R$ is the radius of curvature of optical surface. All parameters and local coordinate systems can have a numerical subscript denoting the surface they correspond to. In addition, a prime symbol indicates that the parameter is in the image space of the surface (i.e., after the light has passed through the surface). Similarly, parameters without the prime symbol are in the object space of the surface (i.e., before the light has passed through the surface). For example, $n_1$ and $n'_1$ are the refractive indexes in the object and image space of the Surface 1. The refractive index, $n$, and the field and pupil plane tilt angles in the image space of one surface are equal to their corresponding quantities in the object space of the next surface.

A local right-handed Cartesian coordinate system is set up for each surface as well as each field or pupil plane with its origin at the point where the OAR intersects the surface, the Z-axis of the local coordinate systems along the surface normal, and the YZ plane in the plane of symmetry. The positive Z direction of each local coordinate system points towards the propagation direction of the OAR before the surface if the OAR has undergone an even number of reflections preceding the surface, otherwise the positive Z direction points in the opposite direction. The axis label of the coordinate systems for field and pupil planes have subscripts, $\rho$ and $h$, to differentiate themselves from other surface coordinate systems. For example, $X_{h1}Y_{h1}Z_{h1}$ in Figure 2.2 represents the object coordinate system for Surface 1, and $X'_{\rho2}Y'_{\rho2}Z'_{\rho2}$ represents the exit pupil coordinate system for Surface 2. This parameterization method can be applied to both reflective and refractive systems. Figure 2.2 shows an example of a reflective case.

All parameters in Figure 2.2 adhere to the following sign conventions: (1) distance along the OAR between two surfaces is positive if measured towards the positive direction of the local Z-axis of the first surface; (2) counterclockwise angles are positive; (3) the radius of curvature of a surface is measured from the vertex to the center of curvature and is positive if measured towards the positive direction of the local Z-axis; (4) the incident and refractive angles of the OAR, $I$ and $I'$, at each surface are measured from the surface normal to the OAR; (5) the field and pupil plane tilt angles, $\theta_h$ and $\theta_\rho$, are measured from the field and pupil planes, respectively, to a plane perpendicular to the OAR; and (6) the refractive index is positive when the OAR is traveling towards the positive direction of the local Z-axis of the previous surface. Consequently, in Figure 3, as a way of example, $\theta_{h1}$ and $\theta'_{h1}$ are the object and image plane tilt angles for the Surface 1 and are both positive; $t_1$ is the distance measured from Surface 1 to Surface 2 along the OAR and is negative; and $I'_2$ is the refractive angle of OAR after Surface 2 and is clockwise, thus negative.

Plane-symmetric imaging spectrometers can be described using this method with the slit along the X-axis of the object plane, one example of which is a Dyson spectrometer shown in Figure 2.3. Note that because of the dispersive element, the OAR is different for different wavelengths. Therefore, the paraxial optics and aberration terms are treated separately for different wavelengths in imaging spectrometers.

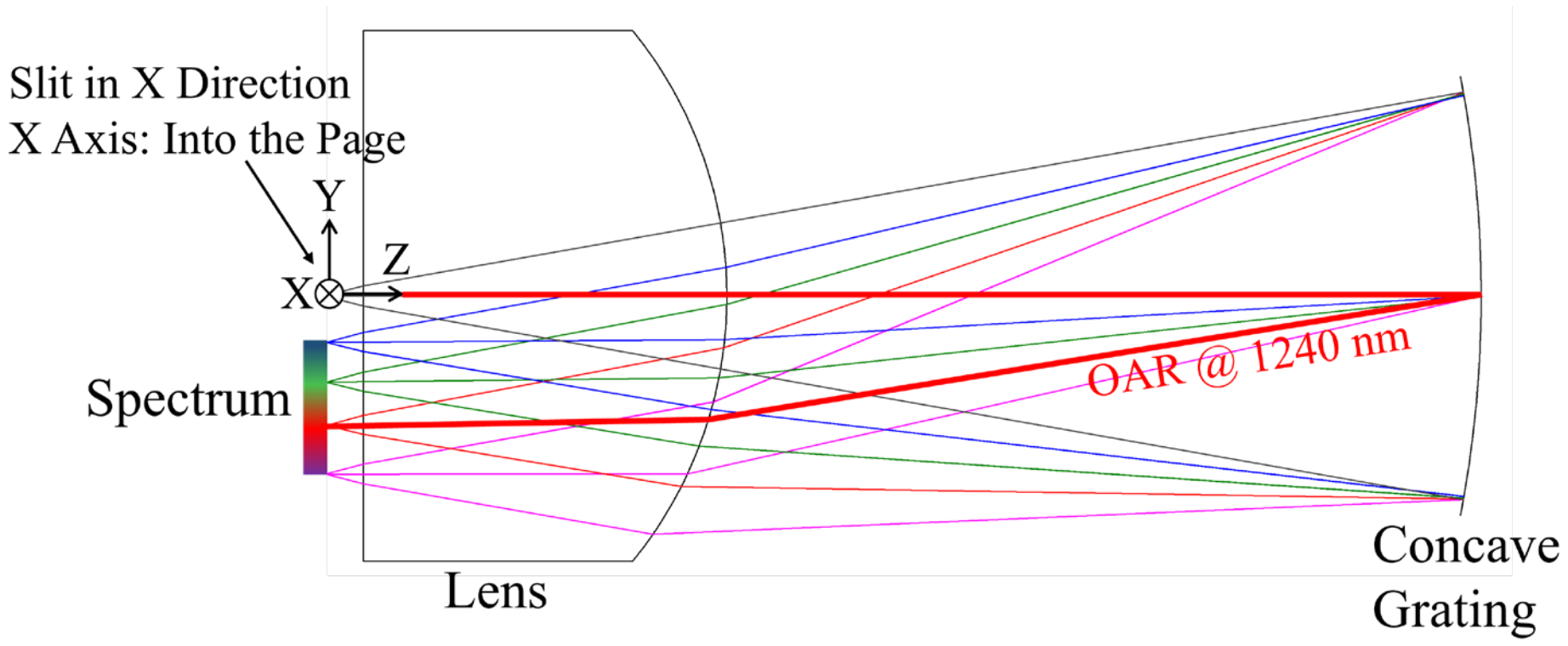

**Figure 2.3. Illustration of a Dyson spectrometer (400-1700 nm) with the slit placed in the X direction of the object coordinate system. The aperture stop is at the concave grating. In red is the OAR at 1240 nm.**

## 2.2 Surface shape description

Using the local surface coordinate systems set in Section 2.1 (blue in Figure 2.2), the surface shape can be described. In general, the surface shape can be defined with a sag function shown as

$$\textit{sag function}: \ z(x,y) \tag{2.1}$$

where the sag, $z$, is the $z$ coordinate of a point on the surface given the $x$ and $y$ coordinates. The origin of the local coordinate system is, by definition, on the surface, which means

$$z(0,0)=0 \tag{2.2}$$

Since the local Z-axis is defined along the surface normal, the first partial derivative of $z$ at $x$=0 and $y$=0 equals zero, which is expressed as

$$\left.\frac{\partial z}{\partial x}\right|_{x=0}=0 \tag{2.3}$$

$$\left.\frac{\partial z}{\partial y}\right|_{y=0}=0 \tag{2.4}$$

In addition, because the YZ plane is the plane of symmetry, the sag remains the same when the $x$ coordinate changes sign, which can be written as

$$z(x,y) = z(-x,y) \tag{2.5}$$

If the surface is spherical, the surface shape can be defined simply with the radius of curvature, as represented by $R$ in Figure 2.2. The radius is signed and is positive when the center of curvature is on the positive side of the local Z-axis. For example, $R_1$ and $R_2$ in Figure 2.2 are both negative because both centers of curvature are on the negative side of the local Z-axis.

Furthermore, under our assumption of plane-symmetry, the surface shape can be freeform if the plane of symmetry is maintained. Freeform surfaces are defined as surfaces that are not part of any rotationally symmetric surfaces [18]. In this work, the freeform shape is modeled as a sag on top of a base spherical surface, and is expressed as

$$z = z_{sph} + z_{freeform} \tag{2.6}$$

where $z_{sph}$ is the sag function for the spherical base surface, expressed as

$$z_{sph}(x,y) = R\left(1-\sqrt{1-\frac{x^2+y^2}{R^2}}\right) \tag{2.7}$$

and $z_{freeform}$ is the freeform sag function. The freeform departure and its derivatives of all orders with respect to $x$ and $y$ are assumed to be continuous over the defined area, which applies to real-life freeform surfaces. Therefore, the freeform departure can be Taylor expanded and expressed in polynomials such as XY or Zernike polynomials. Zernike polynomials are commonly used in freeform optical designs due to their orthogonality, which can be leveraged to find starting-point designs [19] and optimize freeform sag departure [45]. Therefore, in this work, Fringe Zernike polynomials are used to model freeform surfaces, shown as

$$z_{freeform} = \sum_i Z_i \, ZP_i\left(r_n, \theta\right) \tag{2.8}$$

where $ZP_i$ is the $i^{th}$ Fringe Zernike polynomial expressed with polar coordinates $r_n$ and $\theta$, which follow the transformation equations

$$x = R_{zn}\, r_n \cos\theta \tag{2.9}$$

$$y = R_{zn}\, r_n \sin\theta \tag{2.10}$$

and $Z_i$ is the coefficient of $ZP_i$. The parameter $R_{zn}$ in Equations 2.9-2.10 is the normalization radius of the circle over which the Fringe Zernike polynomials are defined. The variable $r_n$ is a normalized radial coordinate that ranges from 0 to 1. Figure 2.4 shows the relation between ($r_n$, $\theta$) and ($x$, $y$).

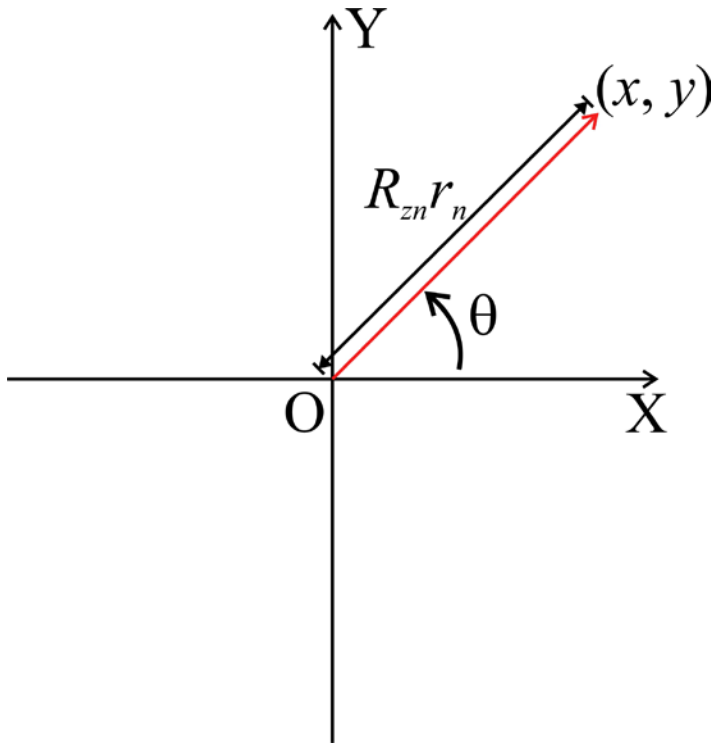

**Figure 2.4. A point in the XY plane defined in terms of ($r_n$, $\theta$). The parameter $R_{zn}$ is the normalization radius of the circle over which the Fringe Zernike polynomials are defined.**

Table 2.1 shows the first 16 terms of Fringe Zernike polynomials. Note that the freeform departure applied onto the base sphere needs to be plane-symmetric about the YZ plane. Therefore, only terms that satisfy the plane-symmetry can be used, which are noted in Table 2.1.

**Table 2.1. First 16 terms of Fringe Zernike polynomials**

| Term ($ZP_i$) | Fringe Zernike polynomial | Plane-symmetric about YZ plane? | Sag gradient plot |
|---|---|---|---|
| 1 | $1$ | Yes | 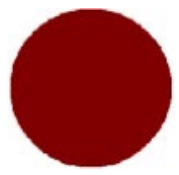 |
| 2 | $r_n \cos\theta$ | No | 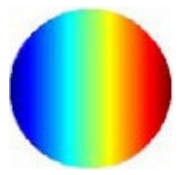 |
| 3 | $r_n \sin\theta$ | Yes | 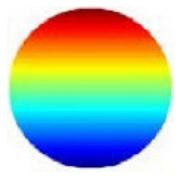 |
| 4 | $2r_n^2-1$ | Yes | 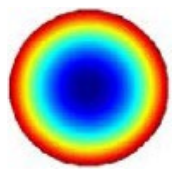 |
| 5 | $r_n^2 \cos2\theta$ | Yes | 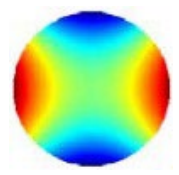 |
| 6 | $r_n^2 \sin2\theta$ | No | 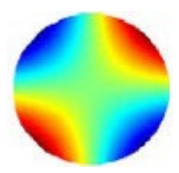 |
| 7 | $(3r_n^3-2r_n) \cos\theta$ | No | 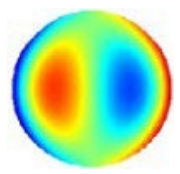 |
| 8 | $(3r_n^3-2r_n) \sin\theta$ | Yes | 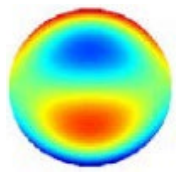 |
| 9 | $6r_n^4-6r_n^2+1$ | Yes | 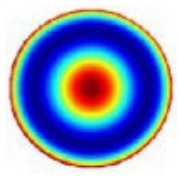 |
| 10 | $r_n^3 \cos3\theta$ | No | 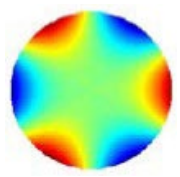 |
| 11 | $r_n^3 \sin3\theta$ | Yes | 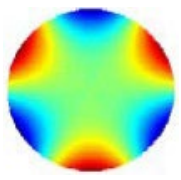 |
| 12 | $(4r_n^4-3r_n^2) \cos2\theta$ | Yes | 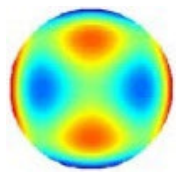 |
| 13 | $(4r_n^4-3r_n^2) \sin2\theta$ | No | 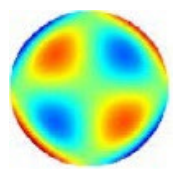 |
| 14 | $(10r_n^5-12r_n^3+3r_n) \cos\theta$ | No | 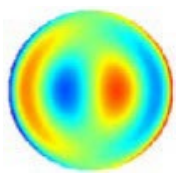 |
| 15 | $(10r_n^5-12r_n^3+3r_n) \sin\theta$ | Yes | 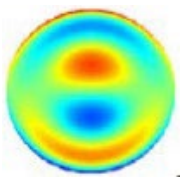 |
| 16 | $20r_n^6-30r_n^4+12r_n^2-1$ | Yes | 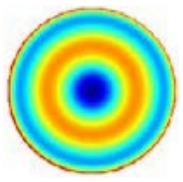 |

In addition, Equations 2.2-2.4 also need to be satisfied for the freeform departure $z_{freeform}$ to follow the definition of surface local coordinate systems. For example, consider a $z_{freeform}$ that consists of terms $ZP_1$, $ZP_3$ and $ZP_4$, shown as

$$example: \ z_{freeform} = Z_1 + Z_3 r_n \sin\theta + Z_4\left(2r_n^2 - 1\right) \qquad (2.11)$$

To satisfy Equation 2.2, the coefficients $Z_1$ and $Z_4$ are related as

$$Z_1 - Z_4 = 0 \qquad (2.12)$$

It can be also seen that $Z_3=0$ to satisfy Equation 2.4.

# Chapter 3. Paraxial optics in plane-symmetric systems

Paraxial optics is an important part of geometric optics and lens design, since it describes the first-order behavior of an optical system and is used to define optical properties such as magnification, image location, and pupil planes. The concept of paraxial optics, often referred to as Gaussian optics, was first introduced by mathematician Gauss [46]. Paraxial optics in plane-symmetric systems is defined as an extension of paraxial optics in rotationally symmetric systems. Therefore, this chapter will start with a brief review of paraxial optics in rotationally symmetric systems.

## 3.1 Review of paraxial optics in rotationally symmetric systems

In rotationally symmetric systems, paraxial optics focuses on the rays that are very close to the optical axis. These rays can be described at any plane normal to the optical axis with two parameters: (1) the ray location at the plane with respect to the optical axis, and (2) the ray angle before (or after) the ray passes through the plane. In a three-dimensional space, both ray location and angle require two numbers to be described and are usually measured in two orthogonal directions defined as the sagittal and tangential directions. However, it can be derived that in the paraxial regime where ray heights and angles approach zero, the ray heights and angles in one direction are independent from those in the other direction. In addition, the rotational symmetry indicates that the system structure is the same in the sagittal and tangential directions. Therefore, tracing paraxial rays in one direction is sufficient to describe the paraxial behavior of the whole rotationally symmetric system.

Figure 3.1 shows how the ray height and angle are defined with respect to the optical axis in the sagittal direction. The parameter $u$ is defined as the tangent of the corresponding ray angle measured from the optical axis.

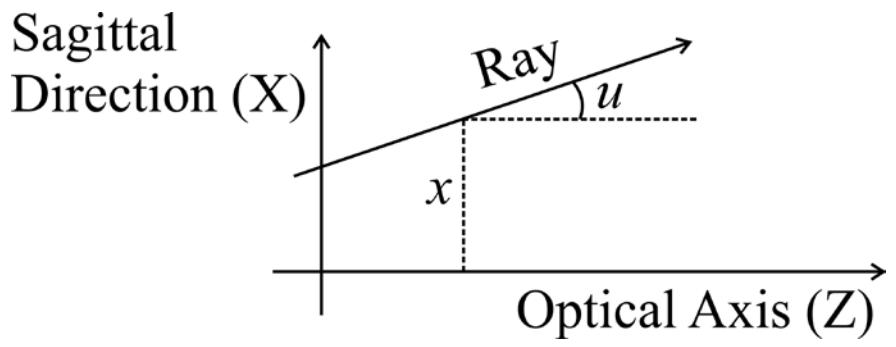

**Figure 3.1. In the sagittal direction (X), ray height (*x*) and the tangent of ray angle (*u*) are defined with respect to the optical axis (Z). *u* is defined as the tangent of the angle between the ray and the optical axis.**

As the ray heights and angles approach zero, the incident angles of the ray on optical surfaces also approach zero. Under this condition, Snell's law for refraction, which is shown as

$$n \sin i = n' \sin i' \tag{3.1}$$

where $i$ and $i'$ are the incident and refractive angles, and $n$ and $n'$ are the refractive indexes before and after the refractive surface (illustrated in Figure 3.2), can be reduced to a linear relation between $i$ and $i'$ as

$$\lim_{i \to 0} (ni) \approx n'i' \tag{3.2}$$

A similar approximation can be shown between tan$i$ and tan$i'$ as

$$\lim_{i \to 0} (n \tan i) \approx n' \tan i' \tag{3.3}$$

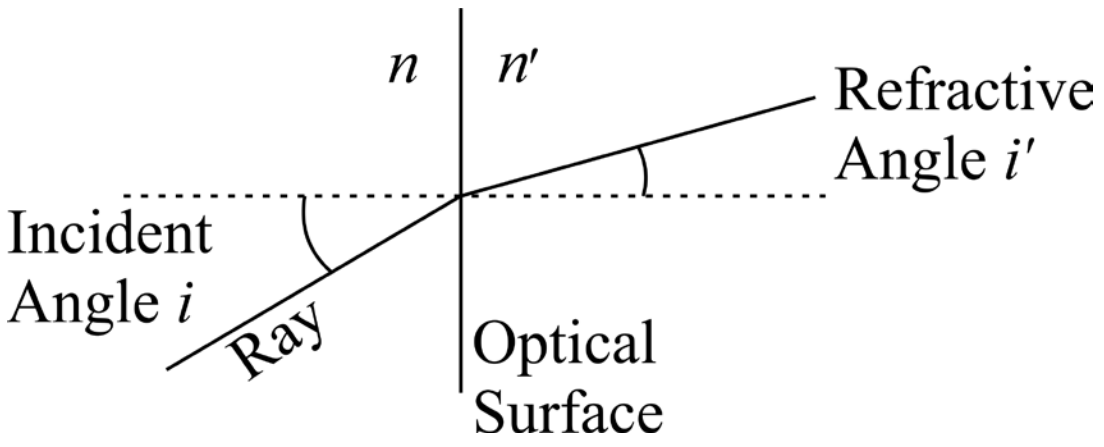

**Figure 3.2. Illustration of a raytracing at an optical surface, with incident and refractive angles, *i* and *i'*, and object and image refractive indexes, *n* and *n'*.**

This linear relationship leads to the derivation of the paraxial raytracing equations shown as

$$x_{i+1} = x_i + u_i' t_i \tag{3.4}$$

$$n_i' u_i' = n_i u_i - x_i \phi_{ri} \tag{3.5}$$

$$u_{i+1} = u_i' \tag{3.6}$$

$$n_{i+1} = n_i' \tag{3.7}$$

where the parameter $u$ (similarly $u'$) was defined in relation to Figure 3.1, and the subscript $i$ corresponds to the surface number, as shown in Figure 3.3. The prime symbol indicates that the quantity is in the image space of the surface, and $\phi_{ri}$ is the optical power of the surface, defined as

$$\phi_{ri} = \frac{n_i' - n_i}{R_i} \tag{3.8}$$

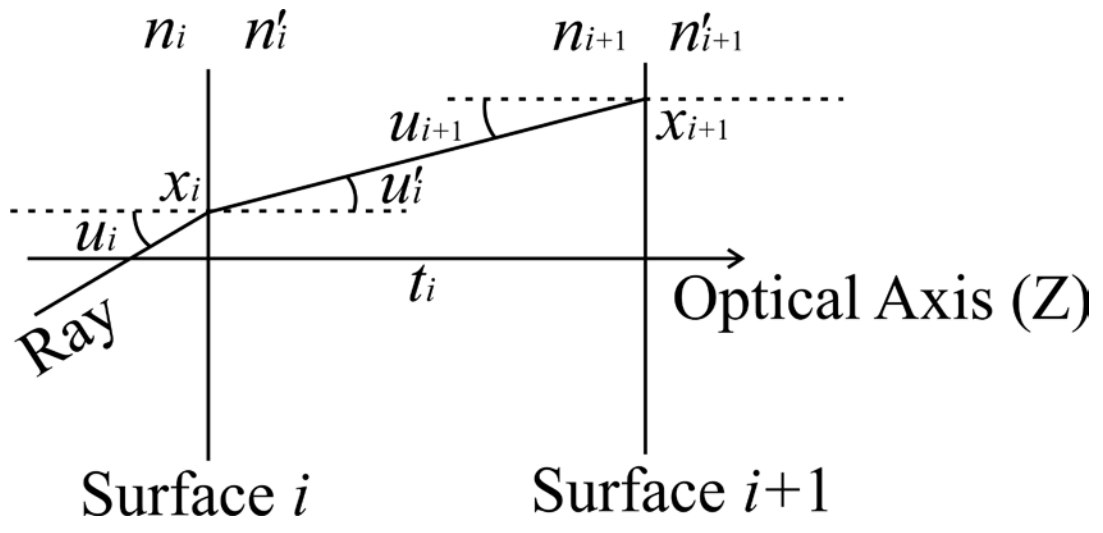

**Figure 3.3. Paraxial raytracing through optical surfaces in a rotationally symmetric system.**

The first-order properties of an optical system can be defined by tracing two characteristic paraxial rays, the chief and marginal rays, as shown in Figure 3.4. The chief ray starts from the edge of the object and passes through the center of the aperture stop, which is the physical aperture that limits the light beam size from an on-axis object. The marginal ray starts on axis at the object plane and passes through the edge of the aperture stop.

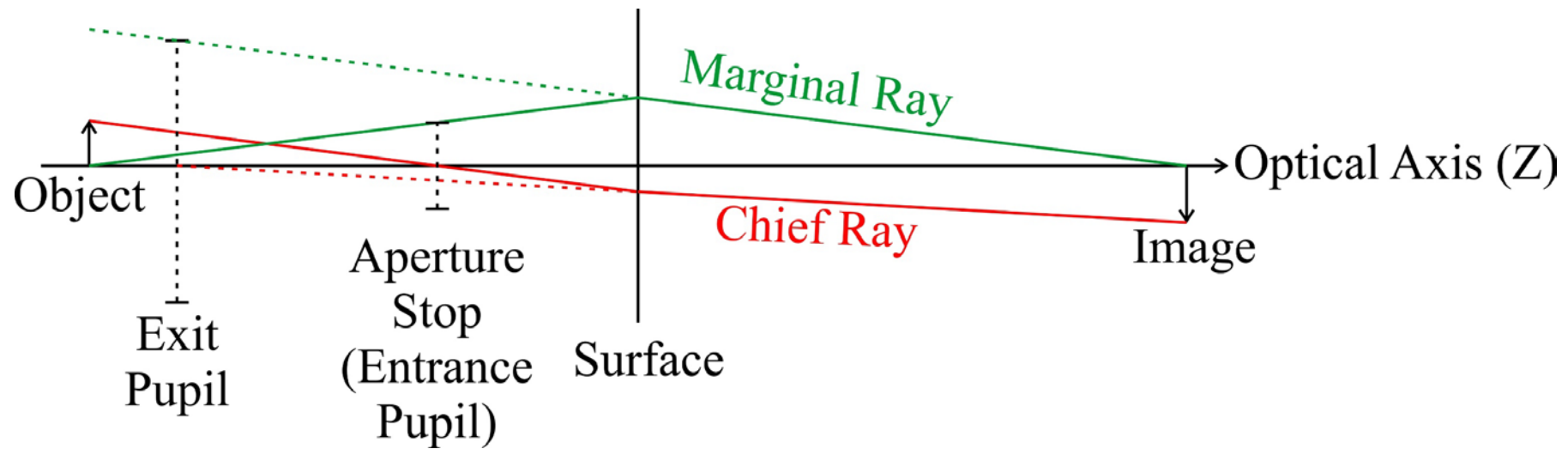

**Figure 3.4. Illustration of the chief and marginal rays in a rotationally symmetric system. Note that the aperture stop is not necessarily in object space.**

As shown in Figure 3.4, the image plane location is defined by the intersection between the marginal ray and the optical axis in image space. The image size is defined by the chief ray height at the image plane, which defines the magnification as well. The entrance and exit pupil plane locations are defined by the intersection between the chief ray and the optical axis in object and image space, respectively. The pupil size is defined by the marginal ray height at the pupil plane. By definition, the image plane is the paraxial image of the object, which is aberration-free in the paraxial regime, and pupil planes are defined similarly as paraxial images of the aperture stop.

## 3.2 Paraxial optics in plane-symmetric systems

Paraxial optics in plane-symmetric systems is based on the same concept as that in rotationally symmetric systems. Instead of approaching an optical axis, the rays are approaching the OAR. Therefore, the ray heights and angles are measured with respect to the OAR.

### 3.2.1 Paraxial raytracing in plane-symmetric systems

Since the OAR can change direction after an optical surface, sagittal and tangential directions may also change. Because the YZ plane is the plane of symmetry in all local coordinate systems, the sagittal direction is defined along the common X direction. Sagittal planes are defined as planes containing the X-axis and the OAR, and can be different for different OAR segments. All sagittal ray heights and angles are defined within sagittal planes. Alternatively, tangential planes are perpendicular to the sagittal planes, intersecting along the OAR, and all tangential ray heights and angles are defined within tangential planes. Figure 3.5 shows an illustration of how ray heights and angles are defined within sagittal and tangential planes. All ray angles are measured between the ray and the OAR. All ray heights at a surface are measured from the OAR intersection with the surface along the local X-axis for sagittal ray heights and local Y-axis for tangential ray heights.

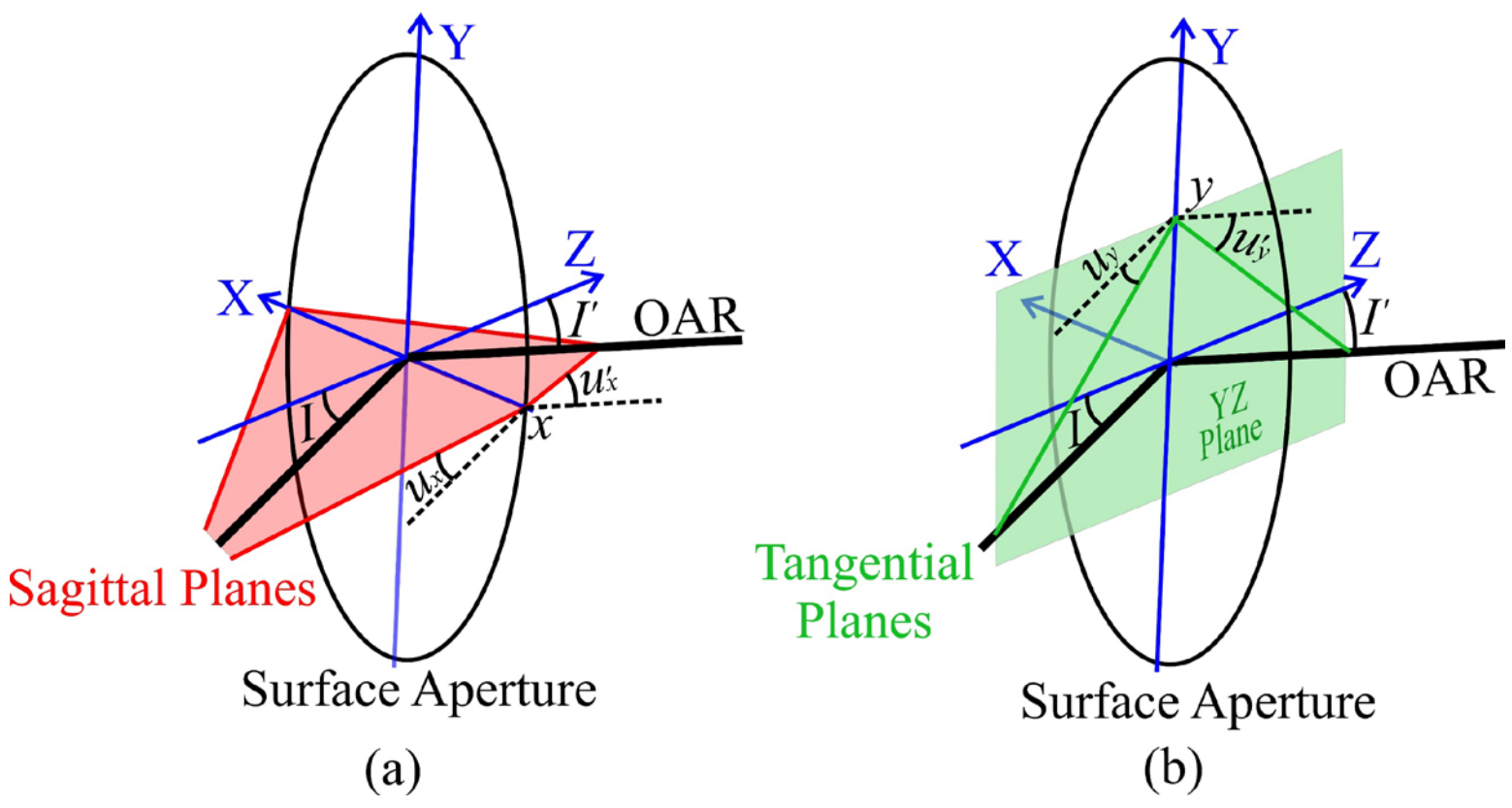

**Figure 3.5. Illustration of (a) sagittal and (b) tangential ray heights and angles.**

The paraxial raytracing equations take similar forms to those in rotationally symmetric systems. However, due to the asymmetry in the sagittal and tangential directions, the raytracing equations are different in the two directions. Note that the $u$ and $u'$ variables are the tangent of the angles shown in Figure 3.5. The sagittal raytracing equations are

$$x_{i+1} = x_i + u'_{xi} t_i \tag{3.9}$$

$$n'_i u'_{xi} = n_i u_{xi} - x_i \phi_{xi} \tag{3.10}$$

$$u_{xi+1} = u'_{xi} \tag{3.11}$$

and the tangential raytracing equations are

$$y_{i+1} \cos I_{i+1} = y_i \cos I'_i + u'_{yi} t_i \tag{3.12}$$

$$n'_i u'_{yi} \cos I'_i = n_i u_{yi} \cos I_i - y_i \phi_{yi} \tag{3.13}$$

$$u_{yi+1} = u'_{yi} \tag{3.14}$$

where $\phi_{xi}$ and $\phi_{yi}$ are the optical powers of the surface in the sagittal and tangential directions respectively, given as

$$\phi_{xi} = \frac{n'_i \cos I'_i - n_i \cos I_i}{R_{xi}} \tag{3.15}$$

$$\phi_{yi} = \frac{n'_i \cos I'_i - n_i \cos I_i}{R_{yi}} \tag{3.16}$$

where $R_{xi}$ and $R_{yi}$ are the radii of curvature in the X and Y directions in the local surface coordinate system for Surface $i$. Recall that, as defined in Chapter 2, the parameter $I_i$ (respectively $I'_i$) is the incident and refractive angles of ORA at Surface $i$. The detailed derivation of the paraxial ray-tracing equations shown in Equations 3.9-3.14 is included in Appendix I. If the surface is spherical, $R_{xi}$ and $R_{yi}$ are equal. In more general cases, $R_{xi}$ and $R_{yi}$ can be different depending on the shape of the surface. Note that when $I_i=I'_i=0$ and $R_{xi}=R_{yi}$ as in the case of rotationally symmetric systems, Equations 3.9-3.16 are reduced to Equations 3.4-3.8.

As opposed to rotationally symmetric systems, paraxial raytracing in plane-symmetric systems is different in the sagittal and tangential directions. The chief and marginal rays are defined the same way as in rotationally symmetric systems. However, due to the asymmetry between the sagittal and tangential directions, chief and marginal ray heights and angles can be different between the two directions at a given plane. The paraxial angles and heights of marginal and chief rays are labeled and illustrated in Table 3.1. Note that in that table, all ray angles and heights are measured from the OAR.

**Table 3.1. Labels of paraxial marginal and chief ray heights and angles**

| Paraxial Ray | Paraxial Quantity | Label |
|---|---|---|
| Sagittal marginal ray | Ray height at $i^{th}$ surface | $x_{ai}$ |
| | Tangent of ray angle before $i^{th}$ surface | $u_{axi}$ |
| | Tangent of ray angle after $i^{th}$ surface | $u'_{axi}$ |
| Sagittal chief ray | Ray height at $i^{th}$ surface | $x_{bi}$ |
| | Tangent of ray angle before $i^{th}$ surface | $u_{bxi}$ |
| | Tangent of ray angle after $i^{th}$ surface | $u'_{bxi}$ |
| Tangential marginal ray | Ray height at $i^{th}$ surface | $y_{ai}$ |
| | Tangent of ray angle before $i^{th}$ surface | $u_{ayi}$ |
| | Tangent of ray angle after $i^{th}$ surface | $u'_{ayi}$ |
| Tangential chief ray | Ray height at $i^{th}$ surface | $y_{bi}$ |
| | Tangent of ray angle before $i^{th}$ surface | $u_{byi}$ |
| | Tangent of ray angle after $i^{th}$ surface | $u'_{byi}$ |

In plane-symmetric systems, an object on the OAR is generally imaged at different locations in the sagittal and tangential planes under the paraxial regime. Therefore, in plane-symmetric systems, there may not be an image plane that is aberration-free in the paraxial regime. In this situation, the image plane needs to be defined differently. In this work, the image and pupil plane locations are defined by the paraxial raytracing in the sagittal direction, as suggested by Sasian[43]. Specifically, the intersection between the image plane and the OAR is defined at the location where the sagittal marginal ray intersects the OAR, as shown in Figure 3.6. The image height is then determined by the sagittal chief ray height in the image plane. Similarly, pupil plane locations are defined by the intersection between the sagittal chief ray and the OAR, and pupil sizes are defined by sagittal marginal ray heights in pupil planes.

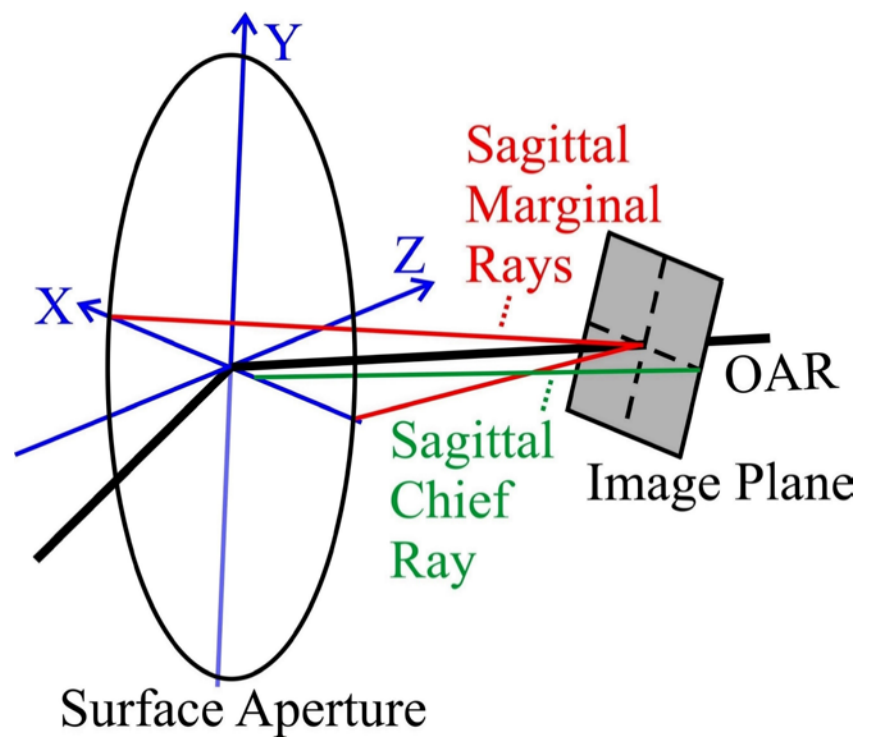

**Figure 3.6. Illustration of the image plane, in light black, defined by the sagittal marginal and chief rays.**

Note that although the image and pupils are defined in the sagittal direction, the ideal image and pupils, which are considered aberration-free, are defined in focus and have the same magnification in both the sagittal and tangential directions. The difference caused by paraxial raytracing in the tangential direction is treated as aberrations (astigmatism and anamorphism).

Furthermore, for a given direction, sagittal or tangential, there are two edges of an object line and also two edges of an aperture stop. Therefore, marginal and chief rays are not singular and can be defined aiming at either edges of the object or the aperture stop. In this work, for all derivations that use marginal or chief ray heights and angles listed in Table 3.1, the marginal and chief rays are chosen to go through the positive edges of the exit pupil and the image, respectively.

### 3.2.2 Tilt angles of field and pupil planes

In rotationally symmetric systems, field and pupil planes are always normal to the optical axis. On the other hand, in plane-symmetric systems, field and pupil planes can have tilt angles as shown in Figure 2.2, which also need to be defined in addition to field and pupil plane locations. In this work, the tilt angles of field and pupil planes are set to zero to simplify the calculation of transverse aberration discussed in Chapter 4. In the following part, we discuss the definitions of field and pupil tilt angles and their impact on aberration calculation.

We will focus on the effect of image plane tilt angle on anamorphism and field-linear defocus, which are related to first order properties: magnification and whether the image plane is in focus. Anamorphism describes the magnification difference between the sagittal and tangential directions, and field-linear defocus is a defocus that is linear to field height in the tangential direction, as shown in Figure 3.7.

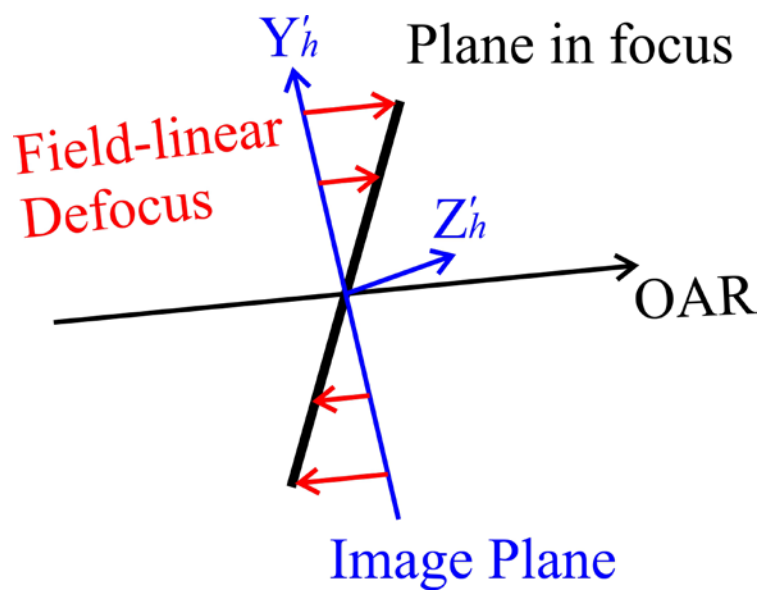

**Figure 3.7. Illustration of field-linear defocus measured along the OAR.**

We list two possible ways to define field plane tilt angles, specifically by the relationship between the object and image tilt angles. The pupil plane tilt angles are defined the same way as field plane tilt angles since pupils can be treated as images of the aperture stop.

(1) The first way is to define field tilt angles following the sagittal paraxial image, which means as the object moves along the tangential direction, the image tilt angle is in the direction of the corresponding sagittal image [47]. The relationship between the object and image plane tilt angles can be derived using the sagittal Coddington equation [48] and Snell's law for the OAR as

$$\Delta\left(\frac{1}{\cos I}\left(\frac{\sin I}{R_x}-\frac{1}{s_0}\tan\theta_h\right)\right)=0 \tag{3.17}$$

where the operator, Δ( ), calculates the change of the quantity inside the bracket on refraction or reflection. For example, Δ($n$)=$n'$–$n$, where $n$ and $n'$ are the refractive index before and after the refraction, respectively. The variables $s_0$ and $s'_0$ are the object and image distances measured from the surface for a pair of on-OAR object and image points, as shown in Figure 3.8. The variables $s_0$ and $s'_0$ are signed and are positive if they are on the positive Z-axis side of the local surface coordinate system.

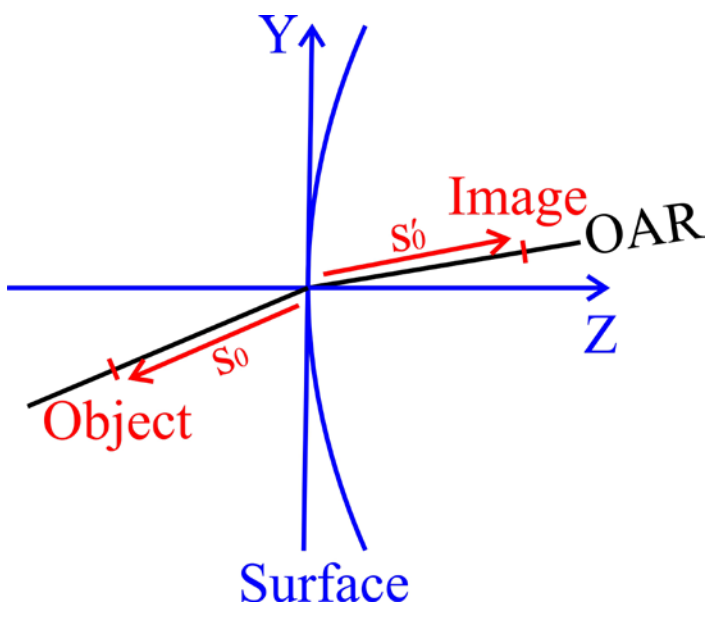

**Figure 3.8. Illustration of on-OAR object and image.**

Note that one of the Coddington equations describes the paraxial imaging in the sagittal direction as

$$\frac{n_i'}{\left(s_0'\right)_i} - \frac{n_i}{\left(s_0\right)_i} = \phi_{xi} \tag{3.18}$$

which can be derived from the sagittal paraxial raytracing equations (Equations 3.9-3.11).

This way of defining field plane tilt angles is a natural generalization of the Scheimpflug condition [49], which is the case when $I$=0 in Equation 3.17. The magnifications of the sagittal paraxial image in the sagittal and tangential directions, $m_x$ and $m_y$, can also be derived as [47]

$$m_x = \frac{n s_0'}{n' s_0} \tag{3.19}$$

$$m_y = \frac{n s_0' \cos\theta_h \cos I}{n' s_0 \cos\theta_h' \cos I'} \tag{3.20}$$

where $n$, $n'$, $\theta_h$, and $\theta'_h$ were defined in Chapter 2. The magnification difference in the sagittal and tangential directions indicates that in order to keep the image focused in the sagittal direction, the image would have an intrinsic anamorphism. In Section 5.2, it is shown this definition also makes the surface contribution of field-linear defocus zero when two conditions are met: (1) the ideal image has an intrinsic anamorphism described in Equation 3.19-3.20, and (2) the stop is at the surface. If the ideal image is defined to have the same magnification, $m_x$, for both sagittal and tangential directions, there will be field-linear defocus in the paraxial regime.

A special case with this tilt angle definition is that the tilt angle of an image plane tends to approach 90 degrees when the image distance approaches infinity. This is consistent with the sagittal Coddington equation, because when the object is close to the focal point, the derivative of the image distance with respect to the object distance approaches infinity as well, causing the tilt angle to be effectively 90 degrees.

(2) The second way to define the tilt angles is to simply define all image planes to be normal to the OAR. In Section 4.3, it is shown that this definition makes the relation between wavefront aberrations and transverse aberrations in plane-symmetric systems the same as that in rotationally symmetric systems, simplifying the calculation of transverse aberrations. It also avoids the 90-degree image plane tilt angle situation discussed previously. In cases where the object, detector, or an internal image plane of interest is placed with a non-zero tilt angle with respect to the OAR, a geometric relation between the transverse aberrations in the image planes with zero and non-zero tilt angles can be found. For the scope of this work, we focus on image planes with zero tilt angles.

On the other hand, if the ideal image has no anamorphism, it is shown in Section 5.2.3 that there may not be an image plane tilt angle that can make anamorphism or field-linear defocus zero, which means the definition of image plane tilt may not be used to eliminate anamorphism or field-linear defocus. In this work, all image and pupil planes are defined normal to the OAR. The magnification is defined by that of the paraxial image in the sagittal direction shown in Equation 3.19.

### 3.2.3 Effect of freeform shapes on the radius of curvature

The radius of curvature of a surface is an important parameter in paraxial raytracing, because it determines the optical power of the surface. While it is easy to find the radius of curvature of a spherical surface, it is more complex to define the radius of curvature for freeform surfaces. Since the radius of curvature is mainly used to describe the impact of surface shape on paraxial raytracing, an effective radius of curvature can be defined as the parameter that determines the surface shape in the paraxial regime.

To identify the surface shape in the paraxial regime, the sag function is expanded into XY polynomials defined in the local surface coordinate system as

$$z(x,y)=\sum_{i,j} A_{i-j}x^i y^j = A_{0-0}+A_{1-0}x+A_{0-1}y+A_{2-0}x^2+A_{0-2}y^2+A_{1-1}xy+\ldots \quad (3.21)$$

where $A_{i-j}$ is the coefficient for the term $x^i y^j$. $i$ and $j$ are integers starting from zero, denoting the power of $x$ and $y$. This sag function also needs to satisfy the requirements discussed in Section 2.2 for the setup of plane-symmetric systems. From Equation 2.2, we can see that the coefficient $A_{0-0}$ is zero. $A_{1-0}$ and $A_{0-1}$ are also zero to satisfy Equations 2.3-2.4. The parameter $A_{1-1}$ is zero to satisfy the plane symmetry requirement shown in Equation 2.5. Therefore, the terms with the lowest power of $x$ or $y$ are $A_{2-0}x^2$ and $A_{0-2}y^2$.

In the paraxial regime, object heights and ray angles with respect to the OAR are approaching zero. Therefore, the ray intersections on the surface are also approaching zero, making $A_{2\text{-}0}x^2$ and $A_{0\text{-}2}y^2$ the dominant terms for the surface shape. The sag function can then be approximated to

$$z\left(x,y\right)\Big|_{x\to 0,y\to 0} \approx A_{2-0}x^2 + A_{0-2}y^2 \tag{3.22}$$

It can be seen that $A_{2\text{-}0}$ and $A_{0\text{-}2}$ are the coefficients that affect the shape of the surface and raytracing in the paraxial regime. In the case of spherical surfaces, the sag function in the paraxial regime is

$$z_{sph}\left(x,y\right)\Big|_{x\to 0,y\to 0} \approx \frac{1}{2R}x^2 + \frac{1}{2R}y^2 \tag{3.23}$$

Therefore, the effective radii of curvature can be defined in the sagittal and tangential directions by comparing Equation 3.22 with Equation 3.23, which can be written as

$$R_{x-eff} = \frac{1}{2A_{2-0}} \tag{3.24}$$

$$R_{y-eff} = \frac{1}{2A_{0-2}} \tag{3.25}$$

The radii of curvature in Equations 3.15-3.16 that define optical powers can be replaced by the effective radii of curvature, given as

$$\phi_x = \frac{n_i' \cos I_i' - n_i \cos I_i}{R_{x-eff}} \tag{3.26}$$

$$\phi_y = \frac{n_i' \cos I_i' - n_i \cos I_i}{R_{y-eff}} \tag{3.27}$$

If the surface shape is described in other ways than XY polynomials, the coefficients $A_{2\text{-}0}$ and $A_{0\text{-}2}$ can be found by converting the surface description to equivalent XY polynomials. For example, if the surface description follows Equation 2.6 with $z_{freeform}$ described with the plane-symmetric terms among the first 16 Fringe Zernike polynomials, they can be converted to XY polynomials following Table 3.2.

**Table 3.2. Conversion from the plane-symmetric terms among the first 16 Fringe Zernike polynomials to XY polynomials**

| Term (ZPi) | Fringe Zernike polynomial | Corresponding XY polynomial |
|---|---|---|
| 1 | $1$ | $1$ |
| 3 | $r_n \sin\theta$ | $y/R_{zn}$ |
| 4 | $2r_n^2-1$ | $-1+2(x^2+y^2)/R^2_{zn}$ |
| 5 | $r_n^2 \cos2\theta$ | $(x^2-y^2)/R^2_{zn}$ |
| 8 | $(3r_n^3-2r_n) \sin\theta$ | $-2y/R_{zn} +3y(x^2+y^2)/R^3_{zn}$ |
| 9 | $6r_n^4-6r_n^2+1$ | $1-6(x^2+y^2)/R^2_{zn}+6(x^2+y^2)^2/R^4_{zn}$ |
| 11 | $r_n^3 \sin3\theta$ | $(3x^2y-y^3)/R^3_{zn}$ |
| 12 | $(4r_n^4-3r_n^2) \cos2\theta$ | $(-3x^2+3y^2)/R^2_{zn}+(4x^4-4y^4)/R^4_{zn}$ |
| 15 | $(10r_n^5-12r_n^3+3r_n) \sin\theta$ | $3y/R_{zn}-12y(x^2+y^2)/R^3_{zn}+10y(x^2+y^2)^2/R^5_{zn}$ |
| 16 | $20r_n^6-30r_n^4+12r_n^2-1$ | $-1+12(x^2+y^2)/R^2_{zn} -30(x^2+y^2)^2/R^4_{zn} +20(x^2+y^2)^3/R^6_{zn}$ |

The corresponding effective radii of curvature can be found as

$$\frac{1}{2R_{x-eff}} = \frac{1}{2R} + \frac{2Z_4 + Z_5 - 6Z_9 - 3Z_{12} + 12Z_{16}}{R_{zn}^2} \tag{3.28}$$

$$\frac{1}{2R_{y-eff}} = \frac{1}{2R} + \frac{2Z_4 - Z_5 - 6Z_9 + 3Z_{12} + 12Z_{16}}{R_{zn}^2} \tag{3.29}$$

where $R$ is the radius of curvature of the base spherical surface.

# Chapter 4. Wavefront aberrations for plane-symmetric systems

Imaging aberrations are defined as the departures from the defined ideal image. In this chapter, we describe how imaging aberrations are quantified using ray or wave models in plane-symmetric systems and the relation between the two models.

The ray model describes aberrations as the difference in location between the ideal image point and a real image point defined by a real ray or real rays. A commonly used approach is to define the aberration for a ray by measuring the corresponding transverse aberration, the distance between the ideal image point and the real ray intersection on the image plane.

The wave model describes the light behavior using wavefronts, which are defined as surfaces with the same phase as the light wave propagates. For incoherent imaging, the wavefront can be simplified to a geometric wavefront, which does not include the diffractive effect of light propagation. Under this model, the phase change between wavefronts is determined by the optical path that a real ray travels between them, as shown in Figure 4.1. Note that the temporal component of the phase is ignored in this work, since the light is assumed to be temporally incoherent. Therefore, the wavefront aberrations can then be defined as the departure of the real geometric wavefront from a spherical wavefront centered at the ideal image point, which is called a reference sphere. Quantitatively, the wavefront aberrations are measured by the optical path difference along the real ray between the real wavefront and the reference sphere.

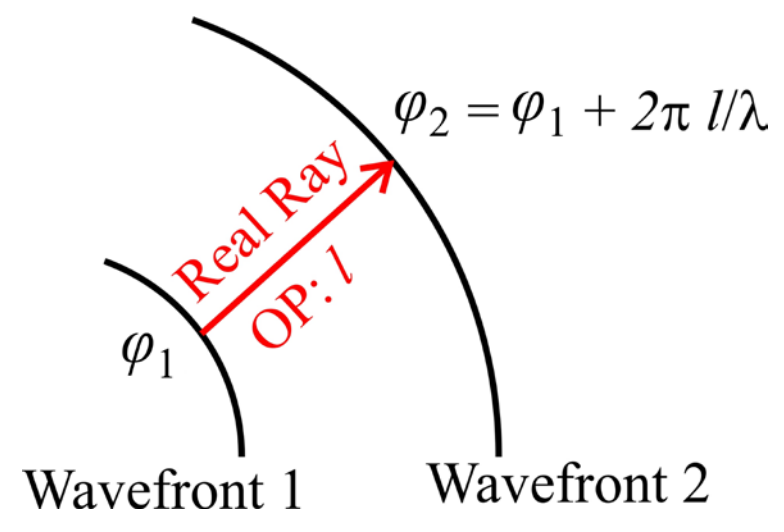

**Figure 4.1. Geometric wavefront propagation is determined by the optical path of a real ray. The phases of wavefront 1 and 2, shown as $\varphi_1$ and $\varphi_2$, are related by the optical path a real ray travels between the two wavefronts. Note that the temporal component of the phase is ignored in this model, since the light is assumed to be temporally incoherent.**

Based on the definition of the geometric wavefront, the rays propagating through a geometric wavefront are always normal to the wavefront surface. Therefore, the wavefront also determines how real rays travel, associating wavefront aberrations with transverse aberrations. For this work, we mainly focus on the wavefront aberrations, for which analytical formulae for intrinsic and induced aberrations are derived. In this chapter, the wavefront aberration function and related expansion are defined. The conversion between wavefront and transverse aberrations is established in Section 4.3.

## 4.1 Wavefront aberrations defined at the exit pupil

The wavefront at the exit pupil of an optical system is of particular interest since it contains the aberration information of a system. In addition, in the paraxial regime, the light beams from all object points fill pupils by definition, making it convenient to describe the wavefronts using exit pupil coordinates. It is also a convenient location to perform interferometric measurements of wavefronts during optical performance testing.

To calculate the wavefront aberrations at the exit pupil, both the reference sphere and real wavefront are set to cross the center of the exit pupil, as shown in Figure 4.2. The reference sphere, with a radius of curvature labeled as $R'_{ref}$, is centered at the ideal image point with coordinates ($H'_{Ix}$, $H'_{Iy}$) on the image plane. The prime symbol in $R'_{ref}$ denotes that the reference sphere is in image space. $R'_{ref}$ is signed and is positive if the center of reference sphere is on the positive $Z'_{\rho}$-axis in the exit pupil coordinate system. A real ray, shown in solid green in Figure 4.2, is normal to the real wavefront and intersects the image plane with the transverse aberration indicated by a vector ($\varepsilon'_x$, $\varepsilon'_y$). The real ray also intersects the exit pupil plane at coordinates ($\rho'_{Ix}$, $\rho'_{Iy}$).

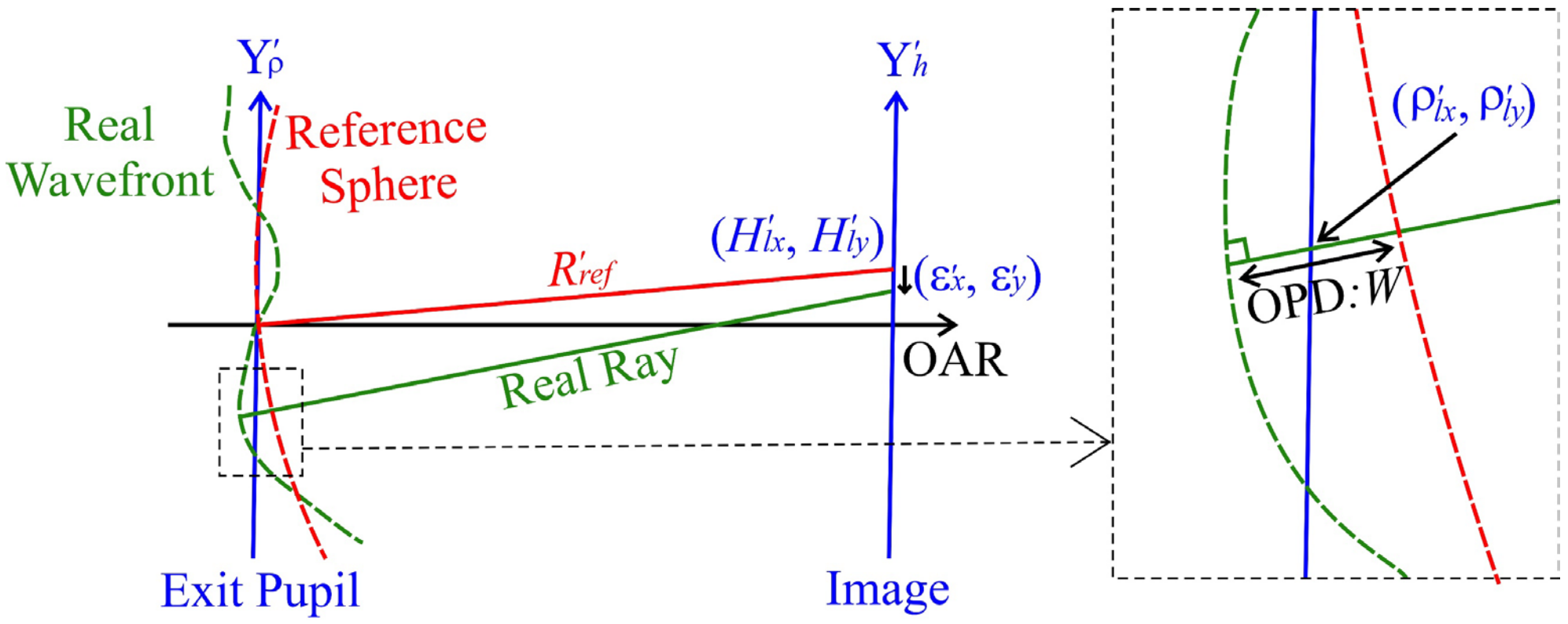

(a) (b)

**Figure 4.2. (a) Illustration of wavefront aberrations defined at the exit pupil. Both the real wavefront in dashed green and the reference sphere in dashed red cross the center of the exit pupil. The reference sphere is centered at the ideal image point ($H'_{Ix}$, $H'_{Iy}$) with a radius of curvature, $R'_{ref}$. (b) is part of (a) enlarged. A real ray in green is normal to the real wavefront and intersects the pupil plane at pupil coordinates ($\rho'_{Ix}$, $\rho'_{Iy}$). The wavefront aberration, $W$, for this ray is the optical path difference between the two wavefronts along the real ray.**

The wavefront aberrations at the exit pupil is the optical path difference between the reference sphere and real wavefront along the real ray, shown as $W$ in Figure 4.2(b). The variable $W$ is a signed quantity and is positive when the real ray reaches the reference sphere before the real wavefront. For example, $W$ is negative for the real ray shown in Figure 15.

It can be seen that $W$ is a function of the ideal image location, ($H'_{lx}$, $H'_{ly}$), and the real ray intersection at the exit pupil plane, ($\rho'_{lx}$, $\rho'_{ly}$). Field and pupil vectors, $\boldsymbol{H'_l}$ and $\boldsymbol{\rho'_l}$, can be defined in the image and exit pupil planes pointing to ($H'_{lx}$, $H'_{ly}$) and ($\rho'_{lx}$, $\rho'_{ly}$) as

$$\boldsymbol{H}'_{\boldsymbol{l}} = \left(H'_{lx}, H'_{ly}\right) \tag{4.1}$$

$$\boldsymbol{\rho}'_{\boldsymbol{l}} = \left(\rho'_{lx}, \rho'_{ly}\right) \tag{4.2}$$

Note that in all equations, vectors are represented by bold symbols. The prime symbol indicates that the quantity is in the image space. The corresponding vectors can also be defined in the object plane and entrance pupil plane as

$$\boldsymbol{H}_{\boldsymbol{l}} = \left(H_{lx}, H_{ly}\right) \tag{4.3}$$

$$\boldsymbol{\rho}_{\boldsymbol{l}} = \left(\rho_{lx}, \rho_{ly}\right) \tag{4.4}$$

With the sizes of the exit pupil and image defined as $L'_p$ and $L'_h$, normalized field and pupil vectors can be defined as

$$\boldsymbol{H} = \left(H'_{lx}, H'_{ly}\right) / L'_h = \left(H_x, H_y\right) \tag{4.5}$$

$$\boldsymbol{\rho} = \left(\rho'_{lx}, \rho'_{ly}\right) / L'_p = \left(\rho_x, \rho_y\right) \tag{4.6}$$

where $L'_p$ and $L'_h$ are always assumed to be positive, and the normalized field and pupil coordinates, $H_x$, $H_y$, $\rho_x$, and $\rho_y$, range from -1 to 1. Because the object and image size are related by the sagittal magnification, $m_x$, the normalized field vector can also be written as

$$\boldsymbol{H} = \left(H'_{lx}, H'_{ly}\right) / L'_h = \left(\frac{H'_{lx}}{m_x}, \frac{H'_{ly}}{m_x}\right) / \frac{L'_h}{m_x} = \left(H_{lx}, H_{ly}\right) / L_h \tag{4.7}$$

where ($H_{lx}$, $H_{ly}$) is the object location corresponding to ($H'_{lx}$, $H'_{ly}$), and $L_h$ is the object size. Similarly, we have

$$\boldsymbol{\rho} = \left(\rho'_{lx}, \rho'_{ly}\right) / L'_p = \left(\frac{\rho'_{lx}}{m_{\rho x}}, \frac{\rho'_{ly}}{m_{\rho x}}\right) / \frac{L'_h}{m_{\rho x}} = \left(\rho_{lx}, \rho_{ly}\right) / L_p \tag{4.8}$$

where $m_{\rho x}$ is the sagittal magnification between the entrance and exit pupils defined similarly to $m_x$; ($\rho_{lx}$, $\rho_{ly}$) is the entrance pupil location, and $L_p$ is the entrance pupil size. It can be seen that $W$ can then be written as a function of $\boldsymbol{H}$ and $\boldsymbol{\rho}$, expressed as $W(\boldsymbol{H}, \boldsymbol{\rho})$ or $W(H_x, H_y, \rho_x, \rho_y)$, and be expanded into aberration terms with different dependencies on $\boldsymbol{H}$ and $\boldsymbol{\rho}$.

## 4.2 Wavefront aberration function expansion in plane-symmetric systems

In a plane-symmetric system, the symmetry requires that the mirror image of a real ray about the plane of symmetry is also a real ray with the same wavefront aberration. Since the plane of symmetry is the YZ plane, the X component of the normalized field and pupil vectors of a ray have the opposite sign as compared to its mirror image. Therefore, the wavefront aberration function has the property given as

$$W\left(H_x, H_y, \rho_x, \rho_y\right) = W\left(-H_x, H_y, -\rho_x, \rho_y\right) \tag{4.9}$$

This property limits the possible aberration terms in the $W$ expansion. In this work, we use the vectorial expansion developed by Sasian [43], given as

$$W\left(\boldsymbol{H}, \boldsymbol{\rho}\right) = \sum_{k,m,n,p,q}^{\infty} W_{2k+n+p,2m+n+q,n,p,q} \left(\boldsymbol{H}\cdot\boldsymbol{H}\right)^k \left(\boldsymbol{\rho}\cdot\boldsymbol{\rho}\right)^m \left(\boldsymbol{H}\cdot\boldsymbol{\rho}\right)^n \left(\boldsymbol{i}\cdot\boldsymbol{H}\right)^p \left(\boldsymbol{i}\cdot\boldsymbol{\rho}\right)^q \tag{4.10}$$

where $W_{2k+n+p,2m+n+q,n,p,q}$ is the aberration coefficient for its associated aberration term. Equation 4.10 expands $W$ in terms of dot products of normalized field and pupil vectors, and there are five possible dot products that satisfy Equation 4.9. The five indexes in the subscript of the aberration coefficients correspond to the sum of the powers of $H_x$ and $H_y$, the sum of the powers of $p_x$ and $p_y$, the power of the dot product $(\boldsymbol{H}\cdot\boldsymbol{\rho})$, the power of $H_y$ (i.e., $(\boldsymbol{i}\cdot\boldsymbol{H})$), and the power of $p_y$ (i.e., $(\boldsymbol{i}\cdot\boldsymbol{\rho})$), respectively. When the dot products are performed between $\boldsymbol{H}$ and $\boldsymbol{\rho}$ vectors in the expansion, the vectors are put into an overlayed coordinate system where all local coordinate systems of surfaces, field and pupil planes coincide with each other, and $\boldsymbol{i}$ is the unit vector pointing towards the Y direction of the overlayed coordinate system. An illustration of the overlayed coordinate system was shown in Figure 3.

Table 4.1 lists four groups of $W$ aberration terms where the terms are grouped by the number of dot products, $k+m+n+p+q$.

**Table 4.1. The first four groups of aberration terms in plane-symmetric systems**

| | | | |
|---|---|---|---|
| First group | $W_{00000}$ | | |
| Second group | $W_{01001}(\boldsymbol{i}\cdot\boldsymbol{\rho})$ | $W_{10010}(\boldsymbol{i}\cdot\boldsymbol{H})$ | $W_{02000}(\boldsymbol{\rho}\cdot\boldsymbol{\rho})$ |
| | $W_{11100}(\boldsymbol{H}\cdot\boldsymbol{\rho})$ | $W_{20000}(\boldsymbol{H}\cdot\boldsymbol{H})$ | |
| Third group | $W_{02002}(\boldsymbol{i}\cdot\boldsymbol{\rho})^2$ | $W_{11011}(\boldsymbol{i}\cdot\boldsymbol{\rho})(\boldsymbol{i}\cdot\boldsymbol{H})$ | $W_{20020}(\boldsymbol{i}\cdot\boldsymbol{H})^2$ |
| | $W_{03001}(\boldsymbol{i}\cdot\boldsymbol{\rho})(\boldsymbol{\rho}\cdot\boldsymbol{\rho})$ | $W_{12101}(\boldsymbol{i}\cdot\boldsymbol{\rho})(\boldsymbol{H}\cdot\boldsymbol{\rho})$ | $W_{12010}(\boldsymbol{i}\cdot\boldsymbol{H})(\boldsymbol{\rho}\cdot\boldsymbol{\rho})$ |
| | $W_{21001}(\boldsymbol{i}\cdot\boldsymbol{\rho})(\boldsymbol{H}\cdot\boldsymbol{H})$ | $W_{21110}(\boldsymbol{i}\cdot\boldsymbol{H})(\boldsymbol{H}\cdot\boldsymbol{\rho})$ | $W_{30010}(\boldsymbol{i}\cdot\boldsymbol{H})(\boldsymbol{H}\cdot\boldsymbol{H})$ |
| | $W_{04000}(\boldsymbol{\rho}\cdot\boldsymbol{\rho})^2$ | $W_{13100}(\boldsymbol{H}\cdot\boldsymbol{\rho})(\boldsymbol{\rho}\cdot\boldsymbol{\rho})$ | $W_{22200}(\boldsymbol{H}\cdot\boldsymbol{\rho})^2$ |
| | $W_{22000}(\boldsymbol{H}\cdot\boldsymbol{H})(\boldsymbol{\rho}\cdot\boldsymbol{\rho})$ | $W_{31100}(\boldsymbol{H}\cdot\boldsymbol{H})(\boldsymbol{H}\cdot\boldsymbol{\rho})$ | $W_{40000}(\boldsymbol{H}\cdot\boldsymbol{H})^2$ |
| Fourth group | $W_{03003}(\boldsymbol{i}\cdot\boldsymbol{\rho})^3$ | $W_{12012}(\boldsymbol{i}\cdot\boldsymbol{H})(\boldsymbol{i}\cdot\boldsymbol{\rho})^2$ | $W_{21021}(\boldsymbol{i}\cdot\boldsymbol{H})^2(\boldsymbol{i}\cdot\boldsymbol{\rho})$ |
| | $W_{30030}(\boldsymbol{i}\cdot\boldsymbol{H})^3$ | $W_{04002}(\boldsymbol{i}\cdot\boldsymbol{\rho})^2(\boldsymbol{\rho}\cdot\boldsymbol{\rho})$ | $W_{13011}(\boldsymbol{i}\cdot\boldsymbol{H})(\boldsymbol{i}\cdot\boldsymbol{\rho})(\boldsymbol{\rho}\cdot\boldsymbol{\rho})$ |
| | $W_{22002}(\boldsymbol{i}\cdot\boldsymbol{\rho})^2(\boldsymbol{H}\cdot\boldsymbol{H})$ | $W_{22020}(\boldsymbol{i}\cdot\boldsymbol{H})^2(\boldsymbol{\rho}\cdot\boldsymbol{\rho})$ | $W_{31011}(\boldsymbol{i}\cdot\boldsymbol{H})(\boldsymbol{i}\cdot\boldsymbol{\rho})(\boldsymbol{H}\cdot\boldsymbol{H})$ |
| | $W_{40020}(\boldsymbol{i}\cdot\boldsymbol{H})^2(\boldsymbol{H}\cdot\boldsymbol{H})$ | $W_{13102}(\boldsymbol{i}\cdot\boldsymbol{\rho})^2(\boldsymbol{H}\cdot\boldsymbol{\rho})$ | $W_{22111}(\boldsymbol{i}\cdot\boldsymbol{H})(\boldsymbol{i}\cdot\boldsymbol{\rho})(\boldsymbol{H}\cdot\boldsymbol{\rho})$ |
| | $W_{31120}(\boldsymbol{i}\cdot\boldsymbol{H})^2(\boldsymbol{H}\cdot\boldsymbol{\rho})$ | $W_{05001}(\boldsymbol{i}\cdot\boldsymbol{\rho})(\boldsymbol{\rho}\cdot\boldsymbol{\rho})^2$ | $W_{14010}(\boldsymbol{i}\cdot\boldsymbol{H})(\boldsymbol{\rho}\cdot\boldsymbol{\rho})^2$ |
| | $W_{14101}(\boldsymbol{i}\cdot\boldsymbol{\rho})(\boldsymbol{H}\cdot\boldsymbol{\rho})(\boldsymbol{\rho}\cdot\boldsymbol{\rho})$ | $W_{23001}(\boldsymbol{i}\cdot\boldsymbol{\rho})(\boldsymbol{H}\cdot\boldsymbol{H})(\boldsymbol{\rho}\cdot\boldsymbol{\rho})$ | $W_{23110}(\boldsymbol{i}\cdot\boldsymbol{H})(\boldsymbol{H}\cdot\boldsymbol{\rho})(\boldsymbol{\rho}\cdot\boldsymbol{\rho})$ |
| | $W_{23201}(\boldsymbol{i}\cdot\boldsymbol{\rho})(\boldsymbol{H}\cdot\boldsymbol{\rho})^2$ | $W_{32010}(\boldsymbol{i}\cdot\boldsymbol{H})(\boldsymbol{H}\cdot\boldsymbol{H})(\boldsymbol{\rho}\cdot\boldsymbol{\rho})$ | $W_{32101}(\boldsymbol{i}\cdot\boldsymbol{\rho})(\boldsymbol{H}\cdot\boldsymbol{H})(\boldsymbol{H}\cdot\boldsymbol{\rho})$ |
| | $W_{32210}(\boldsymbol{i}\cdot\boldsymbol{H})(\boldsymbol{H}\cdot\boldsymbol{\rho})^2$ | $W_{41110}(\boldsymbol{i}\cdot\boldsymbol{H})(\boldsymbol{H}\cdot\boldsymbol{H})(\boldsymbol{H}\cdot\boldsymbol{\rho})$ | $W_{41001}(\boldsymbol{i}\cdot\boldsymbol{\rho})(\boldsymbol{H}\cdot\boldsymbol{H})^2$ |
| | $W_{50010}(\boldsymbol{i}\cdot\boldsymbol{H})(\boldsymbol{H}\cdot\boldsymbol{H})^2$ | $W_{06000}(\boldsymbol{\rho}\cdot\boldsymbol{\rho})^3$ | $W_{15100}(\boldsymbol{H}\cdot\boldsymbol{\rho})(\boldsymbol{\rho}\cdot\boldsymbol{\rho})^2$ |
| | $W_{24000}(\boldsymbol{H}\cdot\boldsymbol{H})(\boldsymbol{\rho}\cdot\boldsymbol{\rho})^2$ | $W_{24200}(\boldsymbol{H}\cdot\boldsymbol{\rho})^2(\boldsymbol{\rho}\cdot\boldsymbol{\rho})$ | $W_{33300}(\boldsymbol{H}\cdot\boldsymbol{\rho})^3$ |
| | $W_{33100}(\boldsymbol{H}\cdot\boldsymbol{H})(\boldsymbol{H}\cdot\boldsymbol{\rho})(\boldsymbol{\rho}\cdot\boldsymbol{\rho})$ | $W_{42200}(\boldsymbol{H}\cdot\boldsymbol{H})(\boldsymbol{H}\cdot\boldsymbol{\rho})^2$ | $W_{42000}(\boldsymbol{H}\cdot\boldsymbol{H})^2(\boldsymbol{\rho}\cdot\boldsymbol{\rho})$ |
| | $W_{51100}(\boldsymbol{H}\cdot\boldsymbol{H})^2(\boldsymbol{H}\cdot\boldsymbol{\rho})$ | $W_{60000}(\boldsymbol{H}\cdot\boldsymbol{H})^3$ | |

The first group only contain one coefficient, $W_{00000}$, representing the term that has no dependency on the five dot products. This constant term contains no information on aberrations and can be ignored. The second group contain five terms, each corresponding to one type of dot product. The $W_{10010}$ and $W_{20000}$ terms are piston terms without dependence on the pupil vector, which can be ignored as well. The $W_{01001}$ term represents the aberration of OAR, which automatically is zero when the OAR is a real ray. The $W_{02000}$ and $W_{11100}$ terms are rotationally symmetric defocus and magnification change. which become zero when the image plane is placed at the sagittal paraxial image plane with the magnification defined by sagittal paraxial raytracing. Therefore, the third group aberration terms represent the main aberration types present in plane-symmetric systems. This work focuses on analytically deriving of the third-group aberration coefficients.

In prior work, Sasian derived approximated formulae for the third-group aberration coefficients. The formulae are limited to intrinsic aberrations with approximations, which will be discussed in Chapter 5. In this work, we provide analytical expressions of third-group coefficients in a complete way that includes accurate intrinsic aberrations and induced aberrations. The detailed derivation process of induced aberrations is discussed in Chapter 6.

### 4.3 Relation between wavefront and transverse aberrations

Consider a real ray in image space as shown in Figure 4.3. The real ray intersects the reference sphere and image plane at $P'$ and $Q'_r$, respectively.

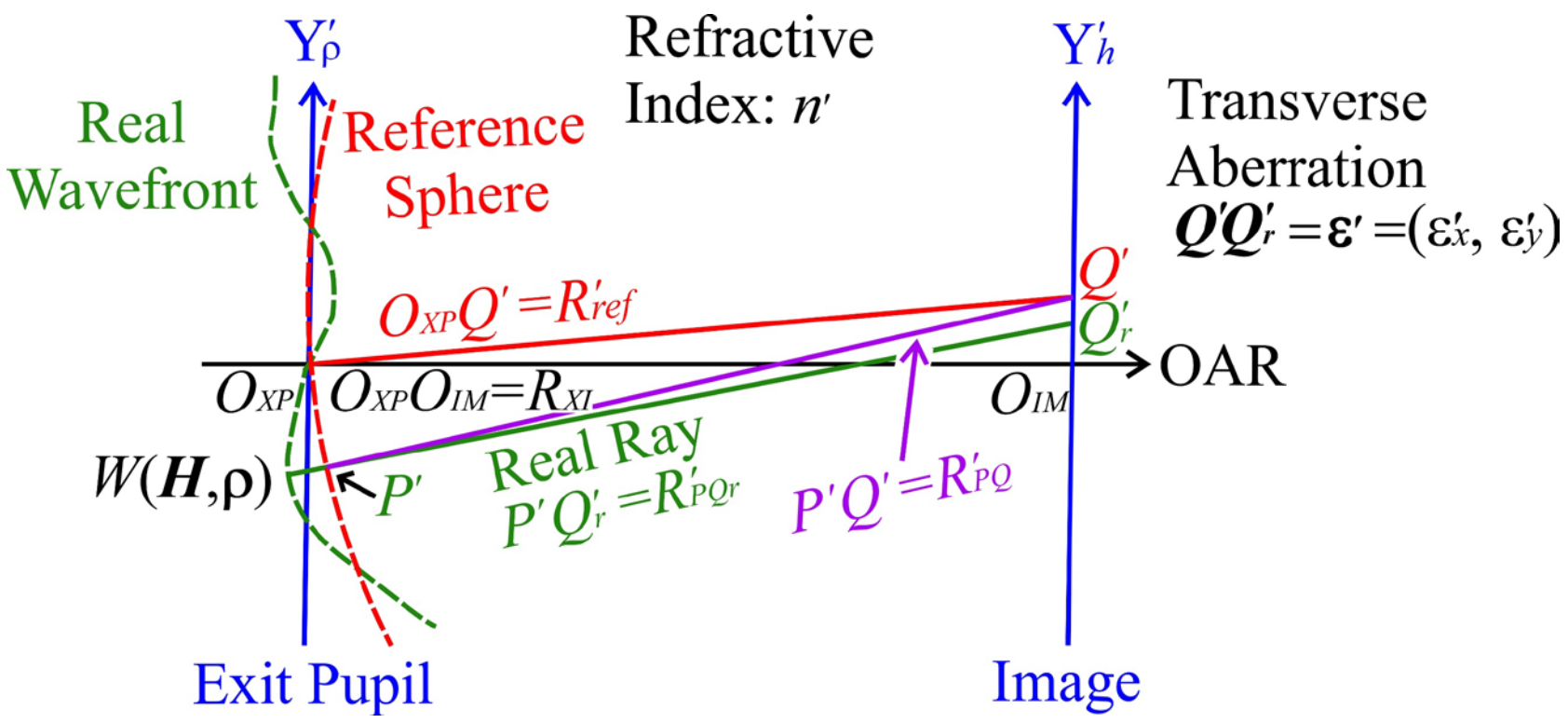

**Figure 4.3. An illustration of a real ray (green line) in image space intersecting the real wavefront and image plane at *P′* and *Q′r*, respectively. *OXP* is the center of the exit pupil. *OIM* is the center of the image plane. *Q′* is the ideal image point.**

Certain vectors are defined. $\boldsymbol{R}_{XI}$ is defined as the vector pointing to $O_{IM}$, the center of image plane, from $O_{XP}$, the center of exit pupil, given as

$$\boldsymbol{R}_{XI} = \boldsymbol{O}_{IM}\boldsymbol{O}_{XP} \tag{4.11}$$

$\boldsymbol{R}'_{PQ}$ is defined as the vector pointing to $Q'$, the ideal image point, from $P'$, given as

$$\boldsymbol{R}'_{PQ} = \boldsymbol{P}'\boldsymbol{Q}' \tag{4.12}$$

Furthermore, $\boldsymbol{R}'_{PQr}$ is defined as the vector pointing to $Q'_r$ from $P'$, given as

$$\boldsymbol{R}'_{PQr} = \boldsymbol{P}'\boldsymbol{Q}'_r \tag{4.13}$$

The lengths of $\boldsymbol{R}_{XI}$, $\boldsymbol{R}'_{PQ}$, and $\boldsymbol{R}'_{PQr}$ are defined as

$$R_{XI} = \left|\boldsymbol{R}_{XI}\right| sign\left(R_{XI}\right) \tag{4.14}$$

$$R'_{PQ} = \left|\boldsymbol{R}'_{PQ}\right| sign\left(R'_{PQ}\right) \tag{4.15}$$

$$R'_{PQr} = \left|\boldsymbol{R}'_{PQr}\right| sign\left(R'_{PQr}\right) \tag{4.16}$$

where *sign*() is the sign of the quantity (1 or -1) inside the bracket. Note that $R_{XI}$, $R'_{PQ}$, and $R'_{PQr}$ are signed quantities, and their signs follow that of $R_{ref}$, which is defined in Section 4.1. Note that since $\boldsymbol{R'_{PQ}}$ is from the reference sphere pointing to the center of the reference sphere, we have

$$R'_{PQ} = R_{ref} \tag{4.17}$$

The transverse aberration vector, labeled as $\boldsymbol{\varepsilon'}$, is represented by the vector pointing from $Q'$ to $Q'_r$, $\boldsymbol{Q'Q'_r}$, in the image plane coordinate system, written as

$$\boldsymbol{\varepsilon'} = \boldsymbol{Q'Q'_r} = \left(\varepsilon'_x, \varepsilon'_y\right) \tag{4.18}$$

It can be derived that wavefront aberrations and the transverse aberrations have the following exact relations [50]:

$$\varepsilon'_x = -\frac{R'_{PQr}}{n'}\frac{\partial W}{\partial \rho'_{lx}} \tag{4.19}$$

$$\varepsilon'_y = -\frac{R'_{PQr}}{n'}\frac{\partial W}{\partial \rho'_{ly}} \tag{4.20}$$

where $n'$ is the refractive index in the image space. Equations 4.19-4.20 are then approximated with $R_{XI}$ in place of $R'_{PQr}$. Also, $\rho'_{lx}$ and $\rho'_{ly}$ can also be replaced with $L'_p\rho_x$ and $L'_p\rho_y$ according to Equation 4.8. The resulting relations become

$$\varepsilon'_x \approx -\frac{R_{XI}}{n'L'_p}\frac{\partial W}{\partial \rho_x} \tag{4.21}$$

$$\varepsilon'_y \approx -\frac{R_{XI}}{n'L'_p}\frac{\partial W}{\partial \rho_y} \tag{4.22}$$

With the relations given in Equations 4.21-4.22, each wavefront aberration term is associated with a pair of transverse aberration terms, $\varepsilon'_x$ and $\varepsilon'_y$. Table 4.2 shows the transverse aberration pairs associated with the wavefront aberration terms in the third group.

**Table 4.2. The transverse aberration pairs associated with wavefront aberration terms in the third group under the $R_{XI}$ approximation**

| $W$ | $\varepsilon'_x/(-R_{XI}/n'L'_p)$ | $\varepsilon'_y/(-R_{XI}/n'L'_p)$ |
|---|---|---|
| $W_{02002}(\boldsymbol{i}\cdot\boldsymbol{\rho})^2$ | 0 | $2W_{02002}\,\rho_y$ |
| $W_{11011}(\boldsymbol{i}\cdot\boldsymbol{\rho})(\boldsymbol{i}\cdot\boldsymbol{H})$ | 0 | $W_{11011}\,H_y$ |
| $W_{20020}(\boldsymbol{i}\cdot\boldsymbol{H})^2$ | 0 | 0 |
| $W_{03001}(\boldsymbol{i}\cdot\boldsymbol{\rho})(\boldsymbol{\rho}\cdot\boldsymbol{\rho})$ | $2W_{03001}\,\rho_x\,\rho_y$ | $W_{03001}\,(\rho_x^2+3\rho_y^2)$ |
| $W_{12101}(\boldsymbol{i}\cdot\boldsymbol{\rho})(\boldsymbol{H}\cdot\boldsymbol{\rho})$ | $W_{12101}\,H_x\,\rho_y$ | $W_{12101}\,(H_x\,\rho_x+2H_y\,\rho_y)$ |
| $W_{12010}(\boldsymbol{i}\cdot\boldsymbol{H})(\boldsymbol{\rho}\cdot\boldsymbol{\rho})$ | $2W_{12010}\,H_y\,\rho_x$ | $2W_{12010}\,H_y\,\rho_y$ |
| $W_{21001}(\boldsymbol{i}\cdot\boldsymbol{\rho})(\boldsymbol{H}\cdot\boldsymbol{H})$ | 0 | $W_{21001}\,(H_x^2+H_y^2)$ |
| $W_{21110}(\boldsymbol{i}\cdot\boldsymbol{H})(\boldsymbol{H}\cdot\boldsymbol{\rho})$ | $W_{21110}\,H_x\,H_y$ | $W_{21110}\,H_y^2$ |
| $W_{30010}(\boldsymbol{i}\cdot\boldsymbol{H})(\boldsymbol{H}\cdot\boldsymbol{H})$ | 0 | 0 |
| $W_{04000}(\boldsymbol{\rho}\cdot\boldsymbol{\rho})^2$ | $2W_{04000}\,(\rho_x^2+\rho_y^2)\,\rho_x$ | $2W_{04000}\,(\rho_x^2+\rho_y^2)\,\rho_y$ |
| $W_{13100}(\boldsymbol{H}\cdot\boldsymbol{\rho})(\boldsymbol{\rho}\cdot\boldsymbol{\rho})$ | $W_{13100}\,(3H_x\,\rho_x^2+H_x\,\rho_y^2+2H_y\,\rho_x\,\rho_y)$ | $W_{13100}\,(3H_y\,\rho_y^2+H_y\,\rho_x^2+2H_x\,\rho_x\,\rho_y)$ |
| $W_{22200}(\boldsymbol{H}\cdot\boldsymbol{\rho})^2$ | $2W_{22200}\,(H_x\,\rho_x+H_y\,\rho_y)\,H_x$ | $2W_{22200}\,(H_x\,\rho_x+H_y\,\rho_y)\,H_y$ |
| $W_{22000}(\boldsymbol{H}\cdot\boldsymbol{H})(\boldsymbol{\rho}\cdot\boldsymbol{\rho})$ | $2W_{22000}\,(H_x^2+H_y^2)\,\rho_x$ | $2W_{22000}\,(H_x^2+H_y^2)\,\rho_y$ |
| $W_{31100}(\boldsymbol{H}\cdot\boldsymbol{H})(\boldsymbol{H}\cdot\boldsymbol{\rho})$ | $W_{31100}\,(H_x^2+H_y^2)\,H_x$ | $W_{31100}\,(H_x^2+H_y^2)\,H_y$ |
| $W_{40000}(\boldsymbol{H}\cdot\boldsymbol{H})^2$ | 0 | 0 |

The transverse aberration pairs are a way to expand transverse aberration similarly to how the wavefront aberration is expanded. It can be seen that the coefficients $W_{11011}$ and $W_{12010}$ correspond to the anamorphism and field-linear defocus discussed in Section 3.2.2.

The following part will illustrate that the approximation used in Equations 4.21-4.22 does not affect the values of the transverse aberration pairs in the third group, as shown in Table 4.2, when the system is free of any aberration in the second group. The elimination of second-group aberrations can be achieved by placing the image plane at the sagittal paraxial image location. To reach this conclusion, we focus on the order of aberration terms, which is the sum of the powers of $H_x$, $H_y$, $\rho_x$ and $\rho_y$. For example, the wavefront aberration term for $W_{02002}$ is second order, and for $W_{31100}$ is fourth order, both of which are in the third group. Note that $R_{XI}$ does not have any dependence on the field and pupil vectors. Therefore, as shown in Table 4.2, the transverse aberration pairs are always one order lower than the corresponding wavefront aberration term due to the derivative operation with respect to normalized pupil coordinates in Equations 4.21-4.22.

Firstly, $R'_{PQr}$ is compared with $R'_{ref}$. Connect three points, $P'$, $Q'$ and $Q'_r$, in Figure 4.3 to form a triangle $P'Q'Q'_r$ as shown in Figure 4.4.

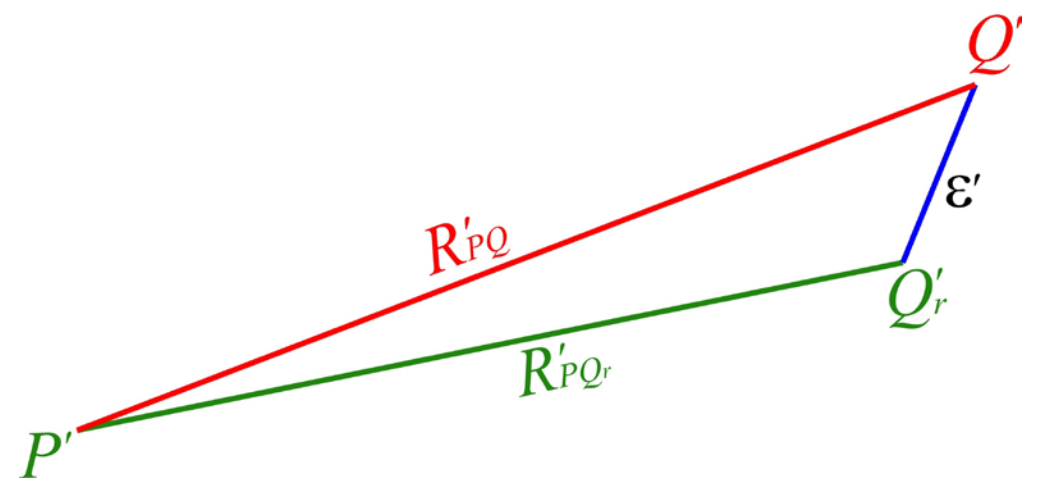

**Figure 4.4. The triangle defined by $P'$, $Q'$, and $Q'_r$. The magnitude of transverse aberration is exaggerated for clear illustration.**

The length of $Q'Q'_r$ is the magnitude of the transverse aberration, $\varepsilon'$, given as

$$\varepsilon' = |\boldsymbol{\varepsilon}'| = \sqrt{\varepsilon_x'^2 + \varepsilon_y'^2} \tag{4.23}$$

Note that the lowest order of $\varepsilon'$ is first order, linear to $H_y$ or $\rho_y$, as in $W_{02002}$ or $W_{11011}$ related terms. The length difference between $R'_{PQr}$ and $R'_{PQ}$ can be written as

$$R'_{PQr} - R'_{PQ} = R'_{PQr} - R'_{ref} = |\boldsymbol{R}'_{PQr}| - R'_{ref} = |\boldsymbol{R}'_{PQ} + \boldsymbol{\varepsilon}'| - R'_{ref} \tag{4.24}$$

Note that $R'_{PQ}$=$R'_{ref}$ from Equation 4.17 is used, so Equation 4.24 also represents the length difference between $R'_{PQr}$ and $R'_{ref}$. For now, we ignore the signs of lengths $R'_{PQr}$, $R'_{PQ}$, $R_{XI}$ and $R'_{ref}$, and treat them all as positive, because they are defined with the same sign, and the signs do not affect the order of related aberration terms. Further expand Equation 4.24 as

$$\begin{aligned} R'_{PQr} - R'_{ref} &= \left|\boldsymbol{R'_{PQ}} + \boldsymbol{\varepsilon'}\right| - R'_{ref} \\ &= \sqrt{\left|\boldsymbol{R'_{PQ}}\right|^2 + \left|\boldsymbol{\varepsilon'}\right|^2 + 2\boldsymbol{R'_{PQ}} \cdot \boldsymbol{\varepsilon'}} - R'_{ref} \\ &= \sqrt{R'^2_{ref} + \left|\boldsymbol{\varepsilon'}\right|^2 + 2\boldsymbol{R'_{PQ}} \cdot \boldsymbol{\varepsilon'}} - R'_{ref} \\ &\approx R'^2_{ref} + \frac{1}{2}\frac{\varepsilon'^2}{R'_{ref}} + \frac{\boldsymbol{R'_{PQ}} \cdot \boldsymbol{\varepsilon'}}{R'_{ref}} - R'_{ref} \\ &= \frac{1}{2}\frac{\varepsilon'^2}{R'_{ref}} + \frac{\boldsymbol{R_{P'Q'}} \cdot \boldsymbol{\varepsilon'}}{R'_{ref}} \end{aligned} \tag{4.25}$$

where $|\boldsymbol{R'_{PQ}}+\boldsymbol{\varepsilon'}|$ is Taylor expanded to include the first three terms with the lowest order. As the field and pupil vectors approach the paraxial regime, $\boldsymbol{R'_{PQ}}$ can be approximated to

$$\begin{aligned} \boldsymbol{R'_{PQ}} &= \boldsymbol{P'O_{XP}} + \boldsymbol{O_{XP}O_{IM}} + \boldsymbol{O_{IM}Q'} \\ &\approx -\boldsymbol{\rho'_l} + \boldsymbol{R_{XI}} + \boldsymbol{H'_l} \end{aligned} \tag{4.26}$$

where $\boldsymbol{R_{XI}}=\boldsymbol{O_{XP}O_{IM}}$ as defined previously, and $\boldsymbol{H'_l}=\boldsymbol{O_{IM}Q'}$ as defined in Section 4.1. In addition, $\boldsymbol{P'O_{XP}}$ is approximated as $-\boldsymbol{\rho'_l}$ as a first-order approximation, which can be seen from Figure 4.3. Equation 4.26 represents the low order terms in the expansion of $\boldsymbol{R'_{PQ}}$, and the lowest order term, $\boldsymbol{R_{XI}}$, is zero order. Substituting Equation 4.26, we have

$$\begin{aligned} R_{P'Q'_r} - R_{ref} &\approx \frac{1}{2}\frac{\varepsilon'^2}{R_{ref}} + \frac{\boldsymbol{R_{P'Q'}} \cdot \boldsymbol{\varepsilon'}}{R_{ref}} \\ &\approx \frac{1}{2}\frac{\varepsilon'^2}{R_{XI}} + \frac{\left(-\boldsymbol{\rho'_l} + \boldsymbol{R_{XI}} + \boldsymbol{H'_l}\right) \cdot \boldsymbol{\varepsilon'}}{R_{XI}} \\ &= \frac{1}{2}\frac{\varepsilon'^2}{R_{XI}} + \frac{\left(-\boldsymbol{\rho'_l} + \boldsymbol{H'_l}\right) \cdot \boldsymbol{\varepsilon'}}{R_{XI}} \end{aligned} \tag{4.27}$$

where $\boldsymbol{R_{XI}} \cdot \boldsymbol{\varepsilon'}=0$ is used since $\boldsymbol{R_{XI}}$ is along the OAR, which is normal to the image plane, and $R'_{ref}$ in the denominators is approximated to its zero-order contribution, $R_{XI}$. Because all approximations made retain low order terms, the result of Equation 4.27 represents the low order contribution to $R'_{PQr} - R'_{ref}$, and it is at least second order since $\varepsilon'$ is at least first order and $(\boldsymbol{H}'_l - \boldsymbol{\rho}'_l)$ is first order.

The length $R'_{ref}$ is then compared to $R_{XI}$. With Equation 4.26, the length difference between $R'_{ref}$ and $R_{XI}$ can be approximated to

$$\begin{aligned} R'_{ref} - R_{XI} &= R'_{PQ} - R_{XI} \\ &= \left|\boldsymbol{R}'_{PQ}\right| - R_{XI} \\ &\approx \left|-\boldsymbol{\rho}'_l + \boldsymbol{R}_{XI} + \boldsymbol{H}'_l\right| - R_{XI} \end{aligned} \tag{4.28}$$

Further expand Equation 4.28 as

$$\begin{aligned} R'_{ref} - R_{XI} &\approx \left|-\boldsymbol{\rho}'_l + \boldsymbol{R}_{XI} + \boldsymbol{H}'_l\right| - R_{XI} \\ &= \sqrt{R_{XI}^2 + \left|\boldsymbol{H}'_l - \boldsymbol{\rho}'_l\right|^2} - R_{XI} \\ &\approx \frac{\left|\boldsymbol{H}'_l - \boldsymbol{\rho}'_l\right|^2}{2R_{XI}} \end{aligned} \tag{4.29}$$

where $\boldsymbol{R}_{XI} \cdot (\boldsymbol{H}'_l - \boldsymbol{\rho}'_l) = 0$ is used. It can be seen from Equation 4.29 that the length difference between $R'_{ref}$ and $R_{XI}$ is at least second order, which together with Equation 4.27, indicates that the length difference between $R'_{PQr}$ and $R_{XI}$ is also at least second order. The length $R'_{PQr}$ can then be written as

$$R'_{PQr} = R_{XI} + \Delta R^{(2)} + \Delta R^{(>2)} \tag{4.30}$$

where $\Delta R^{(2)}$ is the second-order contribution to the difference, and $\Delta R^{(>2)}$ is the contribution higher than second order. Therefore, Equations 4.19-4.20 can be rewritten as

$$\varepsilon'_x = -\frac{R_{XI}}{n'}\frac{\partial W}{\partial \rho'_{lx}} - \frac{\Delta R^{(2)}}{n'}\frac{\partial W}{\partial \rho'_{lx}} - \frac{\Delta R^{(>2)}}{n'}\frac{\partial W}{\partial \rho'_{lx}} \tag{4.31}$$

$$\varepsilon'_y = -\frac{R_{XI}}{n'}\frac{\partial W}{\partial \rho'_{ly}} - \frac{\Delta R^{(2)}}{n'}\frac{\partial W}{\partial \rho'_{ly}} - \frac{\Delta R^{(>2)}}{n'}\frac{\partial W}{\partial \rho'_{ly}} \tag{4.32}$$

where the transverse aberration consists of three terms related to $R_{XI}$, $\Delta R^{(2)}$ and $\Delta R^{(>2)}$. The $R_{XI}$ related term is the approximation used in Equations 4.21-4.22. When $W$ is within the third group, the order of the $R_{XI}$ related term ranges from first order to third order, as shown in Table 4.2. In comparison, the effect of the $\Delta R^{(2)}$ related term is to add two more orders to the transverse aberration pairs, producing terms ranging from third order to fifth order. Similarly, the $\Delta R^{(>2)}$ part produces transverse aberration pairs higher than third order, which do not interfere with the pairs in the $R_{XI}$ related term. Therefore, to determine the effect of $\Delta R^{(2)}$, we need to further evaluate the third-order transverse aberration contribution, which can come from both $R_{XI}$ related and $\Delta R^{(2)}$ related terms.

The third-order transverse aberration pairs in Table 4.2 are rotationally symmetric, meaning if the field and pupil vectors are rotated by an angle, the transverse aberration vector will be rotated by the same angle. However, the only contribution to third-order transverse aberrations from the $\Delta R^{(2)}$ related term is through $W_{11011}$ and $W_{02002}$, which are only in the tangential direction. Therefore, the third-order contribution to the transverse aberration pairs from the $\Delta R^{(2)}$ related term does not interfere with that from the $R_{XI}$ related term. Since the third-group aberration terms are the focus of this work, the approximation in Equations 4.21-4.22 can be used without losing accuracy.

To further simplify Equations 4.21-4.22, the following expression can be derived from paraxial raytracing equations as

$$u'_{ax} = -\frac{R_{XI}}{L'_p} \tag{4.33}$$

where $u'_{ax}$ is the paraxial sagittal marginal ray angle in image space. Note that this marginal ray has a positive ray height at the exit pupil. Therefore, Equations 4.21-4.22 can be rewritten as

$$\varepsilon'_x \approx \frac{1}{n'u'_{ax}}\frac{\partial W}{\partial \rho_x} \tag{4.34}$$

$$\varepsilon'_y \approx \frac{1}{n'u'_{ax}}\frac{\partial W}{\partial \rho_y} \tag{4.35}$$

which, as discussed before, are exact for third-group wavefront aberrations and the corresponding transverse aberrations.

## 4.4 Wavefront aberrations defined at a general reference sphere location

Sections 4.1-4.3 revolve around wavefront aberrations defined at the exit pupil. In general, the reference sphere, where the wavefront aberrations are defined, can be chosen at a general location other than the exit pupil. Consider a general reference sphere shown in Figure 4.5. The reference sphere intersects the OAR at $O_{RI}$ with the distance between $O_{RI}$ and $O_{IM}$ as $R_{RI}$. A coordinate system, $X'_{pr}Y'_{pr}Z'_{pr}$, is set up at $O_{RI}$ with the $Z'_{pr}$-axis aligned with the OAR. This coordinate system is called the reference sphere coordinate system, and its $X'_{pr}Y'_{pr}$ plane will be referred to as the reference plane.

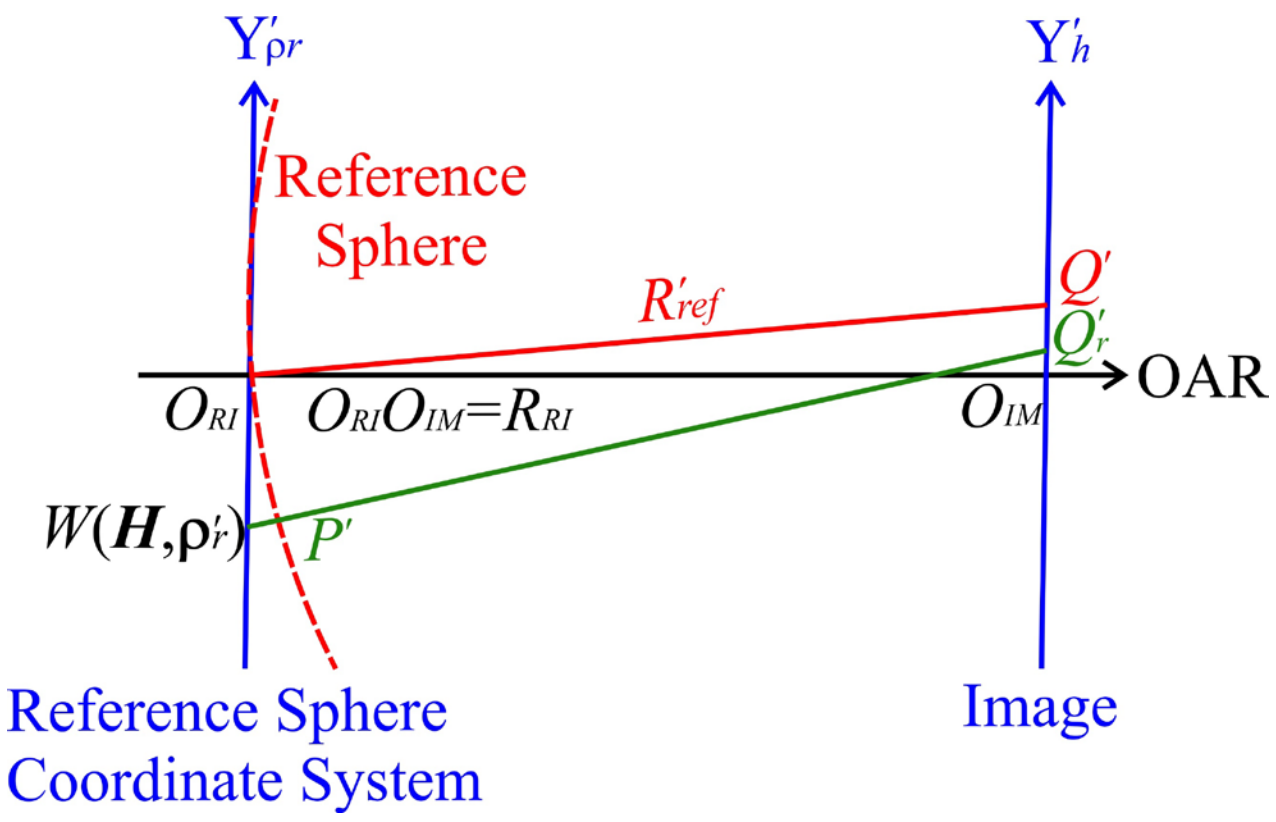

**Figure 4.5. Wavefront aberrations defined at a general reference sphere. A real ray (green line) in image space intersecting the real wavefront and image plane at $P'$ and $Q'_r$, respectively. $O_{RI}$ is the location of the reference sphere on the OAR in the image space. $O_{IM}$ is the center of the image plane. $Q'$ is the ideal image point.**

The intersection between the real ray $P'Q'_r$ and the reference plane can be expressed by a vector $\boldsymbol{\rho}'_r$ in the reference sphere coordinate system, given as

$$\boldsymbol{\rho}'_r = \left(\rho'_{rx}, \rho'_{ry}\right) \tag{4.36}$$

The wavefront aberrations can then be written as a function of $\boldsymbol{H}$ and $\boldsymbol{\rho}'_r$ as $W(\boldsymbol{H}, \boldsymbol{\rho}'_r)$ with $\boldsymbol{\rho}'_r$ serving a similar role as $\boldsymbol{\rho}'_I$. The function $W(\boldsymbol{H}, \boldsymbol{\rho}'_r)$ also can be expanded into similar aberration terms with vector product as shown in Table 4.1. The derivation in Section 4.3 remains valid: within the third group aberrations with $\boldsymbol{\rho}'_r$ in place of $\boldsymbol{\rho}'_I$, the relation can be written as

$$\varepsilon'_x \approx -\frac{R_{RI}}{n'}\frac{\partial W}{\partial \rho_{rx}} \tag{4.37}$$

$$\varepsilon'_y \approx -\frac{R_{RI}}{n'}\frac{\partial W}{\partial \rho_{ry}} \tag{4.38}$$

Because $\boldsymbol{\rho}'_r$ is not necessarily in the exit pupil plane, it cannot be expressed simply by the multiplication of the normalized pupil coordinate and pupil size. Although the exit pupil is normally chosen to be the location of the reference sphere and where the wavefront aberrations are measured, there are other locations with properties beneficial to deriving analytical expressions of wavefront aberrations. One such location is at the surface where approximated rays can be used to derive the optical path between reference spheres without introducing errors to the calculation of third-group aberrations, which is discussed in Section 5.2.

## 4.5 Aberration terms related to distortion in plane-symmetric imaging spectrometers

With the wavefront aberration expansion defined, we can identify the aberration terms related to distortion in plane-symmetric imaging spectrometers. Ideally, the spectrum of an imaging spectrometer can be placed in a spectrum plane with a coordinate system, $X_sY_sZ_s$, defined as shown in Figure 4.6(a). The spectral lines are set parallel to the $X_s$-axis. The $Y_s$ coordinates correspond to different wavelengths, and all spectral lines have their centers on the $Y_s$-axis. The ideal spectral lines are perfectly straight and have equal length for all wavelengths as shown in Figure 4.6(a).

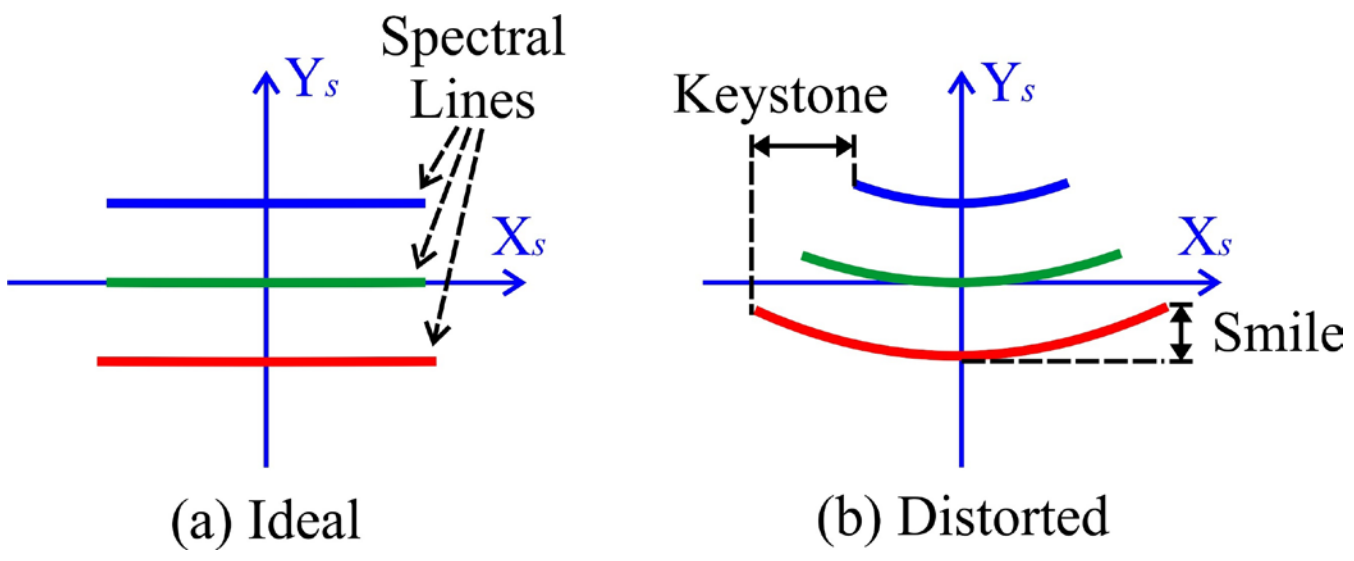

**Figure 4.6. Illustration of distortion of spectral lines on the spectrum plane ($X_sY_s$ plane): (a) Ideal spectrum image, and (b) distorted spectrum image.**

In general, spectral lines are distorted to some degree due to distortion aberrations of the imaging spectrometer. The two commonly used distortion specifications are keystone and smile distortions, which are illustrated in Figure 4.6(b). Keystone distortion measures the maximum $X_s$ displacement of the edge of a spectral line as it moves through all wavelengths, and smile distortion measures the maximum $Y_s$ displacement of points on spectral lines across all wavelengths.

Keystone and smile distortions are related to distortion in $X_s$ and $Y_s$ directions, respectively, but they only measure the maximum distortion of a spectrum. To better describe the distortion behavior, we define X and Y spectral distortions of a given image point on a spectral line based on the $X'_h$ and $Y'_h$ coordinates in its image plane coordinate system, $X'_hY'_hZ'_h$. Note that the coordinate system $X'_hY'_hZ'_h$ is defined related to the OAR of the corresponding wavelength as described in Chapter 2. The $X'_h$-axis is in the same direction as $X_s$-axis, and the $Y'_h$-axis is generally close to being parallel to $Y_s$-axis. Therefore, the X and Y spectral distortions defined in $X'_hY'_h$ plane are used to represent distortion in $X_s$ and $Y_s$ directions.

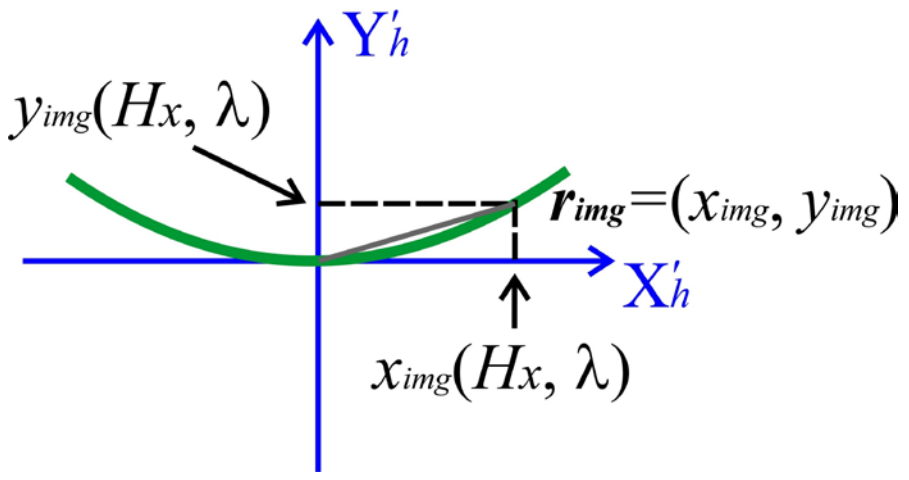

**Figure 4.7. $X'_h$ and $Y'_h$ coordinates of a point on a spectral line as functions of $H_x$ and $\lambda$.**

Figure 4.7 shows the image coordinate system of a wavelength, $\lambda$, and a vector $\boldsymbol{r_{img}}=(x_{img}, y_{img})$ as the real image location in the image coordinate system for the normalized field location $H_x$. Note that $H_y$ is assumed to be zero, because the slit is in the sagittal direction and has a size close to zero in the tangential direction. The real image location $\boldsymbol{r_{img}}$ is determined by the intersection between the real chief ray, the real ray that goes through the center of the exit pupil, and the image plane. Therefore, it is only affected by transverse aberration terms that do not become zero when the normalized pupil vector is zero, ($\rho_x$, $\rho_y$)=(0, 0). From Table 4.1, these aberration terms in the third group are $W_{21001}$, $W_{21110}$ and $W_{31100}$. Because $H_y$=0, the transverse distortion for a spectral line is only affected by $W_{21001}$ and $W_{31100}$ given as

$$\varepsilon'_{x\text{-}D}\left(H_x,\lambda\right)=\frac{1}{n'u'_{ax}}\left[W_{31100}\left(\lambda\right)\right]H_x^3 \tag{4.39}$$

$$\varepsilon'_{y\text{-}D}\left(H_x,\lambda\right)=\frac{1}{n'u'_{ax}}\left[W_{21001}\left(\lambda\right)\right]H_x^2 \tag{4.40}$$

where $\varepsilon'_{x\text{-}D}$ and $\varepsilon'_{y\text{-}D}$ are the transverse distortion aberrations in the third group that are functions of $H_x$ and $\lambda$. Note that the aberration coefficients $W_{21001}$ and $W_{31100}$ are functions of $\lambda$. Therefore, if we ignore the transverse distortion contribution from beyond the third group, the location on the image plane, $\boldsymbol{r_{img}}$, for a point on a spectral line, can be approximated by the sum of two vectors, $\boldsymbol{H'_l}$ and $\boldsymbol{\varepsilon'_D}$, given as

$$\boldsymbol{r_{img}}=\left(x_{img},y_{img}\right)\approx\boldsymbol{H'_l}+\boldsymbol{\varepsilon'_D} \tag{4.41}$$

where $\boldsymbol{\varepsilon'_D}$ is the vector describing transverse distortion in the third group, given as

$$\boldsymbol{\varepsilon'_D}=(\varepsilon'_{x-D},\varepsilon'_{y-D}) \tag{4.42}$$

and $\boldsymbol{H'_l}$ is the field vector defined in the image plane. Note that in this case, the $Y'_h$ component of $\boldsymbol{H'_l}$ is zero. An illustration of $\boldsymbol{r_{img}}$ is shown in Figure 4.8.

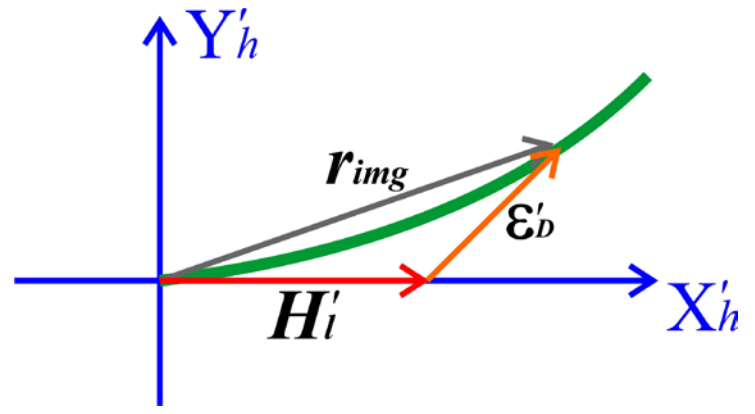

**Figure 4.8. The real chief ray location, $r_{img}$, expressed as a vector sum of the paraxial image location, $H'_l$, and the transverse distortion vector, $\varepsilon'_D$. This expression ignores the contribution of transverse distortion from beyond the third group.**

Therefore, we have

$$x_{img}\left(H_x,\lambda\right)\approx H'_{lx}\left(H_x,\lambda\right)+\varepsilon'_{x\text{-}D}\left(H_x,\lambda\right)=L'_h\left(\lambda\right)H_x+\frac{1}{n'u'_{ax}}\left[W_{31100}\left(\lambda\right)\right]H_x^3 \quad (4.43)$$

$$y_{img}\left(H_x,\lambda\right)\approx \varepsilon'_{y\text{-}D}\left(H_x,\lambda\right)=\frac{1}{n'u'_{ax}}\left[W_{21001}\left(\lambda\right)\right]H_x^2 \quad (4.44)$$

Note that $L'_h$ is also a function of wavelength because the magnification can be different for different wavelengths. Then, we define X spectral distortion, $SD_x$, for a point on a spectral line as the difference in $x_{img}$ between the real chief ray of its wavelength and that of a reference wavelength, $\lambda_{ref}$, given as

$$\begin{aligned}SD_x\left(H_x,\lambda\right)&=x_{img}\left(H_x,\lambda\right)-x_{img}\left(H_x,\lambda_{ref}\right)\\&\approx\left[L'_h\left(\lambda\right)-L'_h\left(\lambda_{ref}\right)\right]H_x+\frac{1}{n'u'_{ax}\left(\lambda\right)}\left[W_{31100}\left(\lambda\right)-W_{31100}\left(\lambda_{ref}\right)\right]H_x^3\end{aligned} \quad (4.45)$$

The Y spectral distortion, $SD_y$, can be defined as the difference in $y_{img}$ between the real chief ray of an $H_x$ position and that of $H_x$=0 position with the same wavelength, given as

$$\begin{aligned}SD_y\left(H_x,\lambda\right)&=y_{img}\left(H_x,\lambda\right)-y_{img}\left(0,\lambda\right)\\&\approx\frac{1}{n'u'_{ax}\left(\lambda\right)}\left[W_{21001}\left(\lambda\right)\right]H_x^2\end{aligned} \quad (4.46)$$

An illustration of $SD_x$ and $SD_y$ is shown in Figure 4.9.

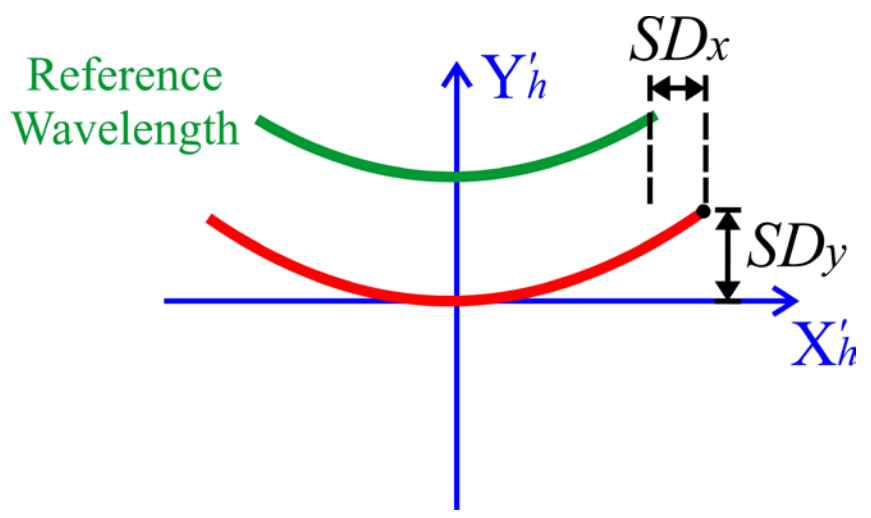

**Figure 4.9. Illustration of $SD_x$ and $SD_y$.**

Keystone and smile distortion can then be calculated from $SD_x$ and $SD_y$. Keystone distortion is the range of $SD_x$ over the whole spectrum, and smile distortion is the maximum $SD_y$ over all $H_x$ and $\lambda$.

Note that because each wavelength is treated independently with different OARs, the paraxial image plane for one wavelength may not coincide with that of a different wavelength. Therefore, strictly speaking, $\boldsymbol{r_{img}}$ is in different image planes for different wavelengths, but in practice, all distortions are measured on the same plane such as a detector plane. With the common detector plane for all wavelengths, $SD_x$ and $SD_y$ can be calculated with $\boldsymbol{r_{img}}$ defined in $X_sY_s$ plane instead of $X'_hY'_h$ plane. Changing the plane where $SD_x$ and $SD_y$ are measured can affect their values. Specifically, $SD_x$ and $SD_y$ can be affected by the defocus from moving the image plane away from the paraxial image location. In practice, the magnitude of this defocus is usually in microns, which does not have a significant impact on $SD_x$ and $SD_y$. On the other hand, $SD_y$ can also be affected by changing the tilt angle from being normal to the OAR, and the difference is approximately a cosine factor of the angle change. Therefore, if the OARs of different wavelengths do not have large angles between them, and the detector is placed close to the paraxial image planes of different wavelengths, $SD_y$ is not greatly affected by the tilt angles.

From Equations 4.45-4.46, it can be seen there are two contributions to keystone distortion. The first contribution is $[L'_h(\lambda)-L'_h(\lambda_{\mathrm{ref}})]H_x$, which is resulting from the difference in magnification for different wavelengths. The second contribution is from the distortion coefficient $W_{31100}$. For smile distortion, the only contribution within the third group is from the distortion coefficient $W_{21001}$. Equations 4.45-4.46 are used to calculate the distortion analytically for plane-symmetric spectrometers, but we still need to evaluate the aberration coefficients $W_{21001}$ and $W_{31100}$. In Chapters 5-6, the aberration coefficients are derived analytically as functions of system parameters.

# Chapter 5. Derivation of third-group aberration coefficients in one-surface systems

The general approach for deriving the aberration coefficients follows the model of geometric wavefront mentioned in Chapter 4. With a known input wavefront, the resulting output wavefront after going through an optical system is determined by the optical paths of the rays between wavefronts. This approach is illustrated in Figure 5.1.

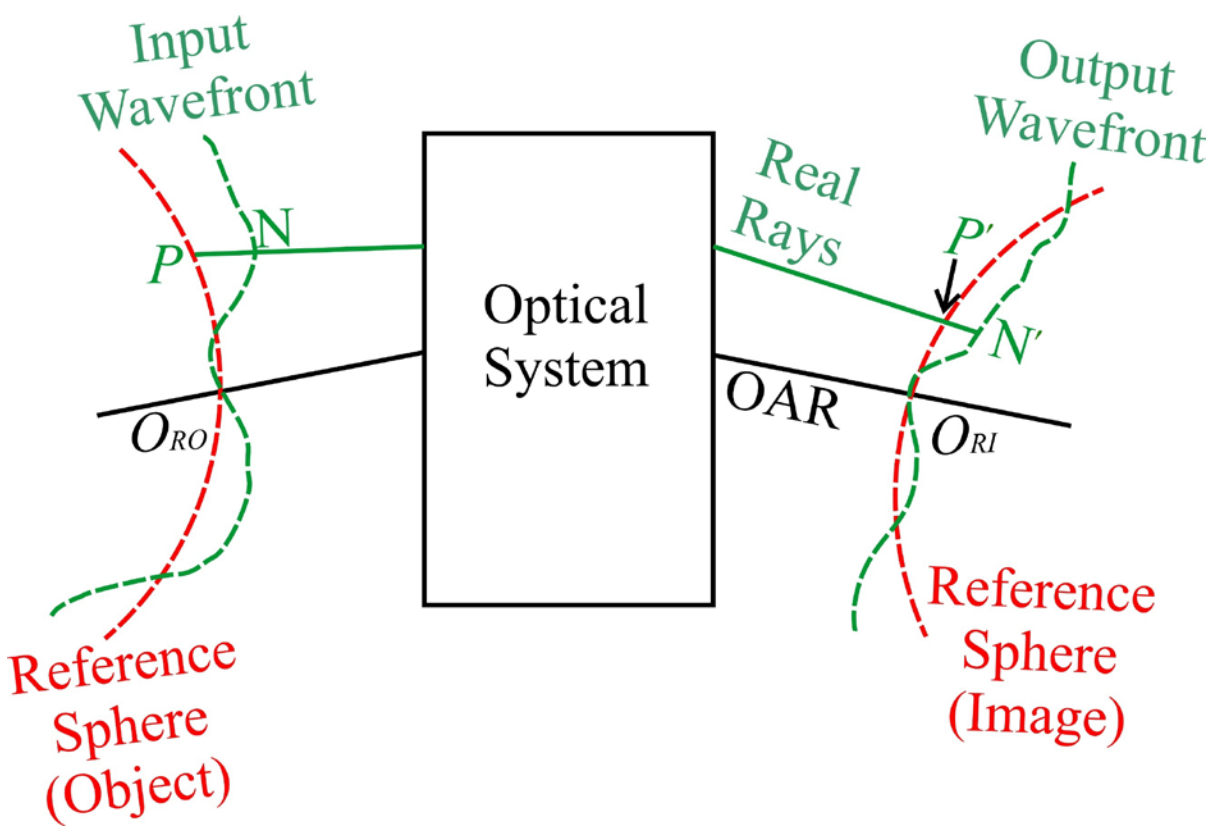

**Figure 5.1. A wavefront propagating through an optical system. In dashed green real wavefronts and in dashed red reference spheres. $O_{RO}$ and $O_{RI}$ are the locations of reference spheres on the OAR in object and image space, respectively. One real ray, represented by green solid lines, connects the input and output wavefronts at points $N$ and $N'$, and intersects the reference spheres at $P$ and $P'$.**

Consider a wavefront propagating through an optical system as shown in Figure 5.1. The two red dashed curves are the reference spheres centered at the object point and ideal image point, which intersect the OAR at $O_{RO}$ and $O_{RI}$, respectively. The dashed green lines denote the input and output wavefronts at the two reference spheres. The optical system is represented by a black box where light rays undergo reflections or refractions.

One real ray, represented by green solid lines, connects the input and output wavefronts at points $N$ and $N'$, and intersects the reference spheres at $P$ and $P'$, as shown in Figure 5.1.

It can be seen that the optical path between $P'$ and $N'$ is the wavefront aberration of the real ray. We use the notation, $P'N'_{(\mathrm{OP})}$, to denote the optical path from $P'$ to $N'$, which is positive if it is in the direction of ray propagation. $P'N'_{(\mathrm{OP})}$ can be expressed as

$$P'N'_{(\mathrm{OP})} = PN_{(\mathrm{OP})} + NN'_{(\mathrm{OP})} - PP'_{(\mathrm{OP})} \tag{5.1}$$

From the definition of wavefront aberrations in Chapter 4, it can be seen that

$$W_{PN} = PN_{(\mathrm{OP})} \tag{5.2}$$

$$W_{P'N'} = P'N'_{(\mathrm{OP})} \tag{5.3}$$

where $W_{PN}$ is the wavefront aberration of the ray $NN'$ in object space, and $W_{P'N'}$ is the corresponding wavefront aberration in image space. Therefore, Equation 5.3 can be rewritten as

$$W_{P'N'} = W_{PN} + NN'_{(\mathrm{OP})} - PP'_{(\mathrm{OP})} \tag{5.4}$$

Note that $NN'_{(\mathrm{OP})}$ is the optical path between the input and output wavefronts and is independent of pupil coordinates. Therefore, $NN'_{(\mathrm{OP})}$ does not contribute to transverse aberrations, and $PP'_{(\mathrm{OP})}$ effectively determines the contribution to wavefront aberrations during the propagation through the optical system.

If $PP'_{(\mathrm{OP})}$ can be expressed analytically, the output wavefront can then be calculated when the input wavefront is known. In the following sections, we focus on $PP'_{(\mathrm{OP})}$ and derive the wavefront aberration coefficients as functions of system parameters such as the radius of curvature of an optical surface, the OAR incident angle, paraxial ray angles and heights of marginal and chief rays, and freeform parameters (Zernike coefficients). The wavefront aberration contribution is divided into two parts. The first part is intrinsic aberration defined as the aberration contribution when the input wavefront is aberration-free. The second part is induced aberration defined as the additional aberration contribution due to an aberrated input wavefront. The derivation starts from one-surface systems, which can simplify the derivation process, and be easily generalized to multiple surfaces by treating the output wavefront of one surface as the input wavefront of the next surface.

## 5.1 Optical path in one-surface systems

Consider a one-surface system shown in Figure 5.2,

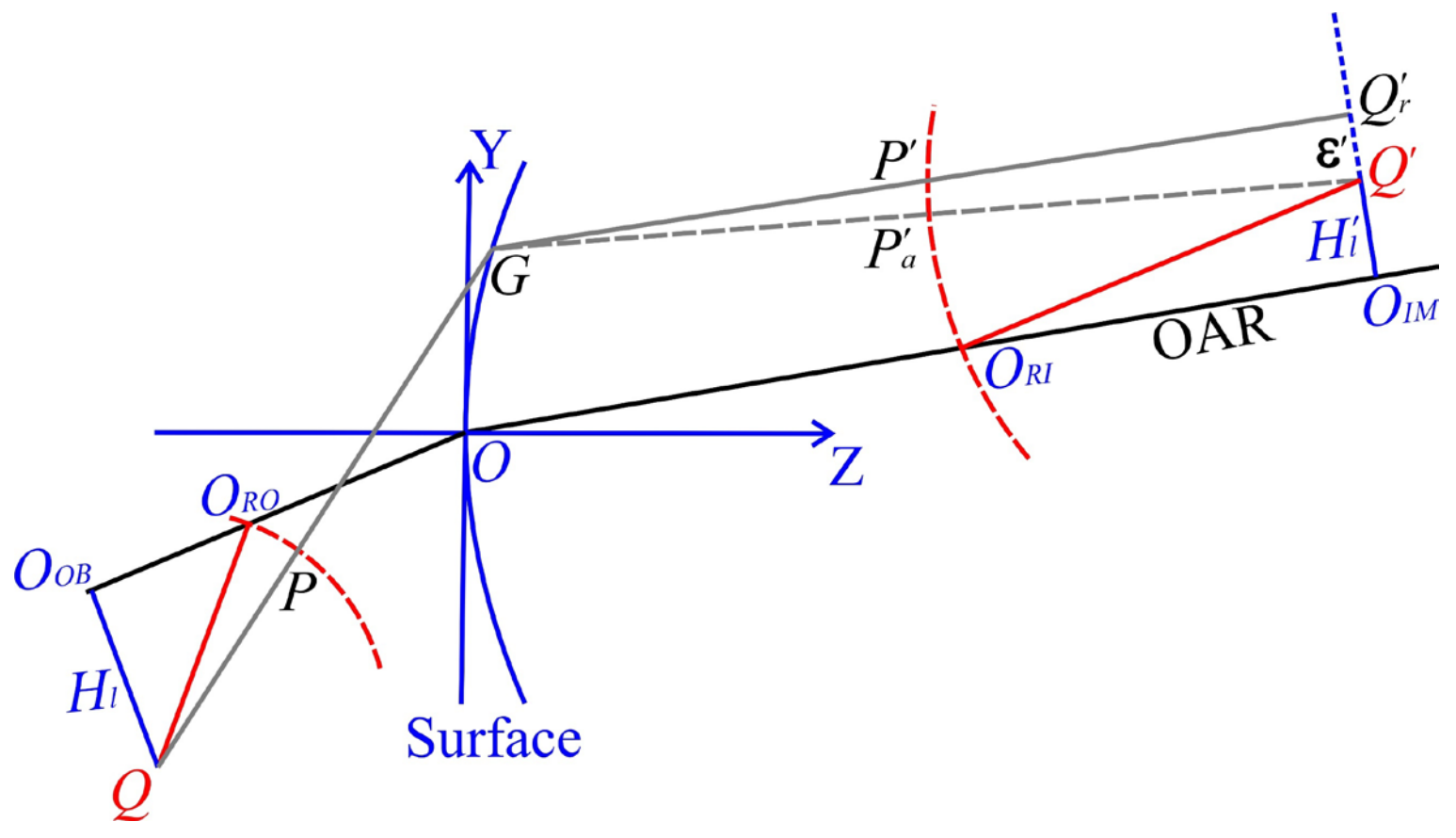

**Figure 5.2. A one-surface system with the object and image at *Q* and *Q′*, respectively, and the OAR in bold black. In solid light black a general real ray emitted from the object and intersecting the image plane at $Q'_r$. In dash light black an approximated ray that goes from *G* to *Q′*. In dash red reference spheres.**

where its local coordinate system is at the vertex of the surface, $O$, where the OAR passes through. An object point and the corresponding ideal image point are located at $Q$ and $Q'$ with the object and image heights, $H_l$ and $H'_l$, respectively. The object and image reference spheres are at $O_{RO}$ and $O_{RI}$ on the OAR, shown as red dashed curves. A real ray, shown in light black in Figure 5.2, emits from the object and intersects the surface at the point $G$. The real ray then hits the image plane at the point $Q'_r$ with the transverse aberration vector denoted by $\boldsymbol{\varepsilon}'$. Note that although the ray is drawn in the YZ plane in Figure 5.2, it is general and can have a ray angle in the sagittal direction. The real ray crosses the object and image reference spheres at points $P$ and $P'$, respectively.

As discussed previously, $PP'_{(OP)}$ is important in the determination of wavefront aberrations. To approach an analytical expression, we first look at the line that connects the points $G$ and $Q'$ and intersects the image reference sphere at $P'_a$, since it is easier to derive the length $GP'_a$ analytically than $GP'$. This approximation has been used in the derivation of Seidel aberration coefficients in rotationally symmetric systems [44]. The accuracy of Seidel coefficients is not affected by this approximation, because it has shown that the approximation only affects higher-order aberration terms [50]. This approximation is also used in the derivation of the third-group aberration coefficients by Sasian, which is mentioned as not accounting for the wavefront deformation during free space propagation [43]. However, for plane-symmetric systems, the length difference between $GP'$ and $GP'_a$ does introduce errors to all third-group coefficients. In the following part, we will investigate and evaluate the error introduced by this approximation.

From Figure 5.2, it is clear that the length difference between $GP'$ and $GP'_a$ depends on the transverse aberration vector $\boldsymbol{\varepsilon}'$, and the difference goes to zero when $\boldsymbol{\varepsilon}'=0$. To derive the length difference, set the following vectors:

$$\boldsymbol{l}' = \boldsymbol{GP}' \tag{5.5}$$

$$\boldsymbol{m'} = \boldsymbol{GQ'_r} \tag{5.6}$$

$$\boldsymbol{m'_a} = \boldsymbol{GQ'} \tag{5.7}$$

$$\boldsymbol{R'_{PQ}} = \boldsymbol{P'Q'} \tag{5.8}$$

Therefore, we have

$$\boldsymbol{R'_{PQ}} = \boldsymbol{m'} - \boldsymbol{l'} - \boldsymbol{\varepsilon'} \tag{5.9}$$

Because $\boldsymbol{m'}$ and $\boldsymbol{l'}$ are aligned, their relation can be expressed as

$$\boldsymbol{m'} - \boldsymbol{l'} = k\,\boldsymbol{m'} \tag{5.10}$$

where $k$ is a real number. Therefore,

$$\boldsymbol{R'_{PQ}} = k\,\boldsymbol{m'} - \boldsymbol{\varepsilon'} \tag{5.11}$$

Similar to how $\boldsymbol{R'_{PQ}}$ is defined in Section 4.3, the length of $\boldsymbol{R'_{PQ}}$ is $R'_{PQ}$ and is equal to $R'_{ref}$, the signed radius of curvature of the reference sphere in image space.

Calculate the length of both sides of Equation 5.11:

$$R'^2_{ref} = k^2\, m'^2 - 2k\left(\boldsymbol{m'}\cdot\boldsymbol{\varepsilon'}\right) + \varepsilon'^2 \tag{5.12}$$

where $m'$ and $\varepsilon'$ are the lengths of $\boldsymbol{m'}$ and $\boldsymbol{\varepsilon'}$, respectively. The length $m'$ is signed and positive when $\boldsymbol{m'}$ is pointing towards positive Z-axis, and $\varepsilon'$ is the length of the transverse aberration and always positive. From Equation 5.12, $k$ can be derived as

$$k = \frac{1}{m'^2}\left[\left(\boldsymbol{m'}\cdot\boldsymbol{\varepsilon'}\right) + m'R'_{ref}\sqrt{\frac{\left(\boldsymbol{m'}\cdot\boldsymbol{\varepsilon'}\right)^2 - m'^2\left(\varepsilon'^2 - R'^2_{ref}\right)}{m'^2 R'^2_{ref}}}\right] \tag{5.13}$$

where the signs of $k$, $m'$ and $R'_{ref}$ are taken into account. From Equation 5.13, we have

$$\boldsymbol{l}' = (1-k)\boldsymbol{m}' = \left[ 1 - \frac{(\boldsymbol{m}'\cdot\boldsymbol{\varepsilon}')}{m'^2} - \frac{R'_{ref}}{m'}\sqrt{\frac{(\boldsymbol{m}'\cdot\boldsymbol{\varepsilon}')^2 - m'^2\left(\varepsilon'^2 - R'^2_{ref}\right)}{m'^2 R'^2_{ref}}} \right]\boldsymbol{m}' \tag{5.14}$$

Therefore, the length of $\boldsymbol{l}'$ is

$$l' = (1-k)m' = \left[ m' - \frac{(\boldsymbol{m}'\cdot\boldsymbol{\varepsilon}')}{m'} - R'_{ref}\sqrt{\frac{(\boldsymbol{m}'\cdot\boldsymbol{\varepsilon}')^2 - m'^2\left(\varepsilon'^2 - R'^2_{ref}\right)}{m'^2 R'^2_{ref}}} \right] \tag{5.15}$$

Also, we have

$$\boldsymbol{m}' = \boldsymbol{m}'_a + \boldsymbol{\varepsilon}' \tag{5.16}$$

$$m' = m'_a\sqrt{1 + \frac{\varepsilon'^2}{m'^2_a} + \frac{2(\boldsymbol{m}'_a\cdot\boldsymbol{\varepsilon}')}{m'^2_a}} \tag{5.17}$$

where $m'$ and $m'_a$ are assumed to have the same sign. Since $\boldsymbol{m}'_a$ and $\boldsymbol{R}'_{ref}$ have no dependence on $\boldsymbol{\varepsilon}'$, Equations 5.15-5.17 describe the relationship between $l'$ and $\boldsymbol{\varepsilon}'$. The transverse aberration vector $\boldsymbol{\varepsilon}'$ is normally small in length compared with $\boldsymbol{m}'_a$. Therefore, $l'$ can be Taylor expanded and written as

$$\begin{aligned} l' &= l'\big|_{\boldsymbol{\varepsilon}'=0} \\ &+ \left.\frac{\partial l'}{\partial \varepsilon'_x}\right|_{\boldsymbol{\varepsilon}'=0}\varepsilon'_x + \left.\frac{\partial l'}{\partial \varepsilon'_y}\right|_{\boldsymbol{\varepsilon}'=0}\varepsilon'_y \\ &+ \frac{1}{2}\left.\frac{\partial^2 l'}{\partial \varepsilon'^2_x}\right|_{\boldsymbol{\varepsilon}'=0}\varepsilon'^2_x + \left.\frac{\partial^2 l'}{\partial \varepsilon'_x \partial \varepsilon'_y}\right|_{\boldsymbol{\varepsilon}'=0}\varepsilon'_x\varepsilon'_y + \frac{1}{2}\left.\frac{\partial^2 l'}{\partial \varepsilon'^2_y}\right|_{\boldsymbol{\varepsilon}'=0}\varepsilon'^2_y + \ldots \end{aligned} \tag{5.18}$$

Equation 5.18 expresses the $l'$ expansion within the image coordinate system so that $\varepsilon'_z$ is zero. It can be derived that

$$l'^{(0)} = l'\big|_{\boldsymbol{\varepsilon}'=0} = m'_a - R'_{ref} \tag{5.19}$$

$$l'^{(1)} = \left.\frac{\partial l'}{\partial \varepsilon'_x}\right|_{\boldsymbol{\varepsilon}'=0} \varepsilon'_x = \left.\frac{\partial l'}{\partial \varepsilon'_y}\right|_{\boldsymbol{\varepsilon}'=0} \varepsilon'_y = 0 \tag{5.20}$$

$$l'^{(2)} = \frac{1}{2}\left.\frac{\partial^2 l'}{\partial \varepsilon'^2_x}\right|_{\boldsymbol{\varepsilon}'=0} \varepsilon'^2_x + \left.\frac{\partial^2 l'}{\partial \varepsilon'_x \partial \varepsilon'_y}\right|_{\boldsymbol{\varepsilon}'=0} \varepsilon'_x \varepsilon'_y + \frac{1}{2}\left.\frac{\partial^2 l'}{\partial \varepsilon'^2_y}\right|_{\boldsymbol{\varepsilon}'=0} \varepsilon'^2_y = \frac{1}{2}\frac{m'_a - R'_{ref}}{m'^2_a R'_{ref}}\left|\boldsymbol{m}'_{\boldsymbol{a}} \times \boldsymbol{\varepsilon}'\right|^2 \tag{5.21}$$

where $l'^{(0)}$, $l'^{(1)}$, and $l'^{(2)}$ are the zero-order, first-order and second-order terms in the Taylor expansion of $l'$. The × operator in Equation 5.21 denotes the cross product between vectors. Note that although the expansion is done within the image coordinate system, the expansion results, Equation 5.19-5.21, do not depend on the selection of coordinate system, which is consistent with how the vector length is determined.

It can be seen that using the length $GP'_a$ as an approximation of $GP'$ is to use the term in Equation 5.19 that is independent of transverse aberrations and ignore other terms dependent on transverse aberrations. The main difference between $GP'$ and $GP'_a$ is the term in Equation 5.21, $l'^{(2)}$. Any terms beyond Equation 5.21 in the expansion do not contribute to wavefront aberrations in the third group.

Now let us investigate the order of $l'^{(2)}$ by looking at the order of $\boldsymbol{m'_a}$ and $\boldsymbol{R'_{ref}}$. Write $\boldsymbol{m'_a}$ and $\boldsymbol{R'_{ref}}$ as

$$\boldsymbol{m}'_{\boldsymbol{a}} = \boldsymbol{GQ}' = \boldsymbol{GO} + \boldsymbol{OO_{IM}} + \boldsymbol{O_{IM}Q}' \tag{5.22}$$

$$\boldsymbol{R}'_{\boldsymbol{ref}} = \boldsymbol{P'Q'} = \boldsymbol{P'O_{RI}} + \boldsymbol{O_{RI}O_{IM}} + \boldsymbol{O_{IM}Q}' \tag{5.23}$$

It can be seen that $\boldsymbol{GO}$, $\boldsymbol{O_{IM}Q'}$, $\boldsymbol{P'O_{RI}}$ and $\boldsymbol{O_{IM}Q'}$ have at least first-order dependence on normalized pupil and field vectors, and $\boldsymbol{OO_{IM}}$ and $\boldsymbol{O_{RI}O_{IM}}$ have no dependence on pupil and field vectors. Vectors $\boldsymbol{OO_{IM}}$ and $\boldsymbol{O_{RI}O_{IM}}$ are the image location on the OAR measured from the surface and the image reference sphere, respectively. Define $\boldsymbol{s'_0}$ and $\boldsymbol{s'_r}$ as

$$\boldsymbol{s}'_{\boldsymbol{0}} = \boldsymbol{OO}_{\boldsymbol{IM}} \tag{5.24}$$

$$\boldsymbol{s}'_{\boldsymbol{r}} = \boldsymbol{OO}_{\boldsymbol{RI}} \tag{5.25}$$

and

$$s'_0 = \left|\boldsymbol{s}'_{\boldsymbol{0}}\right| \operatorname{sign}\left(s'_0\right) \tag{5.26}$$

$$s'_r = \left|\boldsymbol{s}'_{\boldsymbol{r}}\right| \operatorname{sign}\left(s'_r\right) \tag{5.27}$$

where $s'_0$ and $s'_r$ are image and image reference sphere distances from the surface, the signs of which are positive when $\boldsymbol{s'_0}$ and $\boldsymbol{s'_r}$ are pointing towards the positive Z-axis of the local surface coordinate system. Therefore,

$$\boldsymbol{O}_{\boldsymbol{RI}}\boldsymbol{O}_{\boldsymbol{IM}} = \boldsymbol{s}'_{\boldsymbol{0}} - \boldsymbol{s}'_{\boldsymbol{r}} \tag{5.28}$$

Therefore, the low-order contribution to $\boldsymbol{m'_a}$ and $\boldsymbol{R'_{ref}}$ can be written as

$$\boldsymbol{m}'_{\boldsymbol{a}} = \boldsymbol{GO} + \boldsymbol{OO}_{\boldsymbol{IM}} + \boldsymbol{O}_{\boldsymbol{IM}}\boldsymbol{Q}' \approx \boldsymbol{OO}_{\boldsymbol{IM}} = \boldsymbol{s}'_{\boldsymbol{0}} \tag{5.29}$$

$$\boldsymbol{R}'_{\boldsymbol{ref}} = \boldsymbol{P'O}_{\boldsymbol{RI}} + \boldsymbol{O}_{\boldsymbol{RI}}\boldsymbol{O}_{\boldsymbol{IM}} + \boldsymbol{O}_{\boldsymbol{IM}}\boldsymbol{Q}' \approx \boldsymbol{O}_{\boldsymbol{RI}}\boldsymbol{O}_{\boldsymbol{IM}} = \boldsymbol{s}'_{\boldsymbol{0}} - \boldsymbol{s}'_{\boldsymbol{r}} \tag{5.30}$$

The lengths of $\boldsymbol{m'_a}$ and $\boldsymbol{R'_{ref}}$ can then be approximated to

$$m'_a \approx s'_0 \tag{5.31}$$

$$R'_{ref} \approx s'_0 - s'_r \tag{5.32}$$

Note that $\boldsymbol{s'_0}$ and $\boldsymbol{s'_r}$ are both along the OAR, so the length of $\boldsymbol{s'_0}-\boldsymbol{s'_r}$ is simply $s'_0-s'_r$.

Substituting Equations 5.29, 5.31 and 5.32 into Equation 5.21, we can find the low-order contribution of $l'^{(2)}$, labeled as $l'^{(2)}_{(low)}$, given as

$$l' \approx l'^{(2)}_{(low)} = \frac{1}{2}\frac{s'_r \varepsilon'^2}{\left(s'_0 - s'_r\right)} \tag{5.33}$$

In rotationally symmetric systems, $\varepsilon'$ is at least third order if the image is placed at the paraxial image location, making $l'^{(2)}_{(low)}$ at least sixth order. Therefore, the approximation of using $GP'_a$ does not affect Seidel aberrations (i.e., fourth-order in optical path difference) in rotationally symmetric systems, as is discussed by Welford [50]. In plane-symmetric systems where the order of transverse aberrations in the third group ranges from one to three (i.e., from second order to fourth order in optical path difference), all wavefront aberration terms in the third group can be impacted by $l'^{(2)}_{(low)}$. However, this approach can still be useful for plane-symmetric systems. In systems where transverse aberrations are not large, this approximation can still be used without introducing large error. In addition, it can be seen from Equation 5.33 that $l'^{(2)}_{(low)}$ becomes zero when $s'_r$ is zero, which means the image reference sphere is at the surface. In this case, the approximation does not introduce any error to wavefront aberrations in the third group, which is beneficial to deriving the aberration coefficient analytically. Therefore, we first focus on wavefront aberrations of one-surface systems with the image reference sphere at the surface, which serves as a foundation for analyzing the aberration behavior at the exit pupil in multi-surface systems.

## 5.2 Wavefront aberrations for one-surface systems with reference spheres at the surface

A one-surface system only has one optical surface between the object and image planes. Therefore, the incoming wavefront onto the optical surface is aberration-free or perfectly spherical. In this case, the aberration contribution in the image space is defined as intrinsic aberrations, the aberration contribution from an optical surface when the incoming light beams have no aberrations [51]. The intrinsic aberration is a major part of total aberration contribution in an optical system and is the main focus of this section. The aberration contribution caused by an aberrated incoming wavefront will be discussed in Chapter 6.

### 5.2.1 Analytical expression of wavefront aberrations for one-surface systems

Consider a one-surface system similar to Figure 5.2 with the object and image reference sphere locations at the surface as shown in Figure 5.3. The ray *QG* intersects the object reference sphere at *P*, and the approximated ray *GQ′* intersects the image reference sphere at $P'_a$. As discussed in Section 5.1, the approximated ray does not affect wavefront aberrations in the third group when the reference sphere is at the surface. In the following part, we calculate the ray path in the image space using the approximated ray.

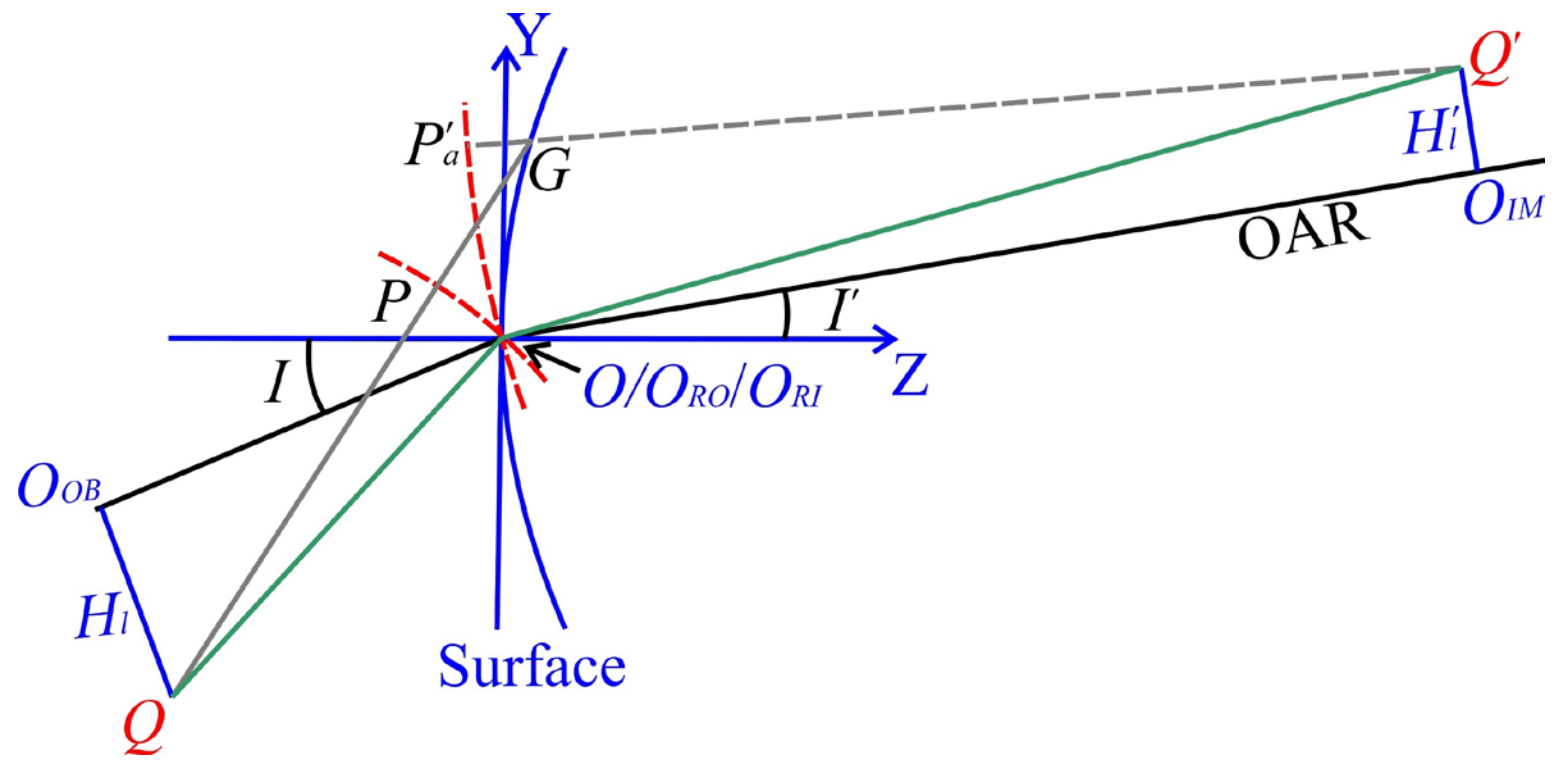

**Figure 5.3. A one-surface system with the object and image reference spheres at the surface. Points *Q* and *Q′* represent the object point and its corresponding ideal image point, respectively. The point *O*, the origin of the surface coordinate system, and the points $O_{RO}$ and $O_{RI}$, the locations of the reference spheres on the OAR in object and image space, coincide. In solid green an approximated ray connecting *Q*, *O* and *Q′*. In light black, solid in object space and dashed in image space, an approximated ray intersecting the surface and the reference spheres in object and image space at *G*, *P* and $P'_a$.**

It can be seen from Equation 5.4 that the wavefront aberration $W^{(S)}$ in this case can be expressed as

$$W^{(S)} = -PP'_{a(\mathrm{OP})} \tag{5.34}$$

where the ($S$) superscript denotes that the wavefront aberration is measured with the reference sphere at the surface. Recall that the ($OP$) subscript denotes the optical path between the two points, and in this case, $PP'_{a(OP)}$ corresponds to the optical path of the approximated ray between $P$ and $P'_a$. From Figure 5.3, $PP'_{a(OP)}$ is

$$PP'_{a(\mathrm{OP})} = n\left|GP\right| - n'\left|GP'_a\right| \tag{5.35}$$

Note that the optical path is negative when the ray is traced backwards. In addition, because

$$\left|PQ\right| = \left|OQ\right| \tag{5.36}$$

$$\left|P'_aQ'\right| = \left|OQ'\right| \tag{5.37}$$

the optical path $PP'_{a(OP)}$ can then be written as

$$\begin{aligned} PP'_{a(\mathrm{OP})} &= n\left|PG\right| + n\left(\left|PQ\right| - \left|OQ\right|\right) - n'\left|GP'_a\right| + n'\left(\left|P'_aQ'\right| - \left|OQ'\right|\right) \\ &= n\left(\left|GQ\right| - \left|OQ\right|\right) + n'\left(\left|GQ'\right| - \left|OQ'\right|\right) \end{aligned} \tag{5.38}$$

Therefore, $W^{(S)}$ can be written as

$$W^{(S)} = -PP'_{a(\mathrm{OP})} = n\left(\left|OQ\right| - \left|GQ\right|\right) + n'\left(\left|OQ'\right| - \left|GQ'\right|\right) \tag{5.39}$$

Define

$$\boldsymbol{s_h} = \boldsymbol{OQ} \tag{5.40}$$

$$\boldsymbol{s'_h} = \boldsymbol{OQ'} \tag{5.41}$$

$$\boldsymbol{s} = \boldsymbol{GQ} \tag{5.42}$$

$$\boldsymbol{s'} = \boldsymbol{GQ'} \tag{5.43}$$

$$\boldsymbol{s_0} = \boldsymbol{OO_{OB}} \tag{5.44}$$

$$\boldsymbol{s'_0} = \boldsymbol{OO_{IM}} \tag{5.45}$$

and

$$s_h = |\boldsymbol{s_h}|\operatorname{sign}(s_h) \tag{5.46}$$

$$s'_h = |\boldsymbol{s'_h}|\operatorname{sign}(s'_h) \tag{5.47}$$

$$s = |\boldsymbol{s}|\operatorname{sign}(s) \tag{5.48}$$

$$s' = |\boldsymbol{s'}|\operatorname{sign}(s') \tag{5.49}$$

$$s_0 = |\boldsymbol{s_0}|\operatorname{sign}(s_0) \tag{5.50}$$

$$s'_0 = |\boldsymbol{s'_0}|\operatorname{sign}(s'_0) \tag{5.51}$$

where the signs of $s_h$, $s'_h$, $s$, $s'$, $s_0$ and $s'_0$ are positive when the corresponding vectors are pointing towards the positive Z-axis of the local surface coordinate system. It can be seen that $s_0$ and $s'_0$ are the object and image distances for the on-OAR conjugates. The vectors, $\boldsymbol{s_0}$ and $\boldsymbol{s'_0}$, can be expressed in the local surface coordinate system as

$$\boldsymbol{s_0} = s_0(0, \sin I, \cos I) \tag{5.52}$$

$$\boldsymbol{s'_0} = s'_0(0, \sin I', \cos I') \tag{5.53}$$

With $s_h$, $s'_h$, $s$, and $s'$ defined in Equations 5.46-5.49, $W^{(S)}$ can be rewritten as

$$\begin{aligned} W^{(S)} &= n(|OQ| - |GQ|) + n'(|OQ'| - |GQ'|) \\ &= n'(s'_h - s') - n(s_h - s) \\ &= \Delta[n(s_h - s)] \end{aligned} \tag{5.54}$$

which indicates that $\boldsymbol{s_h}$ and $\boldsymbol{s}$ vectors are essential to derive the analytical form of $W^{(S)}$. Note that although Equation 5.54 is derived with the reference spheres defined at the surface as shown in Figure 5.3, it is applicable to any reference sphere placements.

To express $\boldsymbol{s_h}$ and $\boldsymbol{s}$ vectors as functions of pupil and field coordinates, firstly express the field vectors in the local surface coordinate system as

$$\boldsymbol{H_{ls}} = \left(H_{lx}, H_{ly}\cos\left(I-\theta_h\right), -H_{ly}\sin\left(I-\theta_h\right)\right) \tag{5.55}$$

$$\boldsymbol{H'_{ls}} = \left(H'_{lx}, H'_{ly}\cos\left(I'-\theta'_h\right), -H'_{ly}\sin\left(I'-\theta'_h\right)\right) \tag{5.56}$$

which results from the angle difference about the X-axis between the surface and image coordinate systems. Note that the object and image tilt angles, $\theta_h$ and $\theta'_h$, are also included in Equations 5.55-5.56. Although their values are set to zero when discussed in Section 3.2.2, they are still written to help evaluate the effect of tilt angles on wavefront aberrations.

Define a $\boldsymbol{g}$ vector pointing towards the intersection between the ray and the surface as

$$\boldsymbol{g} = \boldsymbol{OG} = \left(x, y, z\right) \tag{5.57}$$

Also define an $\boldsymbol{r}$ vector as the projection of $\boldsymbol{g}$ onto the XY plane of the local surface coordinate system as

$$\boldsymbol{r} = \left(x, y, 0\right) \tag{5.58}$$

Therefore, $\boldsymbol{s_h}$ and $\boldsymbol{s}$ vectors can be written as

$$\begin{aligned}\boldsymbol{s_h} &= \boldsymbol{s_0} + \boldsymbol{H_{ls}} \\ &= \left(H_{lx}, s_0\sin I + H_{ly}\cos\left(I-\theta_h\right), s_0\cos I - H_{ly}\sin\left(I-\theta_h\right)\right)\end{aligned} \tag{5.59}$$

$$\begin{aligned}\boldsymbol{s} &= \boldsymbol{s_h} - \boldsymbol{g} = \boldsymbol{s_0} + \boldsymbol{H_{ls}} - \boldsymbol{g} \\ &= \left(H_{lx} - x, s_0\sin I + H_{ly}\cos\left(I-\theta_h\right) - y, s_0\cos I - H_{ly}\sin\left(I-\theta_h\right) - z\right)\end{aligned} \tag{5.60}$$

Similarly, their corresponding vectors in image space, $\boldsymbol{s'_h}$ and $\boldsymbol{s'}$ are

$$\begin{aligned}\boldsymbol{s'_h} &= \boldsymbol{s'_0} + \boldsymbol{H'_{ls}} \\ &= \left(H'_{lx}, s'_0\sin I' + H'_{ly}\cos\left(I'-\theta'_h\right), s'_0\cos I' - H'_{ly}\sin\left(I'-\theta'_h\right)\right)\end{aligned} \tag{5.61}$$

$$\begin{aligned} \boldsymbol{s}' &= \boldsymbol{s}' - \boldsymbol{g} = \boldsymbol{s}_0' + \boldsymbol{H}_{ls}' - \boldsymbol{g} \\ &= \left(H_{lx}' - x, s_0' \sin I' + H_{ly}' \cos\left(I' - \theta_h'\right) - y, s_0' \cos I' - H_{ly}' \sin\left(I' - \theta_h'\right) - z\right) \end{aligned} \tag{5.62}$$

We can then calculate $s^2$ using Equation 5.60 as

$$\begin{aligned} s^2 &= \left(H_{lx} - x\right)^2 + \left[s_0 \sin I + H_{ly} \cos\left(I - \theta_h\right) - y\right]^2 + \left[s_0 \cos I - H_{ly} \sin\left(I - \theta_h\right) - z\right]^2 \\ &= s_h^2 + r^2 - 2\left(\boldsymbol{r} \cdot \boldsymbol{s_h}\right) + z^2 - 2z\left[s_0 \cos I - H_{ly} \sin\left(I - \theta_h\right)\right] \end{aligned} \tag{5.63}$$

As discussed in Section 2.2, $z$ can be written as

$$z = z_{sph} + z_{freeform} \tag{5.64}$$

Based on Equation 2.7, when $|r|<|R|$ which is normally satisfied, $z_{sph}$ can be Taylor expanded and approximated to

$$z_{sph} \approx \frac{r^2}{2R} + \frac{r^4}{8R^3} \tag{5.65}$$

Any terms beyond are higher order terms that do not affect wavefront aberrations in the third group. Similar to Equation 3.21, $z_{freefrom}$ can be expressed as

$$\begin{aligned} z_{freeform} &= F_{2-0}x^2 + F_{0-2}y^2 + F_{2-1}x^2 y + F_{0-3}y^3 \ldots \\ &= F_{2-0}r^2 + \left(F_{0-2} - F_{2-0}\right) y^2 + F_{2-1}x^2 y + F_{0-3}y^3 \ldots \\ &\approx F_{2-0}r^2 + z_{f2} \end{aligned} \tag{5.66}$$

where $F_{i\text{-}j}$ is the coefficient for term $x^i y^j$. The variable $z_{f2}$ represents the freeform departure beyond the term $F_{2\text{-}0}r^2$ that still contributes the wavefront aberrations in the third group. By analyzing the order of each term, $z_{f2}$ can be written as

$$z_{f2} = \left(F_{0-2} - F_{2-0}\right) y^2 + F_{2-1}x^2 y + F_{0-3}y^3 + F_{4-0}x^4 + F_{2-2}x^2 y^2 + F_{0-4}y^4 \tag{5.67}$$

The coordinates of the ray intersection at the surface, $x$ and $y$, are at least first order ($x$=$y$=0 when $\boldsymbol{\rho}$=0 and $\boldsymbol{H}$=0). Furthermore, it can be seen that the first-order terms of $x$ are only related to $\rho_x$ and $H_x$, and the first-order terms of $y$ are only related to $p_y$ and $H_y$. The information about the order of quantities can be valuable to determine which terms contribute to third-group aberrations.

Therefore, $z^2$ can be written as

$$\begin{aligned} z^2 &= \left(z_{sph} + z_{freeform}\right)^2 \\ &\approx \left(\frac{r^2}{2R} + \frac{r^4}{8R^3}\right)^2 + 2\left(\frac{r^2}{2R} + \frac{r^4}{8R^3}\right)\left(F_{2-0}r^2 + z_{f2}\right) + \left(F_{2-0}r^2 + z_{f2}\right)^2 \\ &= \frac{r^4}{4R^2} + \frac{r^6}{8R^4} + \frac{r^8}{64R^6} + F_{2-0}\frac{r^4}{R} + \frac{r^2}{R}z_{f2} + \frac{r^6}{4R^3}F_{2-0} + \frac{r^4}{4R^3}z_{f2} \\ &\quad + F_{2-0}^2 r^4 + 2F_{2-0}r^2 z_{f2} + z_{f2}^2 \end{aligned} \tag{5.68}$$

Because $r$ is at least first order ($r$=0 when $\boldsymbol{\rho}$=0 and $\boldsymbol{H}$=0), any terms with $r$ to a power higher than four are beyond the order of wavefront aberrations in the third group. With the first-order terms of $x$ only related to $\rho_x$ and $H_x$, and the first-order terms of $y$ only related to $p_y$ and $H_y$, it can be seen that $r^2 z_{f2}$, $r^4 z_{f2}$ and $z_{f2}^2$ do not contribute to the wavefront aberrations in the third group. Therefore, for the scope of this derivation, $z^2$ can be approximated to

$$z^2 \approx \frac{r^4}{4R^2} + F_{2-0}\frac{r^4}{R} + F_{2-0}^2 r^4 \tag{5.69}$$

From Section 3.2.3, we know that

$$\frac{1}{2R_{x-eff}} = \frac{1}{2R} + F_{2-0} \tag{5.70}$$

Combine Equations 5.69 and 5.70 to get

$$z^2 \approx \frac{r^4}{4R_{x-eff}^2} \tag{5.71}$$

Substitute Equations 5.64 and 5.71 into Equation 5.63, which results in

$$\begin{aligned}
s^2 &\approx s_h^2 + r^2 - 2(\boldsymbol{r}\cdot\boldsymbol{s_h}) + \frac{r^4}{4R_{x-eff}^2} \\
&\quad -2\left(\frac{r^2}{2R} + \frac{r^4}{8R^3} + F_{2-0}r^2 + z_{f2}\right)\left[s_0\cos I - H_{ly}\sin(I-\theta_h)\right] \\
&\approx s_h^2 + r^2 - 2(\boldsymbol{r}\cdot\boldsymbol{s_h}) + \frac{r^4}{4R_{x-eff}^2} \\
&\quad -2\left(\frac{r^2}{2R_{x-eff}} + \frac{r^4}{8R^3} + z_{f2}\right)s_0\cos I + \frac{r^2}{R_{x-eff}}H_{ly}\sin(I-\theta_h)
\end{aligned} \tag{5.72}$$

where terms that do not contribute to third-group wavefront aberrations (terms related to $r^4H_{ly}$ and $z_{f2}H_{ly}$) are omitted. Therefore, $s$ can be written as

$$\begin{aligned}
s \approx s_h\Bigg[&1 + \frac{r^2}{s_h^2} - 2\frac{(\boldsymbol{r}\cdot\boldsymbol{s_h})}{s_h^2} + \frac{r^4}{4s_h^2R_{x-eff}^2} \\
&-2\frac{s_0}{s_h^2}\left(\frac{r^2}{2R_{x-eff}} + \frac{r^4}{8R^3} + z_{f2}\right)\cos I + \frac{1}{s_h^2}\frac{r^2}{R_{x-eff}}H_{ly}\sin(I-\theta_h)\Bigg]^{1/2}
\end{aligned} \tag{5.73}$$

where $s$ is assumed to have the same sign as $s_h$. In general, $r$, $H_l$ and $z_{f2}$ are small compared to $s_0$ and $R_{x\text{-}eff}$, so Equation 5.73 can be Taylor expanded as

$$\begin{aligned}
s \approx s_h\Bigg[&1 + \frac{r^2}{2s_h^2} - \frac{(\boldsymbol{r}\cdot\boldsymbol{s_h})}{s_h^2} + \frac{r^4}{8s_h^2R_{x-eff}^2} \\
&-\frac{s_0}{s_h^2}\left(\frac{r^2}{2R_{x-eff}} + \frac{r^4}{8R^3} + z_{f2}\right)\cos I + \frac{1}{2s_h^2}\frac{r^2}{R_{x-eff}}H_{ly}\sin(I-\theta_h) \\
&-\frac{1}{4}\left(\frac{r^2}{s_h^2} - 2\frac{(\boldsymbol{r}\cdot\boldsymbol{s_h})}{s_h^2} - \frac{s_0}{s_h^2}\frac{r^2}{R_{x-eff}}\cos I\right)^2\Bigg]
\end{aligned} \tag{5.74}$$

where higher order terms are omitted. The expression $s_h - s$ can then be approximated as

$$
\begin{aligned}
s_h - s \approx & -\frac{r^2}{2s_h} + \frac{(\boldsymbol{r}\cdot\boldsymbol{s_h})}{s_h} - \frac{r^4}{8s_h R_{x-eff}^2} \\
& + \frac{s_0}{s_h}\left(\frac{r^2}{2R_{x-eff}} + \frac{r^4}{8R^3} + z_{f2}\right)\cos I - \frac{1}{2s_h}\frac{r^2}{R_{x-eff}} H_{ly}\sin\left(I-\theta_h\right) \\
& + \frac{1}{4}\frac{1}{s_h^3}\left(r^2 - 2(\boldsymbol{r}\cdot\boldsymbol{s_h}) - \frac{s_0 r^2}{R_{x-eff}}\cos I\right)^2
\end{aligned}
\tag{5.75}
$$

Therefore, the wavefront aberration in Equation 5.54 can be approximated to

$$
\begin{aligned}
W^{(S)} &= \Delta\left[n\left(s_h - s\right)\right] \\
&\approx \Delta\left[-\frac{nr^2}{2s_h} + \frac{n(\boldsymbol{r}\cdot\boldsymbol{s_h})}{s_h} - \frac{nr^4}{8s_h R_{x-eff}^2} + \frac{ns_0}{s_h}\left(\frac{r^2}{2R_{x-eff}} + \frac{r^4}{8R^3} + z_{f2}\right)\cos I\right. \\
&\quad \left. -\frac{n}{2s_h}\frac{r^2}{R_{x-eff}} H_{ly}\sin\left(I-\theta_h\right) + \frac{n}{8s_h^3}\left(r^2 - 2(\boldsymbol{r}\cdot\boldsymbol{s_h}) - \frac{s_0 r^2}{R_{x-eff}}\cos I\right)^2\right]
\end{aligned}
\tag{5.76}
$$

When the image plane is placed at the sagittal paraxial image plane, it can be seen from the sagittal Coddington equation (Equation 3.18) that

$$
\Delta\left(\frac{n}{s_0} - \frac{n\cos I}{R_{x-eff}}\right) = 0 \tag{5.77}
$$

The wavefront aberration function $W^{(S)}$ can then be written as

$$\begin{aligned} W^{(S)} &= \Delta\left[n\left(s-s_h\right)\right] \\ &\approx \underbrace{\Delta\left[\frac{n\left(\boldsymbol{r}\cdot\boldsymbol{s_h}\right)}{s_h}\right]}_{Term\,1} \underbrace{-\frac{1}{2}\left(\frac{n}{s_0}-\frac{n\cos I}{R_{x-eff}}\right)\Delta\left(\frac{s_0}{s_h}\right)r^2}_{Term\,2} \underbrace{+\frac{1}{2}\Delta\left[\frac{n\left(\boldsymbol{r}\cdot\boldsymbol{s_h}\right)^2}{s_h^3}\right]}_{Term\,3} \\ &\underbrace{-\frac{1}{2}\left(\frac{n}{s_0}-\frac{n\cos I}{R_{x-eff}}\right)\Delta\left[\frac{s_0\left(\boldsymbol{r}\cdot\boldsymbol{s_h}\right)}{s_h^3}\right]r^2}_{Term\,4} \underbrace{-\frac{1}{2}\Delta\left[\frac{nH_{ly}}{s_h}\sin\left(I-\theta_h\right)\right]\frac{r^2}{R_{x-eff}}}_{Term\,5} \\ &\underbrace{+\frac{1}{8}\left(\frac{n}{s_0}-\frac{n\cos I}{R_{x-eff}}\right)^2\Delta\left(\frac{s_0^2}{ns_h^3}\right)r^4}_{Term\,6} \underbrace{-\frac{1}{8}\left(\frac{n}{s_0}-\frac{n\cos I}{R_{x-eff}}\right)\Delta\left(\frac{s_0}{s_h}\right)\frac{r^4}{R_{x-eff}^2}}_{Term\,7} \\ &\underbrace{+\left[\left(-F_{2-0}^3-\frac{3}{2}\frac{F_{2-0}^2}{R}-\frac{3}{4}\frac{F_{2-0}}{R^2}\right)r^4+z_{f2}\right]\Delta\left(\frac{ns_0\cos I}{s_h}\right)}_{Term\,8} \end{aligned} \tag{5.78}$$

where $W^{(S)}$ is expressed as the sum of eight terms labeled in Equation 5.78. In Terms 6 and 7, $r^4$ is at least fourth order. Therefore, for the third-group contribution, $s_h$ can be replaced with its zero-order part, $s_0$, which makes Term 7 zero, meaning there is no third group contribution from Term 7. Term 8 represents the freeform contribution to the wavefront aberrations. Similar to Terms 6 and 7, the $s_h$ in Term 8 can be replaced with $s_0$ when only the third-group contribution is considered. In addition, if $F_{2\text{-}0}$ is set to zero to avoid optical power degeneracy [52], Term 8 is reduced to

$$Term\ 8\left(F_{2-0}=0\right)=z_{f2}\Delta\left(n\cos I\right) \tag{5.79}$$

Equation 5.79 can also be used as an approximation when $F_{2\text{-}0}$ is small enough that

$$\left|F_{2-0}^3+\frac{3}{2}\frac{F_{2-0}^2}{R}+\frac{3}{4}\frac{F_{2-0}}{R^2}\right|\ is\ much\ smaller\ than\ \left|F_{4-0}\right|,\ \left|F_{2-2}\right|,\ and\left|F_{0-4}\right| \tag{5.80}$$

In this work, we mainly focus on Equation 5.79 as the freeform contribution to wavefront aberrations. Therefore, $W^{(S)}$ can be written as

$$W^{(S)} \approx \Delta\left[\frac{n\left(\boldsymbol{r}\cdot\boldsymbol{s_h}\right)}{s_h}\right]-\frac{1}{2}\left(\frac{n}{s_0}-\frac{n\cos I}{R_{x-eff}}\right)\Delta\left(\frac{s_0}{s_h}\right)r^2+\frac{1}{2}\Delta\left[\frac{n\left(\boldsymbol{r}\cdot\boldsymbol{s_h}\right)^2}{s_h^3}\right]$$
$$-\frac{1}{2}\left(\frac{n}{s_0}-\frac{n\cos I}{R_{x-eff}}\right)\Delta\left[\frac{s_0\left(\boldsymbol{r}\cdot\boldsymbol{s_h}\right)}{s_h^3}\right]r^2-\frac{1}{2}\Delta\left[\frac{nH_{ly}}{s_h}\sin\left(I-\theta_h\right)\right]\frac{r^2}{R_{x-eff}} \quad (5.81)$$
$$+\frac{1}{8}\left(\frac{n}{s_0}-\frac{n\cos I}{R_{x-eff}}\right)^2\Delta\left(\frac{1}{ns_0}\right)r^4+z_{f2}\Delta\left(n\cos I\right)$$

which serves as the foundation to derive the aberration coefficients.

### 5.2.2 Ray position at the surface as a function of field and pupil coordinates

To express $W$ as a function of field and exit pupil coordinates, a key process is to relate the intersection between the ray and the optical surface to the field and exit pupil coordinates. Consider the same one-surface system discussed in Section 5.2.1 with a general exit pupil location as shown in Figure 5.4. The exit pupil intersects the OAR at $O_{XP}$, and the real ray $\boldsymbol{GQ'_r}$ that intersects the image plane at $Q'_r$, the exit pupil plane at point $E'$, and the XY plane of the image reference sphere coordinate system at point $F'(\rho'_{rx}, \rho'_{ry})$.

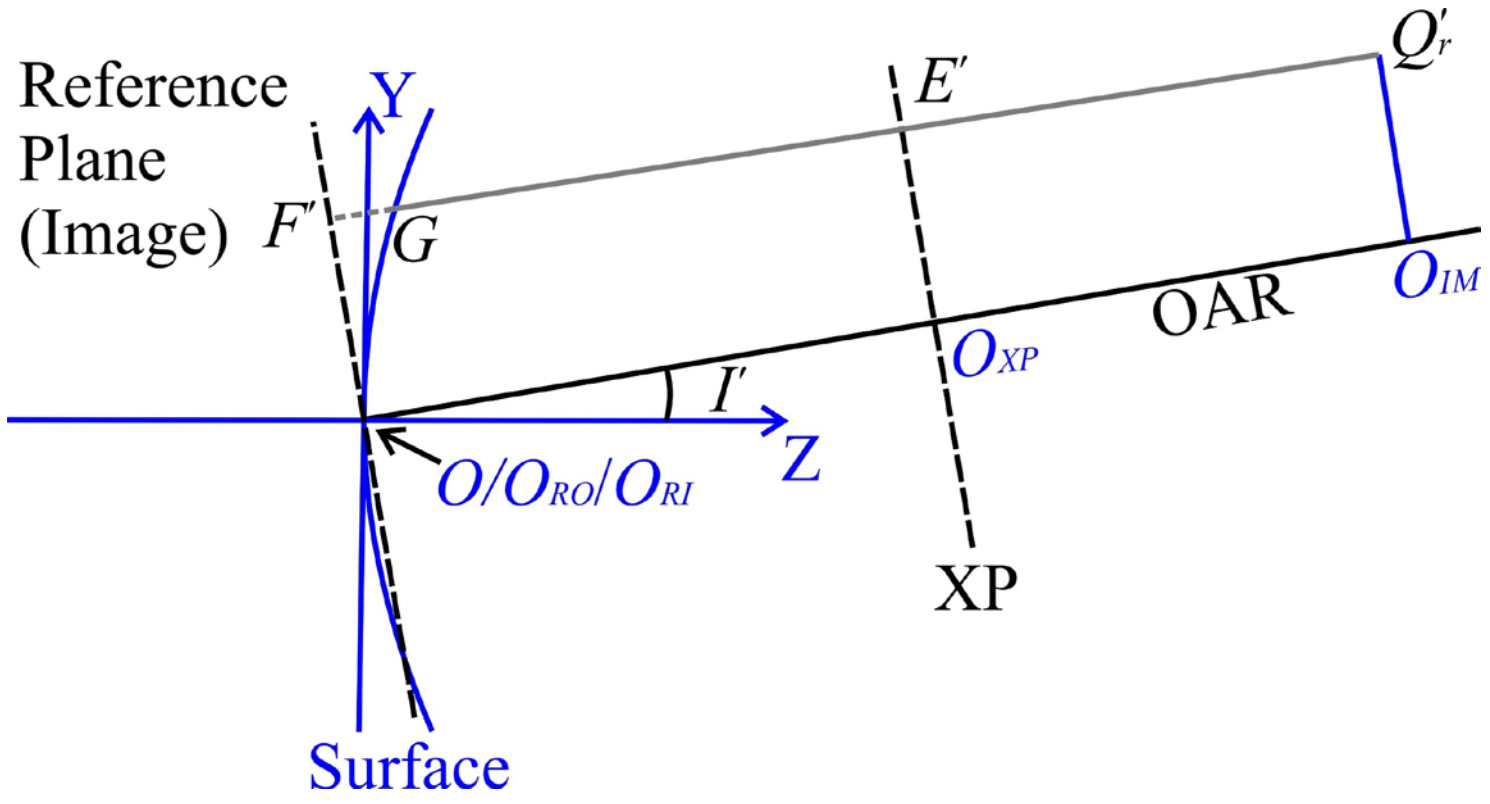

**Figure 5.4. A one-surface system with the real ray $GQ'_r$ in light black intersecting the image plane at $Q'_r$, the exit pupil plane at point $E'$, and the XY plane of the image reference sphere coordinate system at point $F'(\rho'_{rx}, \rho'_{ry})$.**

Define the exit pupil location vector $\boldsymbol{s'_p}$ as

$$\boldsymbol{s}'_p = \boldsymbol{OO}_{XP} \tag{5.82}$$

$$s'_p = \left|\boldsymbol{s}'_p\right| sign\left(\boldsymbol{s}'_p\right) \tag{5.83}$$

where the sign of $s'_p$ is positive if $\boldsymbol{s'_p}$ is pointing towards the positive Z-axis of the local surface coordinate system. Note that a similar quantity can be defined for the entrance pupil location as $\boldsymbol{s_p}$.

Similar to $\boldsymbol{H'_{ls}}$ defined in Section 5.2.1, $Q'_r$ can be expressed in the local surface coordinate system as

$$\begin{aligned} Q'_r &= \left(Q'_{rx}, Q'_{ry}, Q'_{rz}\right) \\ &= \left[H'_{lx} + \varepsilon'_x,\ s'_0 \sin I' + \left(H'_{ly} + \varepsilon'_y\right)\cos\left(I' - \theta'_h\right),\ s'_0 \cos I' - \left(H'_{ly} + \varepsilon'_y\right)\sin\left(I' - \theta'_h\right)\right] \end{aligned} \tag{5.84}$$

which follows a similar form as Equation 5.56. The coordinates of $E'$ in the local surface coordinate system can be expressed as

$$\begin{aligned} E' &= \left(E'_x, E'_y, E'_z\right) \\ &= \left(\rho'_{lx}, s'_p \sin I' + \rho'_{ly} \cos I', s'_p \cos I' - \rho'_{ly} \sin I'\right) \end{aligned} \tag{5.85}$$

where $\rho'_{lx}$ and $\rho'_{ly}$ are the exit pupil coordinates of $E'$ in the exit pupil coordinate system. Note that the transverse aberrations, $\varepsilon'_x$ and $\varepsilon'_y$, are included in the position of $Q'_r$, and for the approximated ray used in Section 5.2.1, $\varepsilon'_x$ and $\varepsilon'_y$ can be set to zero. With $G(x, y, z)$, the following equation needs to be satisfied to ensure the three points are on the same line, given as

$$\frac{E'_x - x}{E'_x - Q'_{rx}} = \frac{E'_y - y}{E'_y - Q'_{ry}} = \frac{E'_z - z}{E'_z - Q'_{rz}} \tag{5.86}$$

Substituting Equations 5.84-5.85 into Equation 5.86, we can express $x$ and $y$ as

$$x = \rho'_{lx} - \frac{\left(s'_p \cos I' - \rho'_{ly} \sin I' - z\right)\left(\rho'_{lx} - H'_{lx} - \varepsilon'_x\right)}{s'_p \cos I' - \rho'_{ly} \sin I' - s'_0 \cos I' + \left(H'_{ly} + \varepsilon'_y\right)\sin\left(I' - \theta'_h\right)} \tag{5.87}$$

$$\begin{aligned} y &= s'_p \sin I' + \rho'_{ly} \cos I' \\ &- \frac{\left(s'_p \cos I' - \rho'_{ly} \sin I' - z\right)\left[s'_p \sin I' + \rho'_{ly} \cos I' - s'_0 \sin I' - \left(H'_{ly} + \varepsilon'_y\right)\cos\left(I' - \theta'_h\right)\right]}{s'_p \cos I' - \rho'_{ly} \sin I' - s'_0 \cos I' + \left(H'_{ly} + \varepsilon'_y\right)\sin\left(I' - \theta'_h\right)} \end{aligned} \tag{5.88}$$

The sag function $z$ is at least second order, $\varepsilon'_x$ is at least second order, and $\varepsilon'_y$ is at least first order. By dropping higher-order terms in Equations 5.87-5.88. It can be derived that the first-order part of $x$ and $y$, denoted by $x^{(1)}$ and $y^{(1)}$, are

$$x^{(1)} = -\frac{s'_0}{s'_p - s'_0}\rho'_{lx} + \frac{s'_p}{s'_p - s'_0}H'_{lx} \tag{5.89}$$

$$y^{(1)} = -\frac{1}{\cos I'}\frac{s'_0}{s'_p - s'_0}\rho'_{ly} + \frac{\cos\theta'_h}{\cos I'}\frac{s'_p}{s'_p - s'_0}\left(H'_{ly} + \varepsilon'^{(1)}_y\right) \tag{5.90}$$

where $\varepsilon'^{(1)}_y$ is the first-order transverse aberration in the tangential direction, which comes from $W_{02002}$ and $W_{11011}$ terms in the wavefront aberration expansion, given as

$$\varepsilon'^{(1)}_y = \frac{1}{\cos\theta'_h n' u'_{ax}}\left(2W^{(XP)}_{02002}\rho_y + W^{(XP)}_{11011}H_y\right) \tag{5.91}$$

Equation 5.91 is derived using the relation between transverse and wavefront aberrations in Equation 4.35. The ($XP$) superscript denotes that the wavefront aberration coefficient is measured with the reference sphere at the exit pupil as discussed in Chapter 3. The factor $\cos\theta'_h$ accounts for the effect of a titled image plane.

For the second-order parts of $x$ and $y$, we focus on terms that have $\rho'^2_{lx}$, $\rho'_{lx}H'_{lx}$, or $H'^2_{lx}$ dependencies, which when described with vector products, contain terms with one vector product. Other dependencies need the multiplication of at least two vector products to be described. For example, $\rho'^2_{lx}$ can be written as

$$\rho'^2_{lx} = L'^2_p \rho_x^2 = L'^2_p \left[ (\boldsymbol{\rho} \cdot \boldsymbol{\rho}) - (\boldsymbol{i} \cdot \boldsymbol{\rho})^2 \right] \tag{5.92}$$

where $(\boldsymbol{\rho} \cdot \boldsymbol{\rho})$ can contribute to third-group wavefront aberrations, while $\rho'^2_{ly}$ is

$$\rho'^2_{ly} = L'^2_p \rho_y^2 = L'^2_p (\boldsymbol{i} \cdot \boldsymbol{\rho})^2 \tag{5.93}$$

which needs the multiplication of two vector products to be described. When $x$ and $y$ need two vector products to be described, by analyzing the resulting terms in Equation 5.81, it can be seen that they only lead to contributions to wavefront aberrations in the fourth or higher groups.

By expanding Equations 5.87-5.88, it can be derived that $x$ does not have terms that are related to $\rho'^2_{lx}$, $\rho'_{lx}H'_{lx}$, or $H'^2_{lx}$, and the second-order part of $y$ with $\rho'^2_{lx}$, $\rho'_{lx}H'_{lx}$, or $H'^2_{lx}$ dependencies, denoted by $y^{(2)}_{W3}$, can be derived as

$$y^{(2)}_{W3} = z^{(2)} \tan I' + \frac{\cos \theta'_h}{\cos I'} \frac{s'_p}{s'_p - s'_0} \varepsilon'^{(2)}_{yW3} \tag{5.94}$$

where $z^{(2)}$ is the part of the surface sag function that has dependencies on $\rho'^2_{lx}$, $\rho'_{lx}H'_{lx}$, or $H'^2_{lx}$, given as

$$\begin{aligned} z^{(2)} &= \frac{x^{(1)2}}{2R_{x-eff}} \\ &= \frac{1}{2\left(s'_p - s'_0\right)^2 R_{x-eff}} \left[ s'^2_0 \left(\boldsymbol{\rho}'_l \cdot \boldsymbol{\rho}'_l\right) - 2 s'_0 s'_p \left(\boldsymbol{H}'_l \cdot \boldsymbol{\rho}'_l\right) + s'^2_p \left(\boldsymbol{H}'_l \cdot \boldsymbol{H}'_l\right) \right] \end{aligned} \tag{5.95}$$

and $\varepsilon'^{(2)}_{yW3}$ is the second-order transverse aberration in the tangential direction that contributes to the third-group wavefront aberrations, given as

$$\varepsilon'^{(2)}_{yW3} = \frac{1}{\cos\theta'_h n' u'_{ax}} \left[ W^{(XP)}_{03001} \left( \boldsymbol{\rho} \cdot \boldsymbol{\rho} \right) + W^{(XP)}_{12101} \left( \boldsymbol{H} \cdot \boldsymbol{\rho} \right) + W^{(XP)}_{21001} \left( \boldsymbol{H} \cdot \boldsymbol{H} \right) \right] \tag{5.96}$$

Therefore, $x$ and $y$ can be approximated to

$$x \approx x^{(1)} = -\frac{s'_0}{s'_p - s'_0} L'_p \rho_x + \frac{s'_p}{s'_p - s'_0} L'_h H_x \tag{5.97}$$

$$\begin{aligned} y \approx y^{(1)} + y^{(2)}_{W3} &= -\frac{1}{\cos I'} \frac{s'_0}{s'_p - s'_0} L'_p \rho_y + \frac{\cos\theta'_h}{\cos I'} \frac{s'_p}{s'_p - s'_0} L'_h H_y \\ &+ z^{(2)} \tan I' + \frac{\cos\theta'_h}{\cos I'} \frac{s'_p}{s'_p - s'_0} \left( \varepsilon'^{(1)}_y + \varepsilon'^{(2)}_{yW3} \right) \end{aligned} \tag{5.98}$$

As defined in Section 4.1, $L'_h$ and $L'_p$ are determined by paraxial raytracing in the sagittal direction, and the sagittal marginal and chief ray heights at the surface, $x_a$ and $x_b$, can be expressed as

$$x_a = -\frac{s'_0}{s'_p - s'_0} L'_p \tag{5.99}$$

$$x_b = \frac{s'_p}{s'_p - s'_0} L'_h \tag{5.100}$$

Another quantity used to simplify the expressions of $x$ and $y$ is the Lagrange invariant, given as

$$\Psi = n \left( u_{bx} x_a - u_{ax} x_b \right) \tag{5.101}$$

It can be seen from paraxial optics that $\Psi$ does not change with refraction or reflection, meaning

$$\Delta \left( \Psi \right) = 0 \tag{5.102}$$

Therefore, Equations 5.97-5.98 can be written as

$$x \approx x_a \rho_x + x_b H_x \tag{5.103}$$

$$\begin{aligned} y \approx \frac{1}{\cos I'} x_a \rho_y + \frac{\cos \theta_h'}{\cos I'} x_b H_y + \frac{\tan I'}{2R_{x-eff}} \left[ x_a^2 (\boldsymbol{\rho} \cdot \boldsymbol{\rho}) + 2 x_a x_b (\boldsymbol{H} \cdot \boldsymbol{\rho}) + x_b^2 (\boldsymbol{H} \cdot \boldsymbol{H}) \right] \\ - \frac{x_b}{\Psi \cos I'} \left[ 2W_{02002}^{(XP)} \rho_y + W_{11011}^{(XP)} H_y + W_{03001}^{(XP)} (\boldsymbol{\rho} \cdot \boldsymbol{\rho}) + W_{12101}^{(XP)} (\boldsymbol{H} \cdot \boldsymbol{\rho}) + W_{21001}^{(XP)} (\boldsymbol{H} \cdot \boldsymbol{H}) \right] \end{aligned} \tag{5.104}$$

For approximated rays, all imaging aberrations are ignored, so $y$ can be further approximated to

$$\begin{aligned} y\left(W^{(XP)} = 0\right) \approx \frac{1}{\cos I'} x_a \rho_y^{(ai)} + \frac{\cos \theta_h'}{\cos I'} x_b H_y \\ + \frac{\tan I'}{2R_{x-eff}} \left[ x_a^2 \left( \boldsymbol{\rho}^{(ai)} \cdot \boldsymbol{\rho}^{(ai)} \right) + 2 x_a x_b \left( \boldsymbol{H} \cdot \boldsymbol{\rho}^{(ai)} \right) + x_b^2 (\boldsymbol{H} \cdot \boldsymbol{H}) \right] \end{aligned} \tag{5.105}$$

where $\boldsymbol{\rho}^{(ai)}$ refers to the normalized exit pupil coordinates where the approximated ray hits, given as

$$\boldsymbol{\rho}^{(ai)} = \left( \rho_x^{(ai)}, \rho_y^{(ai)} \right) \tag{5.106}$$

Note that Equation 5.103 is not affected by transverse aberrations, so the lower-order expression of $x$ is the same for the real ray and the approximated ray.

Equations 5.103-5.105 represent the beam footprint on the surface as functions of field and pupil vectors. An important step during the derivation here is to keep all the terms that can affect third-group aberrations, meaning that any further approximation would add error to the final formulae of third-group aberration coefficients. In Sasian's work, the beam footprint was approximated to be circular, which further approximated Equation 5.104 to be $y \approx x_a \rho_y + x_b H_y$. This approximation was used to simplify the formulae of aberration coefficients. However, it also limits the accuracy of the coefficients with increased error as the OAR incident angle $I$ increases [43].

A similar derivation can be done to find the relationship between $G(x, y, z)$ and $F'(\rho'_{rx}, \rho'_{ry})$ in Figure 5.4 by setting $\boldsymbol{s'_p}$=0 and replacing $\boldsymbol{\rho'_l}$ with $\boldsymbol{\rho'_r}$ in Equations 5.87-5.88, given as

$$x \approx \rho'_{rx} \tag{5.107}$$

$$y \approx \frac{1}{\cos I'}\rho'_{ry} + \frac{(\boldsymbol{\rho'_r} \cdot \boldsymbol{\rho'_r})}{2R_{x-eff}}\tan I' \tag{5.108}$$

For approximated rays, the relation between $\boldsymbol{\rho'_r}$ and $\boldsymbol{\rho}^{(ai)}$ for the same ray can also be found by comparing Equation 5.103 with 5.107, and 5.105 with 5.108, given as

$$\rho'_{rx} = x_a\rho_x^{(ai)} + x_bH_x \tag{5.109}$$

$$\rho'_{ry} = x_a\rho_y^{(ai)} + x_bH_y\cos\theta'_h \tag{5.110}$$

### 5.2.3 Wavefront aberration coefficients with the reference spheres at the surface

With Equations 5.103 and 5.105, the wavefront aberration coefficients can be found by substituting the expression of $x$ and $y$ into the analytical expression of $W$ in Equation 5.81. The results are shown in this section with the contribution from the spherical base surface and freeform departure listed separately.

The contribution from the spherical base surface refers to the contribution from Equation 5.81 with zero freeform departure, $z_{f2}$=0, which will be labeled as $W_{Sph}^{(S)}$, given as

$$W_{Sph}^{(S)} \approx \Delta\left[\frac{n(\boldsymbol{r}\cdot\boldsymbol{s_h})}{s_h}\right] - \frac{1}{2}\left(\frac{n}{s_0} - \frac{n\cos I}{R_{x-eff}}\right)\Delta\left(\frac{s_0}{s_h}\right)r^2 + \frac{1}{2}\Delta\left[\frac{n(\boldsymbol{r}\cdot\boldsymbol{s_h})^2}{s_h^3}\right]$$
$$-\frac{1}{2}\left(\frac{n}{s_0} - \frac{n\cos I}{R_{x-eff}}\right)\Delta\left[\frac{s_0(\boldsymbol{r}\cdot\boldsymbol{s_h})}{s_h^3}\right]r^2 - \frac{1}{2}\Delta\left[\frac{nH_{ly}}{s_h}\sin(I-\theta_h)\right]\frac{r^2}{R_{x-eff}}$$
$$+\frac{1}{8}\left(\frac{n}{s_0} - \frac{n\cos I}{R_{x-eff}}\right)^2\Delta\left(\frac{1}{ns_0}\right)r^4 \tag{5.111}$$
$$= \sum_{k,m,n,p,q}^{\infty} W_{Sph2k+n+p,2m+n+q,n,p,q}^{(S)} (\boldsymbol{H}\cdot\boldsymbol{H})^k \left(\boldsymbol{\rho}^{(ai)}\cdot\boldsymbol{\rho}^{(ai)}\right)^m \left(\boldsymbol{H}\cdot\boldsymbol{\rho}^{(ai)}\right)^n (\boldsymbol{i}\cdot\boldsymbol{H})^p$$
$$\left(\boldsymbol{i}\cdot\boldsymbol{\rho}^{(ai)}\right)^q$$

The process of deriving the aberration coefficients includes steps of expanding, grouping and simplifying. Three quantities are defined to simplify the expressions. We define the quantities $A$ and $B$ as products of the refractive index before a surface, $n$, and the paraxial incident angle of the sagittal marginal and chief rays, $i_a$ and $i_b$, respectively, given as

$$A = n\,i_a = n\left(u_{ax} + \frac{x_a\cos I}{R}\right) \tag{5.112}$$

$$B = n\,i_b = n\left(u_{bx} + \frac{x_b\cos I}{R}\right) \tag{5.113}$$

A quantity $C$ is also defined as the product of the index of refraction before the surface, $n$, and OAR incident angle, $I$, given as

$$C = n\sin I \tag{5.114}$$

$A$, $B$ and $C$ can have corresponding quantities defined in image space by replacing the object-space parameters, such as $n$, $I$ and $u_a$, with the image-space parameters, $n'$, $I'$ and $u'_a$.

However, it can be seen from paraxial optics that $A$ and $B$ are invariants under refraction and reflection, meaning

$$\Delta(A)=0 \tag{5.115}$$

$$\Delta(B)=0 \tag{5.116}$$

It is also easy to see that $C$ can be an invariant under refraction and reflection as well based on Snell's law of the OAR. However, in cases of spectrometers, dispersive surfaces can affect the value of $C$ after the dispersion. Note that this dispersion can only happen in the tangential direction due to the plane of symmetry, so $A$ and $B$ are not affected since they are defined in the sagittal direction. If a prism is used as the dispersive element, $C$ is still an invariant before and after surfaces, because only simple refractions are involved in this case. If a diffractive grating is used as the dispersive element, the additional phase provided by the grating can have significant impact on the wavefront and aberrations. For the scope of this work, we only consider gratings with equal spacing between repetitive diffractive structures, such as a blazed grating with equally spaced rulings, which provides an additional phase linear to the tangential ray height on the surface. In this case, it can be derived from the aberration theory of curved grating [53] that the effect of the grating on paraxial optics and wavefront aberrations is related to the change of $C$ before and after the surface. For a grating with equal spacing, the change of $C$ can be derived from the grating equation as

$$\Delta(C)=\frac{m\lambda}{d} \tag{5.117}$$

where $m$ is the order of the spectrum, $\lambda$ is the wavelength of the light, and $d$ is the length of the spacing. For zero-order spectrum or non-diffractive surfaces ($d\rightarrow\infty$), $C$ is an invariant ($\Delta(C)=0$).

Starting from Equation 5.111, the third-group wavefront aberration coefficients for $W_{Sph}^{(S)}$ are derived with three steps. The first step is to expand the expression in Equation 5.111 as a function of $\boldsymbol{H}$ and $\boldsymbol{\rho}^{(ai)}$. The second step is to group the terms based on their dependence on $\boldsymbol{H}$ and $\boldsymbol{\rho}^{(ai)}$ to find the quantities corresponding to each aberration coefficient. The third step is to simplify the expression using the definitions of $A$, $B$, $C$, and $\Psi$. Recall that $\Psi$ is the Lagrange invariant defined in Equation 5.101, and $A$, $B$, and $C$ are defined in Equations 5.112-5.114. The detailed process is illustrated in Appendix II. The resulting expressions of the third-group wavefront aberration coefficients are given as

$$W_{Sph02002}^{(S)}=-\frac{1}{2}\Delta\left(\frac{u_{ax}C^2}{n}\right)\frac{x_a}{\cos^2 I'} \tag{5.118}$$

$$W_{Sph11011}^{(S)}=\Psi\Delta\left(\cos I\cos\theta_h\right)\frac{1}{\cos I'}-\Delta\left(\frac{u_{ax}C^2}{n}\right)\frac{x_b\cos\theta_h'}{\cos^2 I'} \tag{5.119}$$

$$W_{Sph03001}^{(S)}=-\frac{A}{2}\Delta\left(\frac{u_{ax}C}{n}\right)\frac{x_a}{\cos I'}-\frac{1}{2R_{x-eff}}\Delta\left(\frac{u_{ax}C^2}{n}\right)\frac{x_a^2\tan I'}{\cos I'} \tag{5.120}$$

$$W_{Sph12101}^{(S)}=-B\Delta\left(\frac{u_{ax}C}{n}\right)\frac{x_a}{\cos I'}-\frac{1}{R_{x-eff}}\Delta\left(\frac{u_{ax}C^2}{n}\right)\frac{x_a x_b\tan I'}{\cos I'} \tag{5.121}$$

$$\begin{aligned}W_{Sph12010}^{(S)}=&-\frac{\Psi}{2R_{x-eff}}\Delta\left(\frac{C\cos\theta_h}{n}\right)x_a-\frac{\Psi}{2}\Delta\left(u_{ax}\sin\theta_h\right)-\frac{A}{2}\Delta\left(\frac{u_{ax}C}{n}\right)\frac{x_b\cos\theta_h'}{\cos I'}\\&+\frac{\Psi}{2R_{x-eff}}\Delta\left(\cos I\cos\theta_h\right)x_a\tan I'-\frac{1}{2R_{x-eff}}\Delta\left(\frac{u_{ax}C^2}{n}\right)\frac{x_a x_b\tan I'}{\cos I'}\end{aligned} \tag{5.122}$$

$$W_{Sph21001}^{(S)} = -\frac{B}{2}\Delta\left(\frac{u_{ax}C}{n}\right)\frac{x_b}{\cos I'} - \frac{\Psi}{2}\Delta\left(\frac{u_{bx}C}{n}\right)\frac{1}{\cos I'} - \frac{1}{2R_{x-eff}}\Delta\left(\frac{u_{ax}C^2}{n}\right)\frac{x_b^2 \tan I'}{\cos I'} \tag{5.123}$$

$$\begin{aligned} W_{Sph21110}^{(S)} = & -\frac{\Psi}{R_{x-eff}}\Delta\left(\frac{C\cos\theta_h}{n}\right)x_b - \Psi\Delta\left(u_{bx}\sin\theta_h\right) - B\Delta\left(\frac{u_{ax}C}{n}\right)\frac{x_b\cos\theta_h'}{\cos I'} \\ & + \frac{\Psi}{R_{x-eff}}\Delta\left(\cos I\cos\theta_h\right)x_b\tan I' - \frac{1}{R_{x-eff}}\Delta\left(\frac{u_{ax}C^2}{n}\right)\frac{x_b^2\tan I'}{\cos I'} \end{aligned} \tag{5.124}$$

$$\begin{aligned} W_{Sph04000}^{(S)} = & -\frac{A^2}{8}\Delta\left(\frac{u_{ax}}{n}\right)x_a - \frac{A}{4R_{x-eff}}\Delta\left(\frac{u_{ax}C}{n}\right)x_a^2\tan I' \\ & - \frac{1}{8R_{x-eff}^2}\Delta\left(\frac{u_{ax}C^2}{n}\right)x_a^3\tan^2 I' \end{aligned} \tag{5.125}$$

$$\begin{aligned} W_{Sph13100}^{(S)} = & -\frac{AB}{2}\Delta\left(\frac{u_{ax}}{n}\right)x_a - \frac{Ax_b + Bx_a}{2R_{x-eff}}\Delta\left(\frac{u_{ax}C}{n}\right)x_a\tan I' \\ & - \frac{1}{2R_{x-eff}^2}\Delta\left(\frac{u_{ax}C^2}{n}\right)x_a^2 x_b\tan^2 I' \end{aligned} \tag{5.126}$$

$$\begin{aligned} W_{Sph22200}^{(S)} = & -\frac{B^2}{2}\Delta\left(\frac{u_{ax}}{n}\right)x_a - \frac{B}{R_{x-eff}}\Delta\left(\frac{u_{ax}C}{n}\right)x_a x_b\tan I' \\ & - \frac{1}{2R_{x-eff}^2}\Delta\left(\frac{u_{ax}C^2}{n}\right)x_a x_b^2\tan^2 I' \end{aligned} \tag{5.127}$$

$$\begin{aligned} W_{Sph22000}^{(S)} = & -\frac{AB}{4}\Delta\left(\frac{u_{ax}}{n}\right)x_b - \frac{A\Psi}{4}\Delta\left(\frac{u_{bx}}{n}\right) \\ & - \frac{1}{4R_{x-eff}}\left[\left(Ax_b + Bx_a\right)\Delta\left(\frac{u_{ax}C}{n}\right)x_b + \Psi\Delta\left(\frac{u_{bx}C}{n}\right)x_a\right]\tan I' \\ & - \frac{1}{4R_{x-eff}^2}\Delta\left(\frac{u_{ax}C^2}{n}\right)x_a x_b^2\tan^2 I' \end{aligned} \tag{5.128}$$

$$W_{Sph31100}^{(S)} = -\frac{B^2}{2}\Delta\left(\frac{u_{ax}}{n}\right)x_b - \frac{B\Psi}{2}\Delta\left(\frac{u_{bx}}{n}\right) - \frac{1}{R_{x-eff}}\left[B\Delta\left(\frac{u_{ax}C}{n}\right)x_b + \frac{\Psi}{2}\Delta\left(\frac{u_{bx}C}{n}\right)\right]x_b \tan I' - \frac{1}{2R_{x-eff}^2}\Delta\left(\frac{u_{ax}C^2}{n}\right)x_b^3 \tan^2 I' \tag{5.129}$$

The freeform contribution of wavefront aberrations can also be expanded as given in Equation 5.130 as

$$W_F^{(S)} \approx z_{f2}\Delta\left(n\cos I\right) = \sum_{k,m,n,p,q}^{\infty} W_{F2k+n+p,2m+n+q,n,p,q}^{(S)} \left(\boldsymbol{H}\cdot\boldsymbol{H}\right)^k \left(\boldsymbol{\rho}^{(ai)}\cdot\boldsymbol{\rho}^{(ai)}\right)^m \left(\boldsymbol{H}\cdot\boldsymbol{\rho}^{(ai)}\right)^n \left(\boldsymbol{i}\cdot\boldsymbol{H}\right)^p \left(\boldsymbol{i}\cdot\boldsymbol{\rho}^{(ai)}\right)^q \tag{5.130}$$

The freeform aberration coefficients, $W_{F2k+n+p,2m+n+q,n,p,q}{}^{(S)}$, can be derived similarly to the process of deriving $W_{Sph2k+n+p,2m+n+q,n,p,q}{}^{(S)}$. Here we list the freeform aberration coefficients when $z_{f2}$ is described with the first 16 Fringe Zernike polynomials.

$$W_{F02002}^{(S)} = \Delta\left(n\cos I\right)\left(-2Z_5 + 6Z_{12}\right)\frac{x_a^2}{R_{zn}^2}\frac{1}{\cos^2 I'} \tag{5.131}$$

$$W_{F11011}^{(S)} = \Delta\left(n\cos I\right)\left(-4Z_5 + 12Z_{12}\right)\frac{x_a x_b}{R_{zn}^2}\frac{\cos\theta_h'}{\cos^2 I'} \tag{5.132}$$

$$W_{F03001}^{(S)} = \Delta\left(n\cos I\right)\left[\left(3Z_8 + 3Z_{11} - 12Z_{15}\right)\frac{1}{R_{zn}} + \left(-2Z_5 + 6Z_{12}\right)\frac{\tan I'}{R_{x-eff}}\right]\frac{x_a^3}{R_{zn}^2}\frac{1}{\cos I'} \tag{5.133}$$

$$W_{F12101}^{(S)} = \Delta\left(n\cos I\right)\left[\left(6Z_8 + 6Z_{11} - 24Z_{15}\right)\frac{1}{R_{zn}} + \left(-4Z_5 + 12Z_{12}\right)\frac{\tan I'}{R_{x-eff}}\right]\frac{x_a^2 x_b}{R_{zn}^2}\frac{1}{\cos I'} \tag{5.134}$$

$$W_{F12010}^{(S)} = \Delta\left(n\cos I\right)\left[\left(3Z_8 + 3Z_{11} - 12Z_{15}\right)\frac{1}{R_{zn}} + \left(-2Z_5 + 6Z_{12}\right)\frac{\tan I'}{R_{x-eff}}\right]\frac{x_a^2 x_b}{R_{zn}^2}\frac{\cos\theta_h'}{\cos I'} \tag{5.135}$$

$$W_{F21001}^{(S)} = \Delta\left(n\cos I\right)\left[\left(3Z_8 + 3Z_{11} - 12Z_{15}\right)\frac{1}{R_{zn}} + \left(-2Z_5 + 6Z_{12}\right)\frac{\tan I'}{R_{x-eff}}\right]\frac{x_a x_b^2}{R_{zn}^2}\frac{1}{\cos I'} \quad (5.136)$$

$$W_{F21110}^{(S)} = \Delta\left(n\cos I\right)\left[\left(6Z_8 + 6Z_{11} - 24Z_{15}\right)\frac{1}{R_{zn}} + \left(-4Z_5 + 12Z_{12}\right)\frac{\tan I'}{R_{x-eff}^2}\right]\frac{x_a x_b^2}{R_{zn}^2}\frac{\cos\theta_h'}{\cos I'} \quad (5.137)$$

$$W_{F04000}^{(S)} = \Delta\left(n\cos I\right)\left[\left(6Z_9 + 4Z_{12} - 30Z_{16}\right)\frac{1}{R_{zn}^2} + \left(3Z_8 + 3Z_{11} - 12Z_{15}\right)\frac{\tan I'}{2R_{x-eff}R_{zn}} + \left(-Z_5 + 3Z_{12}\right)\frac{\tan^2 I'}{2R_{x-eff}^2}\right]\frac{x_a^4}{R_{zn}^2} \quad (5.138)$$

$$W_{F13100}^{(S)} = \Delta\left(n\cos I\right)\left[\left(24Z_9 + 16Z_{12} - 120Z_{16}\right)\frac{1}{R_{zn}^2} + \left(6Z_8 + 6Z_{11} - 24Z_{15}\right)\frac{\tan I'}{R_{x-eff}R_{zn}} + \left(-2Z_5 + 6Z_{12}\right)\frac{\tan^2 I'}{R_{x-eff}^2}\right]\frac{x_a^3 x_b}{R_{zn}^2} \quad (5.139)$$

$$W_{F22200}^{(S)} = \Delta\left(n\cos I\right)\left[\left(24Z_9 + 24Z_{12} - 120Z_{16}\right)\frac{1}{R_{zn}^2} + \left(6Z_8 + 6Z_{11} - 24Z_{15}\right)\frac{\tan I'}{R_{x-eff}R_{zn}} + \left(-2Z_5 + 6Z_{12}\right)\frac{\tan^2 I'}{R_{x-eff}^2}\right]\frac{x_a^2 x_b^2}{R_{zn}^2} \quad (5.140)$$

$$W_{F22000}^{(S)} = \Delta\left(n\cos I\right)\left[\left(12Z_9 - 60Z_{16}\right)\frac{1}{R_{zn}^2} + \left(3Z_8 + 3Z_{11} - 12Z_{15}\right)\frac{\tan I'}{R_{x-eff}R_{zn}} + \left(-Z_5 + 3Z_{12}\right)\frac{\tan^2 I'}{R_{x-eff}^2}\right]\frac{x_a^2 x_b^2}{R_{zn}^2} \quad (5.141)$$

$$W_{F31100}^{(S)} = \Delta\left(n\cos I\right)\left[\left(24Z_9 + 16Z_{12} - 120Z_{16}\right)\frac{1}{R_{zn}^2} + \left(6Z_8 + 6Z_{11} - 24Z_{15}\right)\frac{\tan I'}{R_{x-eff}R_{zn}} + \left(-2Z_5 + 6Z_{12}\right)\frac{\tan^2 I'}{R_{x-eff}^2}\right]\frac{x_a x_b^3}{R_{zn}^2} \quad (5.142)$$

Note that the freeform sag function is defined in the local surface coordinate system for Equations 5.131-5.142. An example is given to illustrate the derivation of Equations 5.131-5.142 in Appendix II.

To summarize how each coefficient is affect by Fringe Zernike polynomials, Table *5.1* lists the Fringe Zernike terms associated with each $W_F^{(S)}$ coefficient.

**Table 5.1. Relation between $W_F^{(S)}$ coefficients and Fringe Zernike terms. When a coefficient is affected by a Fringe Zernike term, the corresponding cell will be marked with ● symbol.**

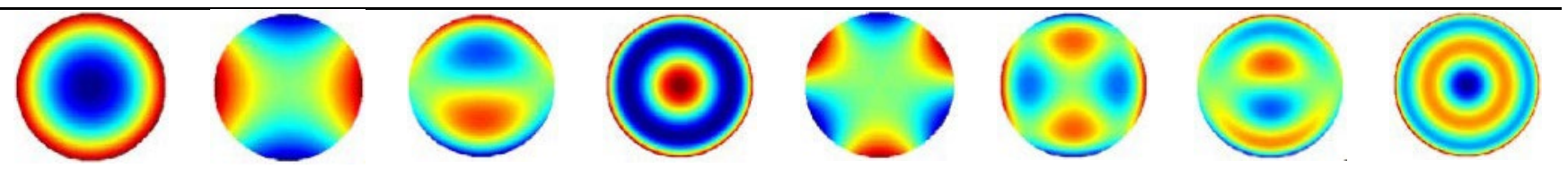

| | $Z_4$ | $Z_5$ | $Z_8$ | $Z_9$ | $Z_{11}$ | $Z_{12}$ | $Z_{15}$ | $Z_{16}$ |
|---|---|---|---|---|---|---|---|---|
| $W_{F02002}^{(S)}$ | - | ● | - | - | - | ● | - | - |
| $W_{F11011}^{(S)}$ | - | ● | - | - | - | ● | - | - |
| $W_{F03001}^{(S)}$ | - | ● | ● | - | ● | ● | ● | - |
| $W_{F12101}^{(S)}$ | - | ● | ● | - | ● | ● | ● | - |
| $W_{F12010}^{(S)}$ | - | ● | ● | - | ● | ● | ● | - |
| $W_{F21001}^{(S)}$ | - | ● | ● | - | ● | ● | ● | - |
| $W_{F21110}^{(S)}$ | - | ● | ● | - | ● | ● | ● | - |
| $W_{F04000}^{(S)}$ | - | ● | ● | ● | ● | ● | ● | ● |
| $W_{F13100}^{(S)}$ | - | ● | ● | ● | ● | ● | ● | ● |
| $W_{F22200}^{(S)}$ | - | ● | ● | ● | ● | ● | ● | ● |
| $W_{F22000}^{(S)}$ | - | ● | ● | ● | ● | ● | ● | ● |
| $W_{F31100}^{(S)}$ | - | ● | ● | ● | ● | ● | ● | ● |

The total wavefront aberration is the sum of the spherical and freeform contributions, given as

$$\begin{aligned} W^{(S)} &= W_{Sph}^{(S)} + W_F^{(S)} \\ &= \sum_{k,m,n,p,q}^{\infty} W_{2k+n+p,2m+n+q,n,p,q}^{(S)} \left(\boldsymbol{H}\cdot\boldsymbol{H}\right)^k \left(\boldsymbol{\rho}^{(ai)}\cdot\boldsymbol{\rho}^{(ai)}\right)^m \left(\boldsymbol{H}\cdot\boldsymbol{\rho}^{(ai)}\right)^n \\ &\quad \left(\boldsymbol{i}\cdot\boldsymbol{H}\right)^p \left(\boldsymbol{i}\cdot\boldsymbol{\rho}^{(ai)}\right)^q \end{aligned} \tag{5.143}$$

Equations 5.118-5.142 describe the third-group wavefront aberration coefficients for a one-surface system when the reference spheres are at the surface, and serve as the foundation for deriving wavefront aberrations at the exit pupil and induced aberrations.

From Equations 5.118-5.142, there are three aberration types related to $\theta_h$ and $\theta'_h$: $W_{11011}^{(S)}$, $W_{12010}^{(S)}$, and $W_{21110}^{(S)}$, which are dependent on $\theta_h$ and $\theta'_h$ for both spherical and freeform contributions. Among them, $W_{11011}^{(S)}$ and $W_{12010}^{(S)}$ are the contribution of anamorphism and field-linear defocus, respectively. The coefficient $W_{21110}^{(S)}$ corresponds to a distortion term related to $\theta_h$ and $\theta'_h$. One may choose to define $\theta_h$ and $\theta'_h$ in a way that one of the three coefficients is reduced to zero. However, there are scenarios where the coefficient cannot be reduced to zero by varying $\theta_h$ and $\theta'_h$. For example, if $\theta'_h$ is set to eliminate $W_{Sph11011}^{(S)}$ for a spherical surface, in a case where the object plane has zero tilt ($\theta_h$=0), and the exit pupil is at the surface ($x_b$=0), $W_{Sph11011}^{(S)}$ is

$$W_{Sph11011}^{(S)}\left(\theta_h = 0, x_b = 0\right) = \Psi\left(\cos\theta'_h - \frac{\cos I}{\cos I'}\right) \tag{5.144}$$

It can be seen that if $\cos I$ is larger than $\cos I'$, $W_{Sph11011}^{(S)}$ cannot be reduced to zero by changing $\theta'_h$. Therefore, it is not universal to define tilt angles based on the elimination of $W_{11011}^{(S)}$. A similar argument can be applied to $W_{12010}^{(S)}$ and $W_{21110}^{(S)}$ as well. As discussed in Section 3.2.2, in this work, $\theta_h$ and $\theta'_h$ are set to zero to simplify the calculation of transverse aberrations. Then, Equations 5.119, 5.122, 5.124, 5.132, 5.135 and 5.137 reduce to

$$W_{Sph11011}^{(S)}\left(\theta_h = 0, \theta'_h = 0\right) = \Delta\left(\cos I\right)\frac{\Psi}{\cos I'} - \Delta\left(\frac{u_{ax}C^2}{n}\right)\frac{x_b}{\cos^2 I'} \tag{5.145}$$

$$\begin{aligned} W_{Sph12010}^{(S)}\left(\theta_h = 0, \theta'_h = 0\right) = &-\frac{1}{2}\frac{\Psi}{R_{x-eff}}\Delta\left(\frac{C}{n}\right)x_a - \frac{1}{2}A\Delta\left(\frac{u_{ax}C}{n}\right)\frac{x_b}{\cos I'} \\ &+\frac{\Psi}{2R_{x-eff}}\Delta\left(\cos I\right)x_a\tan I' - \frac{1}{2R_{x-eff}}\Delta\left(\frac{u_{ax}C^2}{n}\right)\frac{x_a x_b\tan I'}{\cos I'} \end{aligned} \tag{5.146}$$

$$W^{(S)}_{Sph21110}\left(\theta_h=0,\theta'_h=0\right)=-\frac{\Psi}{R_{x-eff}}\Delta\left(\frac{C}{n}\right)x_b-B\Delta\left(\frac{u_{ax}C}{n}\right)\frac{x_b}{\cos I'}+\frac{\Psi}{R_{x-eff}}\Delta\left(\cos I\right)x_b\tan I'-\frac{1}{2R_{x-eff}}\Delta\left(\frac{u_{ax}C^2}{n}\right)\frac{x_a x_b\tan I'}{\cos I'} \tag{5.147}$$

$$W^{(S)}_{F11011}\left(\theta_h=0,\theta'_h=0\right)=\Delta\left(n\cos I\right)\left(-4Z_5+12Z_{12}\right)\frac{x_a x_b}{R^2_{zn}}\frac{1}{\cos^2 I'} \tag{5.148}$$

$$W^{(S)}_{F12010}\left(\theta_h=0,\theta'_h=0\right)=\Delta\left(n\cos I\right)\left[\left(3Z_8+3Z_{11}-12Z_{15}\right)\frac{1}{R_{zn}}+\left(-2Z_5+6Z_{12}\right)\frac{\tan I'}{R_{x-eff}}\right]\frac{x_a^2 x_b}{R^2_{zn}}\frac{1}{\cos I'} \tag{5.149}$$

$$W^{(S)}_{F21110}\left(\theta_h=0,\theta'_h=0\right)=\Delta\left(n\cos I\right)\left[\left(6Z_8+6Z_{11}-24Z_{15}\right)\frac{1}{R_{zn}}+\left(-2Z_5+6Z_{12}\right)\frac{\tan I'}{R^2_{x-eff}}\right]\frac{x_a x_b^2}{R^2_{zn}}\frac{1}{\cos I'} \tag{5.150}$$

Transverse aberrations can then be calculated from wavefront aberrations. From Equation 5.109, the transverse aberration $\varepsilon'_x$ in Equation 4.37 can be rewritten as

$$\varepsilon'_x\approx-\frac{R_{RI}}{n'}\frac{\partial W^{(S)}}{\partial\rho'_{rx}}=-\frac{R_{RI}}{n'}\frac{\partial W^{(S)}}{\partial\rho_x^{(ai)}}\frac{\partial\rho_x^{(ai)}}{\partial\rho'_{rx}}=-\frac{R_{RI}}{x_a n'}\frac{\partial W^{(S)}}{\partial\rho_x^{(ai)}}=\frac{1}{n'u'_{ax}}\frac{\partial W^{(S)}}{\partial\rho_x^{(ai)}} \tag{5.151}$$

which has the same form as Equation 4.34 except that the normalized pupil coordinates follow the approximated ray instead of the real ray. Similarly, $\varepsilon'_y$ follows the form of Equation 4.35, given as

$$\varepsilon'_y\approx\frac{1}{n'u'_{ax}}\frac{\partial W^{(S)}}{\partial\rho_y^{(ai)}} \tag{5.152}$$

### 5.2.4 Wavefront aberrations at the exit pupil in one-surface systems

After the wavefront aberration coefficients are derived with the reference sphere at the surface, they can also be related to wavefront aberrations with the reference sphere at the exit pupil. Consider a real ray and the corresponding approximated ray with the same intersection at the surface as shown in Figure 5.5. The real ray and the approximated ray intersect the exit pupil at the points $E'$ and $E'_a$, respectively. Point $Q'$ is the ideal image point as defined in Section 5.1, and the pupil coordinates of $E'_a$ correspond to the normalized pupil coordinate of the approximated ray, $\boldsymbol{\rho}^{(ai)}$, defined in Section 5.2.2. The normalized exit pupil vector pointing towards $E'$ is, by definition, $\boldsymbol{\rho}$, which can be written as

$$\boldsymbol{\rho} = \boldsymbol{\rho}^{(ai)} + \Delta\boldsymbol{\rho}^{(ai)} \tag{5.153}$$

where $\Delta\boldsymbol{\rho}^{(ai)}$ is the difference between $\boldsymbol{\rho}^{(ai)}$ and $\boldsymbol{\rho}$.

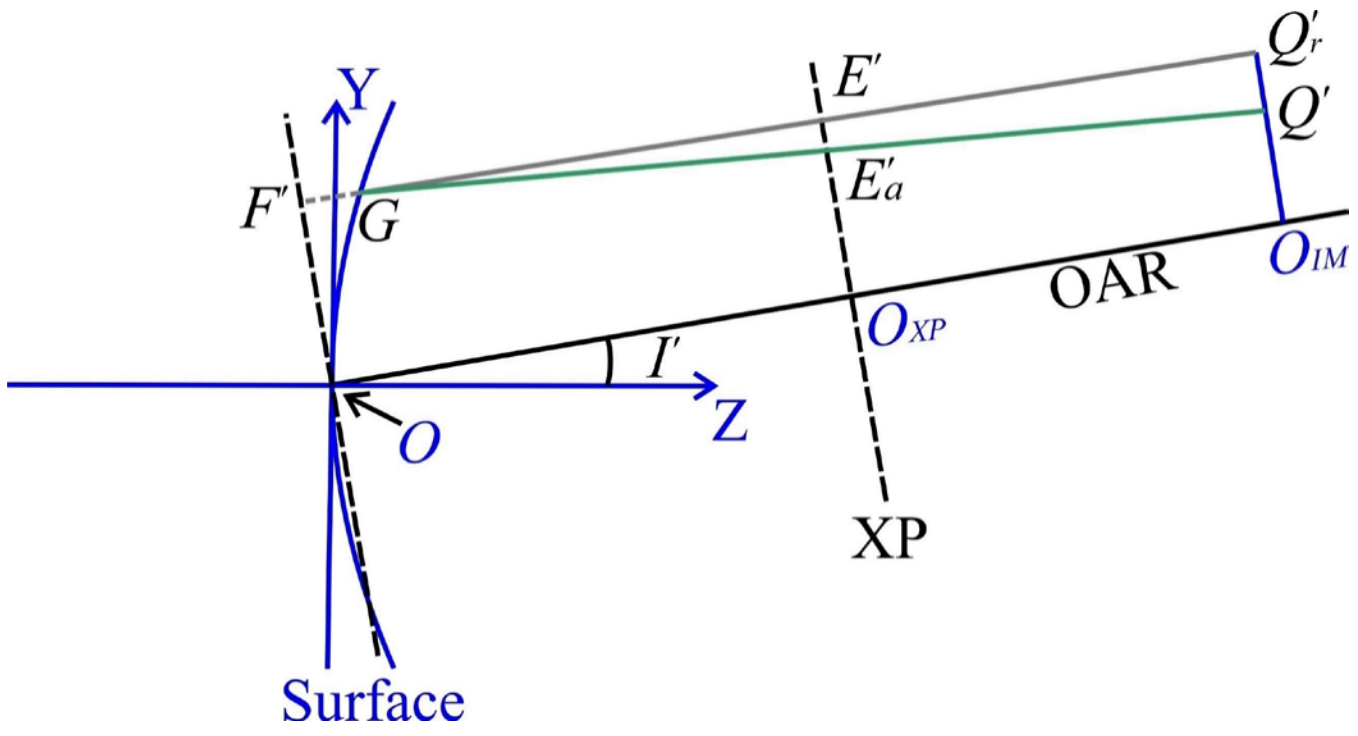

**Figure 5.5. A real ray in light black and its corresponding approximated ray in solid green in the image space of a one-surface system intersecting the image plane at $Q'_r$ and $Q'$. The real ray intersects the exit pupil plane at point $E'$, and the XY plane of the image reference sphere coordinate system of a reference sphere located at the surface at point $F'(\rho'_{rx}, \rho'_{ry})$. The approximated ray intersects the exit pupil plane at point $E'_a$.**

Equations 5.103-5.105 show the coordinates of the ray intersection at the surface as functions of normalized pupil and field coordinates. Since the real ray and the approximated ray pass through the same point on the surface, we have

$$x \approx x_a \rho_x + x_b H_x \approx x_a \rho_x^{(ai)} + x_b H_x \tag{5.154}$$

$$\begin{aligned} y &\approx \frac{1}{\cos I'} x_a \rho_y + \frac{\cos\theta_h'}{\cos I'} x_b H_y + \frac{\tan I'}{2R_{x-eff}} \left[ x_a^2 \left(\boldsymbol{\rho}\cdot\boldsymbol{\rho}\right) + 2x_a x_b \left(\boldsymbol{H}\cdot\boldsymbol{\rho}\right) + x_b^2 \left(\boldsymbol{H}\cdot\boldsymbol{H}\right) \right] \\ &- \frac{x_b}{\Psi\cos I'} \left[ 2W_{02002}^{(XP)} \rho_y + W_{11011}^{(XP)} H_y + W_{03001}^{(XP)} \left(\boldsymbol{\rho}\cdot\boldsymbol{\rho}\right) + W_{12101}^{(XP)} \left(\boldsymbol{H}\cdot\boldsymbol{\rho}\right) + W_{21001}^{(XP)} \left(\boldsymbol{H}\cdot\boldsymbol{H}\right) \right] \\ &\approx \frac{1}{\cos I'} x_a \rho_y^{(ai)} + \frac{\cos\theta_h'}{\cos I'} x_b H_y \\ &+ \frac{\tan I'}{2R_{x-eff}} \left[ x_a^2 \left(\boldsymbol{\rho}^{(ai)}\cdot\boldsymbol{\rho}^{(ai)}\right) + 2x_a x_b \left(\boldsymbol{H}\cdot\boldsymbol{\rho}^{(ai)}\right) + x_b^2 \left(\boldsymbol{H}\cdot\boldsymbol{H}\right) \right] \end{aligned} \tag{5.155}$$

Note that all quantities that have two vector products are omitted, because they only affect wavefront aberration terms beyond the third group. From Equation 5.154, we have

$$\rho_x = \rho_x^{(ai)} \tag{5.156}$$

Set the relation between $\rho_y$ and $\rho_y^{(ai)}$ to be

$$\rho_y = \rho_y^{(ai)} + \Delta\rho_y^{(ai)} \tag{5.157}$$

which leads to

$$\left(\boldsymbol{\rho}^{(ai)}\cdot\boldsymbol{\rho}^{(ai)}\right) = \left(\boldsymbol{\rho}\cdot\boldsymbol{\rho}\right) + 2\rho_y \Delta\rho_y^{(ai)} + \Delta\rho_y^{(ai)2} \tag{5.158}$$

Both $\rho_y\Delta\rho_y^{(ai)}$ and $\Delta\rho_y^{(ai)2}$ contain at least two vector products in terms of $\boldsymbol{\rho}$ and $\boldsymbol{H}$. Therefore, terms related to $\rho_y\Delta\rho_y^{(ai)}$ and $\Delta\rho_y^{(ai)2}$ do not contribute to third-group wavefront aberrations. Within the scope of third-group wavefront aberrations, we can make the approximation given as

$$\left(\boldsymbol{\rho}\cdot\boldsymbol{\rho}\right)\approx\left(\boldsymbol{\rho}^{(ai)}\cdot\boldsymbol{\rho}^{(ai)}\right) \tag{5.159}$$

Similarly, we can write that

$$\left(\boldsymbol{H}\cdot\boldsymbol{\rho}\right)\approx\left(\boldsymbol{H}\cdot\boldsymbol{\rho}^{(ai)}\right) \tag{5.160}$$

By substituting Equations 5.159-5.160 into 5.155, it can then be derived that

$$\begin{aligned}\Delta\rho_y^{(ai)} \approx \frac{x_b}{\Psi x_a}\Big[&2W_{02002}^{(XP)}\rho_y + W_{11011}^{(XP)}H_y + W_{03001}^{(XP)}\left(\boldsymbol{\rho}\cdot\boldsymbol{\rho}\right) + W_{12101}^{(XP)}\left(\boldsymbol{H}\cdot\boldsymbol{\rho}\right)\\ &+W_{21001}^{(XP)}\left(\boldsymbol{H}\cdot\boldsymbol{H}\right)\Big]\end{aligned} \tag{5.161}$$

The transverse aberration of the ray $GQ'_r$ can be related to the wavefront aberration at both the surface and the exit pupil. From Equations 4.34-4.35 and 5.151-5.152, it can be seen that

$$\frac{\partial W^{(XP)}}{\partial \rho_x} = \left.\frac{\partial W^{(S)}}{\partial \rho_x^{(ai)}}\right|_{\boldsymbol{\rho}^{(ai)}=\boldsymbol{\rho}-\Delta\boldsymbol{\rho}^{(ai)}} \tag{5.162}$$

$$\frac{\partial W^{(XP)}}{\partial \rho_y} = \left.\frac{\partial W^{(S)}}{\partial \rho_y^{(ai)}}\right|_{\boldsymbol{\rho}^{(ai)}=\boldsymbol{\rho}-\Delta\boldsymbol{\rho}^{(ai)}} \tag{5.163}$$

Using Equations 5.161, 5.162 and 5.163, the relationship between $W^{(XP)}$ and $W^{(S)}$ can be derived for third-group aberration terms. By comparing the aberration coefficients on both sides of Equations 5.162-5.163, we can find

$$W_{02002}^{(XP)} = \frac{\Psi x_a W_{02002}^{(S)}}{\Psi x_a + 2x_b W_{02002}^{(S)}} \tag{5.164}$$

$$W_{11011}^{(XP)} = \frac{\Psi x_a W_{11011}^{(S)}}{\Psi x_a + 2x_b W_{02002}^{(S)}} \tag{5.165}$$

$$W_{03001}^{(XP)} = \frac{\Psi x_a W_{03001}^{(S)}}{\Psi x_a + 2x_b W_{02002}^{(S)}} \tag{5.166}$$

$$W_{12101}^{(XP)} = \frac{\Psi x_a W_{12101}^{(S)}}{\Psi x_a + 2x_b W_{02002}^{(S)}} \tag{5.167}$$

$$W_{12010}^{(XP)} = W_{12010}^{(S)} - \frac{x_b W_{11011}^{(S)} W_{03001}^{(S)}}{\Psi x_a + 2x_b W_{02002}^{(S)}} \tag{5.168}$$

$$W_{21001}^{(XP)} = \frac{\Psi x_a W_{21001}^{(S)}}{\Psi x_a + 2x_b W_{02002}^{(S)}} \tag{5.169}$$

$$W_{21110}^{(XP)} = W_{21110}^{(S)} - \frac{x_b W_{11011}^{(S)} W_{12101}^{(S)}}{\Psi x_a + 2x_b W_{02002}^{(S)}} \tag{5.170}$$

$$W_{04000}^{(XP)} = W_{04000}^{(S)} - \frac{x_b W_{03001}^{(S)2}}{2\left(\Psi x_a + 2x_b W_{02002}^{(S)}\right)} \tag{5.171}$$

$$W_{13100}^{(XP)} = W_{13100}^{(S)} - \frac{x_b W_{12101}^{(S)} W_{03001}^{(S)}}{\Psi x_a + 2x_b W_{02002}^{(S)}} \tag{5.172}$$

$$W_{22200}^{(XP)} = W_{22200}^{(S)} - \frac{x_b W_{12101}^{(S)2}}{2\left(\Psi x_a + 2x_b W_{02002}^{(S)}\right)} \tag{5.173}$$

$$W_{22000}^{(XP)} = W_{22000}^{(S)} - \frac{x_b W_{21001}^{(S)} W_{03001}^{(S)}}{\Psi x_a + 2x_b W_{02002}^{(S)}} \tag{5.174}$$

$$W_{31100}^{(XP)} = W_{31100}^{(S)} - \frac{x_b W_{12101}^{(S)} W_{21001}^{(S)}}{\Psi x_a + 2x_b W_{02002}^{(S)}} \tag{5.175}$$

The analytical expressions of wavefront aberration coefficients with the reference sphere at the surface in Equations 5.118-5.142, together with the relationship between the wavefront aberrations at the surface and at the exit pupil shown in Equations 5.164-5.175, fully describe the behavior of third-group wavefront aberrations in one-surface systems, where the input wavefront is assumed to be perfect. In the next chapter, we discuss the effect of an aberrated input wavefront on aberrations.

# Chapter 6. Induced aberrations in the third group for plane-symmetric systems

## 6.1 Definition of induced aberration

In the process of deriving wavefront aberration coefficients, the focus is put on one-surface systems that can serve as building blocks for multi-surface systems. The output wavefront of one surface can be treated as the input wavefront of the next surface.

In Chapter 5, we derived wavefront aberration coefficients for one-surface systems when the input wavefront is perfectly spherical and centered at the object point. However, in a multi-surface system, the input wavefront for a surface can be aberrated due to the aberrations from previous surfaces, which can affect the optical path of real rays between the input and output wavefronts. This effect is commonly referred to as the induced aberrations. Hoffman defined induced aberrations as the difference in surface contribution to wavefront aberrations caused by an aberrated input wavefront versus a perfect input wavefront [51], where the input and output wavefronts are located at the entrance and exit pupils. The surface contribution to wavefront aberrations is then defined as the difference between the input and output wavefront aberrations when compared at the same normalized pupil coordinates at the entrance and exit pupils. The same definition is used in this work. In this chapter, we derive the analytical expressions of induced aberrations for third-group aberration types in plane-symmetric systems.

To analytically express induced aberrations, consider the one-surface system shown in Figure 6.1 with the object and image reference spheres located at $O_{EP}$ and $O_{XP}$, the locations of the entrance and exit pupils, respectively.

A real ray in light black emits from the paraxial object point $Q$ and hits the image plane at the point $Q'_r$. The ray intersects the object and image reference spheres at $P$ and $P'$, respectively. The ray intersects the entrance and exit pupil planes at $K$ and $K'$. The location of $K'$ is determined by the normalized pupil vector $\boldsymbol{\rho}$. This ray with no aberration in the object space is referred to as the ray $PP'$.

Consider another ray entering the one-surface system with existing aberrations, indicated by the transverse aberration vector, $\boldsymbol{\varepsilon}$, in the object plane. This ray intersects the exit pupil plane at $K'$, which is the same as the ray $PP'$. The ray also passes through $K_m$ at the entrance pupil plane. Due to pupil aberrations, the location of $K_m$ is generally different from $K'$ in terms of normalized pupil coordinates. Instead of $\boldsymbol{\rho}$, $K_m$ is represented by $\boldsymbol{\rho}+\boldsymbol{\Delta\rho}^{(m)}$ where $\boldsymbol{\Delta\rho}^{(m)}$ is the difference in normalized pupil coordinates between $K_m$ and $K'$. The vector $\boldsymbol{\Delta\rho}^{(m)}$ can be written as

$$\boldsymbol{\Delta\rho}^{(m)} = \left(\Delta\rho_x^{(m)},\ \Delta\rho_y^{(m)}\right) \tag{6.1}$$

This ray is referred to as the ray $P_mP'_m$ with intersections with the object and image reference spheres at $P_m$ and $P'_m$, respectively.

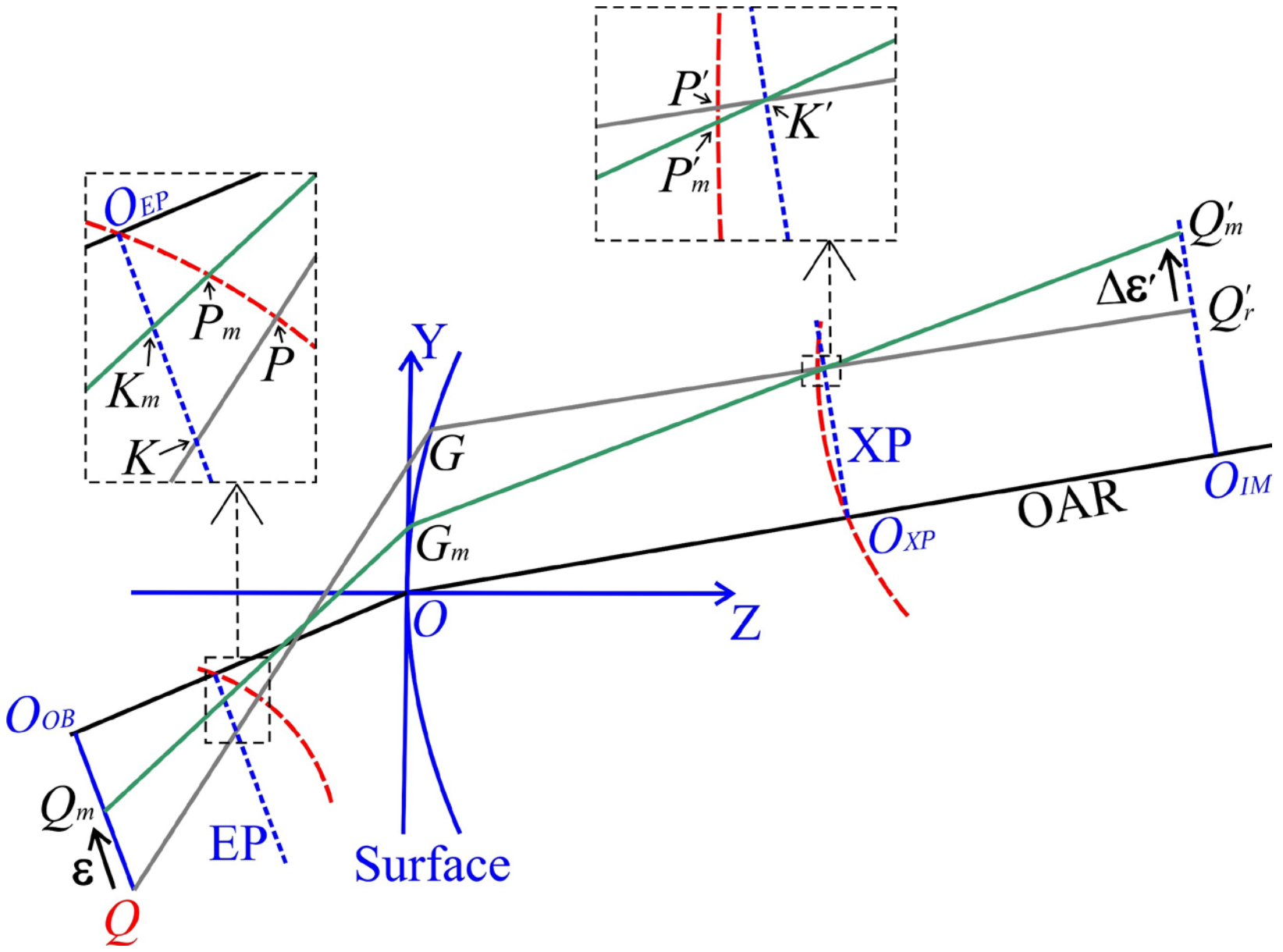

**Figure 6.1. In light black a real ray with no aberrations in the object space propagating through a one-surface system. The light black real ray emits from the object point, *Q*, and intersects the entrance pupil plane at *K*, the object reference sphere at *P*, the surface at *G*, the image reference sphere at *P′*, the exit pupil plane at *K′*, and the image plane at $Q'_r$. In green a real ray with existing aberrations, denoted by *ε*, propagates through a one-surface system. The green ray intersects the object plane at $Q_m$, the entrance pupil plane at $K_m$, the object reference sphere at $P_m$, the surface at $G_m$, the image reference sphere at $P'_m$, the exit pupil plane at *K′*, and the image plane at $Q'_m$.**

For the ray $PP'$, the wavefront aberration at the exit pupil is $W^{(XP)}$ as discussed in Chapter 5. For the ray $P_mP'_m$, the wavefront aberration can be written following Equation 5.4 as

$$W_{P'_mN'} = W_{P_mN} + N_mN'_{m(\mathrm{OP})} - P_mP'_{m(OP)} \tag{6.2}$$

where $N_m$ and $N'_m$ are the locations of the ray $P_mP'_m$ at the input and output wavefronts, respectively. As discussed in Chapter 5, $N_mN'_{m(\mathrm{OP})}$ represents the optical path between the input and output wavefronts, which does not contain information on aberrations. The function $W_{P'_mN'}$ is the wavefront aberration of the ray $P_mP'_m$ at the exit pupil, which can be written as

$$W_{P'_mN'} = W_m^{(XP)}\left(\boldsymbol{H},\boldsymbol{\rho}\right) \tag{6.3}$$

where $W_m^{(XP)}$ is the wavefront aberration at the exit pupil with an aberrated input wavefront. Similarly, $W_{P_mN}$ can be written as

$$W_{P_mN} = W_m^{(EP)}\left(\boldsymbol{H}, \boldsymbol{\rho} + \boldsymbol{\Delta\rho}^{(\boldsymbol{m})}\right) \tag{6.4}$$

where $W_m^{(EP)}$ is the wavefront aberration at the entrance pupil. Note that $\boldsymbol{\Delta\rho}^{(\boldsymbol{m})}$ in Equation 6.4 represents the difference in normalized pupil coordinates between $K_m$ and $K'$. Therefore, we have

$$W_m^{(XP)}\left(\boldsymbol{H}, \boldsymbol{\rho}\right) = W_m^{(EP)}\left(\boldsymbol{H}, \boldsymbol{\rho} + \boldsymbol{\Delta\rho}^{(\boldsymbol{m})}\right) + N_m N'_{m(\mathrm{OP})} - P_m P'_{m(OP)} \tag{6.5}$$

Similarly, $W^{(XP)}$ can be expressed as

$$W^{(XP)}\left(\boldsymbol{H}, \boldsymbol{\rho}\right) = NN'_{(\mathrm{OP})} - PP'_{(OP)} \tag{6.6}$$

With $W_m^{(XP)}$ and $W_m^{(EP)}$ defined, the induced aberration, labeled as $W_{IN}$, can be written as

$$\begin{aligned} W_{IN}\left(\boldsymbol{H}, \boldsymbol{\rho}\right) &= W_m^{(XP)}\left(\boldsymbol{H}, \boldsymbol{\rho}\right) - W^{(XP)}\left(\boldsymbol{H}, \boldsymbol{\rho}\right) - W_m^{(EP)}\left(\boldsymbol{H}, \boldsymbol{\rho}\right) \\ &= W_m^{(EP)}\left(\boldsymbol{H}, \boldsymbol{\rho} + \boldsymbol{\Delta\rho}^{(\boldsymbol{m})}\right) + N_m N'_{m(\mathrm{OP})} - P_m P'_{m(OP)} - NN'_{(\mathrm{OP})} \\ &\quad + PP'_{(OP)} - W_m^{(EP)}\left(\boldsymbol{H}, \boldsymbol{\rho}\right) \\ &\approx W_m^{(EP)}\left(\boldsymbol{H}, \boldsymbol{\rho} + \boldsymbol{\Delta\rho}^{(\boldsymbol{m})}\right) - W_m^{(EP)}\left(\boldsymbol{H}, \boldsymbol{\rho}\right) - P_m P'_{m(OP)} + PP'_{(OP)} \end{aligned} \tag{6.7}$$

where $N_m N'_{m(\mathrm{OP})}$ is approximated to $NN'_{(\mathrm{OP})}$ in the third line of the equation, which does not affect the accuracy of aberration description. It can be seen that induced aberrations can be divided into two parts. The first part, $W_m^{(EP)}(\boldsymbol{H}, \boldsymbol{\rho}+\boldsymbol{\Delta\rho}^{(\boldsymbol{m})}) - W_m^{(EP)}(\boldsymbol{H}, \boldsymbol{\rho})$, is related to pupil aberration or pupil mismatch, which was discussed by Hoffman for rotationally-symmetric systems [51]. The second part, $PP'_{(OP)} - P_m P'_{m(OP)}$, is the change in optical path between reference spheres from the ray $PP'$ to the ray $P_m P'_m$.

## 6.2 Calculation of induced aberrations in the third group

Induced aberrations are mathematically expressed in Equation 6.7. In this section, we derive the analytical expression of induced aberrations in the third group using the first line of Equation 6.7. With $W^{(XP)}$ derived in Chapter 5 and $W_m^{(EP)}$ as the input wavefront at the entrance pupil, $W_m^{(XP)}$ is the key quantity to be calculated in the derivation of $W_{IN}$. The derivation of $W_m^{(XP)}$ follows three steps with each step establishing relationship between wavefront aberrations defined at different references spheres shown in Figure 6.2.

Consider the reference spheres shown in Figure 6.2. The object-side reference sphere at the entrance pupil is labeled as REP, where $W_m^{(EP)}$ is measured. The object-side reference sphere at the surface is labeled as RSO, where $W^{(SO)}$ is measured. Similarly, the image side reference spheres at the exit pupil and the surface are labeled as RXP and RSI, respectively.

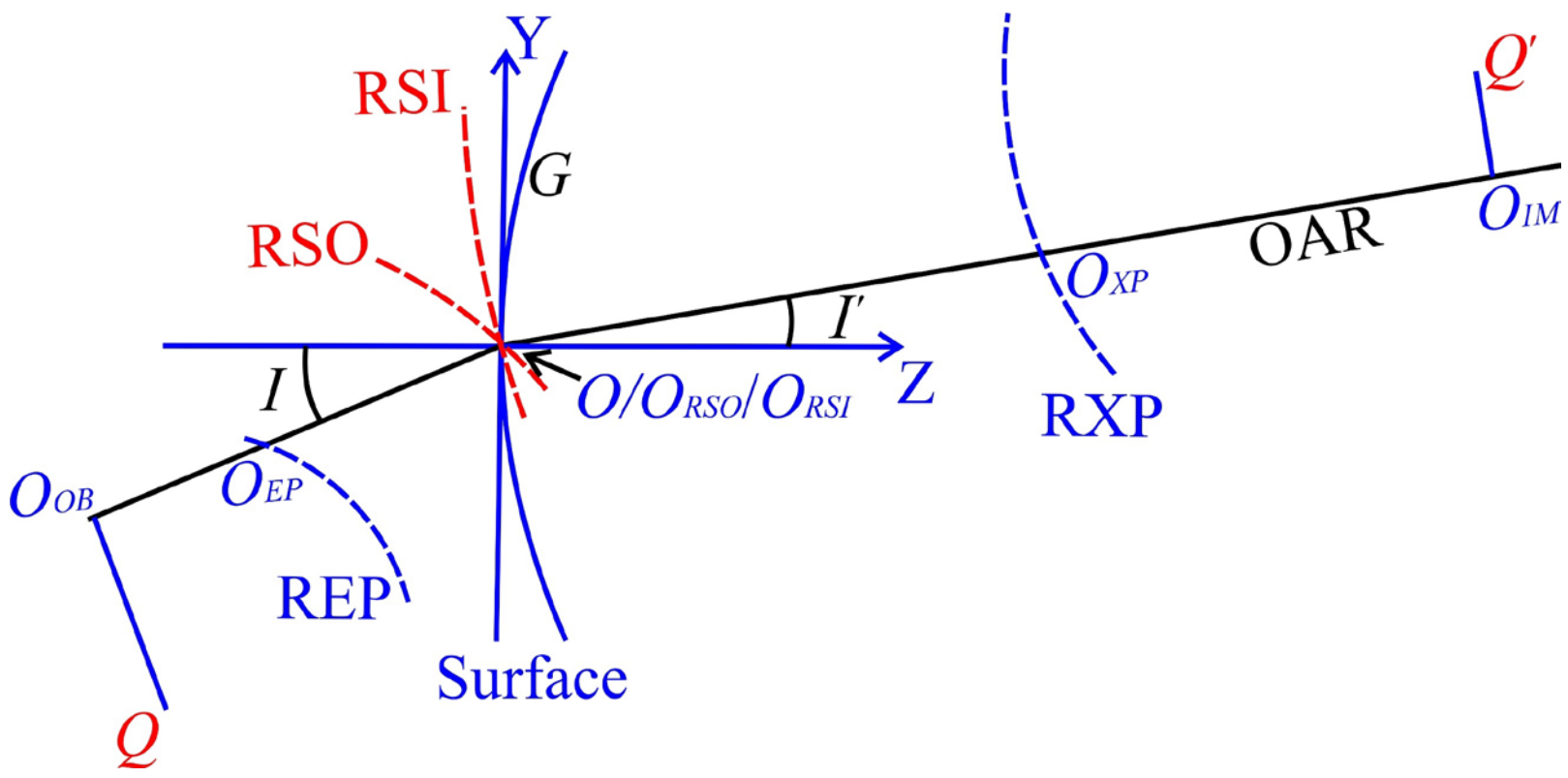

**Figure 6.2. Four reference spheres in a one-surface system: the object-side reference sphere at the entrance pupil, labeled as REP, the object-side reference sphere at the surface, labeled as RSO, the image-side reference sphere at the surface, labeled as RSI, and the image-side reference sphere at the exit pupil, labeled as RXP.**

In Section 5.2.3, the relationship between the wavefront aberrations at RSI and RXP is derived. The same relationship can be applied to the wavefront aberrations at RSO and REP, which can be written as

$$W_{m02002}^{(EP)} = \frac{\Psi x_a W_{02002}^{(SO)}}{\Psi x_a + 2x_b W_{02002}^{(SO)}} \tag{6.8}$$

$$W_{m11011}^{(EP)} = \frac{\Psi x_a W_{11011}^{(SO)}}{\Psi x_a + 2x_b W_{02002}^{(SO)}} \tag{6.9}$$

$$W_{m03001}^{(EP)} = \frac{\Psi x_a W_{03001}^{(SO)}}{\Psi x_a + 2x_b W_{02002}^{(SO)}} \tag{6.10}$$

$$W_{m12101}^{(EP)} = \frac{\Psi x_a W_{12101}^{(SO)}}{\Psi x_a + 2x_b W_{02002}^{(SO)}} \tag{6.11}$$

$$W_{m12010}^{(EP)} = W_{12010}^{(SO)} - \frac{x_b W_{11011}^{(SO)} W_{03001}^{(SO)}}{\Psi x_a + 2x_b W_{02002}^{(SO)}} \tag{6.12}$$

$$W_{m21001}^{(EP)} = \frac{\Psi x_a W_{21001}^{(SO)}}{\Psi x_a + 2x_b W_{02002}^{(SO)}} \tag{6.13}$$

$$W_{m21110}^{(EP)} = W_{21110}^{(SO)} - \frac{x_b W_{11011}^{(SO)} W_{12101}^{(SO)}}{\Psi x_a + 2x_b W_{02002}^{(SO)}} \tag{6.14}$$

$$W_{m04000}^{(EP)} = W_{04000}^{(SO)} - \frac{x_b W_{03001}^{(SO)2}}{2\left(\Psi x_a + 2x_b W_{02002}^{(SO)}\right)} \tag{6.15}$$

$$W_{m13100}^{(EP)} = W_{13100}^{(SO)} - \frac{x_b W_{12101}^{(SO)} W_{03001}^{(SO)}}{\Psi x_a + 2x_b W_{02002}^{(SO)}} \tag{6.16}$$

$$W_{m22200}^{(EP)} = W_{22200}^{(SO)} - \frac{x_b W_{12101}^{(SO)2}}{2\left(\Psi x_a + 2x_b W_{02002}^{(SO)}\right)} \tag{6.17}$$

$$W_{m22000}^{(EP)} = W_{22000}^{(SO)} - \frac{x_b W_{21001}^{(SO)} W_{03001}^{(SO)}}{\Psi x_a + 2x_b W_{02002}^{(SO)}} \tag{6.18}$$

$$W_{m31100}^{(EP)} = W_{31100}^{(SO)} - \frac{x_b W_{12101}^{(SO)} W_{21001}^{(SO)}}{\Psi x_a + 2x_b W_{02002}^{(SO)}} \tag{6.19}$$

where the aberration coefficients follow the wavefront aberration expansion at the entrance pupil and the surface, given as

$$W_{m}^{(EP)} = \sum_{k,j,n,p,q}^{\infty} W_{m2k+n+p,2j+n+q,n,p,q}^{(EP)} \left(\boldsymbol{H}\cdot\boldsymbol{H}\right)^{k} \left(\boldsymbol{\rho}\cdot\boldsymbol{\rho}\right)^{j} \left(\boldsymbol{H}\cdot\boldsymbol{\rho}\right)^{n} \left(\boldsymbol{i}\cdot\boldsymbol{H}\right)^{p} \left(\boldsymbol{i}\cdot\boldsymbol{\rho}\right)^{q} \tag{6.20}$$

$$W^{(SO)} = \sum_{k,j,n,p,q}^{\infty} W_{2k+n+p,2j+n+q,n,p,q}^{(SO)} \left(\boldsymbol{H}\cdot\boldsymbol{H}\right)^{k} \left(\boldsymbol{\rho}^{(\boldsymbol{ao})}\cdot\boldsymbol{\rho}^{(\boldsymbol{ao})}\right)^{j} \left(\boldsymbol{H}\cdot\boldsymbol{\rho}^{(\boldsymbol{ao})}\right)^{n} \left(\boldsymbol{i}\cdot\boldsymbol{H}\right)^{p} \left(\boldsymbol{i}\cdot\boldsymbol{\rho}^{(\boldsymbol{ao})}\right)^{q} \tag{6.21}$$

where $\boldsymbol{\rho}^{(\boldsymbol{ao})}$ is the normalized entrance pupil coordinates where the approximated ray in the object space hits.

From Equations 6.8-6.19, the wavefront aberration coefficients of $W^{(SO)}$ can be expressed as functions of the coefficients of $W_m^{(EP)}$ as shown in Equations 6.22-6.33.

$$W_{02002}^{(SO)} = \frac{\Psi x_a W_{m02002}^{(EP)}}{\Psi x_a - 2x_b W_{m02002}^{(EP)}} \tag{6.22}$$

$$W_{11011}^{(SO)} = \frac{\Psi x_a W_{m11011}^{(EP)}}{\Psi x_a - 2x_b W_{m02002}^{(EP)}} \tag{6.23}$$

$$W_{03001}^{(SO)} = \frac{\Psi x_a W_{m03001}^{(EP)}}{\Psi x_a - 2x_b W_{m02002}^{(EP)}} \tag{6.24}$$

$$W_{12101}^{(SO)} = \frac{\Psi x_a W_{m12101}^{(EP)}}{\Psi x_a - 2x_b W_{m02002}^{(EP)}} \tag{6.25}$$

$$W_{12010}^{(SO)} = W_{m12010}^{(EP)} + \frac{x_b W_{m11011}^{(EP)} W_{m03001}^{(EP)}}{\Psi x_a - 2x_b W_{m02002}^{(EP)}} \tag{6.26}$$

$$W_{21001}^{(SO)} = \frac{\Psi x_a W_{m21001}^{(EP)}}{\Psi x_a - 2x_b W_{m02002}^{(EP)}} \tag{6.27}$$

$$W_{21110}^{(SO)} = W_{m21110}^{(EP)} + \frac{x_b W_{m11011}^{(EP)} W_{m12101}^{(EP)}}{\Psi x_a - 2x_b W_{m02002}^{(EP)}} \tag{6.28}$$

$$W_{04000}^{(SO)} = W_{m04000}^{(EP)} + \frac{x_b W_{m03001}^{(EP)2}}{2\left(\Psi x_a - 2x_b W_{m02002}^{(EP)}\right)} \tag{6.29}$$

$$W_{13100}^{(SO)} = W_{m13100}^{(EP)} + \frac{x_b W_{m03001}^{(EP)} W_{m12101}^{(EP)}}{\Psi x_a - 2x_b W_{m02002}^{(EP)}} \tag{6.30}$$

$$W_{22200}^{(SO)} = W_{m22200}^{(EP)} + \frac{x_b W_{m12101}^{(EP)2}}{2\left(\Psi x_a - 2x_b W_{m02002}^{(EP)}\right)} \tag{6.31}$$

$$W_{22000}^{(SO)} = W_{m22000}^{(EP)} + \frac{x_b W_{m03001}^{(EP)} W_{m21001}^{(EP)}}{\Psi x_a - 2x_b W_{m02002}^{(EP)}} \tag{6.32}$$

$$W_{31100}^{(SO)} = W_{m31100}^{(EP)} + \frac{x_b W_{m12101}^{(EP)} W_{m21001}^{(EP)}}{\Psi x_a - 2x_b W_{m02002}^{(EP)}} \tag{6.33}$$

The next step is to derive $W^{(SI)}$, defined at RSI, from $W^{(SO)}$. Consider the approximated ray shown in Figure 6.3 that connects object and image points, $Q$ and $Q'$, with the ray position on the surface at $G$. The ray intersects the entrance and exit pupils at the normalized pupil vectors, $\boldsymbol{\rho}^{(ao)}$ and $\boldsymbol{\rho}^{(ai)}$. The coordinates of $G$ in the surface coordinate system are $(x, y, z)$.

It has been shown in Section 5.1 that the optical path of rays between the ray intersection at the surface and the image reference sphere can be approximated to that of the approximated rays, when the image reference sphere is at the surface. A similar conclusion can be applied to the object side when the object reference sphere is at the surface. Therefore, when both the object and image reference spheres are at the surface, the optical path between the reference spheres of the approximated rays can be used instead of real rays in the calculation of third-group wavefront aberrations.

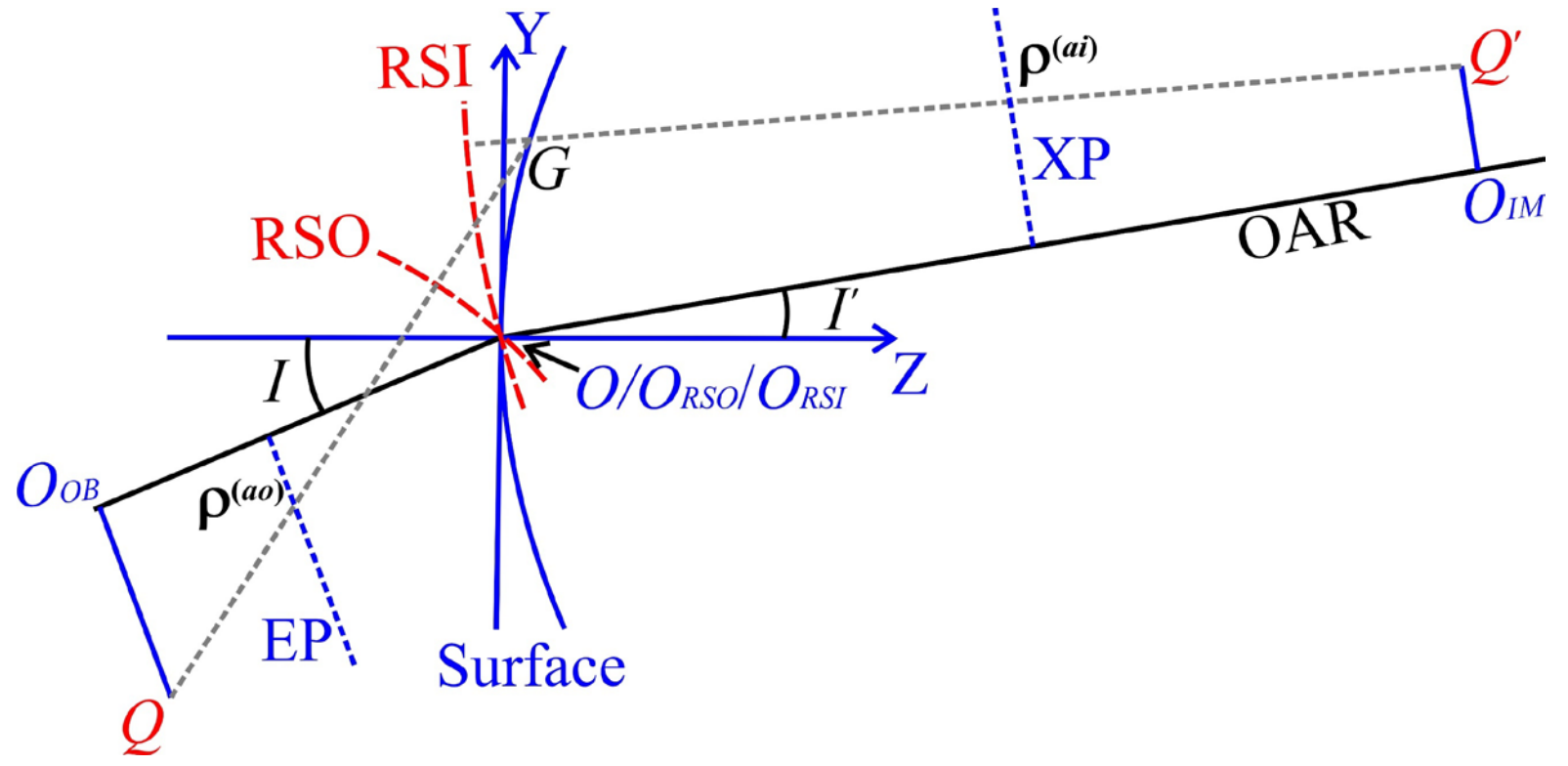

**Figure 6.3. An illustration of an approximated ray in light black propagating through a one-surface system and intersecting the entrance and exit pupils at the normalized pupil vectors, $\boldsymbol{\rho}^{(ao)}$ and $\boldsymbol{\rho}^{(ai)}$, respectively. RSO and RSI are the reference spheres in object and image space located at the surface.**

In Section 5.2.3, $W^{(S)}$ is derived as the surface aberration contribution for intrinsic third-group aberrations when the image reference sphere is RSI. In addition, $W^{(S)}$ can also be treated as the optical path of approximated rays between RSO and RSI, which, as discussed before, is equal to the optical path of real rays in the scope of third-group aberrations. Therefore, in the scope of third-group aberrations, $W^{(SI)}$ can be expressed as

$$W^{(SI)}\left(\boldsymbol{H},\boldsymbol{\rho}^{(ai)}\right)=W^{(SO)}\left(\boldsymbol{H},\boldsymbol{\rho}^{(ao)}\right)+W^{(S)}\left(\boldsymbol{H},\boldsymbol{\rho}^{(ai)}\right) \tag{6.34}$$

To calculate $W^{(SI)}$, the relationship between $\boldsymbol{\rho}^{(ao)}$ and $\boldsymbol{\rho}^{(ai)}$ needs to be derived. From Equations 5.103 in Chapter 5, we have

$$x \approx x_a \rho_x^{(ao)} + x_b H_x = x_a \rho_x^{(ai)} + x_b H_x \tag{6.35}$$

Thus

$$\rho_x^{(ao)} = \rho_x^{(ai)} \tag{6.36}$$

Similarly, from Equation 5.105 in Chapter 5, we have

$$\begin{aligned} y &\approx \frac{1}{\cos I} x_a \rho_y^{(ao)} + \frac{1}{\cos I} x_b H_y \\ &+ \frac{\tan I}{2R_{x-eff}} \left[ x_a^2 \left( \boldsymbol{\rho}^{(ao)} \cdot \boldsymbol{\rho}^{(ao)} \right) + 2x_a x_b \left( \boldsymbol{H} \cdot \boldsymbol{\rho}^{(ao)} \right) + x_b^2 \left( \boldsymbol{H} \cdot \boldsymbol{H} \right) \right] \\ &\approx \frac{1}{\cos I'} x_a \rho_y^{(ai)} + \frac{1}{\cos I'} x_b H_y \\ &+ \frac{\tan I'}{2R_{x-eff}} \left[ x_a^2 \left( \boldsymbol{\rho}^{(ai)} \cdot \boldsymbol{\rho}^{(ai)} \right) + 2x_a x_b \left( \boldsymbol{H} \cdot \boldsymbol{\rho}^{(ai)} \right) + x_b^2 \left( \boldsymbol{H} \cdot \boldsymbol{H} \right) \right] \end{aligned} \tag{6.37}$$

Following a similar derivation process as shown in Equations 5.156-5.160, it can be seen that, within the scope of third-group wavefront aberrations, we have

$$\left( \boldsymbol{\rho}^{(ao)} \cdot \boldsymbol{\rho}^{(ao)} \right) \approx \left( \boldsymbol{\rho}^{(ai)} \cdot \boldsymbol{\rho}^{(ai)} \right) \tag{6.38}$$

$$\left( \boldsymbol{H} \cdot \boldsymbol{\rho}^{(ao)} \right) \approx \left( \boldsymbol{H} \cdot \boldsymbol{\rho}^{(ai)} \right) \tag{6.39}$$

Set $\rho_y{}^{(ao)}$ as

$$\rho_y^{(ao)} = \rho_y^{(ai)} + \Delta \rho_y^{(a)} \tag{6.40}$$

Therefore, it can be derived from Equation 6.37 that

$$\begin{aligned} \Delta \rho_y^{(a)} &\approx \Delta \left( \frac{1}{\cos I} \right) \rho_y^{(ai)} \cos I \\ &+ \frac{\Delta \left( \tan I \right)}{2x_a R_{x-eff}} \left[ x_a^2 \left( \boldsymbol{\rho}^{(ai)} \cdot \boldsymbol{\rho}^{(ai)} \right) + 2x_a x_b \left( \boldsymbol{H} \cdot \boldsymbol{\rho}^{(ai)} \right) + x_b^2 \left( \boldsymbol{H} \cdot \boldsymbol{H} \right) \right] \cos I \end{aligned} \tag{6.41}$$

Therefore, $W^{(SI)}$ can be written as

$$W^{(SI)} \left( \boldsymbol{H}, \boldsymbol{\rho}^{(ai)} \right) = W^{(SO)} \left( \boldsymbol{H}, \boldsymbol{\rho}^{(ai)} + \Delta \boldsymbol{\rho}^{(a)} \right) + W^{(S)} \left( \boldsymbol{H}, \boldsymbol{\rho}^{(ai)} \right) \tag{6.42}$$

where $\boldsymbol{\Delta\rho}^{(a)}$ is defined as

$$\boldsymbol{\Delta\rho}^{(a)} = \left(0,\ \Delta\rho_y^{(a)}\right) \tag{6.43}$$

Write $W^{(SI)}$ as an expansion, given as

$$W^{(SI)} = \sum_{k,j,n,p,q}^{\infty} W^{(SI)}_{2k+n+p,2j+n+q,n,p,q} \left(\boldsymbol{H}\cdot\boldsymbol{H}\right)^k \left(\boldsymbol{\rho}^{(ai)}\cdot\boldsymbol{\rho}^{(ai)}\right)^j \left(\boldsymbol{H}\cdot\boldsymbol{\rho}^{(ai)}\right)^n \left(\boldsymbol{i}\cdot\boldsymbol{H}\right)^p \left(\boldsymbol{i}\cdot\boldsymbol{\rho}^{(ai)}\right)^q \tag{6.44}$$

Therefore, with the expressions of $W^{(S)}$ and $W^{(SO)}$ derived previously in Sections 5.2.3 and 6.2 and the expression of $\boldsymbol{\Delta\rho}^{(a)}$ in terms of $\boldsymbol{\rho}^{(ai)}$ given in Equations 6.41 and 6.43, the coefficients in the $W^{(SI)}$ expansion can be derived from Equation 6.42, as given in Equation 6.45-6.56.

$$W^{(SI)}_{02002} = W^{(S)}_{02002} + \left(\frac{\cos I}{\cos I'}\right)^2 W^{(SO)}_{02002} \tag{6.45}$$

$$W^{(SI)}_{11011} = W^{(S)}_{11011} + \frac{\cos I}{\cos I'} W^{(SO)}_{11011} \tag{6.46}$$

$$W^{(SI)}_{03001} = W^{(S)}_{03001} + \frac{\cos I}{\cos I'} W^{(SO)}_{03001} + \frac{x_a \Delta\left(\tan I\right)\cos I}{R_{x-eff}} W^{(SO)}_{02002} \tag{6.47}$$

$$W^{(SI)}_{12101} = W^{(S)}_{12101} + \frac{\cos I}{\cos I'} W^{(SO)}_{12101} + \frac{2x_b \Delta\left(\tan I\right)\cos I}{R_{x-eff}} W^{(SO)}_{02002} \tag{6.48}$$

$$W^{(SI)}_{12010} = W^{(S)}_{12010} + W^{(SO)}_{12010} + \frac{x_a \Delta\left(\tan I\right)\cos I}{2R_{x-eff}} W^{(SO)}_{11011} \tag{6.49}$$

$$W^{(SI)}_{21001} = W^{(S)}_{21001} + \frac{\cos I}{\cos I'} W^{(SO)}_{21001} + \frac{x_b^2 \Delta\left(\tan I\right)\cos I}{x_a R_{x-eff}} W^{(SO)}_{02002} \tag{6.50}$$

$$W_{21110}^{(SI)} = W_{21110}^{(S)} + W_{21110}^{(SO)} + \frac{x_a \Delta(\tan I)\cos I}{2R_{x-eff}} W_{11011}^{(SO)} \tag{6.51}$$

$$W_{04000}^{(SI)} = W_{04000}^{(S)} + W_{04000}^{(SO)} + \frac{x_a \Delta(\tan I)\cos I}{2R_{x-eff}} W_{03001}^{(SO)} + \frac{x_a^2}{4}\left(\frac{\Delta(\tan I)\cos I}{R_{x-eff}}\right)^2 W_{02002}^{(SO)} \tag{6.52}$$

$$\begin{aligned} W_{13100}^{(SI)} &= W_{13100}^{(S)} + W_{13100}^{(SO)} + \frac{x_b \Delta(\tan I)\cos I}{R_{x-eff}} W_{03001}^{(SO)} + \frac{x_a \Delta(\tan I)\cos I}{2R_{x-eff}} W_{12101}^{(SO)} \\ &+ x_a x_b \left(\frac{\Delta(\tan I)\cos I}{R_{x-eff}}\right)^2 W_{02002}^{(SO)} \end{aligned} \tag{6.53}$$

$$W_{22200}^{(SI)} = W_{22200}^{(S)} + W_{22200}^{(SO)} + \frac{x_b \Delta(\tan I)\cos I}{R_{x-eff}} W_{12101}^{(SO)} + x_b^2\left(\frac{\Delta(\tan I)\cos I}{R_{x-eff}}\right)^2 W_{02002}^{(SO)} \tag{6.54}$$

$$\begin{aligned} W_{22000}^{(SI)} &= W_{22000}^{(S)} + W_{22000}^{(SO)} + \frac{x_b^2 \Delta(\tan I)\cos I}{2x_a R_{x-eff}} W_{03001}^{(SO)} + \frac{x_a \Delta(\tan I)\cos I}{2R_{x-eff}} W_{21001}^{(SO)} \\ &+ \frac{x_b^2}{2}\left(\frac{\Delta(\tan I)\cos I}{R_{x-eff}}\right)^2 W_{02002}^{(SO)} \end{aligned} \tag{6.55}$$

$$\begin{aligned} W_{31100}^{(SI)} &= W_{31100}^{(S)} + W_{31100}^{(SO)} + \frac{x_b^2 \Delta(\tan I)\cos I}{2x_a R_{x-eff}} W_{12101}^{(SO)} + \frac{x_b \Delta(\tan I)\cos I}{R_{x-eff}} W_{21001}^{(SO)} \\ &+ \frac{x_b^3}{x_a}\left(\frac{\Delta(\tan I)\cos I}{R_{x-eff}}\right)^2 W_{02002}^{(SO)} \end{aligned} \tag{6.56}$$

The final step is to derive $W_m^{(XP)}$, defined at RXP, from $W^{(SI)}$. As discussed in Section 5.2.3, $W_m^{(XP)}$ and $W^{(SI)}$ follow Equations 5.164-5.175 with $W_m^{(XP)}$ and $W^{(SI)}$ in place of $W^{(XP)}$ and $W^{(S)}$. Therefore, with $W_m^{(XP)}$ expanded as

$$\begin{aligned} &W_m^{(XP)} = \\ &\sum_{k,j,n,p,q}^{\infty} W_{m2k+n+p,2j+n+q,n,p,q}^{(XP)} \left(\boldsymbol{H}\cdot\boldsymbol{H}\right)^k \left(\boldsymbol{\rho}\cdot\boldsymbol{\rho}\right)^j \left(\boldsymbol{H}\cdot\boldsymbol{\rho}\right)^n \left(\boldsymbol{i}\cdot\boldsymbol{H}\right)^p \left(\boldsymbol{i}\cdot\boldsymbol{\rho}\right)^q \end{aligned} \tag{6.57}$$

we have

$$W_{m02002}^{(XP)} = \frac{\Psi x_a W_{02002}^{(SI)}}{\Psi x_a + 2x_b W_{02002}^{(SI)}} \tag{6.58}$$

$$W_{m11011}^{(XP)} = \frac{\Psi x_a W_{11011}^{(SI)}}{\Psi x_a + 2x_b W_{02002}^{(SI)}} \tag{6.59}$$

$$W_{m03001}^{(XP)} = \frac{\Psi x_a W_{03001}^{(SI)}}{\Psi x_a + 2x_b W_{02002}^{(SI)}} \tag{6.60}$$

$$W_{m12101}^{(XP)} = \frac{\Psi x_a W_{12101}^{(SI)}}{\Psi x_a + 2x_b W_{02002}^{(SI)}} \tag{6.61}$$

$$W_{m12010}^{(XP)} = W_{12010}^{(SI)} - \frac{x_b W_{11011}^{(SI)} W_{03001}^{(SI)}}{\Psi x_a + 2x_b W_{02002}^{(SI)}} \tag{6.62}$$

$$W_{m21001}^{(XP)} = \frac{\Psi x_a W_{21001}^{(SI)}}{\Psi x_a + 2x_b W_{02002}^{(SI)}} \tag{6.63}$$

$$W_{m21110}^{(XP)} = W_{21110}^{(SI)} - \frac{x_b W_{11011}^{(SI)} W_{12101}^{(SI)}}{\Psi x_a + 2x_b W_{02002}^{(SI)}} \tag{6.64}$$

$$W_{m04000}^{(XP)} = W_{04000}^{(SI)} - \frac{x_b W_{03001}^{(SI)2}}{2\left(\Psi x_a + 2x_b W_{02002}^{(SI)}\right)} \tag{6.65}$$

$$W_{m13100}^{(XP)} = W_{13100}^{(SI)} - \frac{x_b W_{12101}^{(SI)} W_{03001}^{(SI)}}{\Psi x_a + 2x_b W_{02002}^{(SI)}} \tag{6.66}$$

$$W_{m22200}^{(XP)} = W_{22200}^{(SI)} - \frac{x_b W_{12101}^{(SI)2}}{2\left(\Psi x_a + 2x_b W_{02002}^{(SI)}\right)} \tag{6.67}$$

$$W_{m22000}^{(XP)} = W_{22000}^{(SI)} - \frac{x_b W_{21001}^{(SI)} W_{03001}^{(SI)}}{\Psi x_a + 2x_b W_{02002}^{(SI)}} \tag{6.68}$$

$$W_{m31100}^{(XP)} = W_{31100}^{(SI)} - \frac{x_b W_{12101}^{(SI)} W_{21001}^{(SI)}}{\Psi x_a + 2 x_b W_{02002}^{(SI)}} \tag{6.69}$$

Induced aberrations can then be calculated using the first line of Equation 6.7, and the distortion related coefficients are

$$W_{21001}^{(IN)} = W_{m21001}^{(XP)} - W_{21001}^{(XP)} - W_{m21001}^{(EP)} \tag{6.70}$$

$$W_{31100}^{(IN)} = W_{m31100}^{(XP)} - W_{31100}^{(XP)} - W_{m31100}^{(EP)} \tag{6.71}$$

With $W_{m21001}{}^{(XP)}$ and $W_{m31100}{}^{(XP)}$ derived together with the related induced aberrations, the X and Y spectral distortion, $SD_x$ and $SD_y$, defined in Section 4.5, can be expressed analytically with system parameters within the scope of wavefront third-group aberrations. In the next chapter, we will use the analytical expressions of $SD_x$ and $SD_y$ on example systems and compare the results from the analytical expressions with results from real raytracing. Note that although we focus on the wavefront aberration coefficients related to distortion in spectrometers, this approach can be applied to all wavefront aberration coefficients in the third group.

# Chapter 7. Freeform hyperspectral imager design in CubeSat format

In this chapter, we present a freeform hyperspectral imager design in CubeSat format that motivates the development of the analytical aberration theory for plane-symmetric systems. The aberration theory helps us understand how the distortion in spectrometers is associated with system parameters. In Chapter 8, the spectrometer component of the hyperspectral imager will be used as an example system for distortion analysis.

## 7.1 Background

Hyperspectral imaging is an imaging method used to acquire both spatial and spectral information of a scene and has been widely used in chemical analysis and environmental monitoring [54]. Utilizing a hyperspectral imager in an aircraft or a satellite enables hyperspectral imaging over a large area using only a strip-field imager due to the relative movement between the ground and the aircraft or satellite. The strip image can then be fed into a spectrometer through a slit where it is dispersed into the spectral information of the strip image. This method of hyperspectral imaging is called pushbroom, which commenced with the development of the Airborne Imaging Spectrometer [55] and has since been used in many hyperspectral imager designs and missions [16, 56-58].

In recent decades, standardized and miniaturized satellites such as CubeSats have gained popularity because they lower the cost of manufacturing and deployment. The size of a CubeSat is standardized to be multiples of a 10×10×10 cm3 cubic unit, referred to as a “U”. The standardized size of CubeSats contributes to the reduced cost of these satellites because standardized components can be leveraged for solar cells, batteries, and other electronic components [59]. A specialized deployment method called the Poly-PicoSatellite Orbital Deployer (P-POD) was also developed for CubeSats [60] which enables CubeSats to be launched as a secondary payload, further reducing the cost. The low-cost characteristic of CubeSats is ideal for industrial and educational research and experimentation, and the unique cost-performance relation also enables CubeSats to fill key gaps in astronomical research [61] and earth imaging [62]. Hyperspectral imagers compatible with CubeSats have been designed and deployed in past space missions [15, 63-65].

CubeSats significantly benefit from the advancements in compact electronics with low power consumption. The combination of freeform optics and the CubeSat format has the potential to enable small satellites with high-performing optical payloads.

## 7.2 First-order specifications for the hyperspectral imager

Recent hyperspectral imager designs have shown a preference for concentric forms for the spectrometer (e.g., Dyson or Offner type), which have low distortion while being compact and fast [13, 15, 66-68]. However, curved gratings or customized prisms are needed in these designs. Curved gratings are effective at reducing the number of optical elements needed, but they are challenging to manufacture. Also, while Offner-Chrisp designs have been shown to provide excellent spectral-spatial uniformity, some limitations have been discussed by Cook and Silny [69]. Chrisp presented a compact spectrometer design that accommodated a planar grating in a configuration similar to an Offner type by adding a catadioptric lens [70]. This creative design with a planar grating is attractive, yet a limitation is the low dispersion that limits the tradeoff between the final wavelength range and the spectral sampling rate. For the design presented in this chapter, we aimed to provide a compact solution with high dispersion and a planar grating for ease of manufacture compared to curved gratings or specialized prisms. The design also leverages the reported advantages of the reflective triplet [69, 71, 72]. The specifications reported in Table 7.1 were chosen in discussion with our industry partners to provide well-rounded performance for earth imaging accounting for prior art, also listed in Table 7.1.

**Table 7.1. Specifications of pushbroom hyperspectral designs**

| Specifications | This Design | Lee (2008) [66] | Mouroulis (2014) [15] | Huang (2021) [73] |
|---|---|---|---|---|
| **Imager Form** | Reflective Triplet | Catadioptric Telescope | Reflective Triplet | TMA |
| **Spectrometer Form** | Double-pass Reflective Triplet | Offner + Prisms | Dyson | Freeform prism |
| **Grating Shape** | Flat | Prisms | Curved | Prism |
| **Entrance Pupil Diameter (mm)** | 30 | 65 | 56 | 40 |
| **Cross-track FOV (degree)** | 15 | 2.5 | 10 | 12 |
| **Instantaneous FOV (arcmin)** | 1.2 | 0.15 | 1.1 | 0.34 / 0.68 |
| **Slit Width (µm)** | 30 | 12 | - | 15 |
| **Slit Length (mm)** | 16 | 12.1 | 17.5 | 30 |
| **System F-number** | F/2.0 (sagittal) F/2.5 (tangential) | F/4.6 | F/1.8 | 3 |
| **Wavelength (nm)** | 400 - 1700 | 400 - 1050 | 350 - 1700 | 400 - 2500 |
| **Spectrum Length (mm)** | 10 | 13 | 7.1 | 3.4 |
| **Pixel Size (µm)** | 20 | 13 | 30 | 15 / 30 |
| **Spectral Sampling (nm)** | 2.6 | 2 – 15 | 5.7 | 5 / 30 |
| **Optics Volume[a] (L)** | 1.7 | ~ 2.7 | ~ 3.5 | ~ 27 |
| **Smile Distortion (%pixel/µm)** | < (25 / 5) | - | < (1 / 0.3) | - |
| **Keystone Distortion (%pixel/µm)** | < (15 / 3) | - | < (1 / 0.3) | - |

[a]The Optics Volume includes only light beams and optical surfaces without necessary mechanical components.

The design covers a wide spectral range, spanning from the visible spectrum to the short-wave infrared while maintaining a high spectral sampling rate compared with prior art. In this design, the smile and keystone distortion are mainly corrected by the optics. Digital post-processing methods can be used for further distortion correction [74, 75].

### 7.3 Design Process

The design process started with selecting the design form for the imager and spectrometer, which were then separately designed according to the specifications. The two designs were adjusted for pupil matching and then combined into one system. A systemwide optimization was then conducted to reach the final design.

#### 7.3.1 The imager design

The structure of the imager is a freeform reflective triplet with the stop at the first mirror. The imager Mirror 1 (IM1) and Mirror 3 (IM3) are positive, while Mirror 2 (IM2) is negative, referred to as a PNP power distribution. PNP has been shown to yield smaller volumes than NPP solutions with equivalent specifications [19, 76]. The presence of a negative mirror is critical for minimizing the Petzval sum to meet overall specifications. Within PNP, the reflective triplet has also been shown to be the best folding geometry for aberration correction among all unobscured PNP three-mirror geometries [19].

The design was started by setting up a spherical three-mirror starting point that is unobscured and satisfies the aperture and field requirements, which occupies a volume of approximately 1U. This design started with spherical surfaces aligned coaxially. The coaxial design roughly met the first-order specifications and had a PNP power distribution. The design was then made unobscured by tilting the mirrors. After re-optimization, we then changed the surface type to Extended Fringe Zernike Polynomial in CODE V, which combines an off-axis conic base surface with a Fringe Zernike overlay and can be strategically used to improve the testability of an optical surface [77]. This surface type was investigated in both the imager and the spectrometer components. To fully leverage off-axis conics to correct aberrations, only the conic constant and the off-axis angle were optimized initially sans freeform departure. CODE V's Global Synthesis (GS) was also utilized for a wider solution search. From the solutions generated by GS, the best-performing off-axis conic solution was selected, and the freeform departures were implemented to further improve the performance. The imager is symmetric about a plane that is perpendicular to the slit. Therefore, during the optimization, only the Fringe Zernike terms that are symmetric about the same plane were used, up to the 25th term. We applied degeneracy constraints to avoid tilt, power, and despace degeneracies [52]. A square sum penalty on the Fringe Zernike coefficients was used during the optimization process to minimize the freeform departure to improve the manufacturability and testability of each surface, which has been shown to be effective [45]. The resulting imager design has a compact size due to the minimum field size in the tangential direction. Figure 7.1 illustrates this design process.

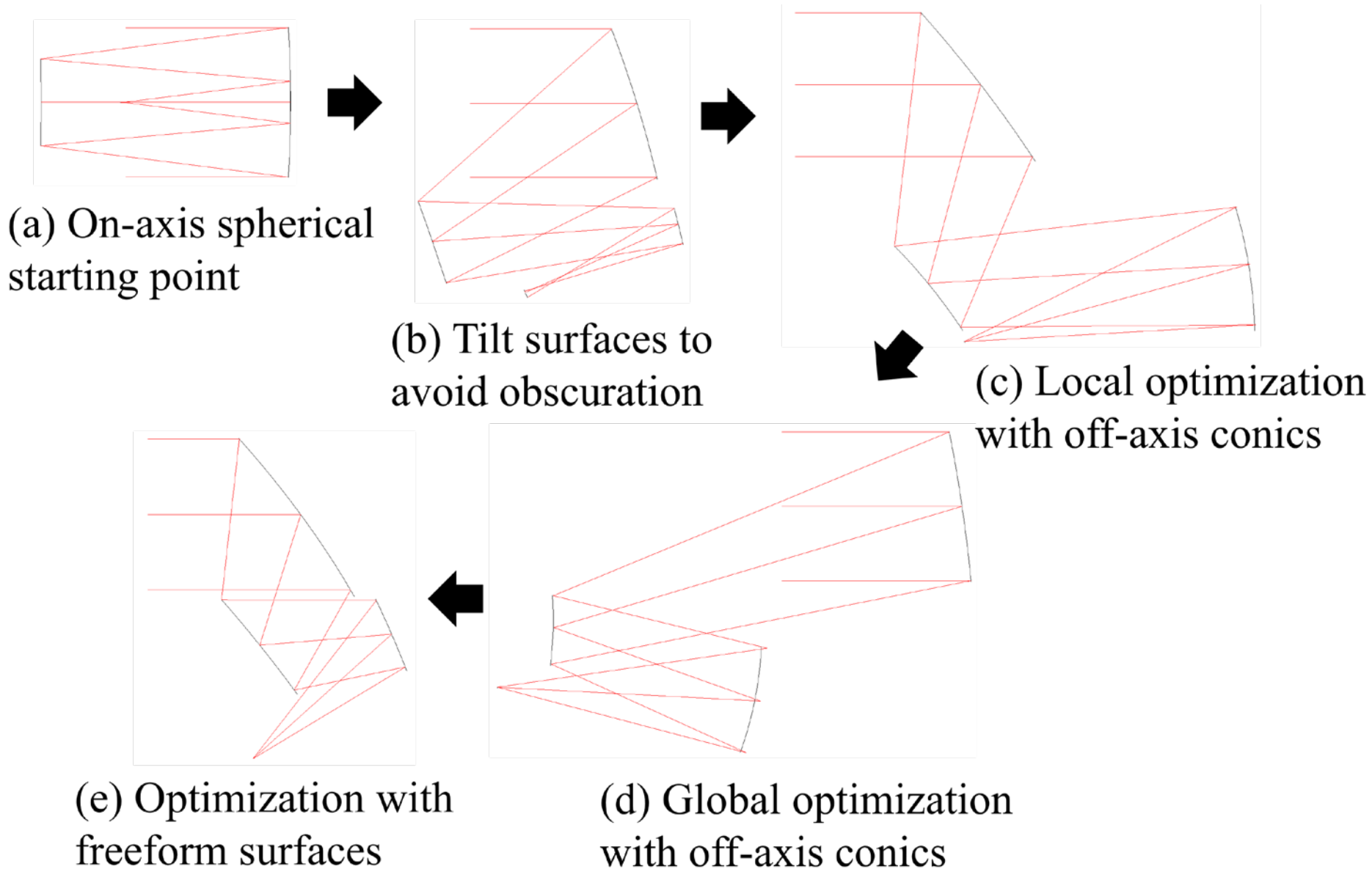

**Figure 7.1. Illustration of imager design process: (a) starting point with on-axis spherical surfaces; (b) tilted spherical surfaces to avoid obscuration; (c) local optimization with the surfaces changed to off-axis conic; (d) global optimization to find better solution spaces; and (e) further optimization with freeform departures added.**

The final design layout is shown in Figure 7.2, which has an average RMS spot size of 7 μm and a smile distortion of less than 2 μm.

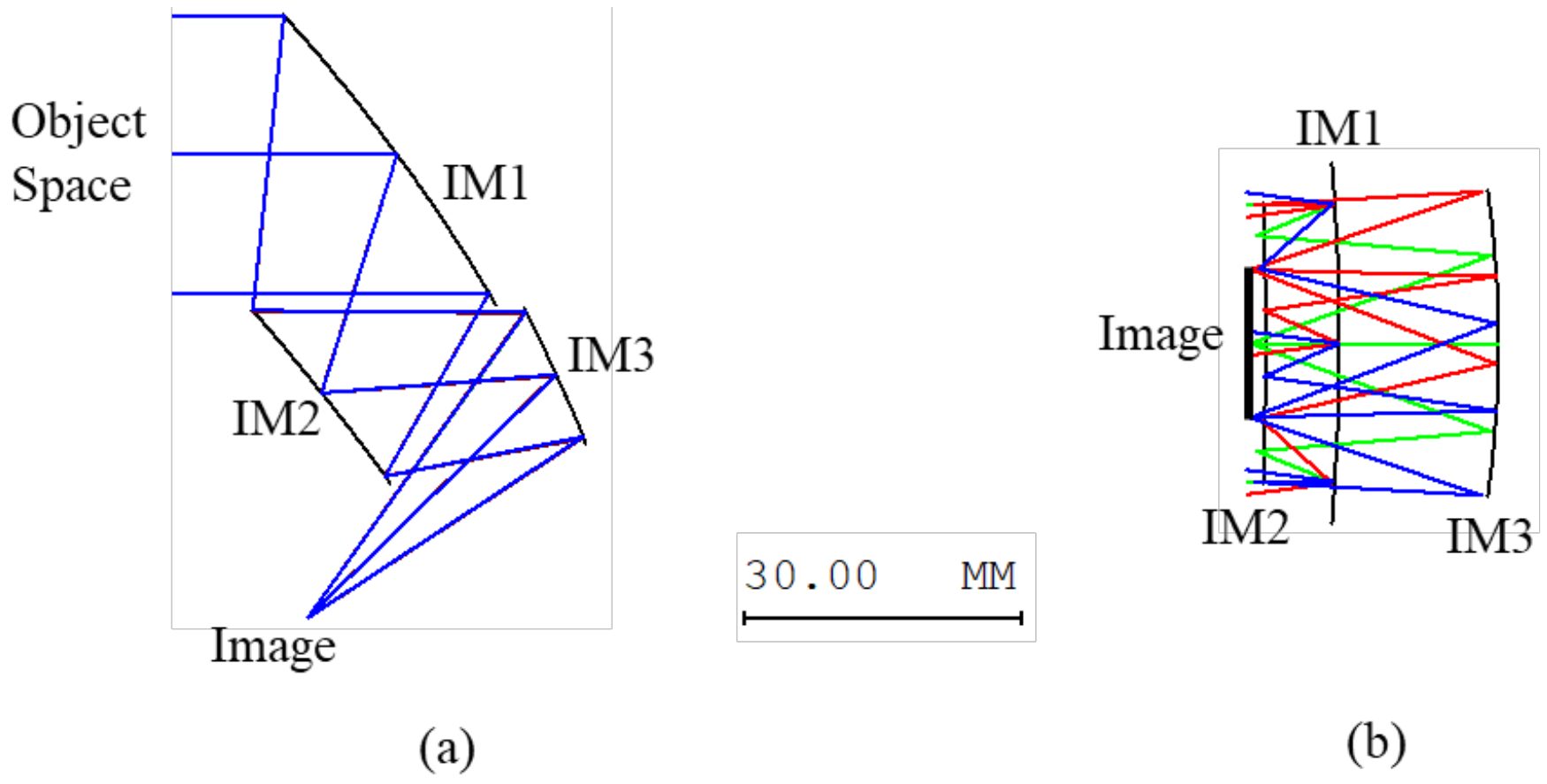

**Figure 7.2. The layout of the final reflective triplet imager design for (a) YZ plane and (b) XZ plane.**

### 7.3.2 The spectrometer design

The design form of the spectrometer is a freeform reflective triplet used in double-pass. The spectrometer utilizes a planar reflective grating as the dispersive element, which also serves as the aperture stop. The light coming from the slit interacts with each powered mirror before being dispersed by the grating, then interacts again with each mirror on the return trip after diffraction. Like the imager design form, the spectrometer leverages the PNP power distribution with the spectrometer Mirror 1 (SM1), Mirrror 2 (SM2), and Mirror 3 (SM3).

Similar to the design process of the imager, the spectrometer design was started by setting up a three-mirror unobscured imager design with sagittal fields only. The double-pass was modeled by placing the slit at the focal point of the imager and using a plane mirror to reflect the outgoing collimated light back into the imager to form an image at the slit location. Then, by changing the plane mirror into a linear grating, the image becomes a spectrum. The reflective grating operates at the -1 order with a constant line spacing.

Similar to the imager design process, the surface type was changed to off-axis conic, and the system was optimized without freeform departure to reach F/2 and 15-mm slit length. The spectrum length specification was met by varying the grating groove density. Additional fields were added at this stage to fully simulate the slit width. Since the incoming beam and the diffracted beam hit different parts of the surfaces, the vertices of the surfaces were kept near the center of the illuminated area for each surface. As the design stabilized (i.e., the design met all constraints and generally remained the same with consecutive optimization cycles), freeform departure was added.

The optimization of the system was performed in CODE V without Global Synthesis since the structure of the design cannot be drastically changed. The error function consisted of the default CODE V error function for imaging quality optimization, combined with weighted constraints on distortion and a square-sum penalty on the surface freeform coefficients. The first-order specifications such as volume, F-number, and image size were prioritized, and constraints such as for distortion, square-sum penalty, and telecentricity were added and tightened gradually.

The final design is shown in Figure 7.3, which has an average RMS spot size of 9.3 μm, smile distortion less than 2.5 μm, and keystone distortion less than 3 μm.

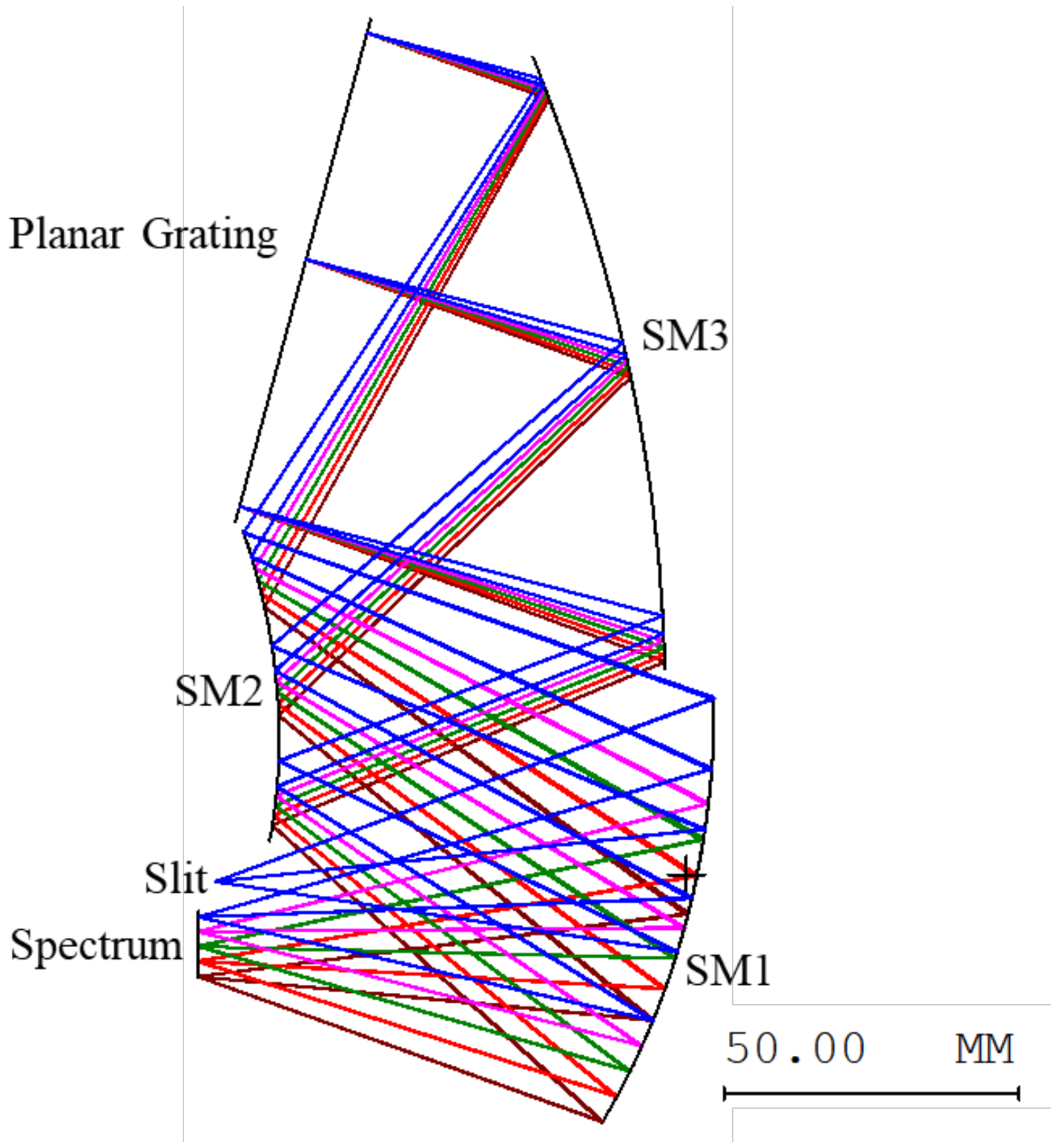

**Figure 7.3. The layout of the double-pass reflective triplet spectrometer designs.**

While the spectrometer was designed with Extended Fringe Zernike Polynomials that use off-axis conic as the base surface, the aberration coefficients in this work are derived using a spherical base surface. Therefore, to make this system applicable to the surface description used in the aberration theory, the surface type was changed from Extended Fringe Zernike Polynomials to Fringe Zernike Polynomials that use spherical base surfaces. This process was done while maintaining the imaging performance and all first-order specifications. The modified system is used as an example system for distortion analysis in Chapter 8. Note that the foundation of the aberration theory does not assume a surface shape description. For future work, Extended Fringe Zernike Polynomials as a surface type can be incorporated into the formulae of aberration coefficients.

### 7.3.3 System matching

To ensure a smooth combination of the imager and spectrometer, the ideal case is when the output rays of the first system can perfectly match the input rays of the second system at the plane of connection. In the ideal case, as illustrated in Figure 7.4(a), the two systems can be combined perfectly without any change to the image quality of each individual design. However, it is not practical to reach the ideal case. In practice, certain first-order quantities are matched as an approximation of a perfect match. For example, if the plane of connection is located in a space where the light beam is collimated in both systems, the two systems can be matched in the first-order by matching the exit pupil of the first system with the entrance pupil of the second system, as shown in Figure 7.4(b). If the plane of connection is located at an internal image plane of the total system and is also telecentric in both systems, the two systems can also be matched in the first-order by matching the F-number or numerical aperture of the two systems, as shown in Figure 7.4(c).

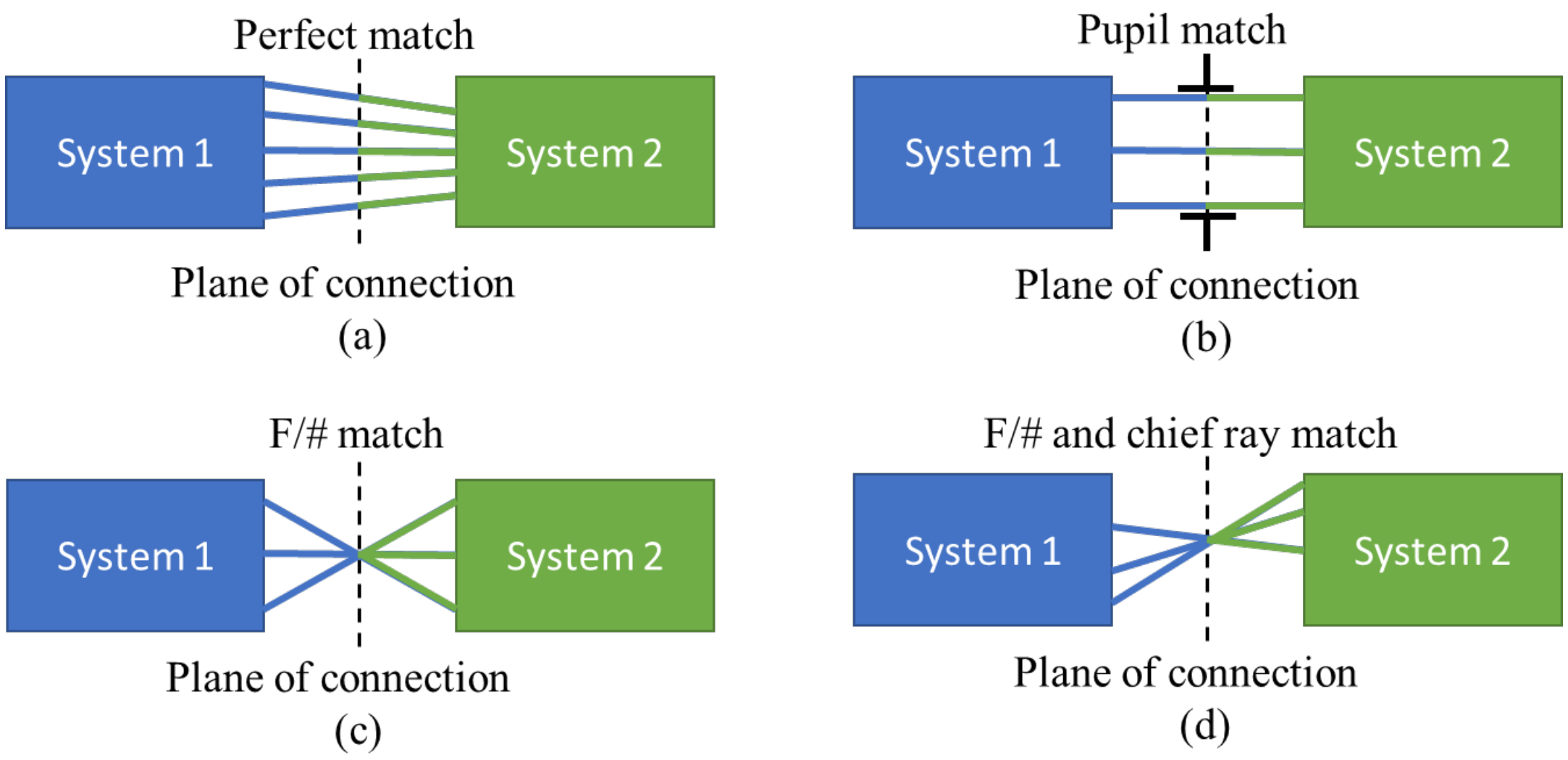

**Figure 7.4. Types of optical system matching: (a) perfect match for every ray, (b) pupil match in the space of collimation, (c) F-number match at an internal image in telecentric systems, and (d) F-number and chief ray match in a general internal image.**

The first-order quantities can be determined by tracing the marginal and chief rays through the system. Therefore, in general, it can be seen that in any plane of connection, two systems can be matched in the first order by matching the marginal and chief rays of two systems. In the case of the hyperspectral imager design in this chapter, the plane of connection is at the slit, which is an internal image plane of the hyperspectral imager. The match of the image planes ensures that the marginal ray and chief ray heights are matched at the plane of connection. In order to fully match the marginal and chief rays, their ray angles also need to be matched. This can be done by matching the F-number and the chief ray angle of the two systems. Because the slit only has a significant length in the sagittal direction, only the sagittal chief ray angle needs to be matched. This matching method is illustrated in Figure 7.4(d). Both the imager and spectrometer were designed at F/2 for a matched F-number. The tangent of the sagittal chief ray angle was adjusted to 0.044 for the two components, which is close to being telecentric.

In addition to matching the rays, the imager can also be combined with the spectrometer in two orientations, as shown in Figure 7.5. The aberration balance is different in each orientation resulting in the variation in the spectrometer geometry shown in Figure 7.5. Results show that both orientations have similar image quality and distortion performance, yet orientation 2 is more compact. Orientation 2 was chosen for the final design due to the larger clearance between the imager and the detector. The compactness gain also supports the choice for this geometry.

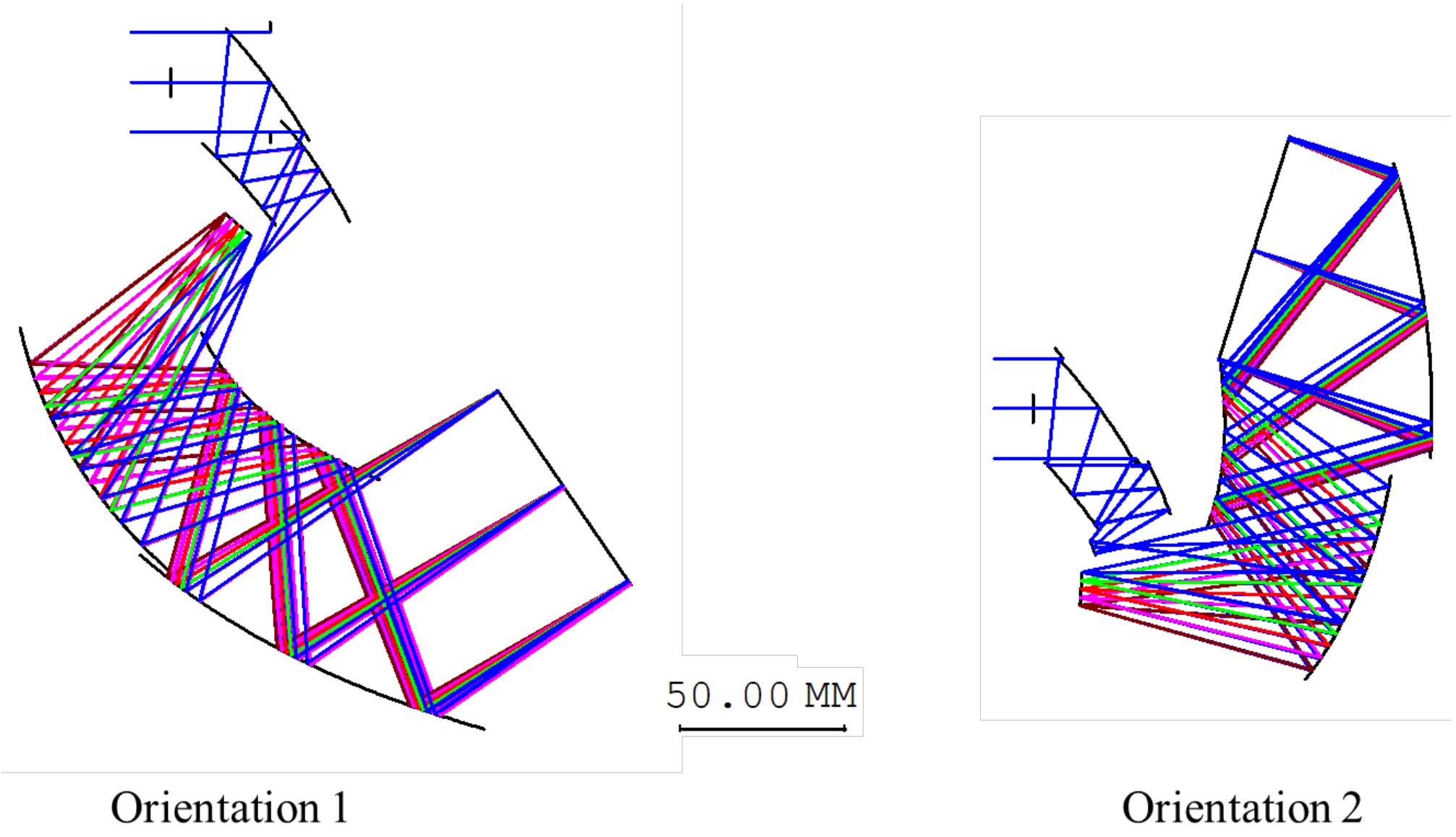

**Figure 7.5. The two orientations of the imager to be combined with the spectrometer.**

### 7.3.4 Combined system optimization

After matching the two systems, a systemwide optimization was conducted to reach the final performance. The system stop location was kept at the first mirror to minimize the stray light level. In the optimization process, the grating was first constrained to be conjugate to the stop at the primary mirror. The conjugate constraint was then relaxed in the final optimization. The telecentricity was constrained to be less than 5 degrees in the image space to lower the incident angle on the detector. During the systemwide optimization, the imaging performance at the slit was also part of the error function besides that at the final spectrum. Therefore, the imager maintained an average RMS spot size of 9.8 μm and smile distortion of less than 2 μm.

## 7.4 Final design and performance of the combined system

### 7.4.1 Design layout and performance

The final optical design layout is shown in Figure 7.6, with a volume of $91\times161\times116$ mm$^3$, which can be fully contained in a 3U format. In a 4U CubeSat, the system can fit comfortably with ample space for additional mechanics or electronics. At the final spectrum, the spot size is generally uniform across all fields and wavelengths, as shown in Figure 7.7, and the average RMS spot size is 5.5 μm, about 28% of the pixel size. The smile distortion is kept under 25% of the pixel, and the keystone distortion is under 15% of the pixel size. The final grating dimension is 90.4 mm along the groove direction and 71 mm perpendicular to the groove direction. The groove density of the grating is 43.67 line/mm.

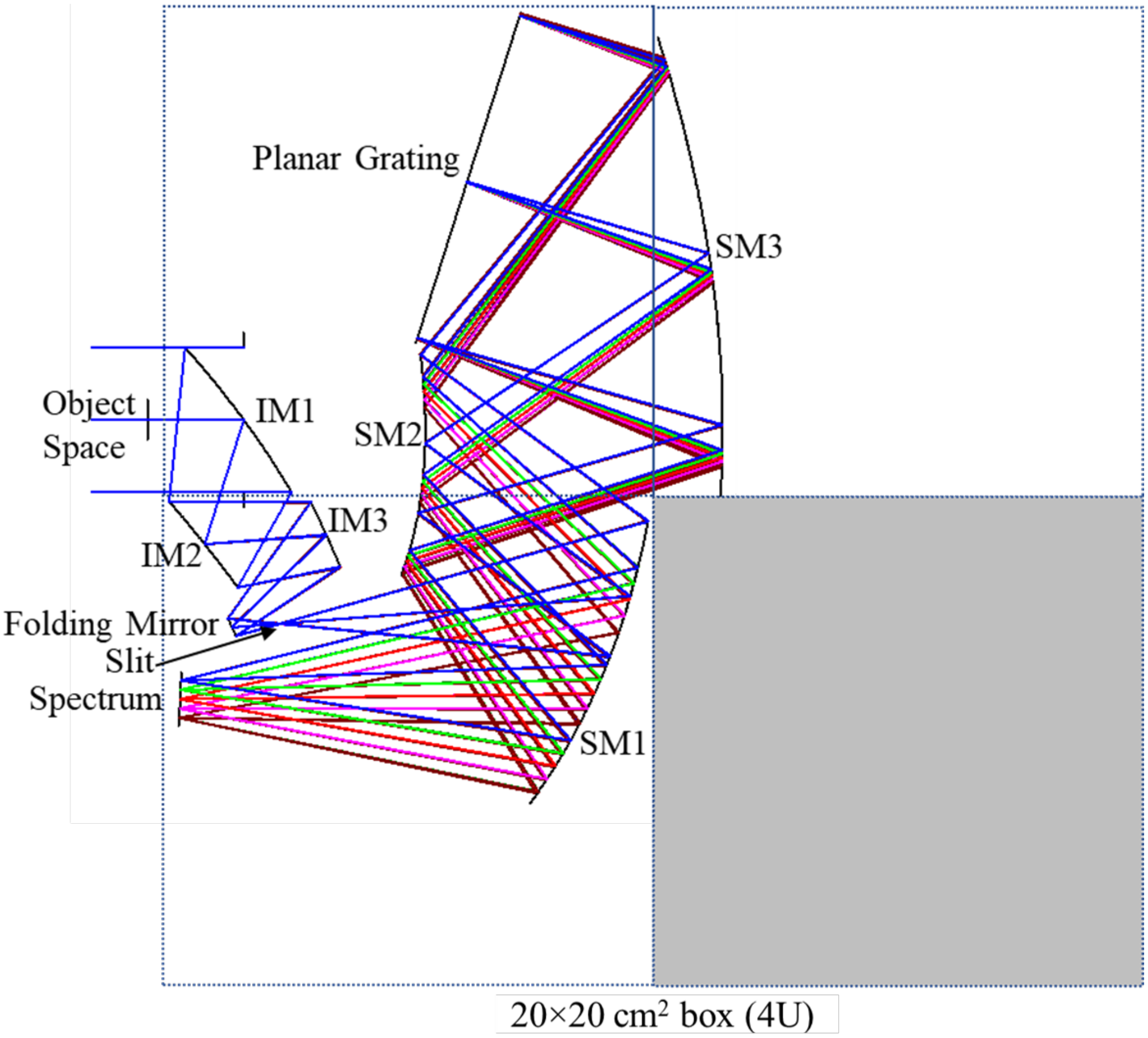

**Figure 7.6. The optical layout of the hyperspectral imager design inside a 10×20×20 cm$^3$ box.**

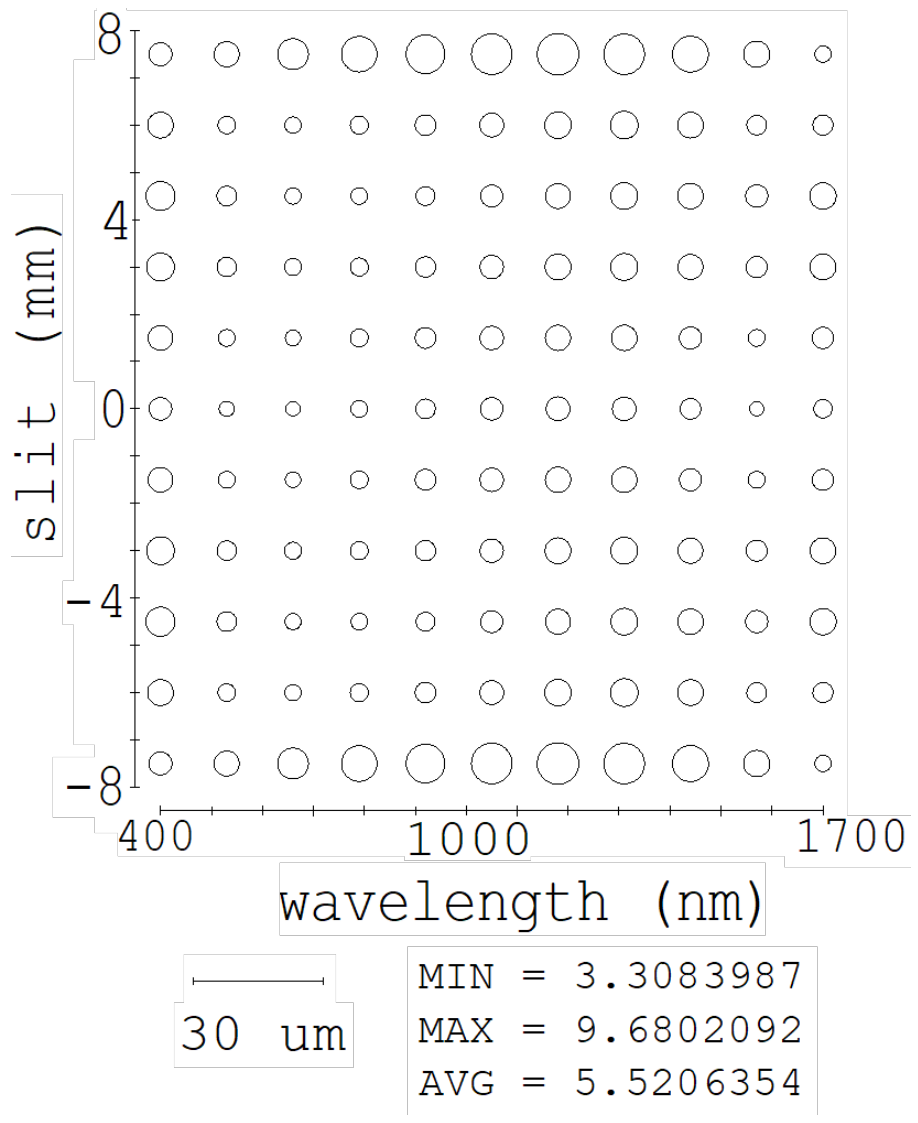

**Figure 7.7. The spectral full-field display for RMS spot size showing the performance over the slit and full spectrum for the combined system.**

### 7.4.2 Sag and slope departure of the optical surfaces

To get a basic understanding of the manufacturability of the resulting optical surfaces, we look at the freeform sag and slope departures. The sag and slope departures were calculated by sampling the surface at a grid of points. The spacing between points was chosen to be 0.05 mm, resulting in a slope value that was invariant within one digit of the value. Finer sampling gave refinement of less than one percent of the value.

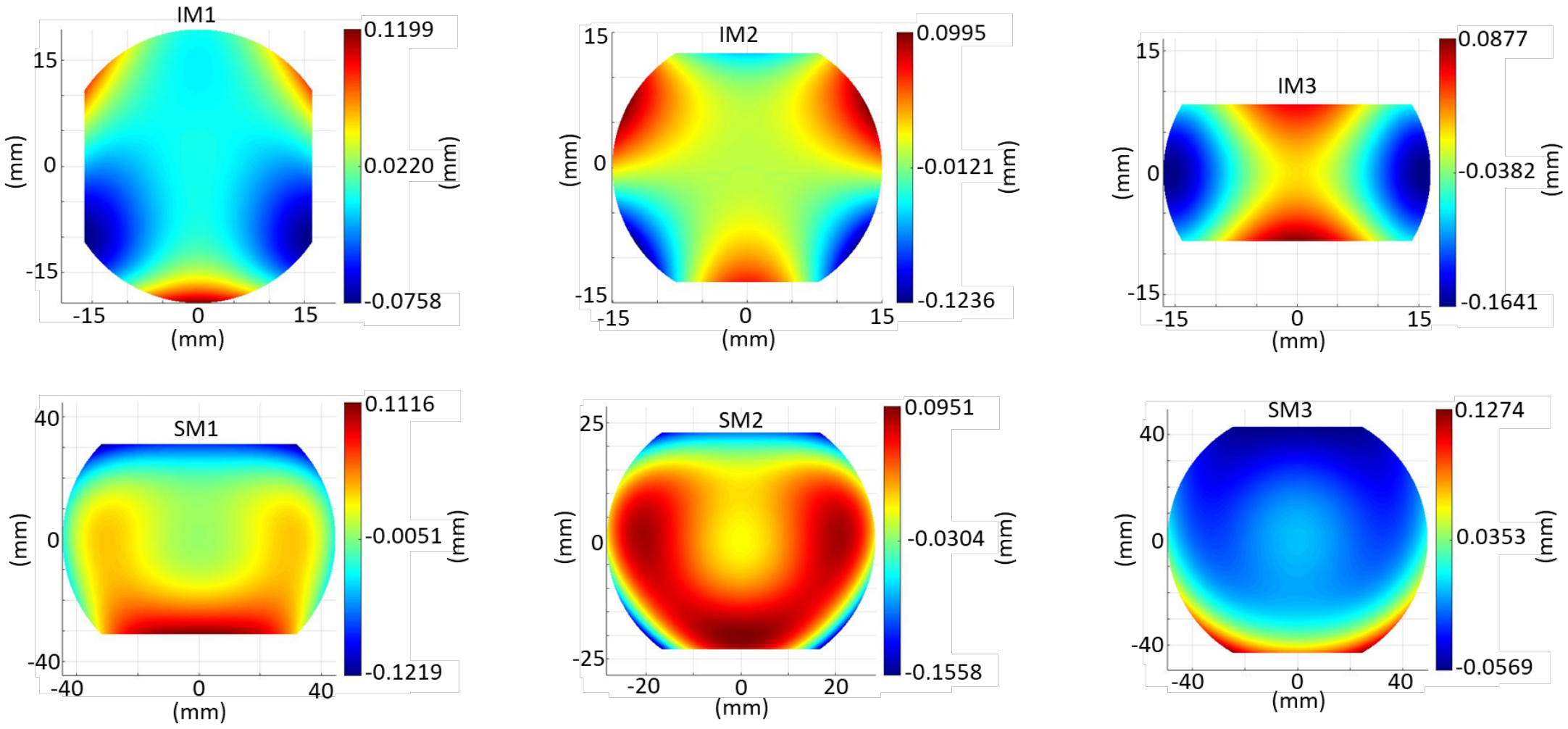

**Figure 7.8. The sag departure map of each mirror from the best-fit off-axis conic.**

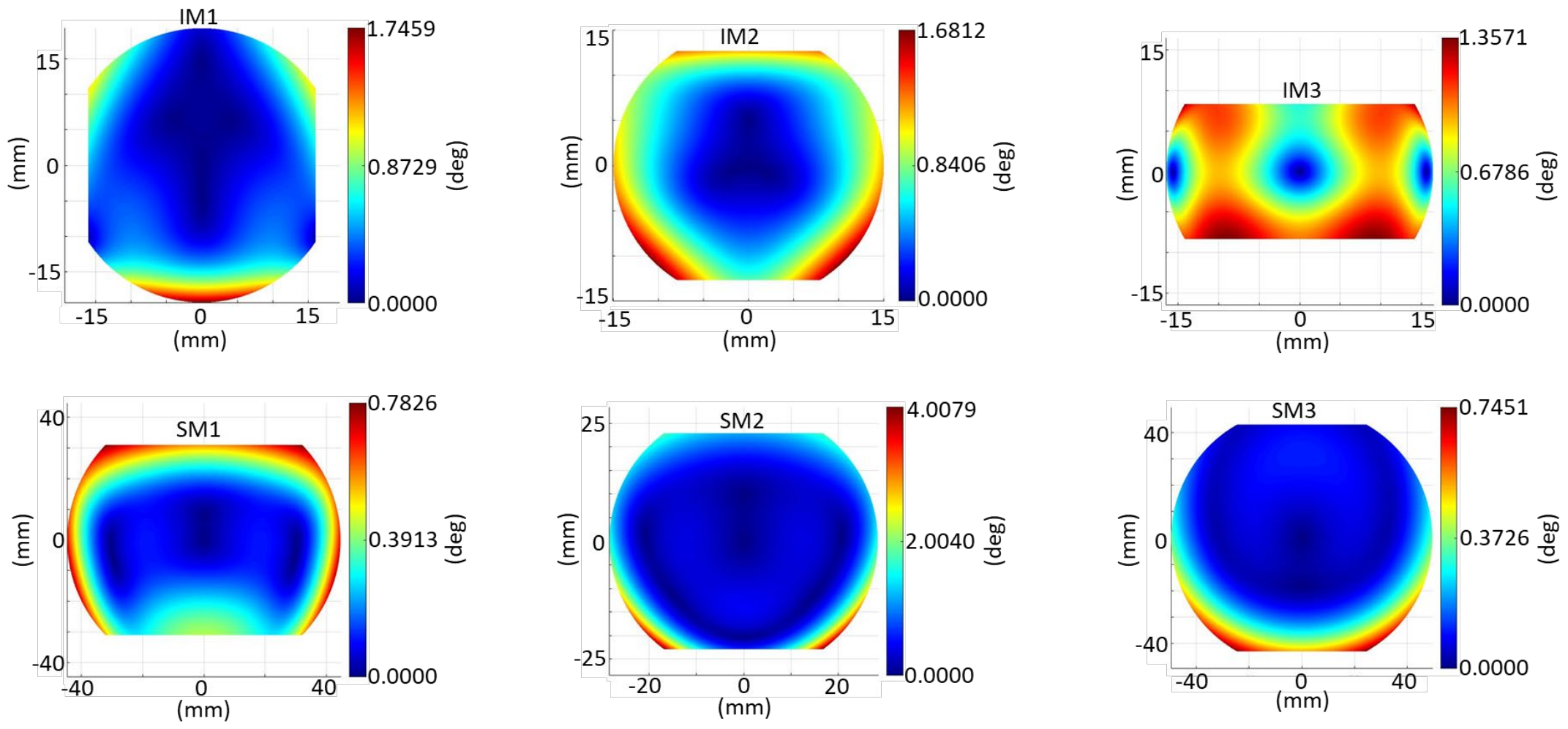

**Figure 7.9. The slope departure map of each mirror from the best-fit off-axis conic.**

Figures 7.8-7.9 show the sag and slope departure maps from the best fit off-axis conic of each mirror surface. The aperture of each mirror is circular, with portions cut out to display only the used area. The maximum peak-to-valley (PV) departure for all mirrors is about 250 μm, and the maximum slope departure is 4.0 degrees.

Table 7.2 shows the PV sag departure and the maximum slope departure for each mirror from the best-fit sphere and the best-fit off-axis conic. One might want to analyze the departure from best-fit off-axis conic because there is the potential to null out the off-axis conic during the surface metrology. The PV sag departure and the maximum slope departure are lower when measured from the best-fit off-axis conic than when measured from the best-fit sphere, especially for the mirrors in the spectrometer. For the current design, however, the residual freeform departure of all mirrors is too large to do a full-field interferometric test after nulling out the best-fit off-axis conic. To make the interferometric test viable, additional nulling optics can be used, such as computer-generated holograms (CGHs) or reconfigurable CGHs [78]. Additionally, other non-interferometric metrology methods can be used to measures the surfaces such as coordinate measuring machines (CMMs).

**Table 7.2. The PV sag departure and the maximum slope departure for each mirror**

| | From the Best-fit Sphere | | From the Best-fit Off-axis Conic | |
|---|---|---|---|---|
| | PV Sag Departure (μm) | Maximum Slope Departure (degree) | PV Sag Departure (μm) | Maximum Slope Departure (degree) |
| IM1 | 293 | 1.4 | 196 | 1.7 |
| IM2 | 323 | 2.2 | 223 | 1.7 |
| IM3 | 318 | 1.7 | 252 | 1.4 |
| SM1 | 908 | 2.3 | 234 | 0.8 |
| SM2 | 1302 | 5.1 | 251 | 4.0 |
| SM3 | 851 | 1.7 | 184 | 0.7 |

## 7.5 Sensitivity analysis

A Monte Carlo process was used to evaluate the as-built performance after random perturbations that emulate fabrication errors. The values of the tolerances were determined so that the as-built performance has above 95% chance to have an average RMS spot size contained within one pixel (20 μm), smile distortion < 8 μm, and keystone distortion < 5 μm.

The tolerances that were investigated for the mirrors and the grating are the decenter, tilt, and the surface figure error modeled by 25 Fringe Zernike polynomials. Image plane tilt (-3.6 – 4.2 arcmin range), decenter (-50 – 50 μm range), and despace (-40 – 30 μm range) were used as compensators.

For the figure error, each Zernike term was assigned a tolerance value that can perturb the surface shape. In this tolerancing process, all Zernike terms were assigned the same value to model a random figure error. If a particular type of figure error is more significant than other types in a known manufacturing process, the tolerance value should be adjusted to reflect the figure error. The resulting tolerance values are summarized in Table 7.3. The tolerance value for each Zernike term is 10 nm, the accumulative effect of all 37 Fringe Zernike terms can reach a maximum PV sag departure of about 0.3 μm by setting all coefficients to be the maximum tolerance. To represent a lower-order figure error, 25 terms of Fringe Zernike terms were also used instead of all 37 terms.

**Table 7.3. Tolerance values to reach the target as-built performance**

| Tolerances on the mirrors and the grating | Unit | Value |
|---|---|---|
| Decenter X | μm | 10 |
| Decenter Y | μm | 10 |
| Despace Z | μm | 10 |
| Tilt X | arcmin | 0.35 |
| Tilt Y | arcmin | 0.35 |
| Tilt Z | arcmin | 0.35 |
| Coefficient Change for 25 Terms of Fringe Zernike Polynomial | nm | 10 nm for each term (maximum PV ~ 0.12 μm) |

Figure 7.10 shows a histogram of PV values for 50000 simulated figure errors, from which it can be seen that the maximum PV value is about 0.12 μm.

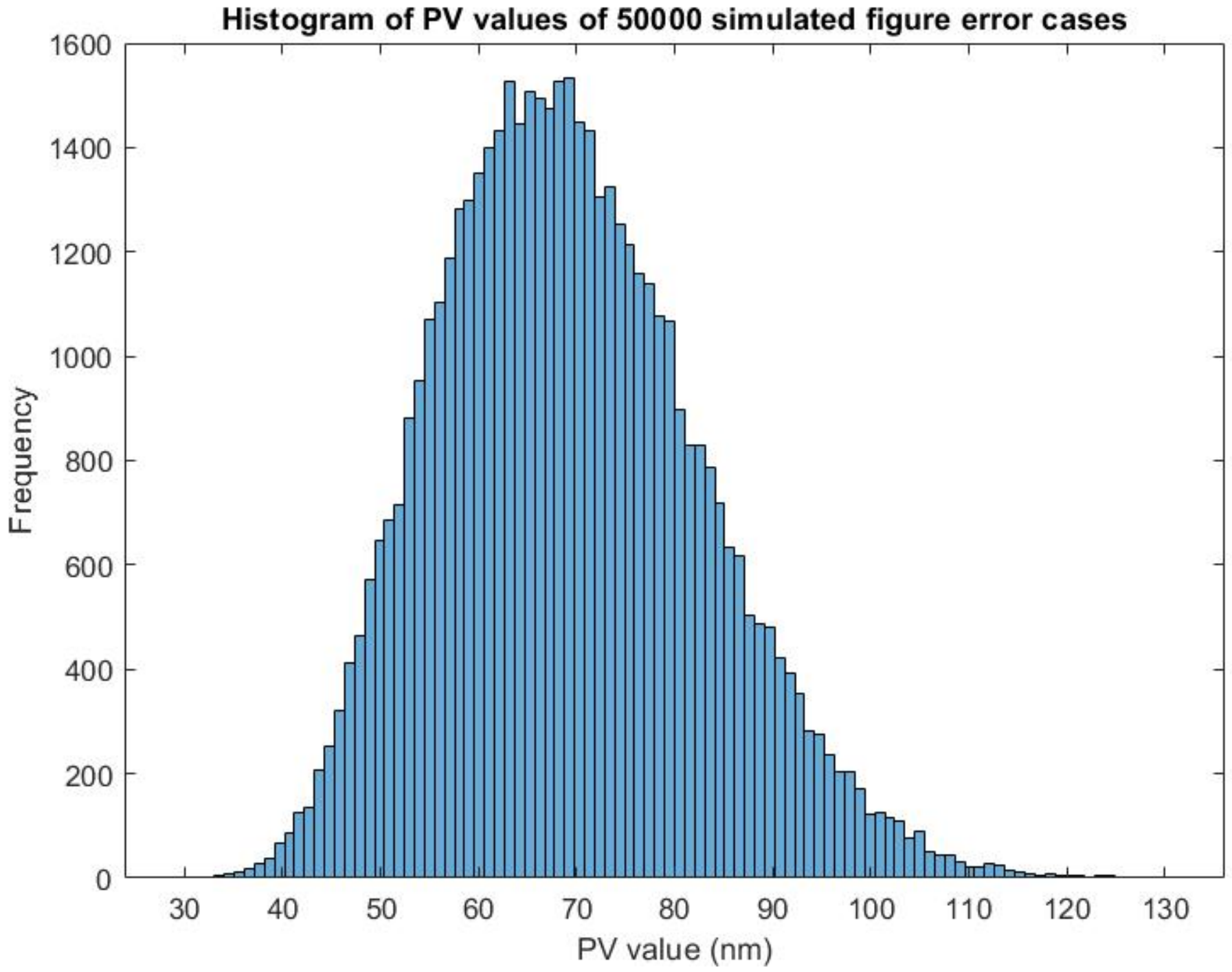

**Figure 7.10. Histogram of PV values of 50000 simulated figure error cases using 25 Fringe Zernike polynomials.**

## 7.6 Stray light analysis and spectrum sorting

Two sources of unwanted light on the detector were investigated: the stray light that interacts with the surfaces in the incorrect order and the light from unwanted diffraction orders of the grating. Note that a full stray light analysis requires the inclusion of mechanical features of the design. Since the mechanical design is beyond the scope of this optical design study, a brief discussion of the stray light sources in the optical design was done to indicate that the parasitic rays coming from the optics themselves and that are amplified when the optical design is compacted, can be managed.

### 7.6.1 Stray light that interacts with the surfaces in the incorrect order

A stray light analysis was performed in LightTools on the solid model of the design, as shown in Figure 7.11. A physical aperture bigger than the beam footprint was created in front of the first mirror, which all light entering the system must pass through. The location and the size of the physical aperture are illustrated in Figure 7.12(a). It was also assumed that the imager housing and the spectrometer housing were separated and only connected with the slit, which essentially acts as a field-stop with regards to limiting the stray light. The analysis assumes smooth optical surfaces (i.e., scatter-free) and that the light missing optical surfaces will be absorbed by the housing. Therefore, any light that goes through the physical aperture, the slit, and lands on the detector in the incorrect optical surface order can be regarded as stray light. With 40 million rays traced backward from the detector with uniform angular distribution across the whole wavelength range, not a single stray light ray path was found.

The main factor for the low stray light from specular reflections is that the slit serves as an effective field stop to block unwanted light. In general, the aperture stop being at the first mirror of the imager generates a lower stray light level compared with the case where the stop is inside the spectrometer. If the stop is set inside the spectrometer, some unwanted light can still enter the spectrometer before blocked by the stop, thus becoming seeds for stray light if not absorbed.

When expanding the stray light analysis to consider surface scatter, particular attention during fabrication should be given to reduce the surface roughness of the optics and the reflectivity of the housing interior. Furthermore, the residual particles inside the system can be minimized to mitigate scatter by assembling the system in a class 100 clean room.

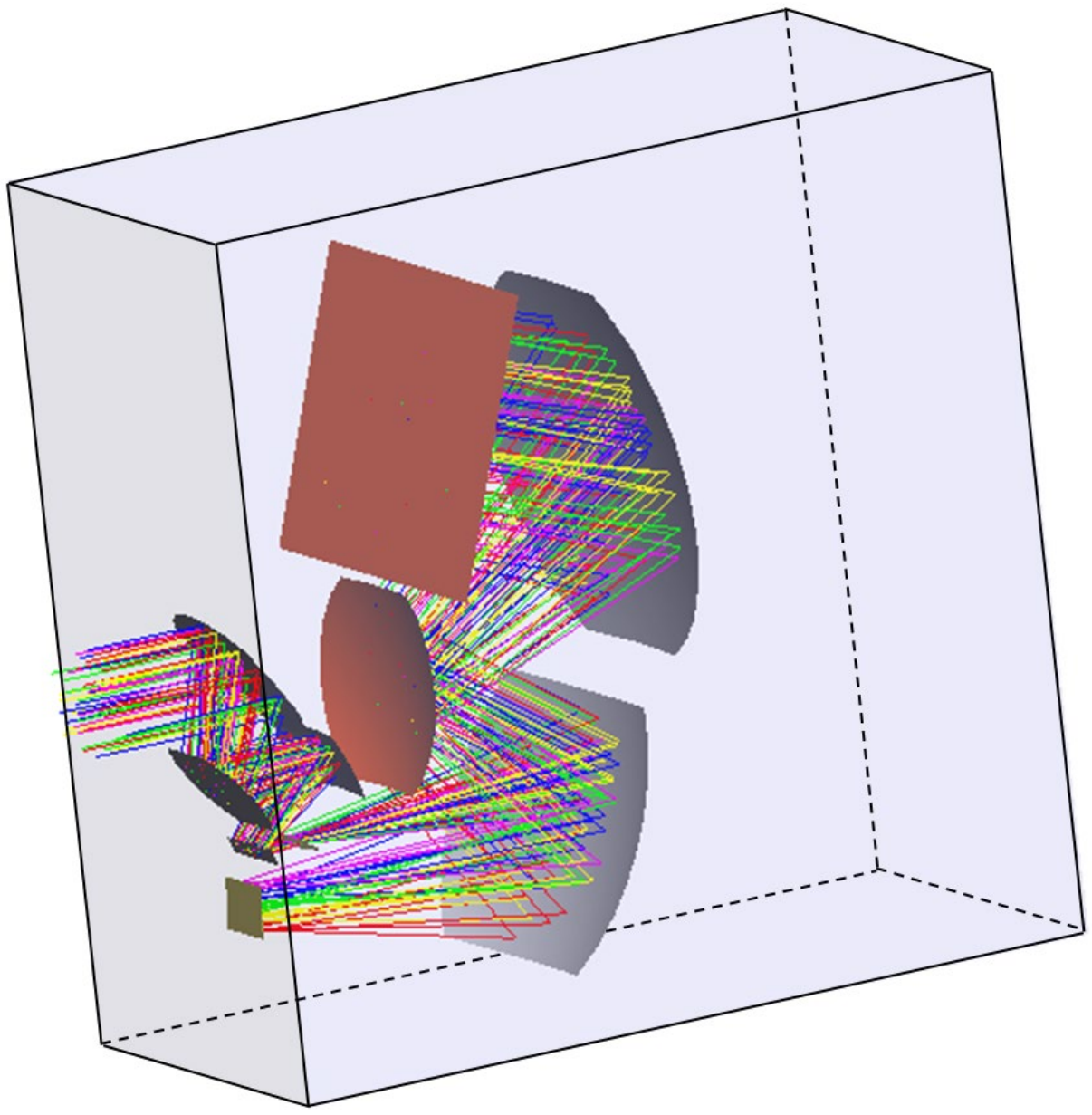

**Figure 7.11. The 3D model of the hyperspectral imager in LightTools (the blue cuboid indicates a 4U CubeSat).**

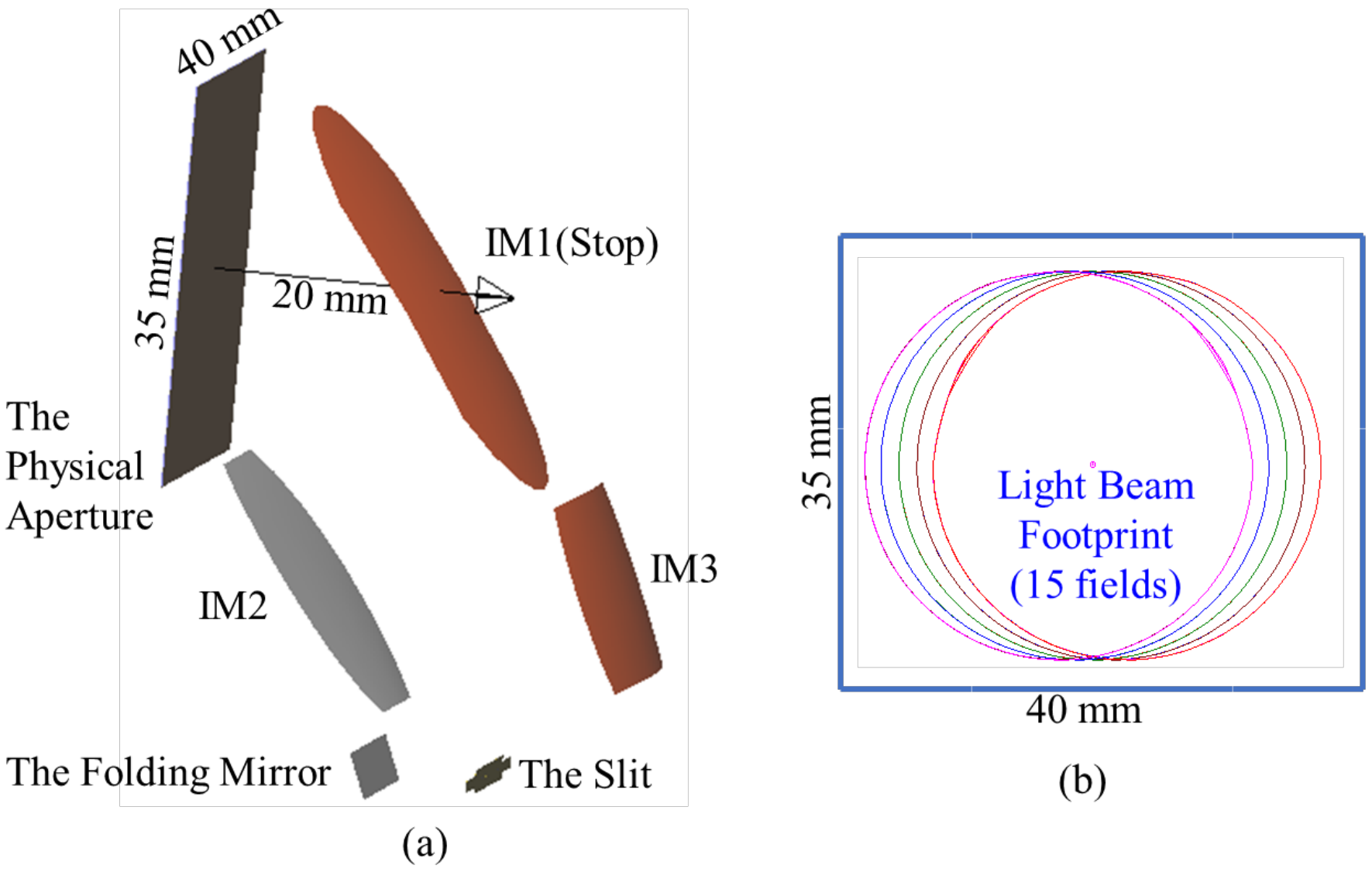

**Figure 7.12. (a) The physical aperture size and location with respect to the first mirror in the imager, (b) the size comparison between the physical aperture and the beam footprint on the aperture.**

### 7.6.2 Spectrum sorting

The hyperspectral imager was designed for a wavelength range of 400 – 1700 nm. The ratio of the upper to lower wavelength is larger than two, which leads to an overlap of higher-order spectra with the first-order spectrum. Since only the first-order spectrum is wanted, spectra of higher orders must be removed. There are two ways to avoid higher-order spectra. The first way is to reduce the efficiency of the grating on higher orders. The second way is to place filters on the detector to filter out the higher-order spectra.

Figure 7.13 illustrates how spectra of different orders are distributed on the detector. The green and orange lines represent two types of high-pass filters that can be used for spectrum sorting. The green step lines represent multiple high-pass filters used in conjunction. Because the wavelength of the second-order spectrum at the end of the detector (850 nm) is higher than the wavelength of the first-order spectrum at where the second-order spectrum starts (800 nm), the spectrum sorting cannot be achieved by using a single high-pass filter. If the upper wavelength was allowed to be adjusted from 1700 nm down to below 1600 nm, a single high-pass filter could be enough to filter out all higher-order spectra. The orange line in Figure 7.13 represents a high-pass linear variable filter that can be used for spectrum sorting. The cutoff wavelength of a linear variable filter changes linearly with spatial positions. By placing the line of cutoff wavelength between the first-order spectrum and higher-order spectra, the spectrum sorting can be achieved with one high-pass linear variable filter. Important factors that affect the manufacturability of linear viable filters are the change rate of the cutoff wavelength and the blocked wavelength range.

Spectrum sorting filters were not integrated during the design process since specific parameters of the filters have not been decided. To estimate the impact of these filters on the imaging qualities, a 2-mm thick NBK7 cover glass was inserted in the final design, 5 mm away from the image plane, to simulate the worst-case scenario of adding the filter after the design was complete. After refocusing, the resulting average RMS spot size increased from 5.5 µm to 6.3 µm, which is not a significant change compared to the 20 µm pixel size. Figure 7.14 shows a spectral full-field display for the RMS spot size of the design with the cover glass inserted. The distortion performance remained the same since the design is nearly telecentric at the image plane.

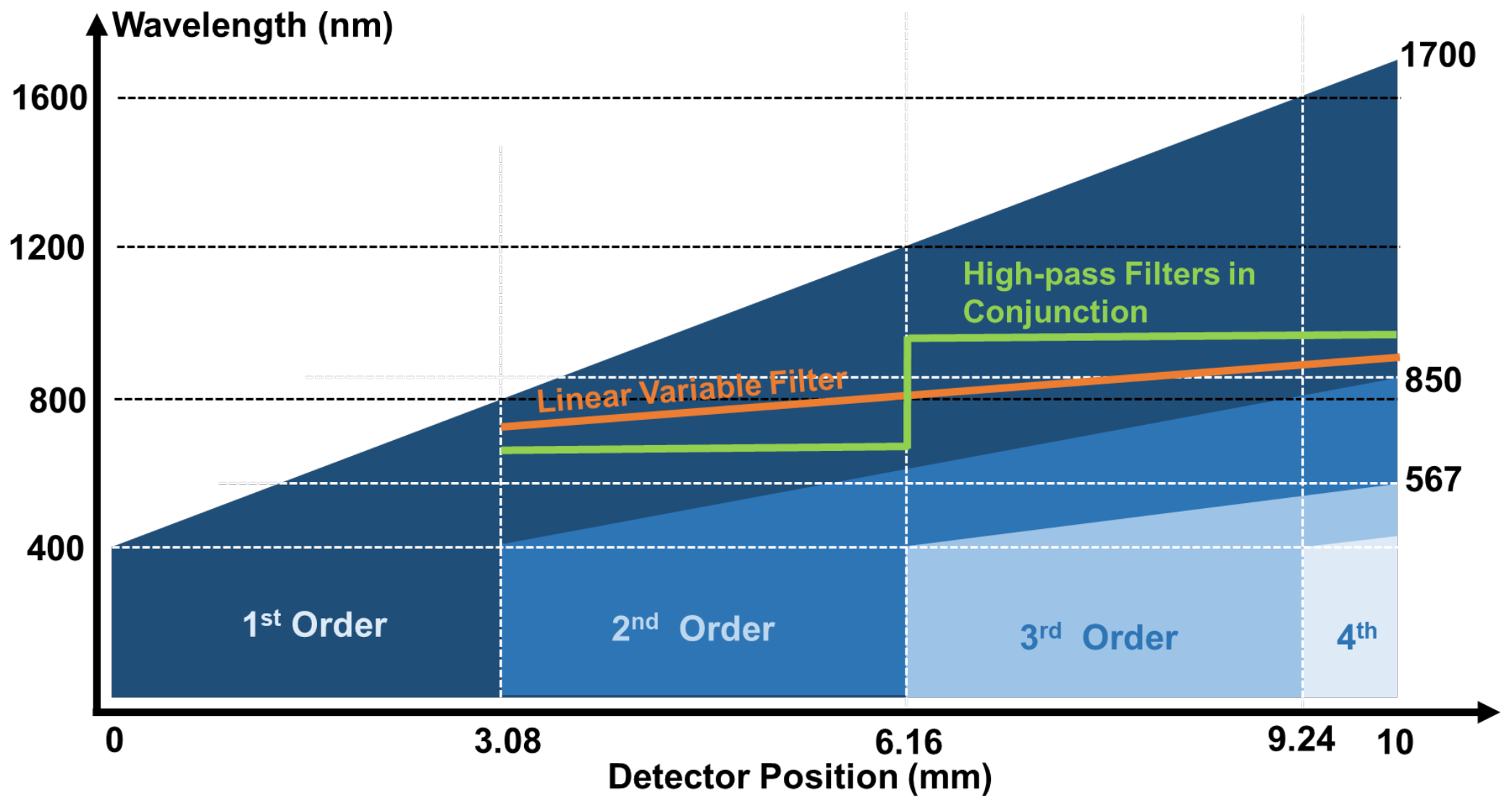

**Figure 7.13. Illustration of the overlap of spectra of different orders and the potential spectrum sorting filters.**

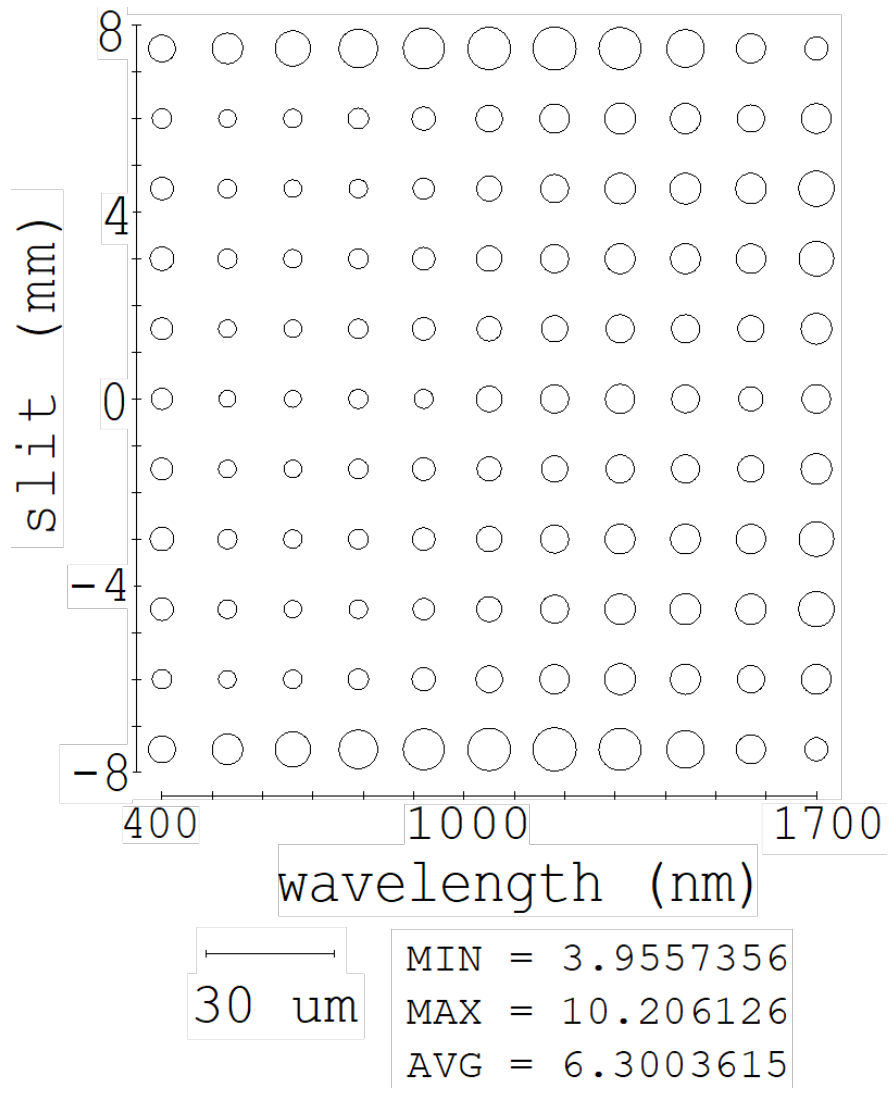

**Figure 7.14. The spectral full-field display for RMS spot size of the design with the cover glass inserted.**

# Chapter 8. Comparison between analytical and raytracing results in spectrometers

In Chapter 4, the distortion behavior in plane-symmetric spectrometers is described by X and Y spectral distortion, $SD_x$ and $SD_y$, defined in Equations 4.45-4.46. Within the scope of third-group wavefront aberrations, $SD_x$ is related to a magnification change between wavelengths and $W_{31100}$, and $SD_y$ is related to $W_{21001}$. In Chapters 5-6, $W_{21001}$ and $W_{31100}$ at the exit pupil of a one-surface system are derived analytically including the effect of induced aberrations. The wavefront aberration of a multi-surface system can then be calculated one surface by one surface, where wavefront aberrations at the exit pupil of one surface is treated as wavefront aberrations at the entrance pupil of the next surface. Therefore, the third-group related $SD_x$ and $SD_y$ in a multi-surface system can be analytically calculated. In this chapter, the results of $SD_x$ and $SD_y$ from analytical calculation and real raytracing are compared for a Dyson spectrometer and three freeform three-mirror double-pass spectrometers.

In addition, since $W_{02002}$ is a second-order aberration and can be calculated from paraxial optics, we also compare $W_{02002}$ calculated from Equation 6.58 and from paraxial optics. Recall that the five indexes in the subscript of the aberration coefficients correspond to the sum of the powers of $H_x$ and $H_y$, the sum of the powers of $p_x$ and $p_y$, the power of the dot product ($\boldsymbol{H}\cdot\boldsymbol{\rho}$), the power of $H_y$ (i.e., ($\boldsymbol{i}\cdot\boldsymbol{H}$)), and the power of $p_y$ (i.e., ($\boldsymbol{i}\cdot\boldsymbol{\rho}$)), respectively.

## 8.1 Calculation of $W_{02002}$ with paraxial optics

The wavefront aberration coefficient $W_{02002}$, referred to as the field-constant astigmatism, does not show dependence over the field and therefore is constant of the field. It can be calculated from the difference in paraxial image locations in the sagittal and tangential directions for the object point on the OAR.

Consider the two paraxial image points shown in Figure 8.1. $O_{IM}$ and $O_{IMt}$ are the sagittal and tangential paraxial image locations for the on-OAR object. $\Delta s'$ denotes the distance between $O_{IM}$ and $O_{IMt}$ and is positive when $O_{IMt}$ is on the positive Z-axis of the image coordinate system at $O_{IM}$. Note that $u'_{ax}$ is the sagittal marginal ray angle in image space, while $u'^{(XP)}_{ay}$ is the ray angle of the tangential ray that passes through the positive edge of the exit pupil from the on-OAR object.

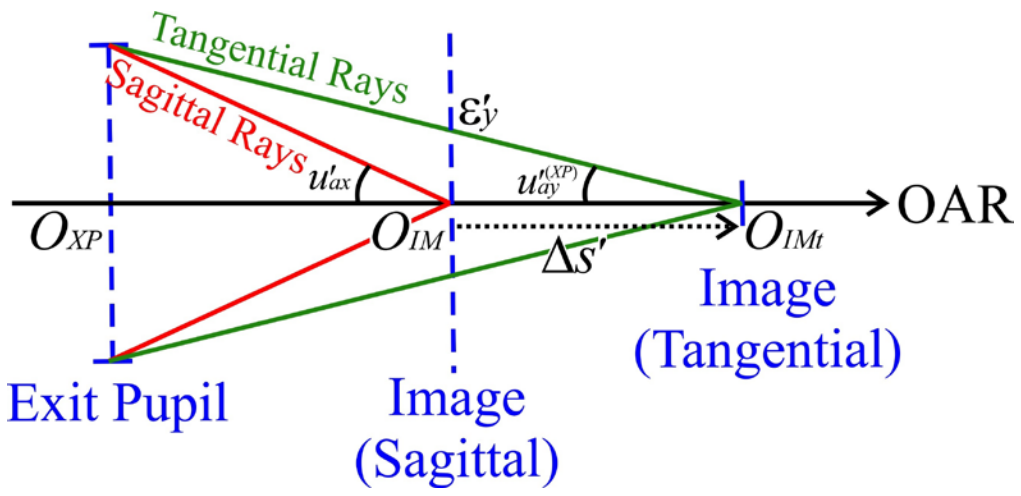

**Figure 8.1. Illustration of paraxial image locations in the sagittal and tangential directions. The sagittal and tangential rays are drawn in the same plane for illustration purpose.**

Since the ray angles in image space are proportional to the normalized pupil coordinate, the transverse aberration in the tangential direction, $\varepsilon'_y$, caused by the paraxial image location difference in sagittal and tangential directions can be expressed as

$$\varepsilon'_y = -u'^{(XP)}_{ay} \Delta s' \rho_y \tag{8.1}$$

Using Equation 4.35, the corresponding wavefront aberration can be derived as

$$W = -\frac{1}{2} n' u'_{ax} u'^{(XP)}_{ay} \Delta s' \rho_y^2 \tag{8.2}$$

Therefore, it can be seen that

$$W_{02002} = -\frac{1}{2} n' u'_{ax} u'^{(XP)}_{ay} \Delta s' \tag{8.3}$$

which can be used to calculate $W_{02002}$ directly from paraxial optics. In following example systems, the result from Equation 8.3 will be compared to the result from Equation 6.58 derived in Chapters 5-6, provided again here for convenience as

$$W^{(XP)}_{m02002} = \frac{\Psi x_a W^{(SI)}_{02002}}{\Psi x_a + 2 x_b W^{(SI)}_{02002}} \tag{6.58}$$

## 8.2 Example system: Dyson spectrometer

First, a relatively simple spectrometer, a Dyson spectrometer, was designed to compare the aberration results acquired from analytical derivation and real raytracing. The system is set up in CODE V, where the real raytracing is performed.

### 8.2.1 System specification and description

The Dyson spectrometer consists of a silica plano-convex lens and a reflective concave grating, as shown in the 2D layout of the Dyson spectrometer in Figure 8.2. The first-order spectrum of the grating ($m$=1) is used to form the spectrum image.

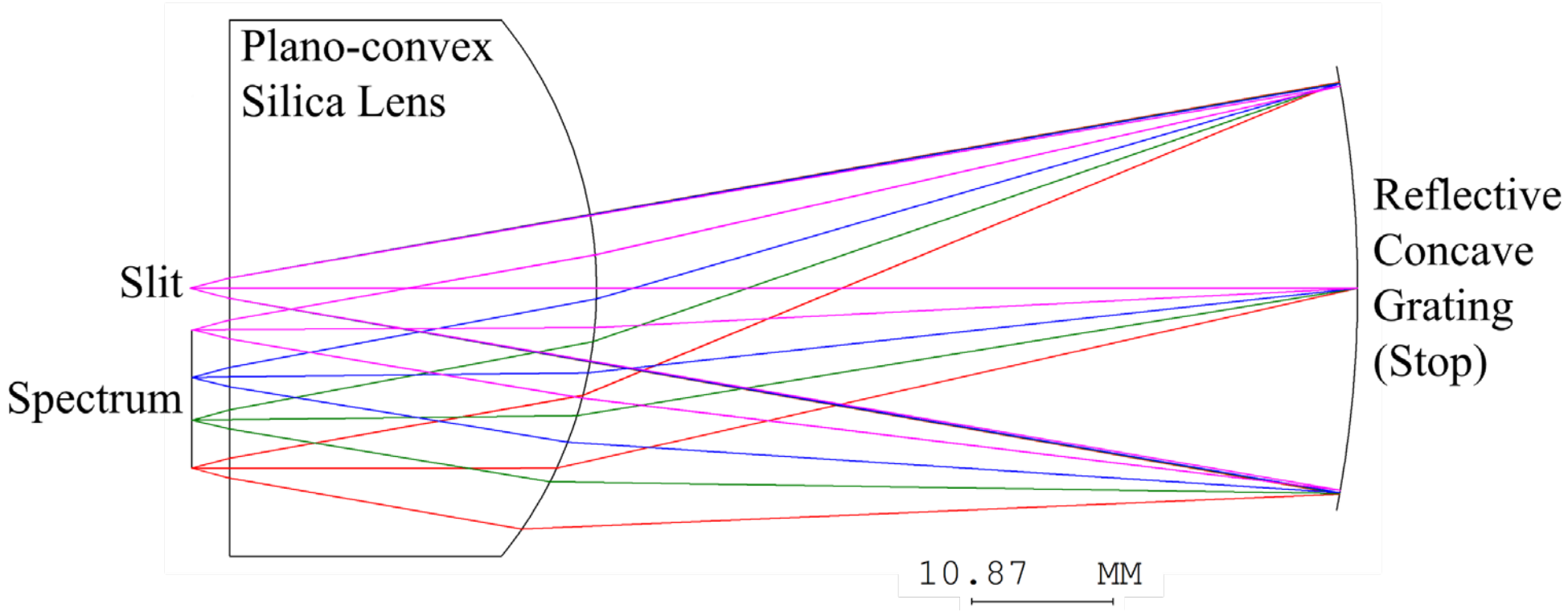

**Figure 8.2. 2D layout of the Dyson spectrometer in the tangential plane.**

The main specifications are described in Table 8.1. Note that the slit width in the tangential direction is not specified, because the width is normally in the magnitude of tens of microns, which does not make the aberration behavior significantly different between the object points in different width positions. Therefore, in this example, we focus on the object points with zero height in the tangential direction.

**Table 8.1. Specifications of the Dyson spectrometer**

| Specifications | Value |
|---|---|
| Image F-number (sagittal direction) | F/2 |
| Wavelength (nm) | 400-1700 |
| Full slit length (sagittal direction) (mm) | 10 mm |
| Grating groove density (lp/mm) | 128.7 |

The system description is detailed in Table 8.2. Note that although during the derivation of the aberration coefficients, the optical system is described following the OAR as discussed in Chapter 2, the system description in Table 8.2 follows how it is set up in CODE V for simplicity [79]. The description methods do not change the structure of the system, and they can be converted from one to another. The optical surfaces of the Dyson spectrometer, including the object and image planes, are perpendicular to an axis shown in

Figure 8.3. Therefore, all surfaces are described sequentially following the axis. The system is double-pass, meaning that the light passes through the silica lens twice, before and after the grating. Therefore, only the optical surfaces in the first pass are needed to describe the whole system.

**Table 8.2. System description of the Dyson spectrometer**

| Surface | Surface Type | Radius of Curvature (mm) | Thickness $t$ (mm) | Material |
|---|---|---|---|---|
| $S_0$ (Object) | Spherical | Infinity | 3.000 | Air |
| $S_1$ | Spherical | Infinity | 27.858 | SILICA_SPECIAL |
| $S_2$ | Spherical | -32.208 | 57.792 | Air |
| $S_3$ (Grating/Stop) | Spherical | -90.000 | - | - |
| Image | Spherical | Infinity | -2.870* | - |

*The thickness value on the Image row is the thickness of the last surface before Image ($S_1$ in this case)

The parameters in Table 8.2 are illustrated in Figure 8.3. Each surface is labeled and has a local surface coordinate system. For this system, all surfaces are spherical with no freeform departure, and the spherical surfaces with an infinite radius of curvature are equivalent to plano surfaces. The slit is in the sagittal direction and passes through the origin of the $S_0$ coordinate system. The radius of curvature is positive if the center of the curvature is on the positive Z-axis. The thicknesses of surfaces before the image, labeled as $t_0$, $t_1$, and $t_2$ in Figure 8.3, are the Z-axis coordinate shift needed for the current surface coordinate system to reach the next surface coordinate system, and is positive if it is towards the positive Z-axis. For the image plane, the thickness number corresponds to the thickness of the last surface before the image plane, which is also $S_1$ in this system due to the second pass. This thickness is labeled as $t_{im}$ in Figure 8.3. The material of a surface is between the surface and the next surface.

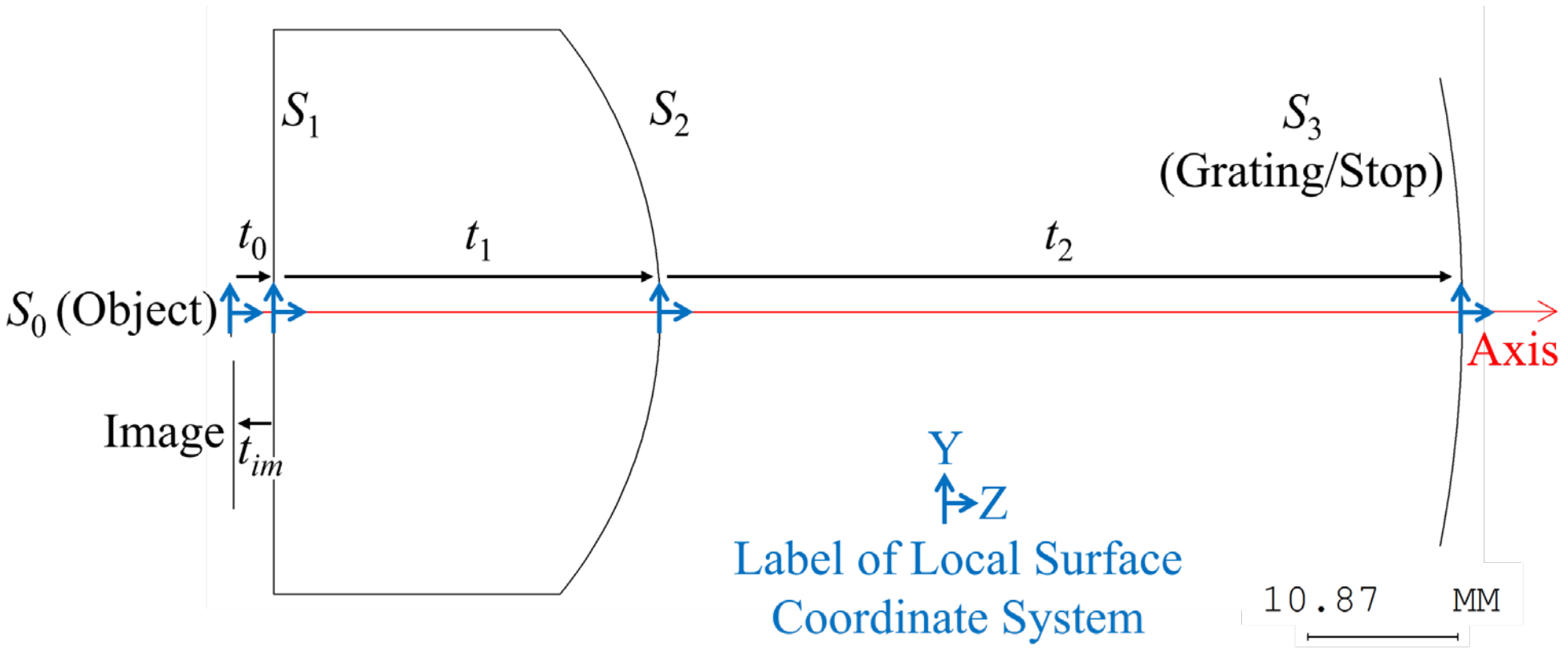

**Figure 8.3. Illustration of surface labels ($S_0$, $S_1$, $S_2$, $S_3$ and Image), surface coordinate systems and surface thicknesses ($t_0$, $t_1$, $t_2$ and $t_{im}$) of the Dyson spectrometer.**

The material code for the silica lens, between $S_1$ and $S_2$, is SILICA_SPECIAL in CODE V. The refractive index of SILICA_SPECIAL can be described with the Standard Sellmeier dispersion formula shown as

$$n^2(\lambda) = 1 + \sum_{i=1}^{6} \frac{B_i \lambda^2}{\lambda^2 - C_i^2} \tag{8.4}$$

where $B_i$ and $C_i$ are coefficients specific to materials. The unit of the wavelength, $\lambda$, is µm. The $B_i$ and $C_i$ coefficients for SILICA_SPECIAL are listed in Table 8.3 [79].

**Table 8.3. The $B_i$ and $C_i$ coefficients for the Standard Sellmeier dispersion formula of SILICA_SPECIAL**

| Coefficient | Value | Coefficient | Value |
|---|---|---|---|
| $B_1$ | 0.1945773949 | $C_1$ | 0.1192472768 |
| $B_2$ | 0.1485676411 | $C_2$ | 0.1.186813304 |
| $B_3$ | 0.7609636424 | $C_3$ | 0.07175045384 |
| $B_4$ | 0.0008818576643 | $C_4$ | 4.641180337 |
| $B_5$ | 0.9357657834 | $C_5$ | 10.12002016 |
| $B_6$ | Not Provided | $C_6$ | Not Provided |

### 8.2.2 Comparison of $W_{02002}$ results

As discussed in Sections 6.2 and 8.1, $W_{02002}$ can be calculated from Equation 6.58 and from paraxial optics. Tables 8.4-8.5 list the results of $W_{m02002}^{(XP)}$ after each surface calculated at the surface exit pupil at 400 nm and 1700 nm wavelength, respectively. Note that during the calculation of $W_{m02002}^{(XP)}$, the system description is converted to follow the description method discussed in Chapter 2, which involves tracing the OAR, shifting each surface coordinate system to the OAR intersection at the surface, and rotating each surface coordinate system to align the Z-axis with the surface normal.

**Table 8.4. $W_{m02002}^{(XP)}$ after each surface of the Dyson spectrometer at 400 nm**

| Surface | $W_{m02002}^{(XP)}$ calculated at the exit pupil of each surface (μm) | |
|---|---|---|
| | From paraxial optics | From Equation 6.58 |
| $S_1$ | 0 | 0 |
| $S_2$ | 0 | 0 |
| $S_3$ (Grating/Stop) | 3.454 | 3.454 |
| $S_2$ (Second Pass) | 0.345 | 0.345 |
| $S_1$ (Second Pass) | 0.348 | 0.348 |

**Table 8.5. $W_{m02002}^{(XP)}$ after each surface of the Dyson spectrometer at 1700 nm**

| Surface | $W_{m02002}^{(XP)}$ calculated at the exit pupil of each surface (μm) | |
|---|---|---|
| | From paraxial optics | From Equation 6.58 |
| $S_1$ | 0 | 0 |
| $S_2$ | 0 | 0 |
| $S_3$ (Grating/Stop) | 66.279 | 66.279 |
| $S_2$ (Second Pass) | 0.170 | 0.170 |
| $S_1$ (Second Pass) | 0.170 | 0.170 |

Tables 8.4-8.5 show that paraxial optics and Equation 6.58 yield the same results for $W_{02002}$. The comparison indicates that the two methods of calculating $W_{02002}$ are equivalent to each other.

### 8.2.3 Comparison of distortion calculated from raytracing and theoretical results

In Section 4.5, spectral distortion $SD_x$ and $SD_y$ defined and related to aberration coefficients $W_{21001}$ and $W_{31100}$ are illustrated in Figure 4.9. $SD_x$ and $SD_y$ can be calculated through real raytracing with CODE V or theoretically through the analytical formulae of $W_{21001}$ and $W_{31100}$ derived in Chapters 5-6. In this section, we compare the raytracing and theoretical results of $SD_x$ and $SD_y$, shown in the spectral full-field displays in Figure 8.4.

Spectral full-field displays are plots that illustrate aberration behavior in imaging spectrometers across slit position and wavelength [80, 81]. In Figure 8.4, the vertical axis represents slit position, and the horizontal axis represent wavelength. Therefore, each point in the plot represents an image position for corresponding slit position and wavelength. The arrow symbols indicate the distortion magnitude and direction for the corresponding image position. The arrows can be viewed as vectors, determined by the magnitude of $SD_x$ and $SD_y$, with $SD_x$ in the vertical direction parallel to the split position axis, and $SD_y$ in the horizontal direction parallel to the wavelength axis. MIN, MAX and AVG represent the minimum, maximum and average values of the magnitude or length of the vectors.

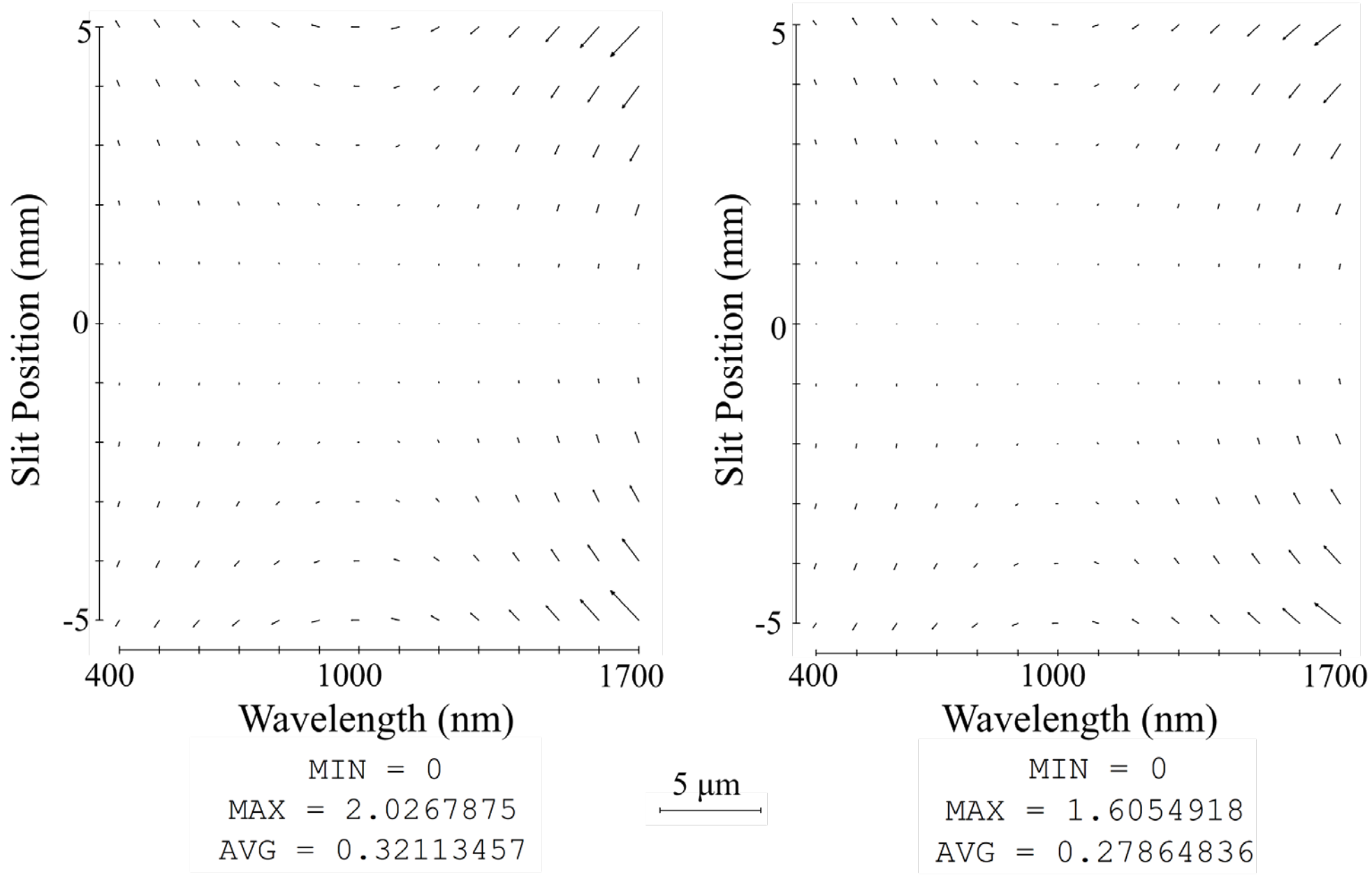

(a) Raytracing Result (b) Theoretical Result

**Figure 8.4. Spectral full-field displays of combined $SD_x$ and $SD_y$ as a vector for the Dyson spectrometer calculated through (a) real raytracing and (b) analytical aberration theory using Equations 4.45-4.46. The reference wavelength for $SD_x$ is 1000 nm.**

The two results from Figure 8.4 have similar distortion magnitude and pattern. The minor differences between the two results can be from several sources. First, the theoretical result represents the distortion behavior up to third order in transverse aberration, while the raytracing result includes the effect of all orders. Secondly, the raytracing result is measured at a designed image plane described in the system description section. As discussed in Section 4.5, the distortion in the theoretical result is measured in different image planes for different wavelengths, because each wavelength is treated independently. However, the independent image planes are located close to each other, thus having minor impact on the results.

Figure 8.5 shows the difference between the raytracing and theoretical results. The distortion differences at all wavelengths and slit positions are under 0.5 μm and are larger at longer wavelengths and larger slit positions. The difference can be from distortion aberrations beyond the third group, since the current theory covers up to third-group aberrations.

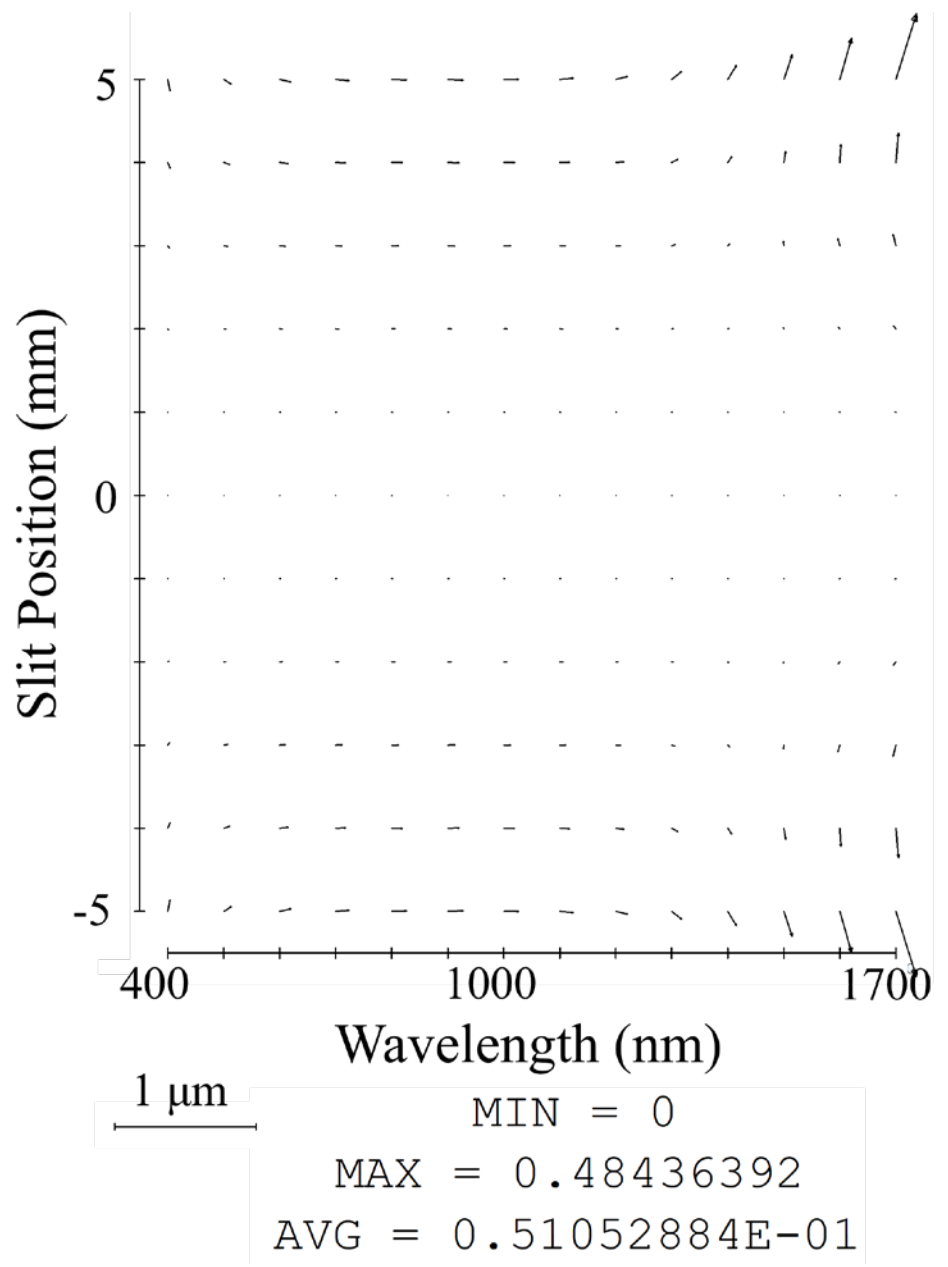

**Figure 8.5. Difference between raytracing and theoretical results in Figure 8.4.**

Figure 8.6 shows the distortion spectral full-field display calculated with Equations 24 and 32 in the work by Sasian [43]. The maximum distortion magnitude in Figure 8.6 is 17.97 µm, while the real raytracing result shows 2.03 µm. This larger difference is due to the approximations used by Sasian during the derivation of third-group aberrations.

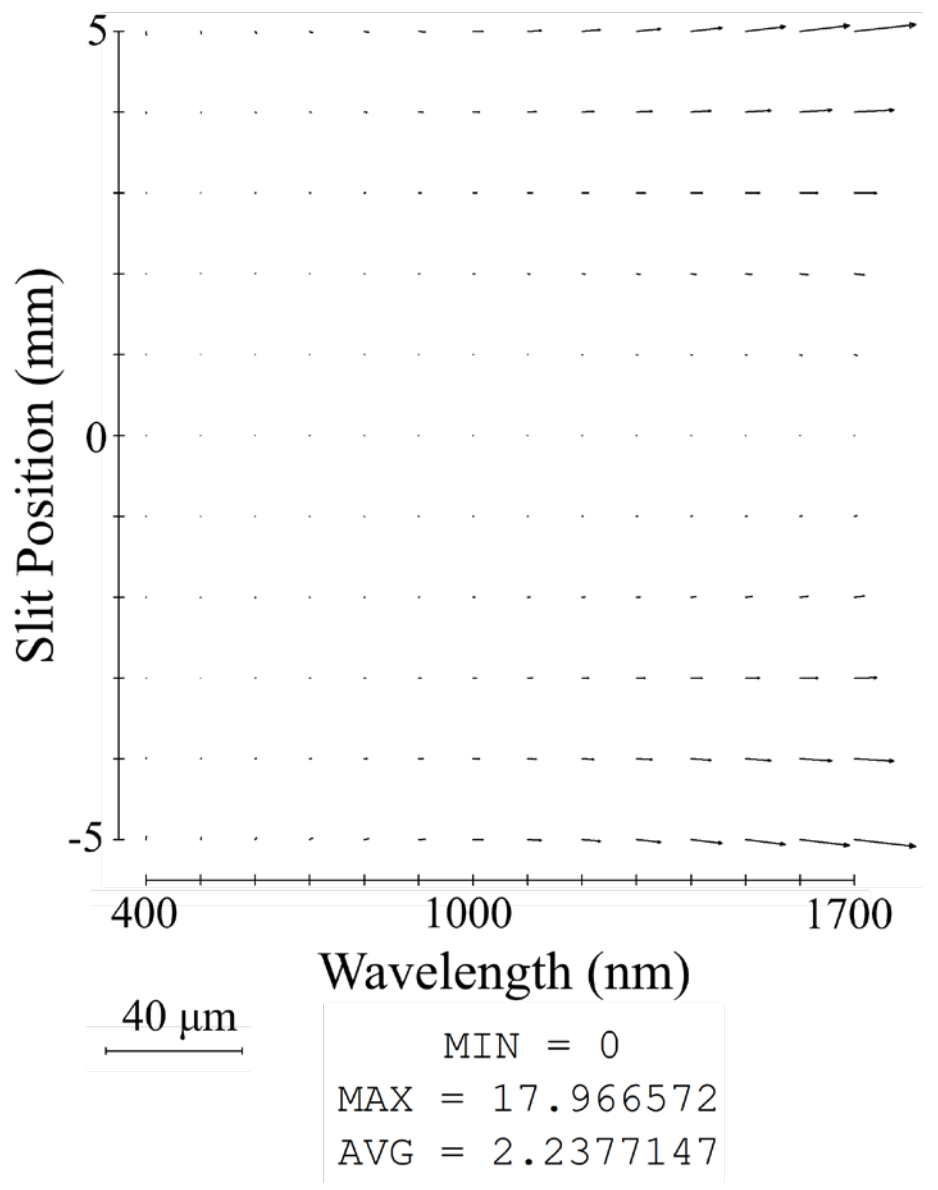

**Figure 8.6. Spectral full-field displays of combined $SD_x$ and $SD_y$ as a vector for the Dyson spectrometer calculated through using Equation 24 and 32 in the work by Sasian [43].**

### 8.2.4 Surface contribution of $W_{21001}$

In Chapters 5-6, the derivation of the system imaging aberrations is through a surface-to-surface process. Therefore, surface contribution to the total aberration is derived and is valuable for analyzing aberration distribution and balancing. As an example, Figure 8.7 shows the surface contribution to smile distortion at 1700 nm wavelength and 5 mm slit position.

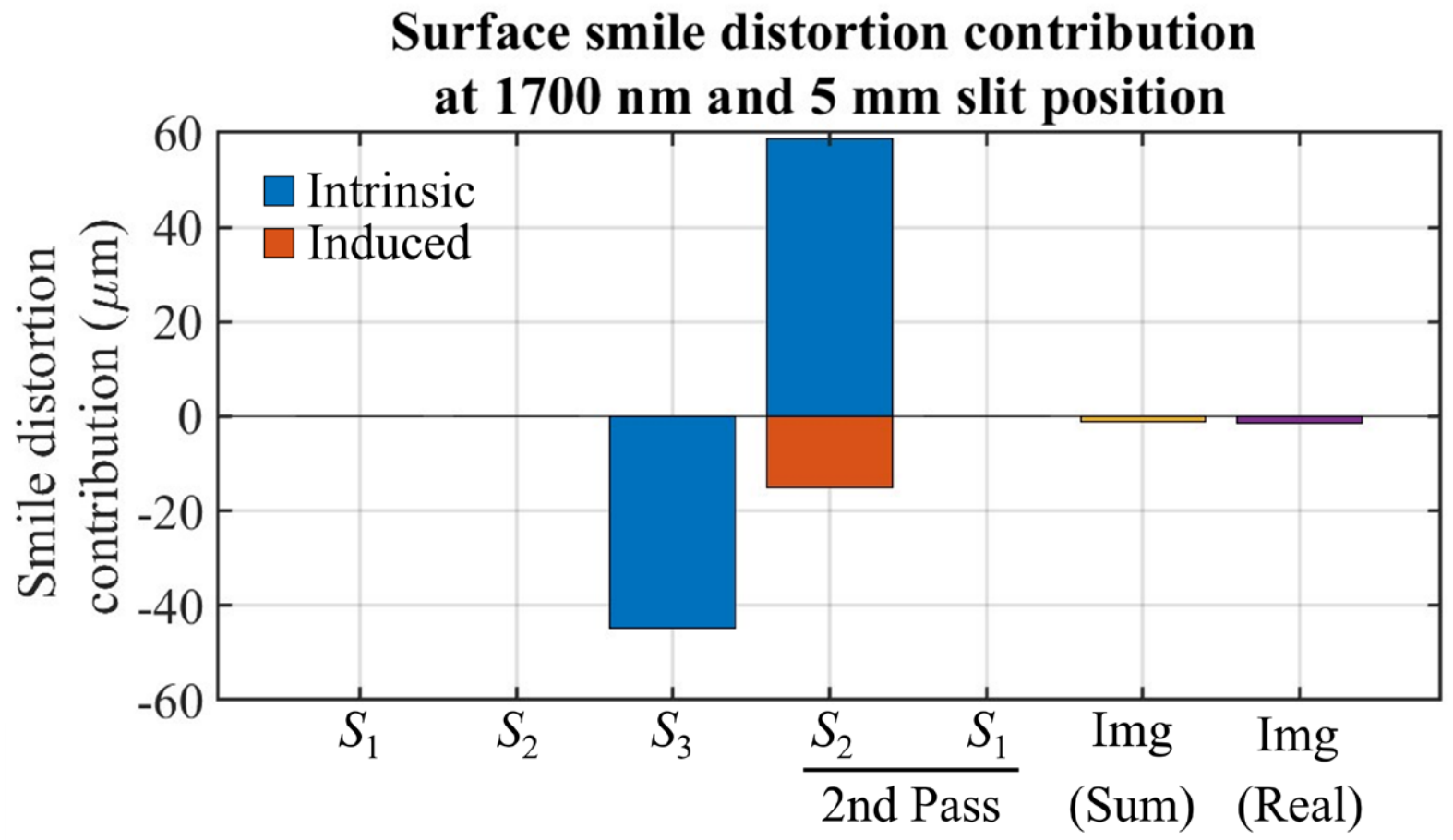

**Figure 8.7. Intrinsic (blue) and Induced (orange) surface contribution to smile distortion computed using Equations 8.5-8.6, respectively, for the Dyson spectrometer at 1700 nm wavelength and 5 mm slit position. The sum of the contributions (yellow) is near the overall contribution validated with real raytracing (purple).**

In Figure 8.7, the surface contribution is the transverse aberration caused in the final image by the surface wavefront aberration. The blue and orange bars represent the surface contribution to the smile distortion caused by intrinsic and induced aberrations, respectively, given as

$$\textit{Intrinsic contribution to the smile distortion}: \frac{1}{n'u'_{ax}} W_{21001}^{(XP)} \tag{8.5}$$

$$\textit{Induced contribution to the smile distortion}: \frac{1}{n'u'_{ax}} W_{21001}^{(IN)} \tag{8.6}$$

where $W_{21001}{}^{(XP)}$ (intrinsic) and $W_{21001}{}^{(IN)}$ (induced) are computed using Equations 5.143 and 6.70, respectively. The "Img (Sum)" bar represents the sum of all surface contributions including all intrinsic and induced contributions, and the "Img (Real)" bar represents the real smile distortion calculated through real raytracing in CODE V.

It can be seen that the smile distortion is balanced between $S_3$ (grating) and $S_2$ in the second pass. Figure 8.7 also shows that the induced aberration at $S_2$ in the second pass is significant in the balancing of smile distortion.

Figure 8.8 shows surface contribution to smile distortion calculated using the $W_{21001}$ formula derived by Sasian (Equation 22 in [43]). The comparison between Figure 8.7 and 8.8 suggests that in this case, most of the prediction error of Sasian's $W_{21001}$ coefficient is from induced aberrations.

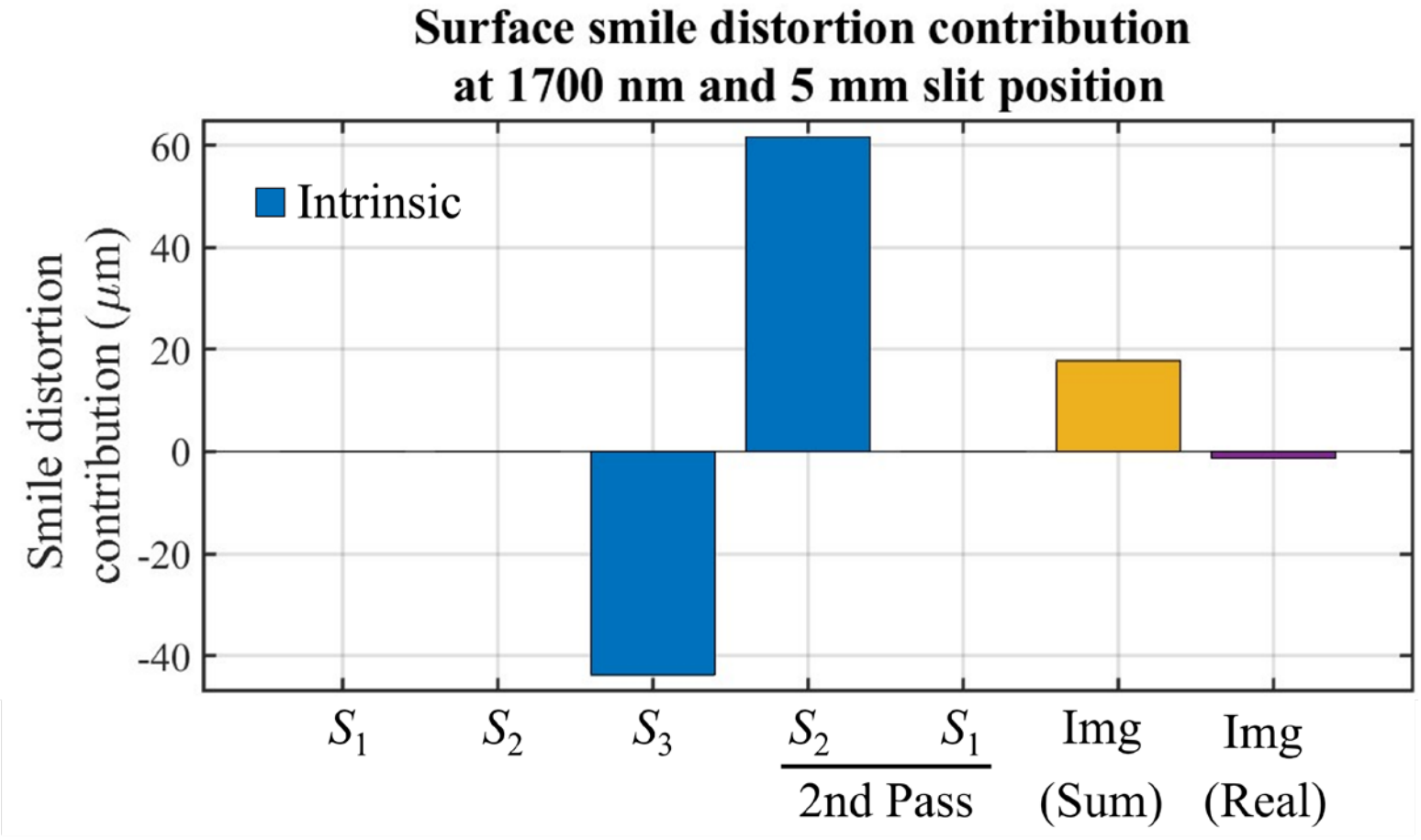

**Figure 8.8. Intrinsic (blue) surface contribution to smile distortion computed using Equation 22 in the work by Sasian [43] for the Dyson spectrometer at 1700 nm wavelength and 5 mm slit position. The sum of the contributions (yellow) is near the overall contribution validated with real raytracing (purple).**

## 8.3 Example systems: freeform three-mirror double-pass spectrometers

As discussed in Chapter 7, a freeform three-mirror double-pass spectrometer was designed for the hyperspectral imager. The following three example systems will follow the design form of the freeform spectrometer. The first two systems have shorter slit length and slower (higher) F-number than the specifications listed in Table 7.1. The two systems are limited by third-group distortion types with one system optimized to correct keystone distortion, labeled as SysK, and the other system optimized to correct smile distortion, labeled as SysS. The third system, labeled as SysF, share the first-order specifications and the performance as the freeform spectrometer described in Chapter 7.

### 8.3.1 System specification and description

The freeform three-mirror double-pass spectrometer consists of three freeform mirrors and a plano reflective grating. The structures of the freeform spectrometers, SysK, SysS and SysF, are shown in Figure 8.9.

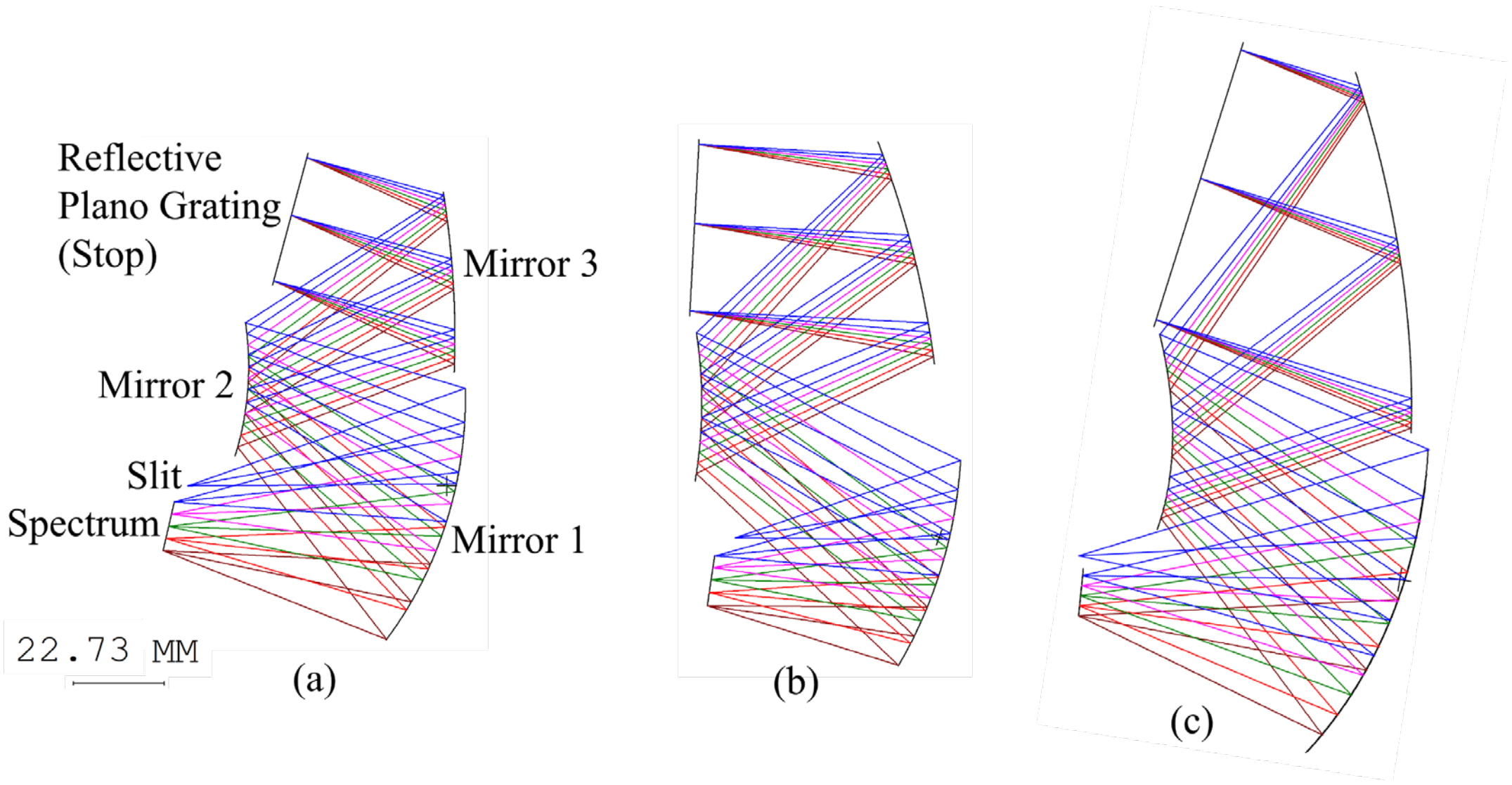

**Figure 8.9. 2D layout (in the tangential plane) of the freeform 3-mirror double-pass spectrometers with similar structures: (a) SysK, (b) SysS, and (c) SysF.**

The specifications for SysK, SysS and SysF are shown in Table 8.6. SysF follows the first-order specifications listed in Table 7.1 and has the performance described in 7.3.2, specifically smile distortion less than 2.5 μm, and keystone distortion less than 3 μm, while SysK and SysS have shorter slit length, slower (higher) F-number and larger residual distortion in comparison.

**Table 8.6. Specifications of the freeform three-mirror double-pass spectrometers**

| Specifications | SysK | SysS | SysF |
|---|---|---|---|
| Image F-number (sagittal direction) | F/2.6 | F/2.6 | F/2.0 |
| Wavelength (nm) | 400-1700 | 400-1700 | 400-1700 |
| Full slit length (sagittal direction) (mm) | 10 | 10 | 15 |
| Grating groove density (lp/mm) | 97.4 | 75.1 | 54.2 |

The freeform spectrometers are described similarly to the Dyson spectrometer described in Section 8.2.1 following the description in CODE V. Note that some surface coordinate systems undergo decenter and tilt operations before the Z-axis shift to the next surface. The decenter and tilt operations are categorized into different types following the definitions in CODE V. In the freeform spectrometers, the two types of decenter and tilt used are called “Decenter and Return” and “Decenter and Bend”, which are illustrated in Figure 8.10. With “Decenter and Return”, after the Z-axis shift from the last surface, decenter and tilt operations are applied to the coordinate system to find the surface coordinate system as shown in Figure 8.10(a). The Z-axis shift to the next surface is measured from the coordinate system before the decenter and tilt operations. For “Decenter and Bend”, the surface coordinate system is found the same way as in “Decenter and Return”. However, the Z-axis shift to the next surface is measured from a coordinate system resulting from performing an additional tilt operation on the surface coordinate system as illustrated in Figure 8.10(b). For the freeform spectrometers, all tilts are about the X-axis of the surface coordinate system, and the tilt angle is positive if the rotation is counter-clockwise. For example, the alpha tilt angles in Figure 8.10 are all positive.

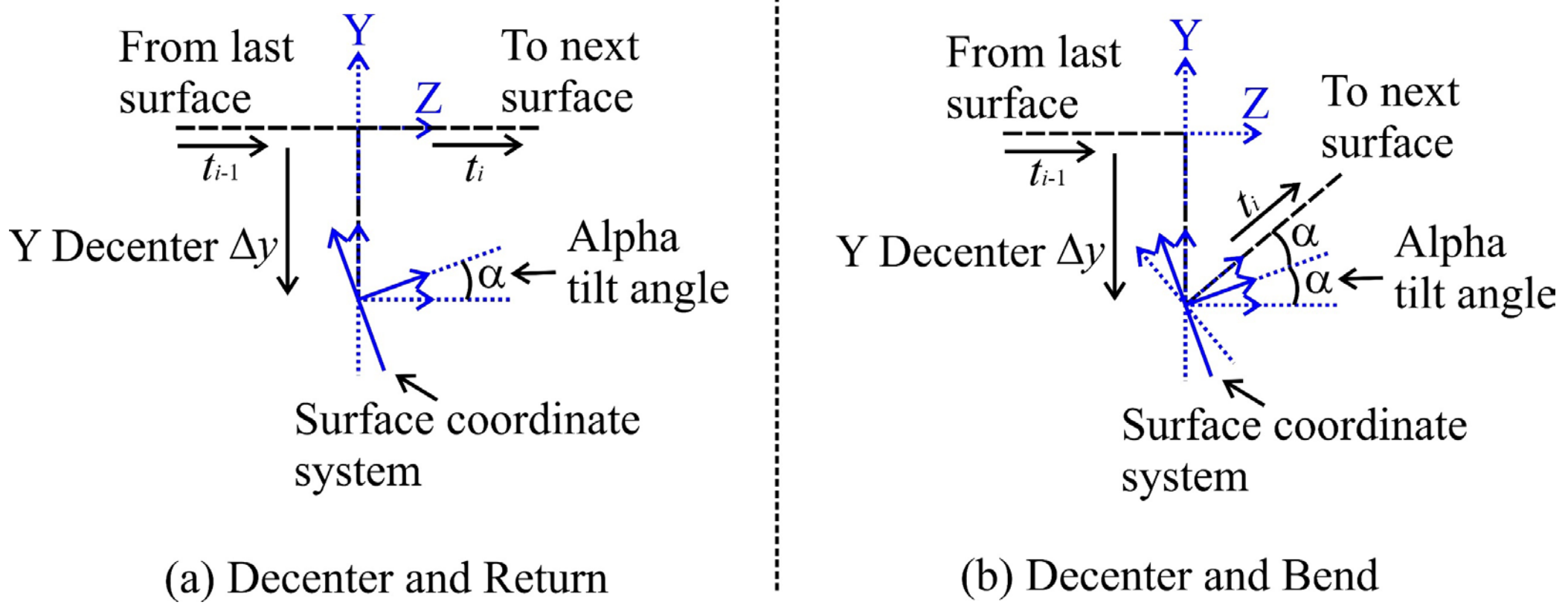

**Figure 8.10. Illustration of decenter and tilt types: (a) Decenter and Return, and (b) Decenter and Bend.**

The system description of SysK is shown in Tables 8.7-8.9, with Table 8.8 showing the decenter and tilt information. The surface type for $S_1$, $S_2$ and $S_3$ is Fringe Zernike Polynomial defined in CODE V as a freeform sag function described with Fringe Zernike polynomials on top of a base spherical surface, and Table 8.9 is showing the freeform coefficients of the three freeform mirrors. SysS is described similarly in Tables 8.10-8.12, and SysF is described in Tables 8.13-8.15. The surface coordinate systems and the thickness parameters in SysK are illustrated in Figure 8.11. The same parameterization also applies to SysS and SysF. Note that the position of the image plane is related not only to the thickness of the surface before, but also the tilt angle of the surface before, which is shown as the Alpha Tilt Angle on the Image-1 row in Tables 8.8, 8.11 and 8.14. This tilt angle, indicated by $\alpha$ in Figure 8.11, corresponds to a Decenter-and-Bend type angle defined in CODE V, but for simplicity it is defined here as the rotation angle of the surface coordinate system before the thickness is calculated along the new Z-axis and is positive for counter-clockwise rotation. The radii of curvature of the base spherical surfaces for $S_1$, $S_2$ and $S_3$ are listed in the Radius of Curvature column.

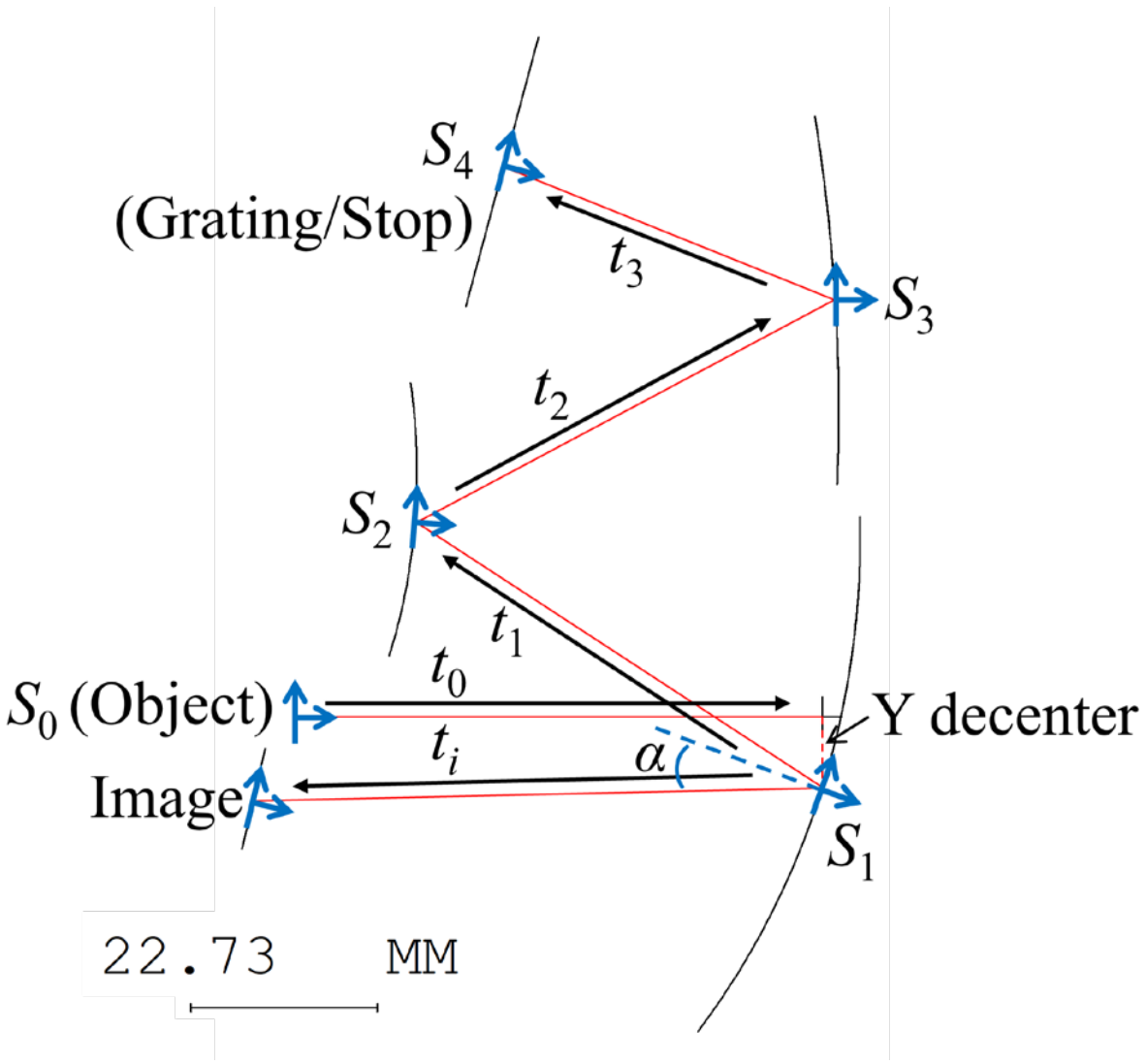

**Figure 8.11. Illustration of surface labels, surface coordinate systems and surface thicknesses of SysK. SysS and SysF follow similar structures. The angle *α* is the rotation angle of the coordinate system for the last surface before image.**

Tables 8.7-8.9 show the system description for SysK.

**Table 8.7. System description of SysK**

| Surface | Surface Type | Radius of Curvature (mm) | Thickness *t* (mm) | Material |
|---|---|---|---|---|
| $S_0$ (Object) | Spherical | Infinity | 64.614 | Air |
| $S_1$ | Fringe Zernike Polynomial | -101.369 | -60.147 | Air |
| $S_2$ | Fringe Zernike Polynomial | -79.576 | 57.813 | Air |
| $S_3$ | Fringe Zernike Polynomial | -218.902 | -42.815 | Air |
| $S_4$ (Grating/Stop) | Spherical | Infinity | - | - |
| Image | Spherical | Infinity | -69.017 | - |

**Table 8.8. Decenter and tilt information for SysK**

| Surface | Decenter and Tilt Type | Y Decenter (mm) | Alpha Tilt Angle (degree) |
|---|---|---|---|
| $S_0$ (Object) | None | - | - |
| $S_1$ | Decenter and Bend | -8.639 | -17.281 |
| $S_2$ | Decenter and Bend | 0 | 31.331 |
| $S_3$ | Decenter and Bend | 0 | -24.087 |
| $S_4$ (Grating/Stop) | Decenter and Return | 0 | 4.925 |
| Image-1(S1) | - | - | 17.281 |
| Image | Decenter and Return | 0 | -12.955 |

**Table 8.9. Freeform coefficients (Fringe Zernike polynomials) for the freeform surfaces in SysK**

| | $S_1$ | $S_2$ | $S_3$ |
|---|---|---|---|
| Unit: µm | | | |
| $Z_1$ | -15.139 | -19.601 | 13.502 |
| $Z_2$ | 0 | 0 | 0 |
| $Z_3$ | 90.454 | 26.148 | 31.967 |
| $Z_4$ | -25.658 | -22.382 | 13.981 |
| $Z_5$ | -7.399 | 29.109 | -25.131 |
| $Z_6$ | 0 | 0 | 0 |
| $Z_7$ | 0 | 0 | 0 |
| $Z_8$ | 47.567 | 12.744 | 16.051 |
| $Z_9$ | -11.204 | -2.910 | 0.477 |
| $Z_{10}$ | 0 | 0 | 0 |
| $Z_{11}$ | 10.211 | 10.374 | 3.475 |
| $Z_{12}$ | 1.179 | 1.842 | 0.210 |
| $Z_{13}$ | 0 | 0 | 0 |
| $Z_{14}$ | 0 | 0 | 0 |
| $Z_{15}$ | 1.560 | -0.220 | 0.045 |
| $Z_{16}$ | -0.685 | -0.129 | -0.002 |

Tables 8.10-8.12 show the system description for SysS.

**Table 8.10. System description of SysS**

| Surface | Surface Type | Radius of Curvature (mm) | Thickness $t$ (mm) | Material |
|---|---|---|---|---|
| $S_0$ (Object) | Spherical | Infinity | 50.924 | Air |
| $S_1$ | Fringe Zernike Polynomial | -114.466 | -73.026 | Air |
| $S_2$ | Fringe Zernike Polynomial | -115.016 | 64.823 | Air |
| $S_3$ | Fringe Zernike Polynomial | -238.929 | -54.793 | Air |
| $S_4$ (Grating/Stop) | Spherical | Infinity | - | - |
| Image | Spherical | Infinity | -56.723 | - |

**Table 8.11. Decenter and tilt information for SysS**

| Surface | Decenter and Tilt Type | Y Decenter (mm) | Alpha Tilt Angle (degree) |
|---|---|---|---|
| $S_0$ (Object) | None | - | - |
| $S_1$ | Decenter and Bend | -9.147 | -17.965 |
| $S_2$ | Decenter and Bend | 0 | 35.758 |
| $S_3$ | Decenter and Bend | 0 | -21.384 |
| $S_4$ (Grating/Stop) | Decenter and Return | 0 | 3.958 |
| Image-1(S1) | - | - | 17.965 |
| Image | Decenter and Return | 0 | -8.573 |

**Table 8.12. Freeform coefficients (Fringe Zernike polynomials) for the freeform surfaces in SysS**

| | $S_1$ | $S_2$ | $S_3$ |
|---|---|---|---|
| Unit: µm | | | |
| $Z_1$ | -0.767 | 0.347 | 1.492 |
| $Z_2$ | 0 | 0 | 0 |
| $Z_3$ | 11.697 | -10.709 | 18.388 |
| $Z_4$ | -1.370 | 0.346 | 2.258 |
| $Z_5$ | 0.087 | 0.232 | 0.152 |
| $Z_6$ | 0 | 0 | 0 |
| $Z_7$ | 0 | 0 | 0 |
| $Z_8$ | 5.283 | -6.351 | 9.049 |
| $Z_9$ | -0.764 | -0.156 | 0.753 |
| $Z_{10}$ | 0 | 0 | 0 |
| $Z_{11}$ | -0.664 | -0.069 | 1.449 |
| $Z_{12}$ | -1.446 | 0.448 | 0.269 |
| $Z_{13}$ | 0 | 0 | 0 |
| $Z_{14}$ | 0 | 0 | 0 |
| $Z_{15}$ | -0.377 | -0.664 | -0.097 |
| $Z_{16}$ | -0.161 | -0.155 | -0.012 |

Tables 8.13-8.15 show the system description for SysF.

**Table 8.13. System description of SysF**

| Surface | Surface Type | Radius of Curvature (mm) | Thickness $t$ (mm) | Material |
|---|---|---|---|---|
| $S_0$ (Object) | Spherical | Infinity | 79.794 | Air |
| $S_1$ | Fringe Zernike Polynomial | -121.671 | -70.894 | Air |
| $S_2$ | Fringe Zernike Polynomial | -87.123 | 70.257 | Air |
| $S_3$ | Fringe Zernike Polynomial | -246.386 | -52.666 | Air |
| $S_4$ (Grating/Stop) | Spherical | Infinity | - | - |
| Image | Spherical | Infinity | -78.426 | - |

**Table 8.14. Decenter and tilt information for SysF**

| Surface | Decenter and Tilt Type | Y Decenter (mm) | Alpha Tilt Angle (degree) |
|---|---|---|---|
| $S_0$ (Object) | None | - | - |
| $S_1$ | Decenter and Bend | -5.449 | -15.780 |
| $S_2$ | Decenter and Bend | 2.790 | 37.316 |
| $S_3$ | Decenter and Bend | -2.233 | -29.808 |
| $S_4$ (Grating/Stop) | Decenter and Return | 0 | 3.352 |
| Image-1(S1) | - | - | 15.780 |
| Image | Decenter and Return | 0 | -1.455 |

**Table 8.15. Freeform coefficients (Fringe Zernike polynomials) for the freeform surfaces in SysF**

| | $S_1$ | $S_2$ | $S_3$ |
|---|---|---|---|
| Unit: μm | | | |
| $Z_1$ | -16.322 | 0.497 | 67.441 |
| $Z_2$ | 0 | 0 | 0 |
| $Z_3$ | 283.344 | 275.979 | 228.005 |
| $Z_4$ | -28.411 | -8.923 | 68.730 |
| $Z_5$ | -12.743 | -36.369 | -129.780 |
| $Z_6$ | 0 | 0 | 0 |
| $Z_7$ | 0 | 0 | 0 |
| $Z_8$ | 150.871 | 143.367 | 114.691 |
| $Z_9$ | -12.583 | -9.784 | 1.308 |
| $Z_{10}$ | 0 | 0 | 0 |
| $Z_{11}$ | -10.055 | 1.978 | 21.673 |
| $Z_{12}$ | 0.295 | 1.463 | 1.608 |
| $Z_{13}$ | 0 | 0 | 0 |
| $Z_{14}$ | 0 | 0 | 0 |
| $Z_{15}$ | 6.133 | 3.591 | 0.459 |
| $Z_{16}$ | -0.494 | -0.366 | 0.019 |

Similar comparisons are done for the two freeform spectrometers as for the Dyson spectrometer in Section 8.2. For the freeform spectrometers, the freeform contribution to $W_{02002}$ and the spectral distortion is included in the comparisons.

### 8.3.2 Comparison of $W_{02002}$ results

Similar to the comparison of $W_{M02002}^{(XP)}$ made in Section 8.2.2 for the Dyson spectrometer, Tables 8.16-8.18 show the $W_{M02002}^{(XP)}$ after each surface of the freeform spectrometers at 1000 nm. The results show high consistency between the data calculated from paraxial optics and from Equation 6.58.

**Table 8.16. $W_{m02002}^{(XP)}$ after each surface of SysK at 1000 nm**

| Surface | $W_{m02002}^{(XP)}$ calculated at the exit pupil of each surface (μm) | |
|---|---|---|
| | From paraxial optics | From Equation 6.58 |
| $S_1$ | 131.249 | 131.286 |
| $S_2$ | -100.294 | -100.273 |
| $S_3$ | -1.436 | -1.437 |
| $S_4$ (Grating/Stop) | -1.662 | -1.663 |
| $S_3$ (Second Pass) | 210.618 | 210.577 |
| $S_2$ (Second Pass) | -223.566 | -223.593 |
| $S_1$ (Second Pass) | -0.014 | -0.039 |

**Table 8.17. $W_{m02002}^{(XP)}$ after each surface of SysS at 1000 nm**

| Surface | $W_{m02002}^{(XP)}$ calculated at the exit pupil of each surface (μm) | |
|---|---|---|
| | From paraxial optics | From Equation 6.58 |
| $S_1$ | 104.508 | 104.508 |
| $S_2$ | -201.718 | -201.718 |
| $S_3$ | 2.299 | 2.299 |
| $S_4$ (Grating/Stop) | 2.056 | 2.056 |
| $S_3$ (Second Pass) | 310.326 | 310.326 |
| $S_2$ (Second Pass) | -130.037 | -130.037 |
| $S_1$ (Second Pass) | -0.074 | -0.074 |

**Table 8.18. $W_{m02002}^{(XP)}$ after each surface of SysF at 1000 nm**

| Surface | $W_{m02002}^{(XP)}$ calculated at the exit pupil of each surface (μm) | |
|---|---|---|
| | From paraxial optics | From Equation 6.58 |
| $S_1$ | 102.611 | 102.404 |
| $S_2$ | -536.407 | -538.321 |
| $S_3$ | -7.311 | -9.179 |
| $S_4$ (Grating/Stop) | -7.240 | -9.115 |
| $S_3$ (Second Pass) | 779.218 | 777.243 |
| $S_2$ (Second Pass) | -474.711 | -473.504 |
| $S_1$ (Second Pass) | 0.526 | -0.829 |

### 8.3.3 Comparison of distortion calculated from raytracing and theoretical results

Similar to the spectral full-field display showed in Section 8.2.3 for the Dyson spectrometer, Figure 8.12-8.16 show the comparison of spectral full-field displays calculated from real raytracing and the aberration theory.

Figure 8.12 shows the comparison between real raytracing and theoretical results for SysK. The distortion vectors are mostly horizontal, indicating residual smile distortion.

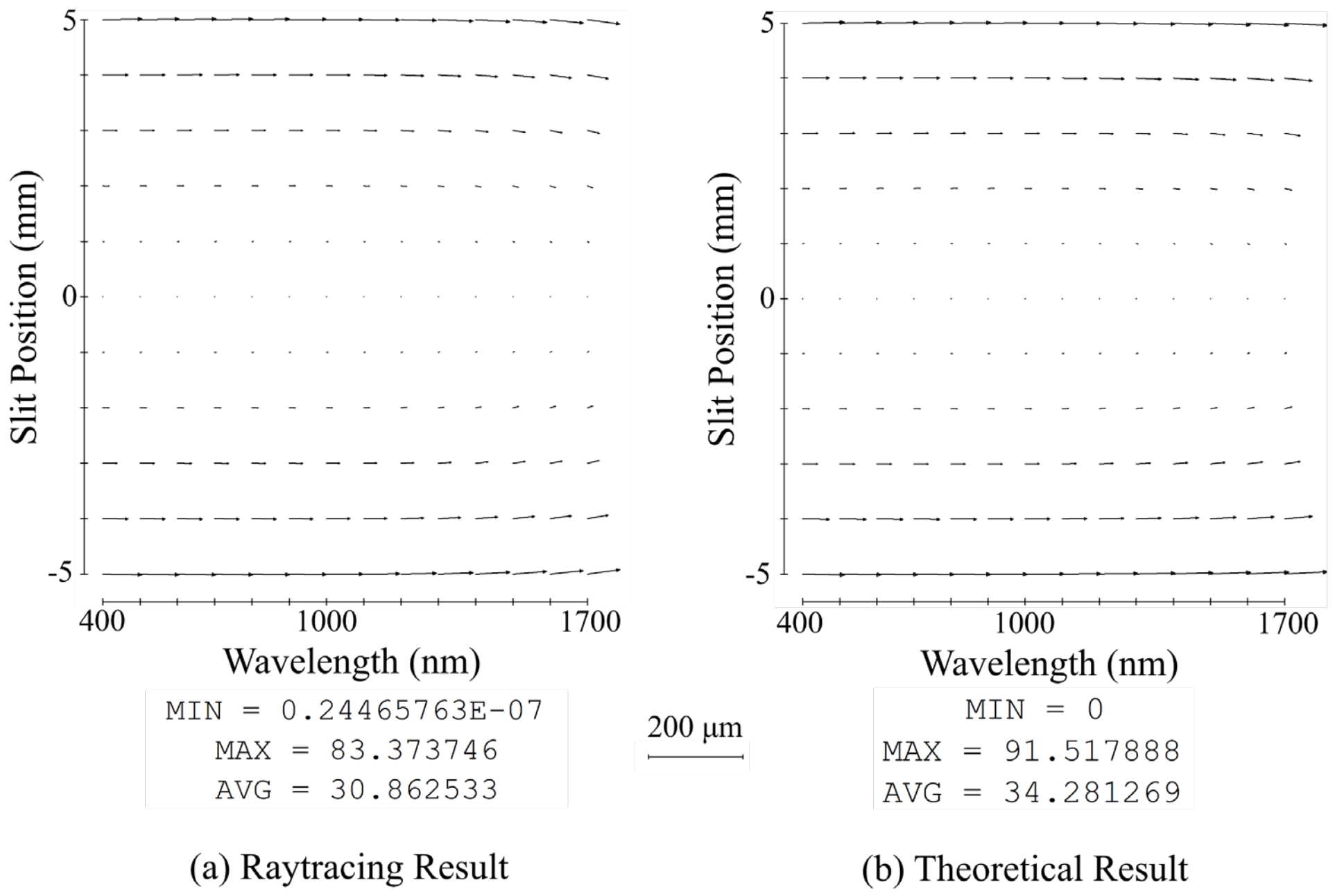

**Figure 8.12. Spectral full-field displays of combined $SD_x$ and $SD_y$ as a vector for SysK calculated through (a) real raytracing and (b) analytical aberration theory using Equations 4.45-4.46. The reference wavelength for $SD_x$ is 1000 nm.**

The difference between Figure 8.12(a) and (b) is shown in Figure 8.13. The difference is mostly small compared to the magnitude of corresponding distortion vectors shown in Figure 8.12. The differences are larger at long wavelengths towards 1700 nm and large slit positions towards 5 mm.

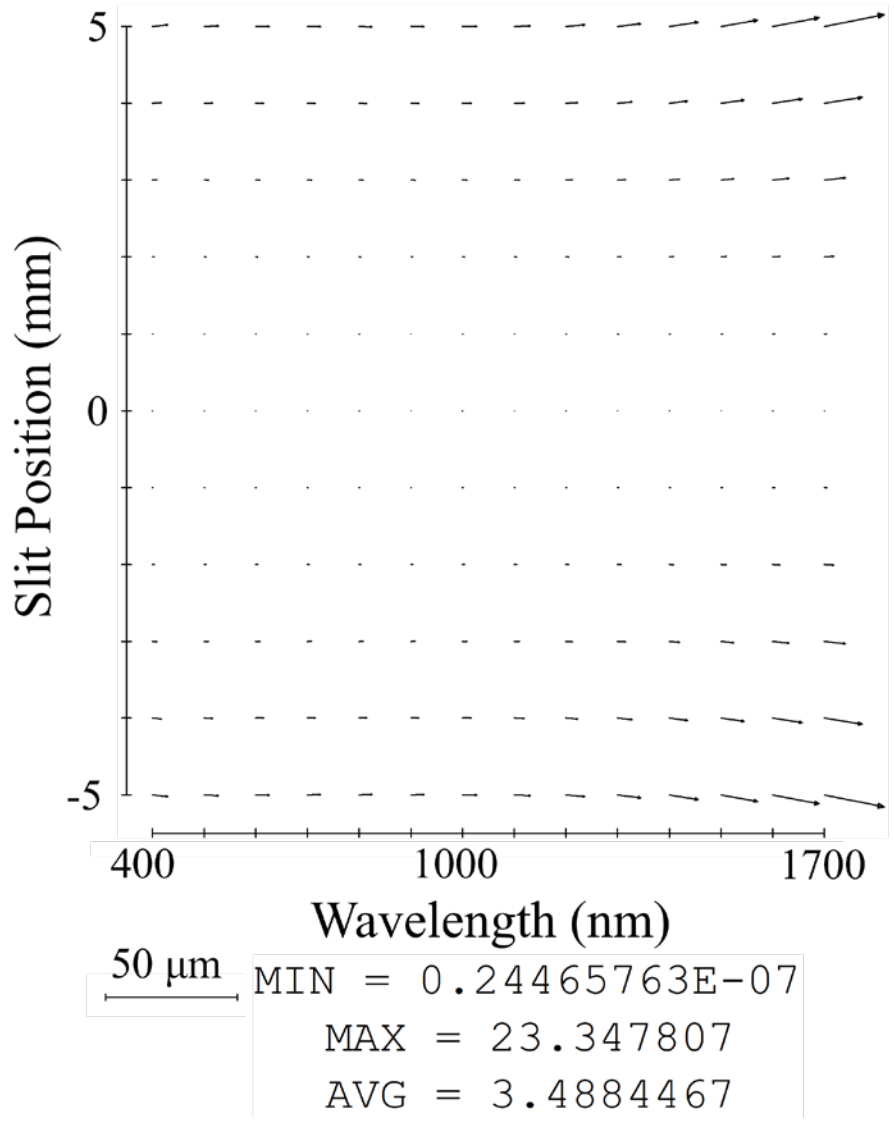

**Figure 8.13. Difference between raytracing and theoretical results in Figure 8.12.**

Figure 8.14 shows the comparison between real raytracing and theoretical results for SysS. In this case, the distortion vectors are mostly vertical, representing keystone distortions with respect to the reference wavelength (1000 nm).

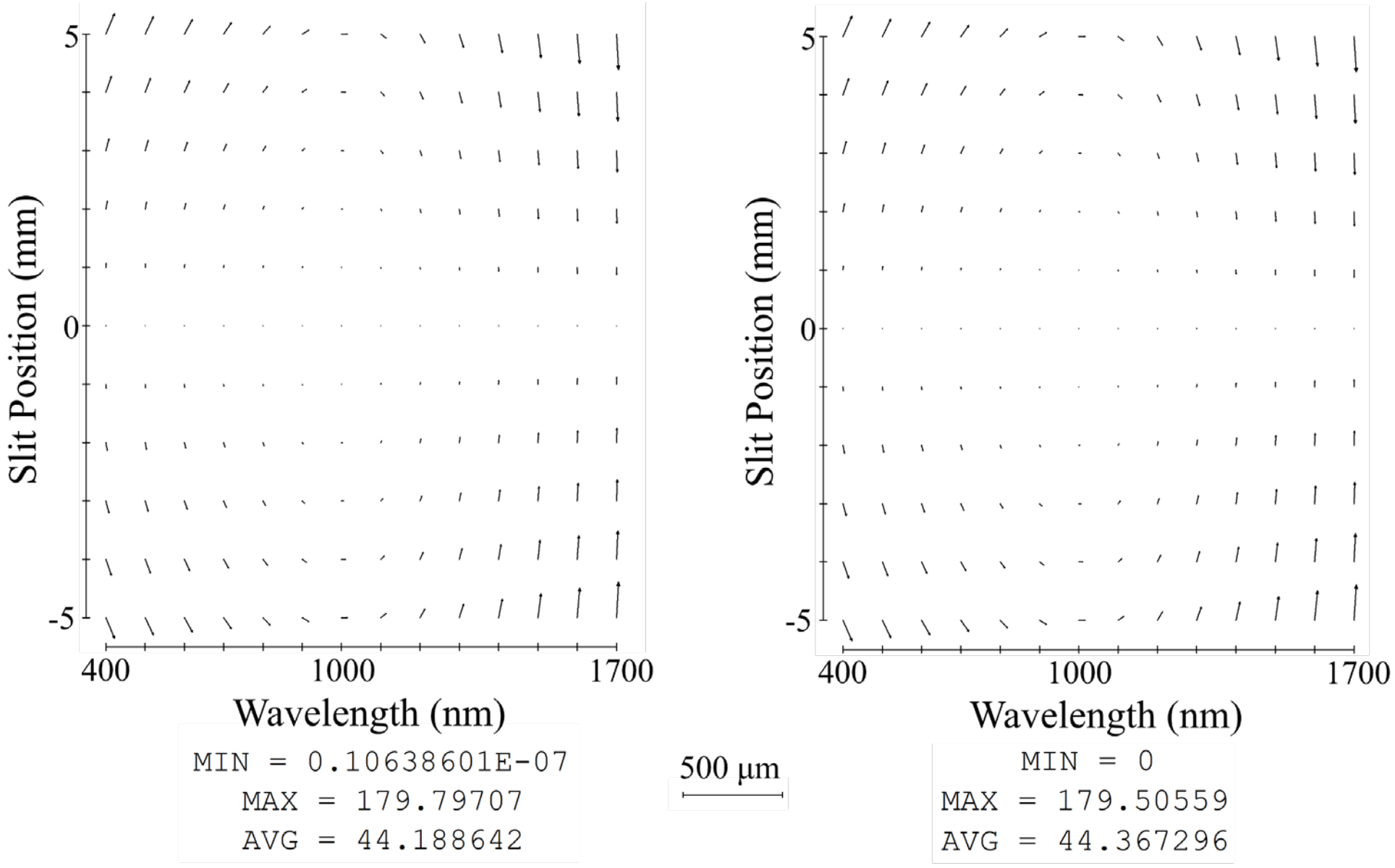

(a) Raytracing Result

(b) Theoretical Result

**Figure 8.14. Spectral full-field displays of combined $SD_x$ and $SD_y$ as a vector for SysS calculated through (a) real raytracing and (b) analytical aberration theory using Equations 4.45-4.46. The reference wavelength for $SD_x$ is 1000 nm.**

The difference between Figure 8.14(a) and (b) is shown in Figure 8.15. It can be seen that the difference is mostly small amount of smile distortion, which can be due to aberrations beyond the third group.

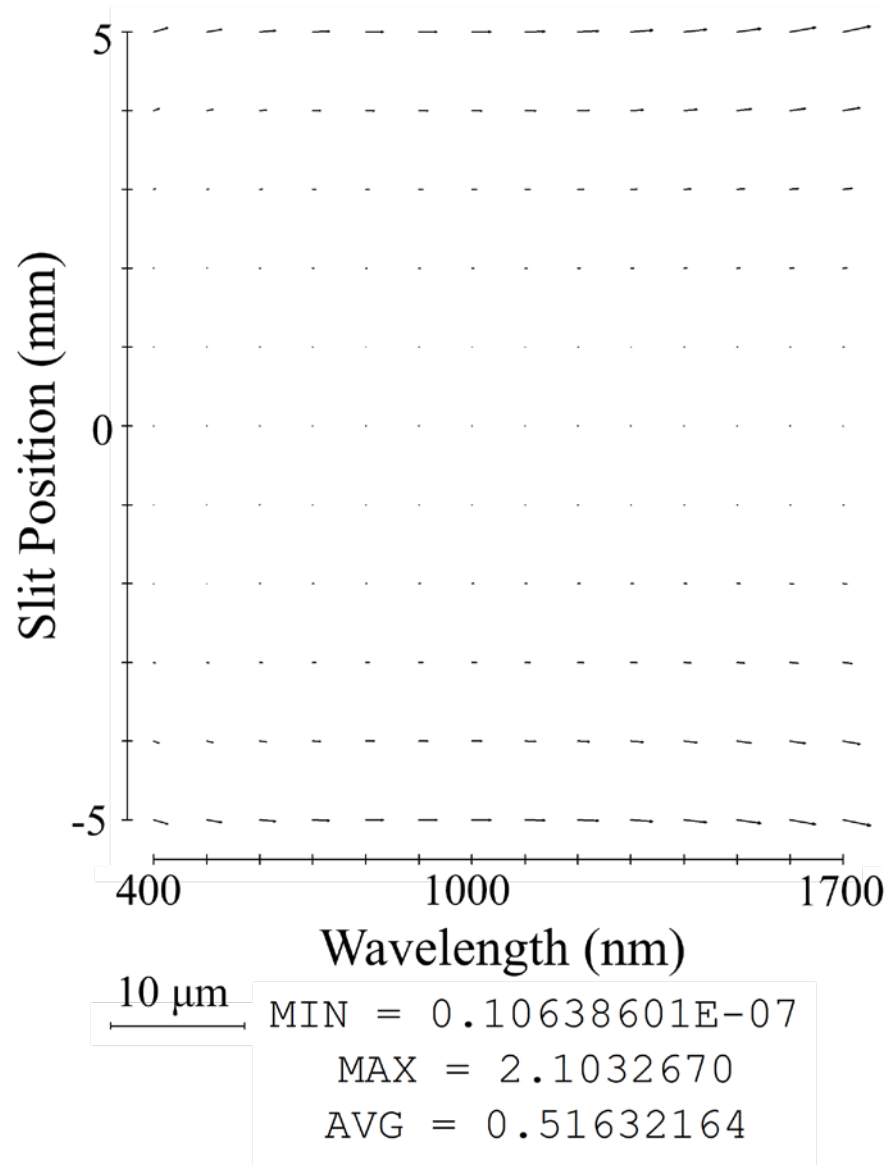

**Figure 8.15. Difference between raytracing and theoretical results in Figure 8.14.**

Figure 8.16 shows the comparison between real raytracing and theoretical results for SysF. The results show significant difference between the distortion calculated from real raytracing and the aberration theory. Note that the scales in Figure 8.16(a) and (b) are different.

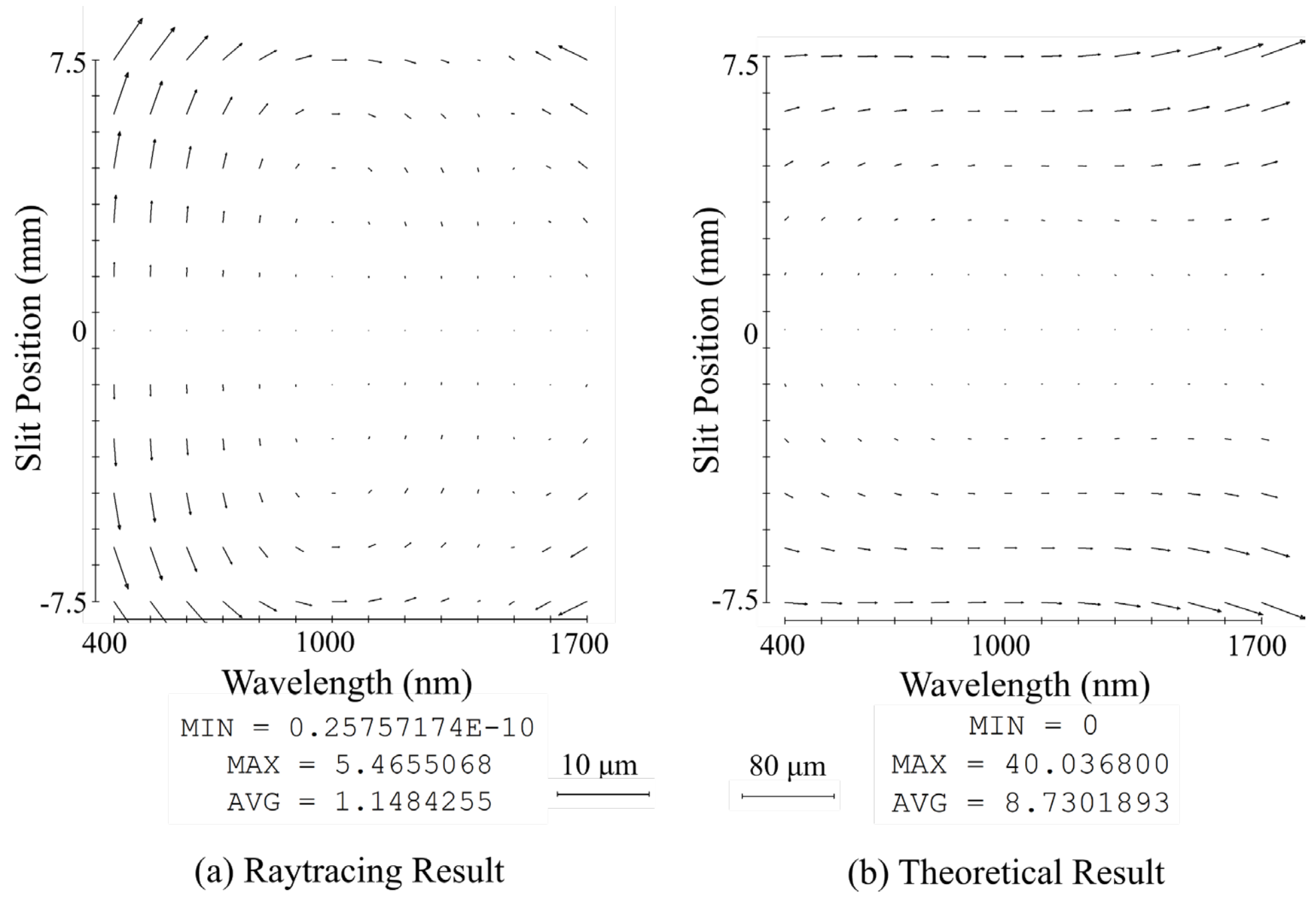

**Figure 8.16. Spectral full-field displays of combined $SD_x$ and $SD_y$ as a vector for SysF calculated through (a) real raytracing and (b) analytical aberration theory using Equations 4.45-4.46. The reference wavelength for $SD_x$ is 1000 nm.**

Figure 8.17 shows the difference between Figure 8.16(a) and (b) within [-2.5 2.5] μm slit position range. At larger slit positions, the difference between real raytracing and theoretical results is relatively large that it follows similar trend as the corresponding distortion vectors in Figure 8.16(b). It can be seen that the difference is dominated by smile distortion.

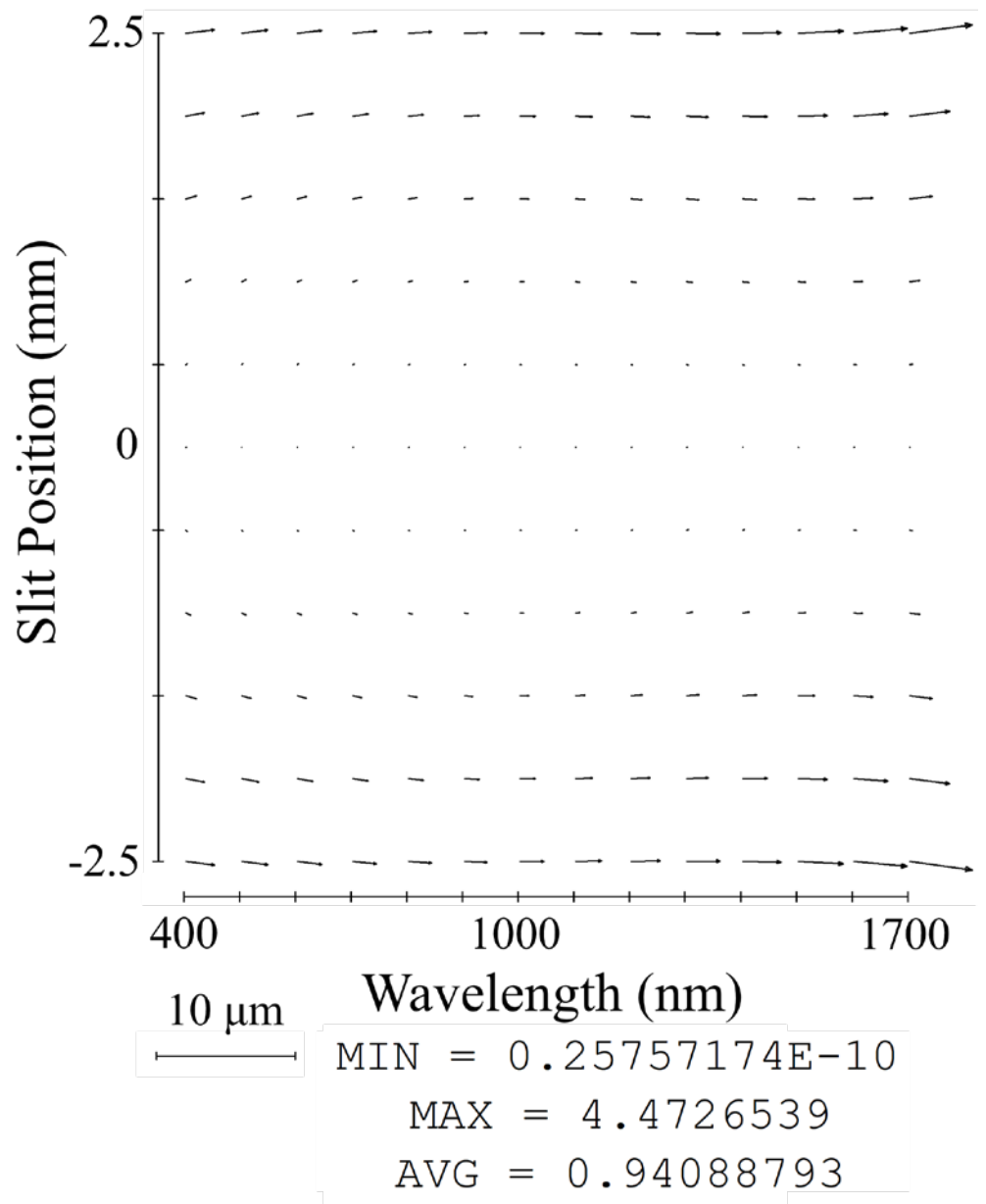

**Figure 8.17. Difference between raytracing and theoretical results in Figure 8.16 within [-2.5 2.5] μm slit position range.**

A possible explanation for the reasonable agreement or significant difference between real raytracing and theoretical results is that the distortion in SysK and SysS is dominated by third-group distortion types related to $W_{21001}$ and $W_{31100}$, while for SysF there may exist a significant aberration balancing between third-group distortion types and distortion types beyond the third group that are not included in the current theory. In the next section, we show the surface contribution plots which also suggest this explanation.

### 8.3.4 Surface contribution of $W_{21001}$

Similar to the surface contribution plot shown in Figure 8.7, the surface contributions to $W_{21001}$ in freeform spectrometers are shown in Figures 8.18-8.19.

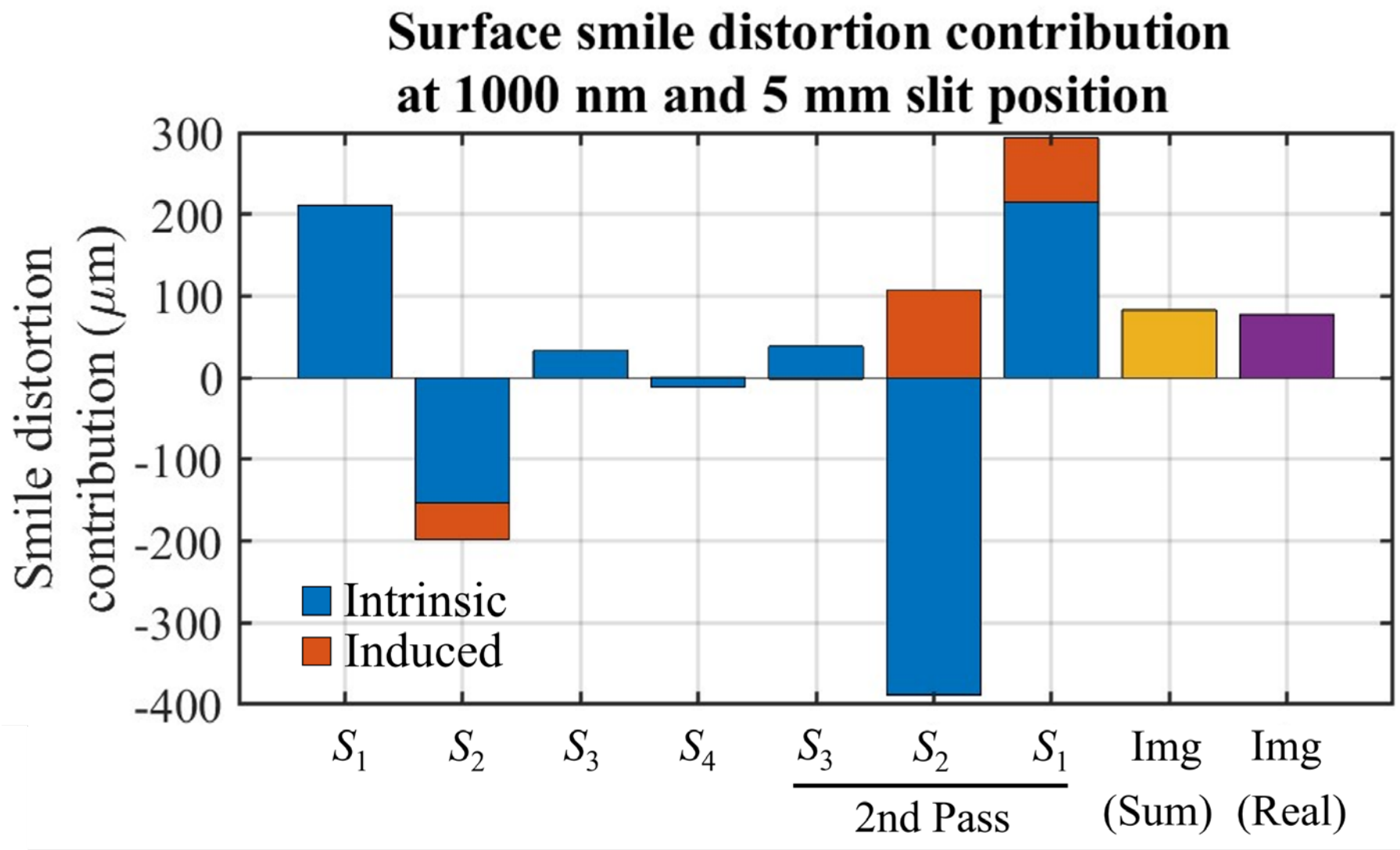

**Figure 8.18. Intrinsic (blue) and Induced (orange) surface contribution to smile distortion computed using Equations 8.5-8.6, respectively, for SysK at 1000 nm wavelength and 5 mm slit position. The sum of the contributions (yellow) is near the overall contribution validated with real raytracing (purple).**

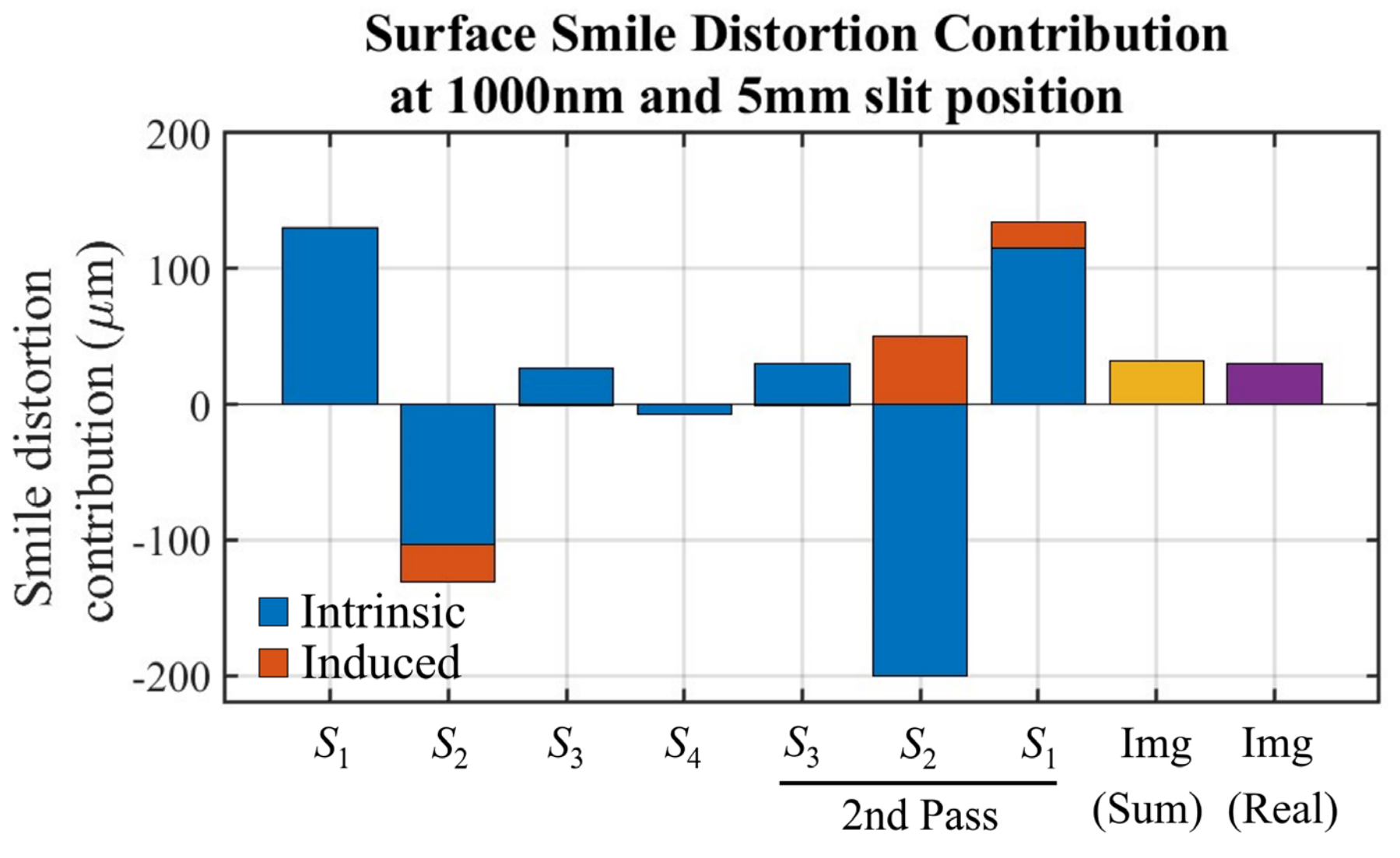

**Figure 8.19. Intrinsic (blue) and Induced (orange) surface contribution to smile distortion computed using Equations 8.5-8.6, respectively, for SysS at 1000 nm wavelength and 5 mm slit position. The sum of the contributions (yellow) is near the overall contribution validated with real raytracing (purple).**

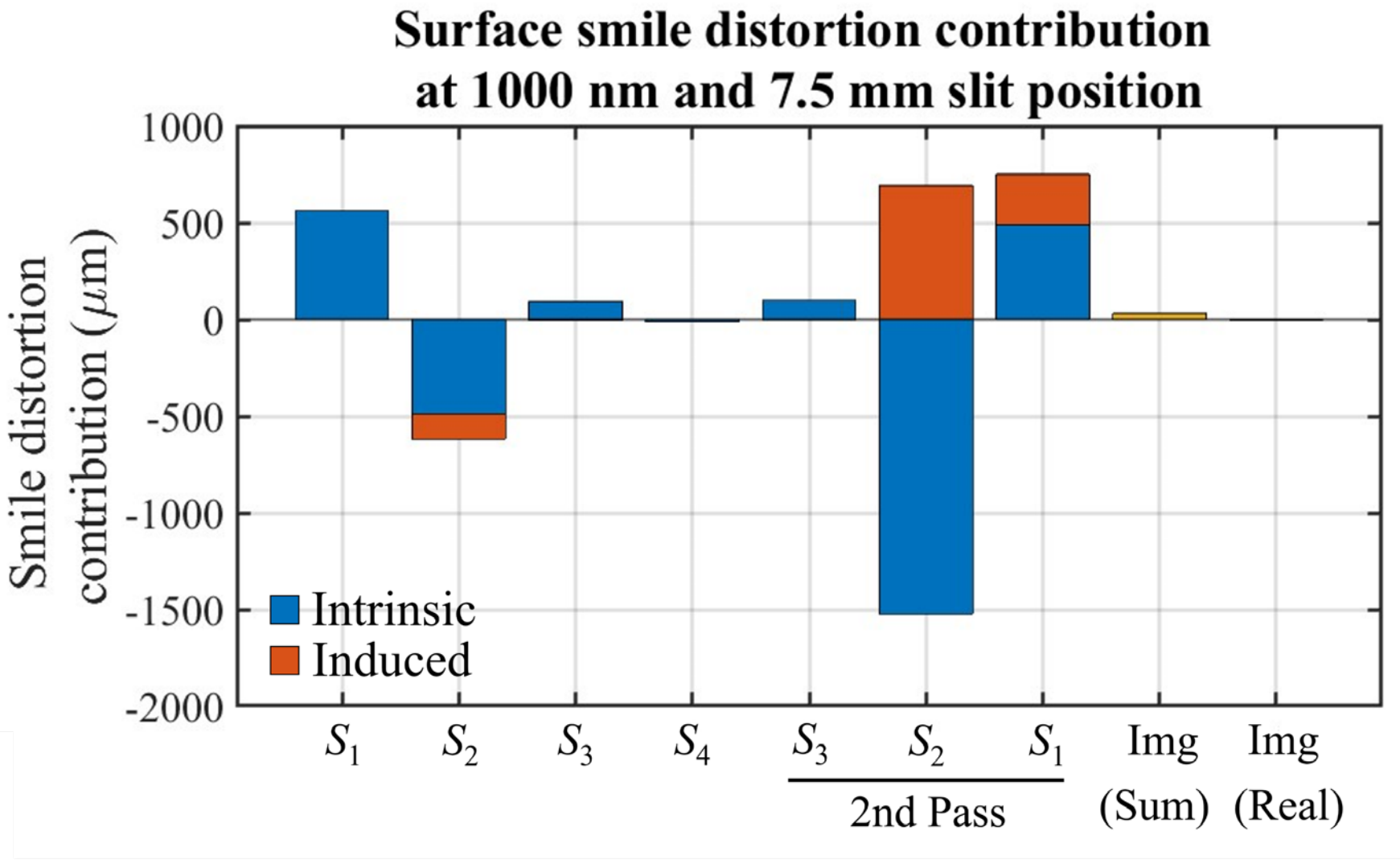

**Figure 8.20. (blue) and Induced (orange) surface contribution to smile distortion computed using Equations 8.5-8.6, respectively, for SysF at 1000 nm wavelength and 7.5 mm slit position. The sum of the contributions (yellow) is near the overall contribution validated with real raytracing (purple).**

Figures 8.18-8.20 provide information on the distortion balancing between surfaces and the contribution of intrinsic and induced aberrations.

It can be seen from Figures 8.18-8.19 that the smile distortion balance in SysK and SysS is mainly between $S_1$ and $S_2$ during both the first pass and the second pass. $S_3$ and $S_4$ are less affected by the induced smile distortion, partly because the aberrations of the incoming beams are smaller, and both surfaces have larger radii of curvature.

In addition, in Figure 8.20 for SysF, the surface contribution is in much larger scale than that in SysK and SysS. At the same time, distortion is well corrected in SysF with much smaller residual distortion (<5 μm) than that in SysK and SysS. These two comparisons suggest a balancing between third-group distortion types and distortion types beyond the third group, since in general, larger surface contribution to low-order aberrations tends to be accompanied by more significant presence of higher-order aberrations.

## Chapter 9. Conclusion and future work

In this work, analytical formulae of aberration coefficients were first derived to shine understanding on the aberration behavior for systems with plane-symmetry. The derivation starts with defining a parameterization method of optical systems following an OAR and related paraxial properties. Wavefront aberration expansion is then defined with the third-group aberration coefficients in plane-symmetric systems identified. To derive the analytical form of the third-group aberration coefficients, the coefficients for one-surface systems with reference spheres at the surface are firstly derived, which serves as a foundation to derive the aberration coefficients at the exit pupil with freeform and induced contribution.

The theory development was motivated to understand the dependence of distortion on system parameters in freeform spectrometers. As such, example systems, a Dyson spectrometer as a benchmark non-freeform and two freeform 3-mirror double-pass spectrometers, are used to test the accuracy of the aberration coefficients, and the results show good consistency between the distortion calculated from real raytracing and from aberration coefficients in the systems limited by aberrations in the third group, which the theory limited itself to.

The advantage of analytical aberration coefficients is to show how aberrations are affected by system parameters such as radii of curvatures, incident angle of the OAR and paraxial ray data. Importantly, for freeform surfaces, the aberration coefficients also show the connection between the aberrations and freeform parameters such as Zernike coefficients. In addition, as the aberration coefficients are derived surface-by-surface, the information of surface aberration contributions can be obtained in analysis of surface aberration balancing.

While this work focuses on the derivation of the aberration coefficients, future efforts can be made in leveraging the knowledge gained to benefit aberration correction in plane-symmetric optical designs. For example, a few qualitative observations can be made from the aberration coefficient formulae. First, it can be seen that the OAR tilt angle, contained in the factor $C$, is a main factor driving all plane-symmetric aberrations. Minimizing surface tilt angle can effectively reduce surface aberration contribution. It can also be seen that since $W_{02002}$ and $W_{11011}$ are low-order aberrations, they are two critical aberration coefficients that have impact on all other plane-symmetric aberrations when calculating $W^{(S)}$, $W^{(XP)}$ and $W_m^{(XP)}$. Managing these two coefficients can be important in plane-symmetric optical design, which also makes $Z_5$ and $Z_{12}$ as important freeform coefficients since they are associated with the freeform contribution to $W_{02002}$ and $W_{11011}$. Finally, lower surface aberration generally means less sensitive to perturbations, so it is good practice to design aberration balancing with low surface contribution for easier tolerancing.

The hyperspectral imager design presented in this work was designed mainly through optimizing the performance at the image plane. With the knowledge of aberration theory, a different approach can be taken by firstly finding a good first-order starting point with decent balancing of $W_{02002}$ and $W_{11011}$. The need for freeform departures can be quickly analyzed with aberration theory to help rank potential starting points. Having a method to locate good starting points can be crucial in the overall design process.

In this work, the description of a spectrometer for a wavelength is independent from the descriptions for other wavelengths. Therefore, to analyze the distortion behavior of a spectrometer across the spectrum, the OARs and aberration coefficients need to be found independently for each wavelength of interest. For future work, a possible way to simplify the calculation process is to incorporate wavelength as a parameter into the paraxial optics in spectrometers.

In paraxial optics for rotationally symmetric systems, wavelength as a parameter has impact on first-order properties such as image distance and magnification. The effects on the first-order properties are usually described with chromatic aberrations. Similar effects are also present in plane-symmetric systems. However, one major difference is that in rotationally symmetric systems, the optical axis of the symmetry remains the same for all wavelengths, while in plane-symmetric systems, the OAR may change upon intersecting gratings or lenses as the wavelength changes. Therefore, the effects of wavelength on first-order properties cannot be simply described as a change of image distance and magnification, since the system description follows a different ray for each wavelength, making the wavelengths, to some extent, independent from each other. For this reason, in this work, each wavelength is treated separately, and in the process, the wavelength as a factor impacting distortion behavior is not reflected in the analytical formulae and can only be analyzed through numerical calculation.

In the context of spectrometers, the idea of incorporating wavelength into the paraxial optics leverages the fact that the fields are only in the direction of the slit in imaging spectrometers with the spectrum in the other direction. A possible way to describe the wavelength dependent OARs is through wavelength-dependent fields in the direction of spectrum, as shown in Figure 9.1. An OAR at a reference wavelength is chosen, and wavelength-dependent fields in the direction of the spectrum are defined with respect to the OAR at the reference wavelength.

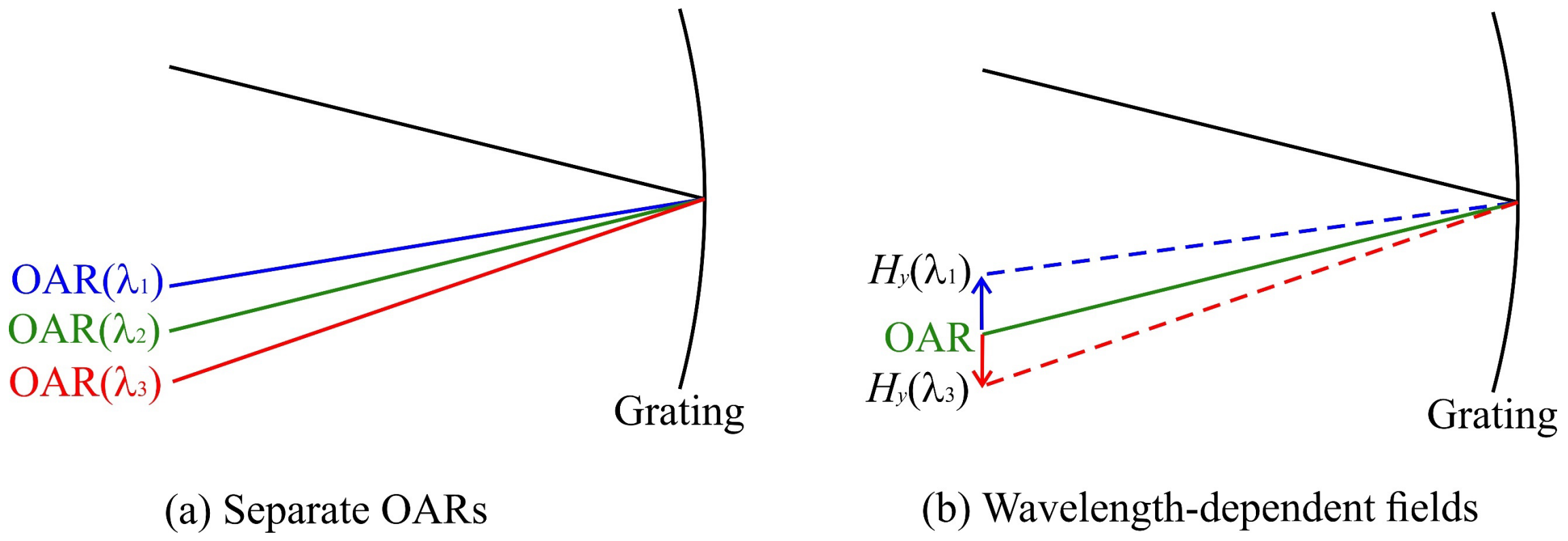

**Figure 9.1. Illustrations of paraxial optics based on (a) separate OARs and (b) wavelength-dependent fields.**

Within a future new framework of paraxial optics, the aberration coefficients derived will be wavelength-dependent and contain all spectral information. In this way, aberration coefficients do not need to be calculated separately for each wavelength, and the system description includes the wavelength as a parameter and remains the same for the whole spectrum. Therefore, the wavelength-dependent paraxial optics can potentially simplify the calculation process and reflect the wavelength-dependent distortion behavior in the analytical aberration coefficients. Note that this approach involves fundamental changes to the paraxial optics, and more investigation is needed in order to fully evaluate the viability of this new framework.

In a scope larger than the distortion in spectrometers, this work presents an general aberration theory for plane-symmetric optical systems. Paraxial optics and third-order aberration theory are important tools to understand the imaging capabilities of optical systems, which have been developed for rotationally symmetric systems over the history of optical design and taught to generations of optical designers. With advancements in optical fabrication, metrology, and assembly, new fields of optical design have been created where designers are pushing the boundaries of system structures, surface shapes, and design strategies. New system structures are proposed with arrangement of optical surfaces not following rotational symmetry. In this new era of optical design, we advance the aberration theory to be applicable to plane-symmetric systems. All third-group aberration coefficients, including freeform contribution and induced aberrations, were derived for plane-symmetric systems as functions of system parameters, which provide valuable design insights and new possibilities on how optical systems may be analyzed and optimized.

One potential development is to apply the aberration theory during the optimization of optical designs. Optimization is an important aspect of optical design in lens design software, where an optical design is gradually changed in search of a better performance. At the beginning and the end of each optimization cycle, the optical performance of the system is evaluated to determine whether the performance has improved. One important aspect of the optical performance is usually the amount of aberrations present in the system, which determines the imaging quality of the system. The evaluation of aberrations often involves tracing tens or hundreds of real rays and measuring their transverse aberrations. Alternatively, the aberration theory can be used to estimate the magnitude of aberrations. Although not all aberrations are included in the theory, aberration coefficients in the third group are often sufficient to cover the main aberration contributions. The advantage of using the aberration theory is that it normally requires much less computation than tracing real rays to gain a general picture of the aberration contributions, since obtaining aberration coefficients only requires the tracing of the OAR and simple calculation through the analytical formulae. Developing an aberration-based optimization method can be beneficial to situations where fast evaluations are desired. For example, it can be especially useful when a solution search is conducted for good starting-point designs, where the emphasis of the search is often to cover a wide range of possible solutions. In this case, using the aberration theory can potentially accelerate the searching process.

In summary, this work establishes a general aberration theory applicable to plane-symmetric systems. It is demonstrated that the theory can be applied to analyze distortion aberrations in spectrometers, and for future work, other potential applications can be explored. In the end, we envision that this work can serve as a useful tool for future optical designers and deepen our understanding of more complicated optical systems.

## Appendix I. Derivation of the paraxial ray-tracing equations in plane-symmetric systems

This derivation is based on the system description method presented in Chapter 2 with the establishment of the OAR and surface coordinate systems. The goal of this appendix is to derive the paraxial ray-tracing equations shown in Equations 3.9-3.14.

Firstly, we start by noting that Equations 3.11 and 3.14 are self-evident. Since rays do not change directions between optical surfaces under the system setup of this work, the ray angles in the object space of one surface are equal to the corresponding angles of the same ray in the image space of the previous surface.

Next, we derive the paraxial equations for the propagation between surfaces, namely Equations 3.9 and 3.12.

Consider two sequential surfaces, Surface $i$ and Surface $i$+1, as shown in Figure I.1. A real ray is propagating between the two surfaces and intersecting the two surfaces at points $P_i$ and $P_{i+1}$. Points $O_i$ and $O_{i+1}$ are the origins of the surface coordinate systems of Surface $i$ and Surface $i$+1, respectively. The two surface coordinate systems are separated by the distance $t_i$ along the OAR. Vectors $\boldsymbol{O_iP_i}$ and $\boldsymbol{O_{i+1}P_{i+1}}$ are defined in their corresponding surface coordinate systems as

$$\boldsymbol{O_iP_i} = \left(x_i, y_i, z_i\right) \;\; in \;\; \mathrm{X}_i\mathrm{Y}_i\mathrm{Z}_i \tag{I.1}$$

$$\boldsymbol{O_{i+1}P_{i+1}} = \left(x_{i+1}, y_{i+1}, z_{i+1}\right) \;\; in \;\; \mathrm{X}_{i+1}\mathrm{Y}_{i+1}\mathrm{Z}_{i+1} \tag{I.2}$$

where $z_i$ and $z_{i+1}$ are the surface sag which follow the general sag function as shown in Equation 3.21.

Another coordinate system, labeled as $X'_{qi}Y'_{qi}Z'_{qi}$, is defined at $O_i$ with the Z-axis pointing towards the direction of the OAR in the image space of Surface *i*. It can be seen that the sagittal and tangential planes, as defined in Section 3.2.1, are $X'_{qi}Z'_{qi}$ and $Y'_{qi}Z'_{qi}$ planes, respectively. The vector connecting the two surface coordinate systems, $\boldsymbol{O_iO_{i+1}}$, can be written in $X'_{qi}Y'_{qi}Z'_{qi}$ as

$$\boldsymbol{O_iO_{i+1}} = (0, 0, t_i) \;\; in \;\; X'_{qi}Y'_{qi}Z'_{qi} \tag{I.3}$$

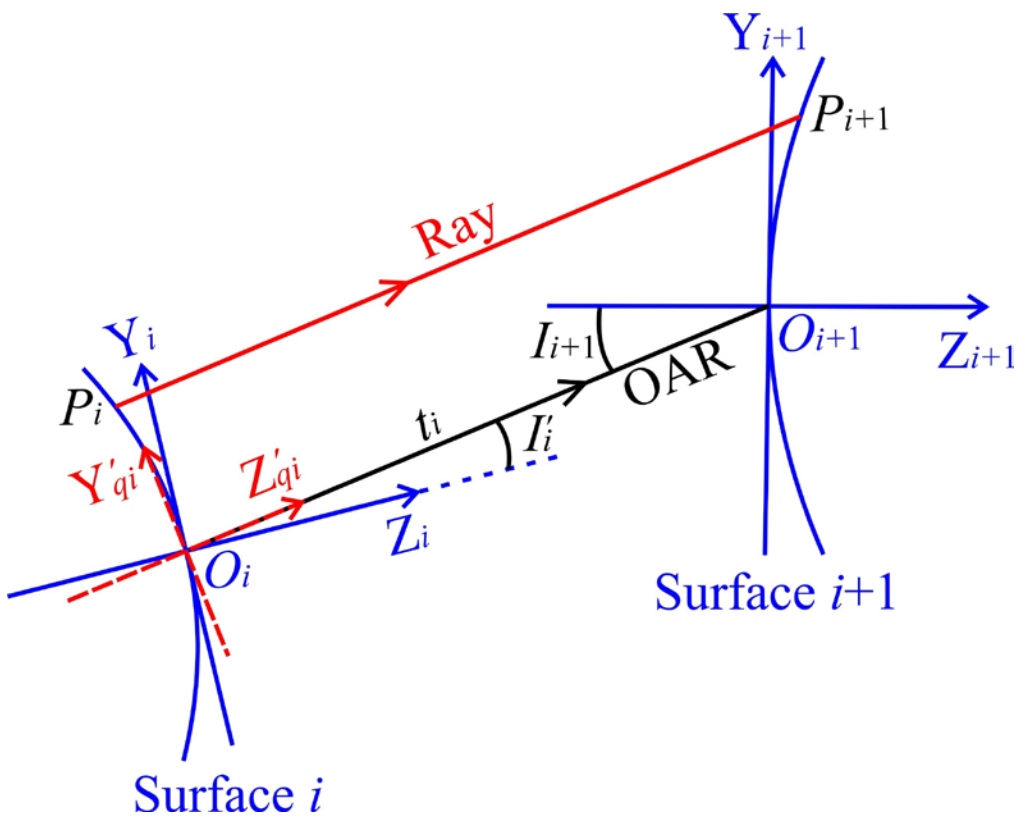

**Figure I.1. Illustration of a real ray in solid red propagating between two surfaces, Surface *i* and Surface *i*+1, and intersecting the surfaces at $P_i$ and $P_{i+1}$. In solid black the OAR.**

From Figure I.1, it can be seen that the coordinate systems $X_iY_iZ_i$ and $X_{i+1}Y_{i+1}Z_{i+1}$ are related to $X'_{qi}Y'_{qi}Z'_{qi}$ by the OAR tilt angles, $I'_i$ and $I_{i+1}$, respectively. With coordinate transformation, vectors $\boldsymbol{O_iP_i}$ and $\boldsymbol{O_{i+1}P_{i+1}}$ can be expressed in $X'_{qi}Y'_{qi}Z'_{qi}$ as

$$\boldsymbol{O_iP_i} = (x_{qi}, y_{qi}, z_{qi}) \;\; in \;\; X'_{qi}Y'_{qi}Z'_{qi} \tag{I.4}$$

$$\boldsymbol{O_{i+1}P_{i+1}} = (x_{qi+1}, y_{qi+1}, z_{qi+1}) \;\; in \;\; X'_{qi}Y'_{qi}Z'_{qi} \tag{I.5}$$

where

$$\begin{cases} x_{qi} = x_i \\ y_{qi} = y_i \cos I'_i - z_i \sin I'_i \\ z_{qi} = y_i \sin I'_i + z_i \cos I'_i \end{cases} \tag{I.6}$$

$$\begin{cases} x_{qi+1} = x_{i+1} \\ y_{qi+1} = y_{i+1} \cos I_{i+1} - z_{i+1} \sin I_{i+1} \\ z_{qi+1} = y_{i+1} \sin I_{i+1} + z_{i+1} \cos I_{i+1} \end{cases} \tag{I.7}$$

Therefore, the vector along the ray, $\boldsymbol{P_iP_{i+1}}$, can be written in $X'_{qi}Y'_{qi}Z'_{qi}$ as

$$\begin{aligned} \boldsymbol{P_iP_{i+1}} &= -\boldsymbol{O_iP_i} + \boldsymbol{O_iO_{i+1}} + \boldsymbol{O_{i+1}P_{i+1}} \\ &= \left(x_{qi+1} - x_{qi}, y_{qi+1} - y_{qi}, z_{qi+1} - z_{qi} + t_i\right) \quad in\ X'_{qi}Y'_{qi}Z'_{qi} \end{aligned} \tag{I.8}$$

The projections of $\boldsymbol{P_iP_{i+1}}$ onto the sagittal and tangential planes, or $X'_{qi}Z'_{qi}$ and $X'_{qi}Z'_{qi}$ planes, are defined as $\boldsymbol{P_iP_{i+1(s)}}$ and $\boldsymbol{P_iP_{i+1(t)}}$, which are given as

$$\boldsymbol{P_iP_{i+1(s)}} = \left(x_{qi+1} - x_{qi}, 0, z_{qi+1} - z_{qi} + t_i\right) \quad in\ X'_{qi}Y'_{qi}Z'_{qi} \tag{I.9}$$

$$\boldsymbol{P_iP_{i+1(t)}} = \left(0, y_{qi+1} - y_{qi}, z_{qi+1} - z_{qi} + t_i\right) \quad in\ X'_{qi}Y'_{qi}Z'_{qi} \tag{I.10}$$

$\boldsymbol{P_iP_{i+1(s)}}$ and $\boldsymbol{P_iP_{i+1(t)}}$ can also be used to represent real rays in the sagittal and tangential planes. Therefore, according to the definitions of $u_x$ and $u_y$ defined in Section 3.2.1, $u'_{xi}$ is the tangent of the angle between $\boldsymbol{P_iP_{i+1(s)}}$ and $\boldsymbol{O_iO_{i+1}}$ and is positive if $x_{qi+1}$-$x_{qi}$ is positive. Similarly, $u'_{yi}$ is the tangent of the angle between $\boldsymbol{P_iP_{i+1(t)}}$ and $\boldsymbol{O_iO_{i+1}}$ and is positive if $y_{qi+1}$-$y_{qi}$ is positive. With the properties of vector operations, $u'_{xi}$ and $u'_{yi}$ can be expressed as

$$u'_{xi} = -\frac{\left(\boldsymbol{P_iP_{i+1(s)}} \times \boldsymbol{O_iO_{i+1}}\right)_y}{\boldsymbol{P_iP_{i+1(s)}} \cdot \boldsymbol{O_iO_{i+1}}} \tag{I.11}$$

$$u'_{yi} = \frac{\left(\boldsymbol{P_iP_{i+1(t)}} \times \boldsymbol{O_iO_{i+1}}\right)_x}{\boldsymbol{P_iP_{i+1(t)}} \cdot \boldsymbol{O_iO_{i+1}}} \tag{I.12}$$

where the notation (***Vector***)$_x$ means the $x$ component of ***Vector***, and the symbol × denotes vector cross products.

Substituting Equations I.3, I.9 and I.10 in Equations I.11-I.12, we have

$$u'_{xi} = \frac{t_i \left( x_{qi+1} - x_{qi} \right)}{t_i \left( z_{qi+1} - z_{qi} + t_i \right)} \tag{I.13}$$

$$u'_{yi} = \frac{t_i \left( y_{qi+1} - y_{qi} \right)}{t_i \left( z_{qi+1} - z_{qi} + t_i \right)} \tag{I.14}$$

Further expanding Equations I.13 and I.14, we have

$$x_{qi+1} - x_{qi} = \left( z_{qi+1} - z_{qi} + t_i \right) u'_{xi} \tag{I.15}$$

$$y_{qi+1} - y_{qi} = \left( z_{qi+1} - z_{qi} + t_i \right) u'_{yi} \tag{I.16}$$

Use the transformation equations in Equations I.6-I.7, Equations I.15-I.16 can be rewritten as

$$x_{i+1} - x_i = \left( y_{i+1} \sin I_{i+1} + z_{i+1} \cos I_{i+1} - y_i \sin I'_i - z_i \cos I'_i + t_i \right) u'_{xi} \tag{I.17}$$

$$\begin{gathered} y_{i+1} \cos I_{i+1} - z_{i+1} \sin I_{i+1} - y_i \cos I'_i + z_i \sin I'_i = \\ \left( y_{i+1} \sin I_{i+1} + z_{i+1} \cos I_{i+1} - y_i \sin I'_i - z_i \cos I'_i + t_i \right) u'_{yi} \end{gathered} \tag{I.18}$$

As the ray approach the OAR, ray angles and heights, $u'_{xi}$, $u'_{yi}$, $x_i$, $x_{i+1}$, $y_i$, and $y_{i+1}$, are all approaching zero. In the paraxial regime, the dominant terms are linear to the ray heights and angles. Since $z_i$ and $z_{i+1}$ are at least second-order in terms of ray heights as discussed in Section 3.2.3, $z_i$ and $z_{i+1}$ are reduced to zero as we are focused on linear terms. Therefore, in the paraxial regime, Equations I.17-I.18 are reduced to

$$x_{i+1} = x_i + u'_{xi} t_i \tag{I.19}$$

$$y_{i+1}\cos I_{i+1} = y_i \cos I_i' + u_{yi}' t_i \tag{I.20}$$

which are Equations 3.9 and 3.12 presented in Section 3.2.1.

Then, we derive the paraxial equations for refraction or reflection at optical surfaces, which are shown in Equations 3.10 and 3.13.

Consider a ray undergoing refraction at a point $P_i(x_i, y_i, z_i)$ on Surface $i$ as shown in Figure I.2. The signed refractive indexes in object and image space are $n_i$ and $n'_i$, respectively. Vector $\boldsymbol{N}$ represents the surface normal vector at the point $P_i$ and can be expressed in the surface coordinate system $X_iY_iZ_i$ as

$$\boldsymbol{N} = \left(\frac{\partial z_i}{\partial x_i}, \frac{\partial z_i}{\partial y_i}, -1\right) \quad in\ X_iY_iZ_i \tag{I.21}$$

Note that in this derivation, $\boldsymbol{N}$ only needs to be along the surface normal. The magnitude of $\boldsymbol{N}$ and whether $\boldsymbol{N}$ is pointing towards object space or image space is of no importance.

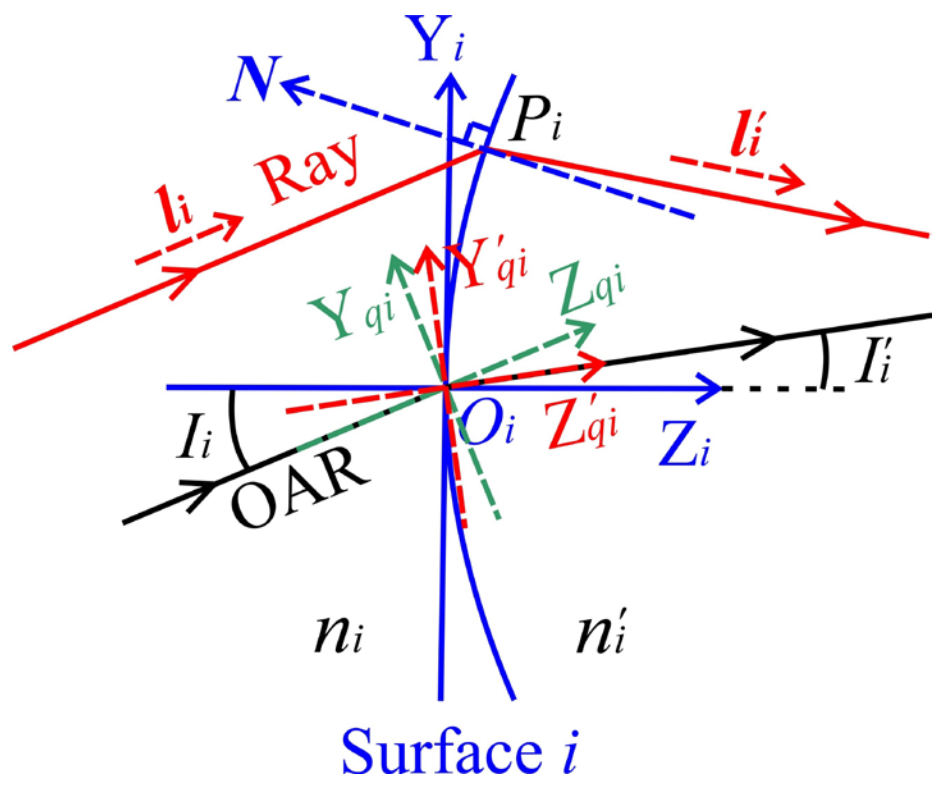

**Figure I.2. Illustration of a real ray in solid red refracting at $P_i$ on Surface $i$. In black the OAR. The vector $N$ is the surface normal at $P_i$.**

Similar to how $X'_{qi}Y'_{qi}Z'_{qi}$ is defined in Figure I.1, two coordinate systems, $X_{qi}Y_{qi}Z_{qi}$ and $X'_{qi}Y'_{qi}Z'_{qi}$, are set up with their Z-axes pointing along the direction of the OAR in object and image space, respectively. Therefore, $X_{qi}Y_{qi}Z_{qi}$ and $X'_{qi}Y'_{qi}Z'_{qi}$ are connected to the surface coordinate system $X_iY_iZ_i$ by the OAR tilt angles, $I_i$ and $I_{i+1}$.

Define vectors $\boldsymbol{l_i}$ and $\boldsymbol{l'_i}$ as vectors pointing towards the ray direction before and after refraction, respectively. With $u_x$ and $u_y$ defined, $\boldsymbol{l_i}$ and $\boldsymbol{l'_i}$ can be expressed in $X_{qi}Y_{qi}Z_{qi}$ and $X'_{qi}Y'_{qi}Z'_{qi}$ as

$$\boldsymbol{l_i} = l_{qzi}\left(u_{xi}, u_{yi}, 1\right) \quad in\ \mathrm{X}_{qi}\mathrm{Y}_{qi}\mathrm{Z}_{qi} \tag{I.22}$$

$$\boldsymbol{l'_i} = l'_{qzi}\left(u'_{xi}, u'_{yi}, 1\right) \quad in\ \mathrm{X}'_{qi}\mathrm{Y}'_{qi}\mathrm{Z}'_{qi} \tag{I.23}$$

where $l_{qzi}$ and $l'_{qzi}$ are real numbers without specific values assigned, because only the directions of $\boldsymbol{l_i}$ and $\boldsymbol{l'_i}$ are important for this derivation. Vectors $\boldsymbol{l_i}$ and $\boldsymbol{l'_i}$ can also be expressed in $X_iY_iZ_i$ as

$$\boldsymbol{l_i} = \left(l_{xi}, l_{yi}, l_{zi}\right) \quad in\ \mathrm{X}_i\mathrm{Y}_i\mathrm{Z}_i \tag{I.24}$$

$$\boldsymbol{l'_i} = \left(l'_{xi}, l'_{yi}, l'_{zi}\right) \quad in\ \mathrm{X}_i\mathrm{Y}_i\mathrm{Z}_i \tag{I.25}$$

with the transformation equations given as

$$\begin{cases} l_{xi} = l_{qzi}u_{xi} \\ l_{yi} = l_{qzi}\left(u_{yi}\cos I_i + \sin I_i\right) \\ l_{zi} = l_{qzi}\left(-u_{yi}\sin I_i + \cos I_i\right) \end{cases} \tag{I.26}$$

$$\begin{cases} l'_{xi} = l'_{qzi}u'_{xi} \\ l'_{yi} = l'_{qzi}\left(u'_{yi}\cos I'_i + \sin I'_i\right) \\ l'_{zi} = l'_{qzi}\left(-u'_{yi}\sin I'_i + \cos I'_i\right) \end{cases} \tag{I.27}$$

Let us define the wave vectors $\boldsymbol{k_i}$ and $\boldsymbol{k'_i}$ in $X_iY_iZ_i$ along the ray directions in object and image space as

$$\boldsymbol{k_i} = \frac{2\pi}{\lambda_i}\boldsymbol{l_{ui}} \quad in\ \mathrm{X}_i\mathrm{Y}_i\mathrm{Z}_i \tag{I.28}$$

$$\boldsymbol{k}_i' = \frac{2\pi}{\lambda_i'}\boldsymbol{l}_{ui}' \quad in\ \mathrm{X}_i\mathrm{Y}_i\mathrm{Z}_i \tag{I.29}$$

where $\lambda_i$ and $\lambda'_i$ are wavelengths of the ray in object and image space, respectively, and $\boldsymbol{l}_{ui}$ and $\boldsymbol{l}'_{ui}$ are unit vectors pointing in the directions of $\boldsymbol{l}_i$ and $\boldsymbol{l}'_i$. Although only rays are considered in this derivation, their corresponding wave vectors still need to satisfy the boundary conditions at the surface [82], which are given as

$$\boldsymbol{k}_{i\parallel} = \boldsymbol{k}'_{i\parallel} \tag{I.30}$$

where $\boldsymbol{k}_{i\parallel}$ and $\boldsymbol{k}'_{i\parallel}$ are the components of $\boldsymbol{k}_i$ and $\boldsymbol{k}'_i$ that are parallel to the optical surface at $P_i$. The directions of $\boldsymbol{k}_i$ and $\boldsymbol{k}'_i$ components normal to the optical surface at $P_i$ remain the same on refraction and change signs on reflection, which are given as

$$\boldsymbol{k}_{ui\perp} = \boldsymbol{k}'_{ui\perp} \quad (refraction) \tag{I.31}$$

$$\boldsymbol{k}_{ui\perp} = -\boldsymbol{k}'_{ui\perp} \quad (reflection) \tag{I.32}$$

where $\boldsymbol{k}_{ui\perp}$ and $\boldsymbol{k}'_{ui\perp}$ are the unit vectors along the directions of $\boldsymbol{k}_i$ and $\boldsymbol{k}'_i$ components normal to the optical surface at $P_i$, respectively.

The paraxial equations for refraction or reflection can be obtained by expanding Equations I.30-I.32 and focusing on the low-order terms in terms of ray angles and heights. First, we focus on Equations I.31-I.32. Vectors $\boldsymbol{k}_{ui\perp}$ and $\boldsymbol{k}'_{ui\perp}$ can be written as

$$\boldsymbol{k}_{ui\perp} = \frac{\boldsymbol{k}_i \cdot \boldsymbol{N}}{|\boldsymbol{k}_i \cdot \boldsymbol{N}|}\boldsymbol{N}_u \tag{I.33}$$

$$\boldsymbol{k}'_{ui\perp} = \frac{\boldsymbol{k}'_i \cdot \boldsymbol{N}}{|\boldsymbol{k}'_i \cdot \boldsymbol{N}|}\boldsymbol{N}_u \tag{I.34}$$

where $\boldsymbol{N}_u$ is the unit vector in the direction of $\boldsymbol{N}$.

By expressing $\boldsymbol{k_i}$ and $\boldsymbol{N}$ in $X_iY_iZ_i$ with Equations I.21 and I.24-I.29, $\boldsymbol{k_i}\cdot\boldsymbol{N}$ can be written as

$$\boldsymbol{k_i}\cdot\boldsymbol{N}=\frac{2\pi}{\lambda_i\left|\boldsymbol{l_i}\right|}l_{qzi}\left[u_{xi}\frac{\partial z_i}{\partial x_i}+\left(u_{yi}\cos I_i+\sin I_i\right)\frac{\partial z_i}{\partial y_i}-\left(-u_{yi}\sin I_i+\cos I_i\right)\right] \quad (I.35)$$

In the paraxial regime where all ray angles and heights approach zero, we have

$$\frac{\partial z_i}{\partial x_i}\approx 2A_{2-0}x_i \quad (I.36)$$

$$\frac{\partial z_i}{\partial y_i}\approx 2A_{0-2}y_i \quad (I.37)$$

which is based on the surface sag function discussed in Section 3.2.3, and $\boldsymbol{k_i}\cdot\boldsymbol{N}$ is dominated by its zero-order term in terms of ray angles and heights, given as

$$\boldsymbol{k_i}\cdot\boldsymbol{N}\approx\frac{2\pi}{\lambda_i}l_{qzi}\cos I_i \quad (I.38)$$

Since $\lambda$ and $\cos I'$ are always positive, in the paraxial regime, Equation I.33-I.34 can be written as

$$\boldsymbol{k_{ui\perp}}\approx sign\left(l_{qzi}\right)\boldsymbol{N_u} \quad (I.39)$$

$$\boldsymbol{k'_{ui\perp}}\approx sign\left(l'_{qzi}\right)\boldsymbol{N_u} \quad (I.40)$$

Therefore, Equations I.31-I.32 become

$$sign\left(l_{qzi}\right)=sign\left(l'_{qzi}\right)\quad (refraction) \quad (I.41)$$

$$sign\left(l_{qzi}\right)=-sign\left(l'_{qzi}\right)\quad (reflection) \quad (I.42)$$

Equations I.41-I.42 will help on determining the signs of refractive indexes in the following derivation focusing on expanding Equation I.30.

Equation I.30 can be rewritten as

$$\boldsymbol{k}_i - \left(\boldsymbol{k}_i \cdot \boldsymbol{N}_u\right)\boldsymbol{N}_u = \boldsymbol{k}_i' - \left(\boldsymbol{k}_i' \cdot \boldsymbol{N}_u\right)\boldsymbol{N}_u \tag{I.43}$$

further expanding Equation I.43 using Equations I.28-I.29 yields

$$\frac{2\pi}{\lambda_i}\frac{\boldsymbol{l}_i}{|\boldsymbol{l}_i|} - \frac{2\pi}{\lambda|\boldsymbol{l}_i||\boldsymbol{N}|^2}\left(\boldsymbol{l}_i \cdot \boldsymbol{N}\right)\boldsymbol{N} = \frac{2\pi}{\lambda_i'}\frac{\boldsymbol{l}_i'}{|\boldsymbol{l}_i'|} - \frac{2\pi}{\lambda'|\boldsymbol{l}_i'||\boldsymbol{N}|^2}\left(\boldsymbol{l}_i' \cdot \boldsymbol{N}\right)\boldsymbol{N} \tag{I.44}$$

we then rewrite Equation I.44 as

$$\frac{\lambda_i'}{\lambda_i}\frac{|\boldsymbol{l}_i'|}{|\boldsymbol{l}_i|}\left[|\boldsymbol{N}|^2\boldsymbol{l}_i - \left(\boldsymbol{l}_i \cdot \boldsymbol{N}\right)\boldsymbol{N}\right] = |\boldsymbol{N}|^2\boldsymbol{l}_i' - \left(\boldsymbol{l}_i' \cdot \boldsymbol{N}\right)\boldsymbol{N} \tag{I.45}$$

The relationship between ray angles before and after refraction or reflection can be obtained by substituting Equations I.21 and I.26-I.27 into Equation I.45.

To simplify the expression, certain quantities can be reduced to terms linear to ray angles and heights in the paraxial regime, which are given as

$$\frac{|\boldsymbol{l}_i'|}{|\boldsymbol{l}_i|} = \frac{|l_{qzi}'|\sqrt{u_{xi}'^2 + u_{yi}'^2 + 1}}{|l_{qzi}|\sqrt{u_{xi}^2 + u_{yi}^2 + 1}} \approx \frac{|l_{qzi}'|}{|l_{qzi}|} \tag{I.46}$$

$$|\boldsymbol{N}|^2 = \left(\frac{\partial z_i}{\partial x_i}\right)^2 + \left(\frac{\partial z_i}{\partial y_i}\right)^2 + 1 \approx 1 \tag{I.47}$$

$$\begin{aligned}\boldsymbol{l}_i \cdot \boldsymbol{N} &= l_{qzi}\left[u_{xi}\frac{\partial z_i}{\partial x_i} + \left(u_{yi}\cos I_i + \sin I_i\right)\frac{\partial z_i}{\partial y_i} - \left(-u_{yi}\sin I_i + \cos I_i\right)\right] \\ &\approx l_{qzi}\left(2A_{0-2}y_i\sin I_i + u_{yi}\sin I_i - \cos I_i\right)\end{aligned} \tag{I.48}$$

Similarly,

$$\boldsymbol{l}_i' \cdot \boldsymbol{N} \approx l_{qzi}' \left(2A_{0-2} y_i \sin I_i' + u_{yi}' \sin I_i' - \cos I_i'\right) \tag{I.49}$$

Therefore, comparing the $x$-component on both sides of Equation I.45 in the paraxial regime, we have

$$\begin{aligned} &\frac{\lambda_i'}{\lambda_i} \frac{\left|l_{qzi}'\right|}{\left|l_{qzi}\right|} \left[ l_{qzi} u_{xi} - l_{qzi} \left(2A_{0-2} y_i \sin I_i + u_{yi} \sin I_i - \cos I_i\right) 2A_{2-0} x_i \right] = \\ &\left[ l_{qzi}' u_{xi}' - l_{qzi}' \left(2A_{0-2} y_i' \sin I_i' + u_{yi}' \sin I_i' - \cos I_i'\right) 2A_{2-0} x_i \right] \end{aligned} \tag{I.50}$$

Further simplifying Equation I.50 and dropping terms with second order or higher order dependence on ray angles and heights, we have

$$\frac{\lambda_i'}{\lambda_i} \frac{\left|l_{qzi}'\right|}{\left|l_{qzi}\right|} \frac{l_{qzi}}{l_{qzi}'} \left(u_{xi} + 2A_{2-0} x_i \cos I_i\right) = u_{xi}' + 2A_{2-0} x_i \cos I_i' \tag{I.51}$$

In addition, the relationship between wavelengths and refractive indexes is given as [82]

$$\frac{\lambda_i'}{\lambda_i} = \left|\frac{n_i}{n_i'}\right| \tag{I.52}$$

where $n_i$ and $n'_i$ are signed refractive indexes in object and image space. Since wavelengths are always positive, the absolute value of ratio between $n_i$ and $n'_i$ is taken in Equation I.52. Furthermore, according to Equations I.41-I.42, we have

$$\frac{\left|l_{qzi}'\right|}{\left|l_{qzi}\right|} \frac{l_{qzi}}{l_{qzi}'} = \frac{sign\left(l_{qzi}\right)}{sign\left(l_{qzi}'\right)} = \begin{cases} 1 & (refraction) \\ -1 & (reflection) \end{cases} \tag{I.53}$$

Based on how the sign of refractive index is defined in Section 2.1, reflection always induces a sign change on the refractive index from object space to image space. Therefore, we can write

$$\frac{\lambda_i'}{\lambda_i}\frac{\left|l_{qzi}'\right|}{\left|l_{qzi}\right|}\frac{l_{qzi}}{l_{qzi}'}=\left|\frac{n_i}{n_i'}\right|\frac{sign\left(l_{qzi}\right)}{sign\left(l_{qzi}'\right)}=\frac{n_i}{n_i'} \tag{I.54}$$

Then, Equation I.51 becomes

$$\frac{n_i}{n_i'}\left(u_{xi}+\frac{x_i\cos I_i}{R_{x-eff}}\right)=u_{xi}'+\frac{x_i\cos I_i'}{R_{x-eff}} \tag{I.55}$$

where the effective radius of curvature defined in Equation 3.24 is used. A simple rearrangement of terms in Equation I.55 yields the paraxial raytracing equation for refraction or reflection in the sagittal direction, given as

$$\begin{aligned} n_i'u_{xi}' &= n_i u_{xi}-\left(\frac{n_i'\cos I_i'-n_i\cos I_i}{R_{x-effi}}\right)x_i \\ &= nu_{xi}-x_i\phi_{xi} \end{aligned} \tag{I.56}$$

The paraxial raytracing equation for refraction or reflection in the tangential direction, shown in Equation 3.13, can be derived the same way by comparing the $y$-component on both sides of Equation I.45 in the paraxial regime.

## Appendix II. Illustration of the process of deriving aberration coefficients from $W^{(S)}$

In Section 5.2.3, aberration coefficients for one-surface systems with reference spheres at the optical surface were obtained from the general expression of $W^{(S)}$ in Equation 5.81, which is given for convenience as

$$W^{(S)} \approx \Delta\left[\frac{n(\boldsymbol{r}\cdot\boldsymbol{s_h})}{s_h}\right]-\frac{1}{2}\left(\frac{n}{s_0}-\frac{n\cos I}{R_{x-eff}}\right)\Delta\left(\frac{s_0}{s_h}\right)r^2+\frac{1}{2}\Delta\left[\frac{n(\boldsymbol{r}\cdot\boldsymbol{s_h})^2}{s_h^3}\right]$$
$$-\frac{1}{2}\left(\frac{n}{s_0}-\frac{n\cos I}{R_{x-eff}}\right)\Delta\left[\frac{s_0(\boldsymbol{r}\cdot\boldsymbol{s_h})}{s_h^3}\right]r^2-\frac{1}{2}\Delta\left[\frac{nH_{ly}}{s_h}\sin(I-\theta_h)\right]\frac{r^2}{R_{x-eff}} \tag{5.81}$$
$$+\frac{1}{8}\left(\frac{n}{s_0}-\frac{n\cos I}{R_{x-eff}}\right)^2\Delta\left(\frac{1}{ns_0}\right)r^4+z_{f2}\Delta(n\cos I)$$

The process of obtaining the analytical expression of aberration coefficients include steps of expanding $W^{(S)}$, grouping terms based on their dependencies, and simplifying the expressions of aberration coefficients. In this appendix, we illustrate this process on two terms from the $W^{(S)}$ expression in Equation 5.81. The rest of terms follow the same process.

Define $W_{T1}{}^{(S)}$ as the first term in the $W^{(S)}$ expression, given as

$$W_{T1}^{(S)}=\Delta\left[\frac{n(\boldsymbol{r}\cdot\boldsymbol{s_h})}{s_h}\right] \tag{II.1}$$

We will first expand $W_{T1}{}^{(S)}$ to see how it contribute to each aberration coefficient. Recall that vector $\boldsymbol{r}$ starts from the origin of the surface coordinate system and points towards the XY-plane projection in the surface coordinate system of the intersection between the ray and the surface. Vector $\boldsymbol{r}$ is defined in Equation 5.58 as

$$\boldsymbol{r} = (x, y, 0) \tag{5.58}$$

Vector $\boldsymbol{s_h}$ starts from the origin of the surface coordinate system and points towards the object point. Vector $\boldsymbol{s_h}$ is expressed in Equation 5.59 as

$$\begin{aligned} \boldsymbol{s_h} &= \boldsymbol{s_0} + \boldsymbol{H_{ls}} \\ &= \left(H_{lx}, s_0 \sin I + H_{ly} \cos\left(I - \theta_h\right), s_0 \cos I - H_{ly} \sin\left(I - \theta_h\right)\right) \end{aligned} \tag{5.59}$$

where $\boldsymbol{s_0}$, defined in Section 5.2.1, is the vector from the origin of the surface coordinate system to the on-OAR object point, and $\boldsymbol{H_{ls}}$, defined in Section 5.2.1, is the vector from the on-OAR object point to the off-OAR object point expressed in the surface coordinate system. Vectors $\boldsymbol{s_h}$, $\boldsymbol{s_0}$, and $\boldsymbol{H_{ls}}$, are all defined in the object space and have their corresponding vectors defined in the image space.

Quantities $s_h$ in Equation II.1 and $s_0$ in Equation 5.59 are the signed lengths of vectors $\boldsymbol{s_h}$ and $\boldsymbol{s_0}$, respectively, and are positive if the corresponding vector is pointing towards the positive Z-axis of the local surface coordinate system.

From Equation 5.59, we can calculate $s_h$ as

$$s_h = \left|\boldsymbol{s_h}\right| sign\left(\boldsymbol{s_h}\right) = s_0 \left[1 + \frac{\left(\boldsymbol{H_l} \cdot \boldsymbol{H_l}\right)}{s_0^2} + \frac{2H_{ly}}{s_0} \sin\theta_h\right]^{\frac{1}{2}} \tag{II.2}$$

where $\boldsymbol{H_l}=(H_{lx}, H_{ly})$, defined in Section 4.1, is the same vector as $\boldsymbol{H_{ls}}$ but expressed in the object coordinate system. In Equation II.2, $s_h$ and $s_0$ are assumed to have the same sign, which is true when $\boldsymbol{H_{ls}}$ does not greatly impact the directions of $\boldsymbol{s_h}$. With $s_h$ expressed in Equation II.2, $1/s_h$ can be written as

$$\begin{aligned}\frac{1}{s_h} &= \frac{1}{s_0}\left(1+\frac{(\boldsymbol{H_l}\cdot\boldsymbol{H_l})}{s_0^2}+\frac{2H_{ly}}{s_0}\sin\theta_h\right)^{-\frac{1}{2}} \\ &\approx \frac{1}{s_0}-\frac{1}{2s_0}\left(\frac{(\boldsymbol{H_l}\cdot\boldsymbol{H_l})}{s_0^2}+\frac{2H_{ly}}{s_0}\sin\theta_h\right)+\frac{3}{8s_0}\left(\frac{(\boldsymbol{H_l}\cdot\boldsymbol{H_l})}{s_0^2}+\frac{2H_{ly}}{s_0}\sin\theta_h\right)^2\end{aligned} \tag{II.3}$$

where $1/s_h$ is Taylor expanded to the second order in the second line. Higher order terms do not contribute to third-group aberrations. The Taylor expansion is valid when $\boldsymbol{H_l}$ is smaller than $\boldsymbol{s_0}$ in length.

With Equations 5.58, 5.59 and II.3, $W_{T1}^{(S)}$ can be expressed as

$$\begin{aligned}W_{T1}^{(S)} &= \Delta\left[\frac{n(\boldsymbol{r}\cdot\boldsymbol{s_h})}{s_h}\right] \\ &\approx \Delta\Bigg\{\frac{n}{s_0}\left[H_{lx}x+ys_0\sin I+yH_{ly}\cos(I-\theta_h)\right] \\ &\quad\left[1-\frac{1}{2}\left(\frac{(\boldsymbol{H_l}\cdot\boldsymbol{H_l})}{s_0^2}+\frac{2H_{ly}}{s_0}\sin\theta_h\right)+\frac{3}{8}\left(\frac{(\boldsymbol{H_l}\cdot\boldsymbol{H_l})}{s_0^2}+\frac{2H_{ly}}{s_0}\sin\theta_h\right)^2\right]\Bigg\} \\ &= \Delta\Bigg\{\frac{n}{s_0}\left[H_{lx}x+ys_0\sin I+yH_{ly}\cos(I-\theta_h)\right] \\ &\quad\left(1-\frac{(\boldsymbol{H_l}\cdot\boldsymbol{H_l})}{2s_0^2}-\frac{H_{ly}}{s_0}\sin\theta_h+\frac{3(\boldsymbol{H_l}\cdot\boldsymbol{H_l})^2}{8s_0^4}+\frac{3H_{ly}^2}{2s_0^2}\sin^2\theta_h \right. \\ &\quad\left.+\frac{3(\boldsymbol{H_l}\cdot\boldsymbol{H_l})H_{ly}}{2s_0^3}\sin\theta_h\right)\Bigg\}\end{aligned} \tag{II.4}$$

From Equation II.4, $W_{T1}^{(S)}$ can be viewed as a function of $x$, $y$, $H_{lx}$ and $H_{ly}$, which are components of vectors $\boldsymbol{r}$ and $\boldsymbol{H_l}$. We then expand $W_{T1}^{(S)}$ as

$$W_{T1}^{(S)} = \sum_{k,m,n,p,q}^{\infty} W_{T1r-2k+n+p,2m+n+q,n,p,q}^{(S)}(\boldsymbol{H_l}\cdot\boldsymbol{H_l})^k(\boldsymbol{r}\cdot\boldsymbol{r})^m(\boldsymbol{H_l}\cdot\boldsymbol{r})^n(\boldsymbol{i}\cdot\boldsymbol{H_l})^p(\boldsymbol{i}\cdot\boldsymbol{r})^q \tag{II.5}$$

where $W_{T1r\text{-}2k+n+p,2m+n+q,n,p,q}^{(S)}$ is the coefficient for each term. Ignoring terms beyond the third group, $W_{T1}^{(S)}$ can be approximated as

$$\begin{aligned} W_{T1}^{(S)} &\approx \Delta\left\{\frac{n}{s_0}\left[H_{lx}x + ys_0\sin I + yH_{ly}\cos\left(I-\theta_h\right)\right]\right. \\ &\left.\left(1-\frac{\left(\boldsymbol{H}_l\cdot\boldsymbol{H}_l\right)}{2s_0^2}-\frac{H_{ly}}{s_0}\sin\theta_h\right)\right\} \\ &= \Delta\left\{\frac{n}{s_0}\left[\left(\boldsymbol{H}_l\cdot\boldsymbol{r}\right) + ys_0\sin I + yH_{ly}\left(\cos\left(I-\theta_h\right)-1\right)\right]\right. \\ &\left.\left(1-\frac{\left(\boldsymbol{H}_l\cdot\boldsymbol{H}_l\right)}{2s_0^2}-\frac{H_{ly}}{s_0}\sin\theta_h\right)\right\} \end{aligned} \tag{II.6}$$

Therefore, by expanding Equation II.6, we can find

$$W_{T1r-11100}^{(S)}\left(\boldsymbol{H}_l\cdot\boldsymbol{r}\right) = \Delta\left[\frac{n}{s_0}\left(\boldsymbol{H}_l\cdot\boldsymbol{r}\right)\right] = 0 \tag{II.7}$$

where the following equations are used to simplify the expression:

$$\Delta\left(\frac{n\boldsymbol{H}_l}{s_0}\right) = 0 \tag{II.8}$$

$$\Delta\left(\boldsymbol{r}\right) = 0 \tag{II.9}$$

Equation II.8 can be derived from the sagittal magnification shown in Equation 3.19, and Equation II.9 is always true due to the fact that the ray intersection at the surface does not change with refraction or reflection.

In addition, other terms in the $W_{T1r}^{(S)}$ expansion can be found as

$$\begin{aligned} W_{T1r-31100}^{(S)}\left(\boldsymbol{H}_l\cdot\boldsymbol{H}_l\right)\left(\boldsymbol{H}_l\cdot\boldsymbol{r}\right) &= \Delta\left[-\frac{n}{2s_0^3}\left(\boldsymbol{H}_l\cdot\boldsymbol{H}_l\right)\left(\boldsymbol{H}_l\cdot\boldsymbol{r}\right)\right] \\ &= -\frac{n^3}{2s_0^3}\Delta\left(\frac{1}{n^2}\right)\left(\boldsymbol{H}_l\cdot\boldsymbol{H}_l\right)\left(\boldsymbol{H}_l\cdot\boldsymbol{r}\right) \end{aligned} \tag{II.10}$$

$$W_{T1r-21110}^{(S)}\left(\boldsymbol{H}_l\cdot\boldsymbol{r}\right)\left(\boldsymbol{i}\cdot\boldsymbol{H}_l\right)=\Delta\left[-\frac{n}{s_0}\left(\boldsymbol{H}_l\cdot\boldsymbol{r}\right)\frac{H_{ly}}{s_0}\sin\theta_h\right]$$
$$=-\frac{n^2}{s_0^2}\Delta\left[\frac{1}{n}\sin\theta_h\right]\left(\boldsymbol{H}_l\cdot\boldsymbol{r}\right)\left(\boldsymbol{i}\cdot\boldsymbol{H}_l\right) \tag{II.11}$$

$$W_{T1r-01001}^{(S)}\left(\boldsymbol{i}\cdot\boldsymbol{r}\right)=\Delta\left(\frac{n}{s_0}ys_0\sin I\right)=\Delta\left(C\right)\left(\boldsymbol{i}\cdot\boldsymbol{r}\right) \tag{II.12}$$

$$W_{T1r-21001}^{(S)}\left(\boldsymbol{H}_l\cdot\boldsymbol{H}_l\right)\left(\boldsymbol{i}\cdot\boldsymbol{r}\right)=\Delta\left[-\frac{n}{s_0}ys_0\sin I\frac{\left(\boldsymbol{H}_l\cdot\boldsymbol{H}_l\right)}{2s_0^2}\right]$$
$$=-\frac{n^2}{2s_0^2}\Delta\left(\frac{C}{n^2}\right)\left(\boldsymbol{H}_l\cdot\boldsymbol{H}_l\right)\left(\boldsymbol{i}\cdot\boldsymbol{r}\right) \tag{II.13}$$

$$W_{T1r-11011}^{(S)}\left(\boldsymbol{i}\cdot\boldsymbol{H}_l\right)\left(\boldsymbol{i}\cdot\boldsymbol{r}\right)=-\Delta\left(\frac{n}{s_0}ys_0\sin I\frac{H_{ly}}{s_0}\sin\theta_h\right)$$
$$+\Delta\left(\frac{n}{s_0}yH_{ly}\left(\cos\left(I-\theta_h\right)-1\right)\right)$$
$$=\frac{n}{s_0}\Delta\left(\cos I\cos\theta_h\right)\left(\boldsymbol{i}\cdot\boldsymbol{H}_l\right)\left(\boldsymbol{i}\cdot\boldsymbol{r}\right) \tag{II.14}$$

The terms that are beyond the third group are ignored, since they do not contribute to third-group aberrations. The term related to $W_{T1-01001}^{(S)}$, shown in Equation II.12, can be further written as

$$W_{T1r-01001}^{(S)}\left(\boldsymbol{i}\cdot\boldsymbol{r}\right)=\Delta\left(C\right)\left(\boldsymbol{i}\cdot\boldsymbol{r}\right)=\frac{m\lambda y}{d} \tag{II.15}$$

which cancels the additional linear phase that the grating diffractive elements provide. Therefore, this term does not contribute to aberrations and is ignored.

The next step is to express $\boldsymbol{H}_l$ and $\boldsymbol{r}$ vectors in terms of normalized field and pupil vectors. From Equation 4.7, $\boldsymbol{H}_l$ can be written as

$$\boldsymbol{H}_l = L_h \boldsymbol{H} = L_h \left( H_x, H_y \right) \tag{II.16}$$

In Section 5.2.2, for approximated rays, the components of $\boldsymbol{r}$ vector, $x$ and $y$, can be approximated, shown in Equations 5.103 and 5.105 as

$$x \approx x_a \rho_x + x_b H_x \tag{5.103}$$

and

$$\begin{aligned} y\left(W^{(XP)} = 0\right) \approx & \frac{1}{\cos I'} x_a \rho_y^{(ai)} + \frac{\cos \theta_h'}{\cos I'} x_b H_y \\ & + \frac{\tan I'}{2R_{x-eff}} \left[ x_a^2 \left( \boldsymbol{\rho}^{(ai)} \cdot \boldsymbol{\rho}^{(ai)} \right) + 2 x_a x_b \left( \boldsymbol{H} \cdot \boldsymbol{\rho}^{(ai)} \right) + x_b^2 \left( \boldsymbol{H} \cdot \boldsymbol{H} \right) \right] \end{aligned} \tag{5.105}$$

which are the parts of $x$ and $y$ that contribute to third-group aberrations. Note that $\rho_x$ in Equation 5.103 is equal to $\rho_x^{(ai)}$ in the scope of third-group aberrations as discussed in Section 5.2.2.

Therefore, using Equations II.16, 5.103 and 5.105, dot products, $(\boldsymbol{H}_l \cdot \boldsymbol{H}_l)$, $(\boldsymbol{H}_l \cdot \boldsymbol{r})$, $(\boldsymbol{r} \cdot \boldsymbol{r})$, $(\boldsymbol{i} \cdot \boldsymbol{H}_l)$ and $(\boldsymbol{i} \cdot \boldsymbol{r})$ can be expressed as

$$\left( \boldsymbol{H}_l \cdot \boldsymbol{H}_l \right) = L_h^2 \left( \boldsymbol{H} \cdot \boldsymbol{H} \right) \tag{II.17}$$

$$\begin{aligned} \left( \boldsymbol{H}_l \cdot \boldsymbol{r} \right) &= L_h x H_x + L_h y H_y \\ &\approx L_h H_x \left( x_a \rho_x + x_b H_x \right) + L_h H_y \left( \frac{1}{\cos I'} x_a \rho_y^{(ai)} + \frac{\cos \theta_h'}{\cos I'} x_b H_y \right) \\ &\quad + \frac{L_h H_y \tan I'}{2R_{x-eff}} \left[ x_a^2 \left( \boldsymbol{\rho}^{(ai)} \cdot \boldsymbol{\rho}^{(ai)} \right) + 2 x_a x_b \left( \boldsymbol{H} \cdot \boldsymbol{\rho}^{(ai)} \right) + x_b^2 \left( \boldsymbol{H} \cdot \boldsymbol{H} \right) \right] \\ &\approx L_h H_x \left( x_a \rho_x^{(ai)} + x_b H_x \right) + L_h H_y \left( x_a \rho_y^{(ai)} + x_b H_y \right) \\ &= L_h x_a \left( \boldsymbol{H} \cdot \boldsymbol{\rho}^{(ai)} \right) + L_h x_b \left( \boldsymbol{H} \cdot \boldsymbol{H} \right) \end{aligned} \tag{II.18}$$

$$
\begin{aligned}
(\boldsymbol{r}\cdot\boldsymbol{r}) &= x^2 + y^2 \\
&\approx \left(x_a\rho_x + x_b H_x\right)^2 + \left\{\left(\frac{1}{\cos I'}x_a\rho_y^{(ai)} + \frac{\cos\theta_h'}{\cos I'}x_b H_y\right)\right. \\
&\quad \left. + \frac{\tan I'}{2R_{x-eff}}\left[x_a^2\left(\boldsymbol{\rho}^{(ai)}\cdot\boldsymbol{\rho}^{(ai)}\right) + 2x_a x_b\left(\boldsymbol{H}\cdot\boldsymbol{\rho}^{(ai)}\right) + x_b^2\left(\boldsymbol{H}\cdot\boldsymbol{H}\right)\right]\right\}^2 \\
&\approx \left(x_a\rho_x^{(ai)} + x_b H_x\right)^2 + \left(x_a\rho_y^{(ai)} + x_b H_y\right)^2 \\
&= x_a^2\left(\boldsymbol{\rho}^{(ai)}\cdot\boldsymbol{\rho}^{(ai)}\right) + 2x_a x_b\left(\boldsymbol{H}\cdot\boldsymbol{\rho}^{(ai)}\right) + x_b^2\left(\boldsymbol{H}\cdot\boldsymbol{H}\right)
\end{aligned}
\tag{II.19}
$$

$$
\left(\boldsymbol{i}\cdot\boldsymbol{H}_l\right) = L_h\left(\boldsymbol{i}\cdot\boldsymbol{H}\right) \tag{II.20}
$$

$$
\begin{aligned}
(\boldsymbol{i}\cdot\boldsymbol{r}) &= y \\
&\approx \frac{1}{\cos I'}x_a\rho_y^{(ai)} + \frac{\cos\theta_h'}{\cos I'}x_b H_y \\
&\quad + \frac{\tan I'}{2R_{x-eff}}\left[x_a^2\left(\boldsymbol{\rho}^{(ai)}\cdot\boldsymbol{\rho}^{(ai)}\right) + 2x_a x_b\left(\boldsymbol{H}\cdot\boldsymbol{\rho}^{(ai)}\right) + x_b^2\left(\boldsymbol{H}\cdot\boldsymbol{H}\right)\right] \\
&= \frac{1}{\cos I'}x_a\left(\boldsymbol{i}\cdot\boldsymbol{\rho}^{(ai)}\right) + \frac{\cos\theta_h'}{\cos I'}x_b\left(\boldsymbol{i}\cdot\boldsymbol{H}\right) \\
&\quad + \frac{\tan I'}{2R_{x-eff}}\left[x_a^2\left(\boldsymbol{\rho}^{(ai)}\cdot\boldsymbol{\rho}^{(ai)}\right) + 2x_a x_b\left(\boldsymbol{H}\cdot\boldsymbol{\rho}^{(ai)}\right) + x_b^2\left(\boldsymbol{H}\cdot\boldsymbol{H}\right)\right]
\end{aligned}
\tag{II.21}
$$

Note that in Equations II.17-II.21, approximations are made where the terms that contain two vector products are ignored, since they do not contribute to third-group aberrations. Therefore, the aberration terms in Equations II.10-II.11 and II.13-II.14 can be rewritten as

$$
\begin{aligned}
W_{T1r-31100}^{(S)}\left(\boldsymbol{H}_l\cdot\boldsymbol{H}_l\right)\left(\boldsymbol{H}_l\cdot\boldsymbol{r}\right) &= W_{T1r-31100}^{(S)}L_h^2\left(\boldsymbol{H}\cdot\boldsymbol{H}\right)\left[L_h x_a\left(\boldsymbol{H}\cdot\boldsymbol{\rho}^{(ai)}\right)\right. \\
&\quad \left. + L_h x_b\left(\boldsymbol{H}\cdot\boldsymbol{H}\right)\right] \\
&= W_{T1r-31100}^{(S)}L_h^3 x_a\left(\boldsymbol{H}\cdot\boldsymbol{H}\right)\left(\boldsymbol{H}\cdot\boldsymbol{\rho}^{(ai)}\right) \\
&\quad + W_{T1r-31100}^{(S)}L_h^3 x_b\left(\boldsymbol{H}\cdot\boldsymbol{H}\right)^2
\end{aligned}
\tag{II.22}
$$

$$W^{(S)}_{T1r-21110}\left(\boldsymbol{H_l}\cdot\boldsymbol{r}\right)\left(\boldsymbol{i}\cdot\boldsymbol{H_l}\right)=W^{(S)}_{T1r-21110}\left[L_h x_a\left(\boldsymbol{H}\cdot\boldsymbol{\rho}^{(\boldsymbol{ai})}\right)+L_h x_b\left(\boldsymbol{H}\cdot\boldsymbol{H}\right)\right]L_h\left(\boldsymbol{i}\cdot\boldsymbol{H}\right)$$
$$=W^{(S)}_{T1r-21110}L_h^2 x_a\left(\boldsymbol{H}\cdot\boldsymbol{\rho}^{(\boldsymbol{ai})}\right)\left(\boldsymbol{i}\cdot\boldsymbol{H}\right)\qquad\text{(II.23)}$$
$$+W^{(S)}_{T1r-21110}L_h^2 x_b\left(\boldsymbol{H}\cdot\boldsymbol{H}\right)\left(\boldsymbol{i}\cdot\boldsymbol{H}\right)$$

$$W^{(S)}_{T1r-21001}\left(\boldsymbol{H_l}\cdot\boldsymbol{H_l}\right)\left(\boldsymbol{i}\cdot\boldsymbol{r}\right)=W^{(S)}_{T1r-21001}L_h^2\left(\boldsymbol{H}\cdot\boldsymbol{H}\right)\left\{\frac{1}{\cos I'}x_a\left(\boldsymbol{i}\cdot\boldsymbol{\rho}^{(\boldsymbol{ai})}\right)\right.$$
$$+\frac{\cos\theta_h'}{\cos I'}x_b\left(\boldsymbol{i}\cdot\boldsymbol{H}\right)+\frac{\tan I'}{2R_{x-eff}}\left[x_a^2\left(\boldsymbol{\rho}^{(\boldsymbol{ai})}\cdot\boldsymbol{\rho}^{(\boldsymbol{ai})}\right)\right.$$
$$\left.\left.+2x_a x_b\left(\boldsymbol{H}\cdot\boldsymbol{\rho}^{(\boldsymbol{ai})}\right)+x_b^2\left(\boldsymbol{H}\cdot\boldsymbol{H}\right)\right]\right\}$$
$$=W^{(S)}_{T1r-21001}\frac{L_h^2 x_a}{\cos I'}\left(\boldsymbol{H}\cdot\boldsymbol{H}\right)\left(\boldsymbol{i}\cdot\boldsymbol{\rho}^{(\boldsymbol{ai})}\right)$$
$$+W^{(S)}_{T1r-21001}\frac{L_h^2 x_b\cos\theta_h'}{\cos I'}\left(\boldsymbol{H}\cdot\boldsymbol{H}\right)\left(\boldsymbol{i}\cdot\boldsymbol{H}\right)$$
$$+W^{(S)}_{T1r-21001}\frac{L_h^2 x_a^2\tan I'}{2R_{x-eff}}\left(\boldsymbol{H}\cdot\boldsymbol{H}\right)\left(\boldsymbol{\rho}^{(\boldsymbol{ai})}\cdot\boldsymbol{\rho}^{(\boldsymbol{ai})}\right)$$
$$+W^{(S)}_{T1r-21001}\frac{L_h^2 x_a x_b\tan I'}{R_{x-eff}}\left(\boldsymbol{H}\cdot\boldsymbol{H}\right)\left(\boldsymbol{H}\cdot\boldsymbol{\rho}^{(\boldsymbol{ai})}\right)$$
$$+W^{(S)}_{T1r-21001}\frac{L_h^2 x_b^2\tan I'}{2R_{x-eff}}\left(\boldsymbol{H}\cdot\boldsymbol{H}\right)\left(\boldsymbol{H}\cdot\boldsymbol{H}\right)\qquad\text{(II.24)}$$

$$\begin{aligned}
W_{T1r-11011}^{(S)}\left(\boldsymbol{i}\cdot\boldsymbol{H}_{l}\right)\left(\boldsymbol{i}\cdot\boldsymbol{r}\right) = {} & W_{T1r-11011}^{(S)}L_{h}\left(\boldsymbol{i}\cdot\boldsymbol{H}\right)\Bigg\{\frac{1}{\cos I'}x_{a}\left(\boldsymbol{i}\cdot\boldsymbol{\rho}^{(ai)}\right) \\
& +\frac{\cos\theta_{h}'}{\cos I'}x_{b}\left(\boldsymbol{i}\cdot\boldsymbol{H}\right)+\frac{\tan I'}{2R_{x-eff}}\Big[x_{a}^{2}\left(\boldsymbol{\rho}^{(ai)}\cdot\boldsymbol{\rho}^{(ai)}\right) \\
& +2x_{a}x_{b}\left(\boldsymbol{H}\cdot\boldsymbol{\rho}^{(ai)}\right)+x_{b}^{2}\left(\boldsymbol{H}\cdot\boldsymbol{H}\right)\Big]\Bigg\} \\
& = W_{T1r-11011}^{(S)}\frac{L_{h}x_{a}}{\cos I'}\left(\boldsymbol{i}\cdot\boldsymbol{H}\right)\left(\boldsymbol{i}\cdot\boldsymbol{\rho}^{(ai)}\right) \\
& +W_{T1r-11011}^{(S)}\frac{L_{h}x_{b}\cos\theta_{h}'}{\cos I'}\left(\boldsymbol{i}\cdot\boldsymbol{H}\right)\left(\boldsymbol{i}\cdot\boldsymbol{H}\right) \\
& +W_{T1r-11011}^{(S)}\frac{L_{h}x_{a}^{2}\tan I'}{2R_{x-eff}}\left(\boldsymbol{i}\cdot\boldsymbol{H}\right)\left(\boldsymbol{\rho}^{(ai)}\cdot\boldsymbol{\rho}^{(ai)}\right) \\
& +W_{T1r-11011}^{(S)}\frac{L_{h}x_{a}x_{b}\tan I'}{R_{x-eff}}\left(\boldsymbol{i}\cdot\boldsymbol{H}\right)\left(\boldsymbol{H}\cdot\boldsymbol{\rho}^{(ai)}\right) \\
& +W_{T1r-11011}^{(S)}\frac{L_{h}x_{b}^{2}\tan I'}{2R_{x-eff}}\left(\boldsymbol{i}\cdot\boldsymbol{H}\right)\left(\boldsymbol{H}\cdot\boldsymbol{H}\right)
\end{aligned} \tag{II.25}$$

Note that our goal is to find the aberration coefficients of the $W^{(S)}$ expansion based on $\boldsymbol{H}$ and $\boldsymbol{\rho}^{(ai)}$ dependencies, shown in Equation 5.143 as

$$\begin{aligned}
W^{(S)} &= W_{Sph}^{(S)}+W_{F}^{(S)} \\
&= \sum_{k,m,n,p,q}^{\infty} W_{2k+n+p,2m+n+q,n,p,q}^{(S)}\left(\boldsymbol{H}\cdot\boldsymbol{H}\right)^{k}\left(\boldsymbol{\rho}^{(ai)}\cdot\boldsymbol{\rho}^{(ai)}\right)^{m}\left(\boldsymbol{H}\cdot\boldsymbol{\rho}^{(ai)}\right)^{n} \\
&\quad \left(\boldsymbol{i}\cdot\boldsymbol{H}\right)^{p}\left(\boldsymbol{i}\cdot\boldsymbol{\rho}^{(ai)}\right)^{q}
\end{aligned} \tag{5.143}$$

Equations II.22-II.25 show how $W_{T1}^{(S)}$ contributes to each term in Equation 5.143. Since $W_{T1}^{(S)}$ only represents part of $W_{Sph}^{(S)}$, and other parts $W^{(S)}$, including $W_{F}^{(S)}$, can be expanded in the same process described above to find their contribution. The contributions with the same $\boldsymbol{H}$ and $\boldsymbol{\rho}^{(ai)}$ dependence are grouped together to find the coefficients in the $W^{(S)}$ expansion.

For example, in order to find the expression of $W_{Sph11011}^{(S)}$, we need to group contributions with $(\boldsymbol{i}\cdot\boldsymbol{H})(\boldsymbol{i}\cdot\boldsymbol{\rho}^{(ai)})$ dependence. One contribution can be found in Equation II.25 given as

$$W_{T1r-11011}^{(S)}\frac{L_h x_a}{\cos I'}\left(\boldsymbol{i}\cdot\boldsymbol{H}\right)\left(\boldsymbol{i}\cdot\boldsymbol{\rho}^{(ai)}\right) \tag{II.26}$$

Other contribution can be found in $W_{T3}^{(S)}$ which is given as

$$W_{T3}^{(S)}=\frac{1}{2}\Delta\left[\frac{n\left(\boldsymbol{r}\cdot\boldsymbol{s_h}\right)^2}{s_h^3}\right] \tag{II.27}$$

Following the same expansion process described above, we can find that the contribution from $W_{T3}^{(S)}$ that has $(\boldsymbol{i}\cdot\boldsymbol{H})(\boldsymbol{i}\cdot\boldsymbol{\rho}^{(ai)})$ dependence is

$$W_{T3r-02002}^{(S)}\frac{2x_a x_b\cos\theta_h'}{\cos^2 I'}\left(\boldsymbol{i}\cdot\boldsymbol{H}\right)\left(\boldsymbol{i}\cdot\boldsymbol{\rho}^{(ai)}\right) \tag{II.28}$$

where $W_{T3r\text{-}02002}^{(S)}$ is

$$W_{T3r-02002}^{(S)}=\frac{1}{2}\Delta\left(\frac{C^2}{ns_0}\right) \tag{II.29}$$

Therefore, we have

$$\begin{aligned} W_{Sph11011}^{(S)} &= W_{T1r-11011}^{(S)}\frac{L_h x_a}{\cos I'}+W_{T3r-02002}^{(S)}\frac{x_a^2}{\cos^2 I'} \\ &= \frac{n}{s_0}\Delta\left(\cos I\cos\theta_h\right)\frac{L_h x_a}{\cos I'}+\Delta\left(\frac{C^2}{ns_0}\right)\frac{x_a x_b\cos\theta_h'}{\cos^2 I'} \end{aligned} \tag{II.30}$$

The next step is simplification. From paraxial ray tracing, we have

$$u_{ax}=-\frac{x_a}{s_0} \tag{II.31}$$

$$u'_{ax} = -\frac{x_a}{s'_0} \tag{II.32}$$

$$u_{bx} = \frac{L_h - x_b}{s_0} \tag{II.33}$$

$$u'_{bx} = \frac{L'_h - x_b}{s'_0} \tag{II.34}$$

Therefore, we also have

$$\begin{aligned} \Psi &= n\left(u_{bx}x_a - u_{ax}x_b\right) \\ &= n\left(\frac{L_h - x_b}{s_0}x_a + \frac{x_a}{s_0}x_b\right) \\ &= \frac{nL_h x_a}{s_0} \end{aligned} \tag{II.35}$$

With Equations II.31-II.32 and II.35, Equation II.30 can be simplified as

$$W^{(S)}_{Sph11011} = \Psi\Delta\left(\cos I \cos\theta_h\right)\frac{1}{\cos I'} - \Delta\left(\frac{u_a C^2}{n}\right)\frac{x_b \cos\theta'_h}{\cos^2 I'} \tag{II.36}$$

which is the same expression shown in Equation 5.119.

The example discussed above focuses on aberration contributions from a base spherical sphere. Next, we provide another example illustrating the expansion of freeform contributions. The freeform contribution is part of $W^{(S)}$ and is defined in Equation 5.130 as

$$\begin{aligned} &W^{(S)}_F \approx z_{f2}\Delta\left(n\cos I\right) \\ &= \sum_{k,m,n,p,q}^{\infty} W^{(S)}_{F2k+n+p,2m+n+q,n,p,q}\left(\boldsymbol{H}\cdot\boldsymbol{H}\right)^k\left(\boldsymbol{\rho}^{(ai)}\cdot\boldsymbol{\rho}^{(ai)}\right)^m\left(\boldsymbol{H}\cdot\boldsymbol{\rho}^{(ai)}\right)^n\left(\boldsymbol{i}\cdot\boldsymbol{H}\right)^p\left(\boldsymbol{i}\cdot\boldsymbol{\rho}^{(ai)}\right)^q \end{aligned} \tag{5.130}$$

where $z_{f2}$ is defined as the freeform sag departure from the spherical surface with the effective radius of curvature in the sagittal direction, which is defined in Equation 5.67 as

$$z_{f2} = \left(F_{0-2} - F_{2-0}\right)y^2 + F_{2-1}x^2y + F_{0-3}y^3 + F_{4-0}x^4 + F_{2-2}x^2y^2 + F_{0-4}y^4 \qquad (5.67)$$

Coefficients, $F_{i\text{-}j}$, are defined in the XY-polynomial expansion of freeform sag function, $z_{freeform}$, shown in Equation 5.66 as

$$\begin{aligned} z_{freeform} &= F_{2-0}x^2 + F_{0-2}y^2 + F_{2-1}x^2y + F_{0-3}y^3 \ldots \\ &= F_{2-0}r^2 + \left(F_{0-2} - F_{2-0}\right)y^2 + F_{2-1}x^2y + F_{0-3}y^3 \ldots \\ &\approx F_{2-0}r^2 + z_{f2} \end{aligned} \qquad (5.66)$$

Consider a freeform sag function described with the fifth Fringe Zernike polynomial, $Z_5$, described in Section 2.2. The freeform sag function can be written as

$$z_{freeform} = Z_5 r_n^2 \cos 2\theta = Z_5\left(x^2 - y^2\right)/R_{zn}^2 \qquad (II.37)$$

In this case, $F_{0\text{-}2}=Z_5/R^2_{zn}$ and $F_{0\text{-}2}=-Z_5/R^2_{zn}$, so we can write $z_{f2}$ as

$$z_{f2} = -2Z_5y^2/R_{zn}^2 \qquad (II.38)$$

With the $y$ expression in Equation 5.105, the freeform aberration contribution, $W_F^{(S)}$, can be written as

$$\begin{aligned} W_F^{(S)} &\approx z_{f2}\Delta\left(n\cos I\right) \\ &= -\frac{2Z_5}{R_{zn}^2}\left\{\frac{1}{\cos I'}x_a\rho_y^{(ai)} + \frac{\cos\theta_h'}{\cos I'}x_bH_y \right. \\ &\quad + \frac{\tan I'}{2R_{x-eff}}\left[x_a^2\left(\boldsymbol{\rho}^{(ai)}\cdot\boldsymbol{\rho}^{(ai)}\right) + 2x_ax_b\left(\boldsymbol{H}\cdot\boldsymbol{\rho}^{(ai)}\right)\right. \\ &\quad \left.\left. + x_b^2\left(\boldsymbol{H}\cdot\boldsymbol{H}\right)\right]\right\}^2\Delta\left(n\cos I\right) \end{aligned} \qquad (II.39)$$

Then, the freeform aberration coefficients defined in Equation 5.130 that are related to $Z_5$ can be found as

$$W_{F02002}^{(S)}\left(\boldsymbol{i}\cdot\boldsymbol{\rho}^{(ai)}\right)^2 = \Delta\left(n\cos I\right)\left(-2Z_5\right)\frac{x_a^2}{R_{zn}^2}\frac{1}{\cos^2 I'}\left(\boldsymbol{i}\cdot\boldsymbol{\rho}^{(ai)}\right)^2 \qquad (II.40)$$

$$W^{(S)}_{F11011}\left(\boldsymbol{i}\cdot\boldsymbol{H}\right)\left(\boldsymbol{i}\cdot\boldsymbol{\rho}^{(ai)}\right)=\Delta\left(n\cos I\ \right)\left(-4Z_5\right)\frac{x_a x_b}{R_{zn}^2}\frac{\cos\theta_h'}{\cos^2 I'}\left(\boldsymbol{i}\cdot\boldsymbol{H}\right)\left(\boldsymbol{i}\cdot\boldsymbol{\rho}^{(ai)}\right) \quad \text{(II.41)}$$

$$W^{(S)}_{F03001}\left(\boldsymbol{\rho}^{(ai)}\cdot\boldsymbol{\rho}^{(ai)}\right)\left(\boldsymbol{i}\cdot\boldsymbol{\rho}^{(ai)}\right)=\Delta\left(n\cos I\ \right)\left(-2Z_5\right)\frac{\tan I'}{R_{x-eff}}\frac{x_a^3}{R_{zn}^2}\frac{1}{\cos I'}\left(\boldsymbol{\rho}^{(ai)}\cdot\boldsymbol{\rho}^{(ai)}\right)\left(\boldsymbol{i}\cdot\boldsymbol{\rho}^{(ai)}\right) \quad \text{(II.42)}$$

$$W^{(S)}_{F12101}\left(\boldsymbol{H}\cdot\boldsymbol{\rho}^{(ai)}\right)\left(\boldsymbol{i}\cdot\boldsymbol{\rho}^{(ai)}\right)=\Delta\left(n\cos I\ \right)\left(-4Z_5\right)\frac{\tan I'}{R_{x-eff}}\frac{x_a^2 x_b}{R_{zn}^2}\frac{1}{\cos I'}\left(\boldsymbol{H}\cdot\boldsymbol{\rho}^{(ai)}\right)\left(\boldsymbol{i}\cdot\boldsymbol{\rho}^{(ai)}\right) \quad \text{(II.43)}$$

$$W^{(S)}_{F21001}\left(\boldsymbol{H}\cdot\boldsymbol{H}\right)\left(\boldsymbol{i}\cdot\boldsymbol{\rho}^{(ai)}\right)=\Delta\left(n\cos I\ \right)\left(-2Z_5\right)\frac{\tan I'}{R_{x-eff}}\frac{x_a x_b^2}{R_{zn}^2}\frac{1}{\cos I'}\left(\boldsymbol{H}\cdot\boldsymbol{H}\right)\left(\boldsymbol{i}\cdot\boldsymbol{\rho}^{(ai)}\right) \quad \text{(II.44)}$$

$$W^{(S)}_{F12010}\left(\boldsymbol{\rho}^{(ai)}\cdot\boldsymbol{\rho}^{(ai)}\right)\left(\boldsymbol{i}\cdot\boldsymbol{H}\right)=\Delta\left(n\cos I\ \right)\left(-2Z_5\right)\frac{\tan I'}{R_{x-eff}}\frac{x_a^2 x_b}{R_{zn}^2}\frac{\cos\theta_h'}{\cos I'}\left(\boldsymbol{\rho}^{(ai)}\cdot\boldsymbol{\rho}^{(ai)}\right)\left(\boldsymbol{i}\cdot\boldsymbol{H}\right) \quad \text{(II.45)}$$

$$W^{(S)}_{F21110}\left(\boldsymbol{H}\cdot\boldsymbol{\rho}^{(ai)}\right)\left(\boldsymbol{i}\cdot\boldsymbol{H}\right)=\Delta\left(n\cos I\ \right)\left(-4Z_5\right)\frac{\tan I'}{R_{x-eff}}\frac{x_a x_b^2}{R_{zn}^2}\frac{\cos\theta_h'}{\cos I'}\left(\boldsymbol{H}\cdot\boldsymbol{\rho}^{(ai)}\right)\left(\boldsymbol{i}\cdot\boldsymbol{H}\right) \quad \text{(II.46)}$$

$$W^{(S)}_{F04000}\left(\boldsymbol{\rho}^{(ai)}\cdot\boldsymbol{\rho}^{(ai)}\right)^2=\Delta\left(n\cos I\ \right)\left(-Z_5\right)\frac{\tan^2 I'}{2R_{x-eff}^2}\frac{x_a^4}{R_{zn}^2}\left(\boldsymbol{\rho}^{(ai)}\cdot\boldsymbol{\rho}^{(ai)}\right)^2 \quad \text{(II.47)}$$

$$W^{(S)}_{F13100}\left(\boldsymbol{\rho}^{(ai)}\cdot\boldsymbol{\rho}^{(ai)}\right)\left(\boldsymbol{H}\cdot\boldsymbol{\rho}^{(ai)}\right)=\Delta\left(n\cos I\ \right)\left(-2Z_5\right)\frac{\tan^2 I'}{R_{x-eff}^2}\frac{x_a^3 x_b}{R_{zn}^2}\left(\boldsymbol{\rho}^{(ai)}\cdot\boldsymbol{\rho}^{(ai)}\right)\left(\boldsymbol{H}\cdot\boldsymbol{\rho}^{(ai)}\right) \quad \text{(II.48)}$$

$$W_{F22000}^{(S)}\left(\boldsymbol{\rho}^{(ai)}\cdot\boldsymbol{\rho}^{(ai)}\right)\left(\boldsymbol{H}\cdot\boldsymbol{H}\right)=\Delta\left(n\cos I\ \right)\left(-Z_5\right)\frac{\tan^2 I'}{R_{x-eff}^2}\frac{x_a^2x_b^2}{R_{zn}^2}\left(\boldsymbol{\rho}^{(ai)}\cdot\boldsymbol{\rho}^{(ai)}\right)\left(\boldsymbol{H}\cdot\boldsymbol{H}\right) \tag{II.49}$$

$$W_{F22200}^{(S)}\left(\boldsymbol{H}\cdot\boldsymbol{\rho}^{(ai)}\right)^2=\Delta\left(n\cos I\ \right)\left(-2Z_5\right)\frac{\tan^2 I'}{R_{x-eff}^2}\frac{x_a^2x_b^2}{R_{zn}^2}\left(\boldsymbol{H}\cdot\boldsymbol{\rho}^{(ai)}\right)^2 \tag{II.50}$$

$$W_{F31100}^{(S)}\left(\boldsymbol{H}\cdot\boldsymbol{H}\right)\left(\boldsymbol{H}\cdot\boldsymbol{\rho}^{(ai)}\right)=\Delta\left(n\cos I\ \right)\left(-2Z_5\right)\frac{\tan^2 I'}{R_{x-eff}^2}\frac{x_a x_b^3}{R_{zn}^2}\left(\boldsymbol{H}\cdot\boldsymbol{\rho}^{(ai)}\right)\left(\boldsymbol{H}\cdot\boldsymbol{H}\right) \tag{II.51}$$

Other Fringe Zernike terms can undergo the same process to find their contribution to each third-group aberration coefficient in Equation 5.130.